\documentclass[a4paper,11pt]{article}
\pdfoutput=1 

\usepackage{jheppub} 

\usepackage{graphicx}
\usepackage{hyperref}
\usepackage{pdfpages}
\numberwithin{figure}{section}
\usepackage[section]{placeins}
\usepackage{epstopdf}

\usepackage[colorinlistoftodos]{todonotes}
\usepackage{epsfig}
\usepackage{graphicx,color}
\usepackage{cancel}
\usepackage[utf8]{inputenc}

\newcommand{\be}{\begin{equation}}
\newcommand{\ee}{\end{equation}}

\newcommand{\bea}{\begin{aligned}}
\newcommand{\eea}{\end{aligned}}

\newcommand{\beq}{\begin{eqnarray}}
\newcommand{\eeq}{\end{eqnarray}}

\definecolor{red}{rgb}{1,0,0}
\definecolor{gray}{rgb}{0.5,0.5,0.5}

\usepackage{bbold}
\usepackage{tensor}
\usepackage[normalem]{ulem} 
\renewcommand\sout{\bgroup  \ULdepth=-.5ex \ULset}

\usepackage{pifont}
\newcommand{\cmark}{\ding{51}}%

\title{Dijet photoproduction at low $x$ at next-to-leading order and its back-to-back limit}
\author[a,b]{Pieter Taels,}
\author[c]{Tolga Altinoluk,}
\author[c]{Guillaume Beuf,}
\author[a]{and Cyrille Marquet}

\affiliation[a]{Centre de Physique Th\'eorique, \'Ecole polytechnique, CNRS, I.P. Paris, F-91128 Palaiseau, France}
\affiliation[b]{Universiteit Antwerpen, Departement fysica, Groenenborgerlaan
171, 2020 Antwerpen, Belgium}
\affiliation[c]{National Centre for Nuclear Research, 02-093 Warsaw, Poland}

\emailAdd{pieter.taels@uantwerpen.be}
\emailAdd{tolga.altinoluk@ncbj.gov.pl}
\emailAdd{guillaume.beuf@ncbj.gov.pl}
\emailAdd{cyrille.marquet@polytechnique.edu}

\date{\today}

\preprint{???}

\abstract{We compute the cross section for inclusive photoproduction of a pair of jets at next-to-leading order accuracy in the Color Glass Condensate (CGC) effective theory. The aim is to study the back-to-back limit, to investigate whether transverse momentum dependent (TMD) factorization can be recovered at this perturbative order. In particular, we focus on large Sudakov double logarithms, which are dominant terms in the TMD evolution kernel. Interestingly, the kinematical improvement of the low-$x$ resummation scheme turns out to play a crucial role in our analysis.}

\begin{document}

\maketitle

\section{Introduction}

A key factor in the success of perturbative Quantum Chromodynamics (pQCD) is the resummation of large logarithms that would otherwise spoil the perturbative expansion. Generally speaking, such logarithms are sensitive to the available phase space for gluon radiation. In one of the most common approaches, known as collinear factorization, the cross section is, up to power corrections, written as a convolution of perturbative hard parts and nonperturbative parton distribution functions (PDFs), and large logarithms in $\ln(\mu^{2}/\Lambda_{\mathrm{QCD}}^{2})$ are absorbed into the latter with the help of the Dokshitzer-Gribov-Lipatov-Altarelli-Parisi (DGLAP)~\cite{Gribov:1972ri,Dokshitzer:1977sg,Altarelli:1977zs} evolution equations. A hard scale $\mu^{2}$ is required to be present in a particle collision for pQCD to be applicable, and is typically provided by the photon virtuality in deep-inelastic scattering (DIS) or by the mass of the produced boson in proton-proton collisions.

In collinear factorization, it is tacitly assumed that the center-of-mass energy $\sqrt{s}$ is of the same order as the hard scale. At very high energies, or equivalently very small parton longitudinal momentum fractions $x$, this approximation breaks down, and large `rapidity' logarithms in $\ln(s/\mu^{2})\sim\ln(1/x)$ become equally or even more important than the collinear ones. Their resummation is often performed using the Balitsky-Fadin-Kuraev-Lipatov (BFKL) evolution equations~\cite{Kuraev:1977fs,Balitsky:1978ic}, which are embedded in another factorization framework known as $k_{\scriptscriptstyle{T}}$- or High-Energy Factorization (HEF)~\cite{Catani:1990xk,Catani:1990eg,Catani:1994sq}. BFKL predicts a steep rise of the unintegrated or $k_{\scriptscriptstyle{T}}$-dependent gluon density which eventually violates unitarity~\cite{Froissart:1961ux}. However, below a dynamically generated saturation scale $Q_s(x)$, nonlinear gluon recombination effects will counteract this unphysical exponential growth~\cite{Gribov:1983ivg}. At this point, High-Energy Factorization needs to be generalized to include nonlinear low-$x$ evolution. This is done by the Color Glass Condensate (CGC) effective theory~\cite{McLerran:1993ni,McLerran:1993ka,McLerran:1994vd,Jalilian-Marian:1997qno,Jalilian-Marian:1997jhx,Jalilian-Marian:1997ubg,Kovner:2000pt,Weigert:2000gi,Iancu:2000hn,Iancu:2001ad,Ferreiro:2001qy} employed in this paper. In the CGC, the highly dense gluon distribution is treated as a large semiclassical field sharply localized on the light cone (a `shockwave'), whose rapidity evolution of the gluon distribution is governed by the Balitsky-Kovchegov/Jalilian-Marian-Iancu-McLerran-Weigert-Leonidov-Kovner (BK-JIMWLK) evolution equations~\cite{Balitsky:1995ub, Balitsky:1998kc, Balitsky:1998ya, Kovchegov:1999yj, Jalilian-Marian:1997qno, Jalilian-Marian:1997jhx, Jalilian-Marian:1997ubg, Kovner:2000pt, Weigert:2000gi, Iancu:2000hn, Iancu:2001ad, Ferreiro:2001qy}, which can be regarded as a nonlinear generalization of BFKL.

Another situation where collinear factorization breaks down is when the process is sensitive to a second scale $\mu_b^2$ that is much smaller than the hard scale: $\mu^2\gg\mu_b^2\gtrsim\Lambda_\mathrm{QCD}^2$. Large `Sudakov' logarithms $\ln(\mu^2/\mu_b^2)$~\cite{Dokshitzer:1978hw} need to be resummed in addition to the collinear ones through the Collins-Soper-Sterman (CSS) evolution equations~\cite{Collins:1984kg,Collins:1988ig} and can be absorbed~\cite{Collins:2011zzd,Echevarria:2011epo} into the parton distributions, which need to be extended to transverse momentum dependent parton distributions functions (TMD PDFs~\cite{Angeles-Martinez:2015sea}). Classic examples are Z-boson hadroproduction (Drell-Yan) or semi-inclusive deep-inelastic scattering (SIDIS~\cite{Ji:2004wu}) at low transverse momenta, where the large scale is provided by the boson mass or the photon virtuality, and the small scale by the transverse momentum of the produced boson or hadron.

In kinematics with a hierarchy of scales $s\!\gg\!\mu^2\!\gg\!\mu_b^2\!\gtrsim\!\Lambda_\mathrm{QCD}^2$, the necessity arises to simultaneously treat large $\ln(s/\mu^{2})$ and $\ln(\mu^2/\mu_b^2)$ logarithms. Such a combined low-$x$ and Sudakov resummation is the subject of intensive research, some of which very recent, either based on the HEF approach~\cite{Hautmann:2008vd,Deak:2009xt,Jung:2010si,Deak:2011ga,Dooling:2014kia,Hentschinski:2016wya,Bury:2017jxo,Hentschinski:2017ayz,Blanco:2019qbm,vanHameren:2019ysa,vanHameren:2020rqt}, BFKL~\cite{Hentschinski:2021lsh,Nefedov:2021vvy}, BK~\cite{Zheng:2019zul}, the CGC~\cite{Sun:2011iw,Mueller:2012uf,Mueller:2013wwa,Xiao:2017yya,Stasto:2018rci,Marquet:2019ltn}, or TMD factorization~\cite{Boer:2017xpy,Zhou:2018lfq,Boer:2022njw}.

We should stress, however, that TMD factorization goes beyond Sudakov resummation. Indeed, the quark- and gluon TMD PDFs parameterize the transverse-momentum \emph{and} spin dependence of the partons inside the proton or nucleus, and moreover come in different types according to the underlying hard process~\cite{Bomhof:2006dp}. By contrast, the HEF and BFKL frameworks depend on a single unintegrated gluon distribution and, because of the particular `nonsense' polarization tensor used, seem to be incompatible with the full structure of the TMD PDFs except at large transverse momenta~\cite{Metz:2011wb,Altinoluk:2021ygv}. 

It turns out that the CGC, which generalizes BFKL, at leading order (LO) in perturbation theory also encompasses TMD factorization. Therefore, one might hope that the CGC provides a unified framework for both TMD factorization and (non)linear low-$x$ evolution. Even at leading order this is not trivial, because a generic CGC cross section involves a complicated intertwining of perturbative coefficients with nonperturbative correlators of semiclassical fields. In the seminal papers~\cite{Dominguez:2010xd, Dominguez:2011wm} it was demonstrated that, for a large class of $2\to2$ processes, the CGC and TMD LO calculations do result in the same cross sections, given a proper identification of the correlators of semiclassical gluon fields and gluon TMD PDFs~\cite{Mulders:2000sh, Meissner:2007rx,Bomhof:2006dp}. This triggered a series of studies, demonstrating the sensitivity to the linearly polarized gluon TMD PDF when masses are included~\cite{Metz:2011wb,Dominguez:2011br,Akcakaya:2012si}, applying JIMWLK evolution to gluon TMDs~\cite{Dumitru:2015gaa,Marquet:2016cgx,Marquet:2017xwy}, and extending the CGC-TMD correspondence to $2\to3$ processes~\cite{Altinoluk:2018byz,Altinoluk:2020qet,Bury:2018kvg}. In parallel, the so-called small-$x$ improved transverse momentum dependent (ITMD) factorization framework~\cite{Kotko:2015ura,vanHameren:2016ftb,Altinoluk:2019fui,Altinoluk:2019wyu,Boussarie:2020vzf,Altinoluk:2021ygv,vanHameren:2021sqc,Boussarie:2021ybe} was developed as a way to combine the applicability of TMD factorization with the resummation of powers $(Q_s/\mu_b)^n$ and $(\mu_b/\mu)^n$, where $Q_s$ is the saturation scale, of the CGC. 

The aim of this paper is twofold. First, we contribute to the effort to bring CGC calculations to higher perturbative accuracy by calculating the full NLO cross section of inclusive dijet photoproduction, i.e. the process $\gamma+A\to \mathrm{dijet}+X$, using light-cone perturbation theory (LCPT)~\cite{Kogut:1969xa,Bjorken:1970ah,Brodsky:1997de}. This process could be measured at low photon virtualities in the future Electron-Ion Collider~\cite{Accardi:2012qut}, the proposed LHeC~\cite{LHeC:2020van}, or in ultraperipheral lead-proton collisions at the LHC. Moreover, our calculation provides an important cross-check of the $\gamma^*_T+A\to \mathrm{dijet}+X$ impact factor recently obtained in~\cite{Caucal:2021ent} using the covariant formulation of the CGC. Second, we want to address the important open question whether the compatibility of the CGC with TMD factorization is preserved beyond leading order. To do so, we study the limit in which the dijets are back-to-back in the transverse plane, thus creating a scale hierarchy $s\gg\mathbf{P}_\perp^2\gg\mathbf{k}_\perp^2$, where $\mathbf{P}_\perp$ is the typical large transverse momentum of each jet and $\mathbf{k}_\perp$ is their small momentum imbalance. We can reproduce the large Sudakov double logarithms that are essential ingredients in the CSS evolution of the gluon TMD, obtaining the same result as what was predicted in ref.~\cite{Mueller:2012uf,Mueller:2013wwa,Sun:2014gfa,Sun:2015doa}. However, we show that the usual subtraction of low-$x$ logarithms and their absorption into JIMWLK leads to an oversubtraction incompatible with the extraction of the Sudakov logarithms performed in~\cite{Mueller:2012uf,Mueller:2013wwa}, and demonstrate that one must rather employ the kinematically-improved JIMWLK equation~\cite{Ciafaloni:1987ur,Andersson:1995ju,Kwiecinski:1996td,Salam:1998tj,Motyka:2009gi,Beuf:2014uia,Iancu:2015vea,Hatta:2016ujq,Ducloue:2019ezk}. Finally, we observe that, at least at first sight, a class of virtual diagrams which contribute to the finite NLO corrections seem to break factorization. The analysis of these contributions and thus the answer to whether the CGC-TMD correspondence holds at full NLO accuracy is left for future work.

Note that the central role of the large semiclassical gluon field, as well as the nonlinearity of the evolution equations, introduce additional complications into CGC computations. Therefore, we are still far away from the next-to-next-to-leading-order precision reached for some collinear observables. In the last decade, however, a huge effort has been made to bring CGC calculations to NLO accuracy. Prominent examples are the cross sections for inclusive hadron production in proton-nucleus collisions~\cite{Chirilli:2011km,Chirilli:2012jd}, inclusive deep-inelastic-scattering (DIS)~\cite{Balitsky:2010ze, Balitsky:2012bs, Beuf:2011xd, Beuf:2016wdz, Beuf:2017bpd, Hanninen:2017ddy}, DIS with massive quarks~\cite{Beuf:2021qqa, Beuf:2021srj,Beuf:2022ndu}, exclusive vector meson production in DIS~\cite{Boussarie:2016bkq,Mantysaari:2022bsp,Mantysaari:2021ryb}, photon+dijet production in DIS~\cite{Roy:2019hwr}, diffractive dijet production in DIS~\cite{Boussarie:2016ogo} and very recently inclusive dijet production in DIS~\cite{Caucal:2021ent}. Moreover, the next-to-leading logarithmic extension of the BK-JIMWLK equations was studied in refs.~\cite{Balitsky:2013fea,Kovner:2013ona,Kovner:2014lca,Lublinsky:2016meo}.

The paper is organized as follows. In section~\ref{sec:NLO}, we use light-cone perturbation theory to calculate the loop corrections to $\gamma + A\to q+\bar{q}+X$. One set of diagrams: the initial-state loop corrections, were already calculated in ref.~\cite{Beuf:2016wdz} and are not recomputed here. In section~\ref{sec:223}, we revisit the calculation of the real NLO corrections, i.e. the process $\gamma + A\to q+\bar{q}+g+X$, which was calculated earlier in~\cite{Altinoluk:2020qet}.

As in any NLO calculation, individual diagrams can be plagued by ultraviolet (UV) divergences, while squared diagrams or interferences might exhibit soft and collinear divergences. We demonstrate their cancellation in sections~\ref{sec:UV},~\ref{sec:soft}, and~\ref{sec:coll}, respectively.
Large rapidity logarithms are absorbed in the JIMWLK equation for the LO cross section, as is shown in section~\ref{sec:JIMWLK}. The full NLO cross section is then presented in section~\ref{sec:crosssection}, after which we investigate the back-to-back or `correlation' limit in section~\ref{sec:corrlimit}.

\section{\label{sec:LO}Leading-order calculation}
Throughout this work, we will work in LCPT
in the conventions of Bjorken-Kogut-Soper~\cite{Kogut:1969xa,Bjorken:1970ah}.
In this picture, the dynamics of the \textquoteleft projectile', i.e.
the photon splitting into the quark-antiquark pair, take place on
a much longer timescale than the partonic dynamics of the \textquoteleft target'
proton or nucleus~\cite{Mueller:1989st,Nikolaev:1990ja}. The Color Glass Condensate effective theory then
asserts that the target effectively behaves as a localized \textquoteleft shockwave'
of semiclassical gluon fields, which may be described by an external
potential built from Wilson lines.

Taking the incoming photon to travel along the $+$ light-cone (LC)
direction, colliding head on with the hadronic target which travels
along the $-$ LC direction, the differential cross section for the
process $\gamma A\to q\bar{q}X$ (see fig.~\ref{fig:virtual}) is
given by:
\begin{equation}
\begin{aligned}\mathrm{d}\sigma & =\frac{1}{2q^{+}}\frac{\mathrm{d}p_{1}^{+}\mathrm{\mathrm{d}}^{D-2}\mathbf{p}_{1}\theta(p^+_1)}{(2\pi)^{D-1}2p_{1}^{+}}\frac{\mathrm{d}p_{2}^{+}\mathrm{\mathrm{d}}^{D-2}\mathbf{p}_{2}\theta(p^+_2)}{(2\pi)^{D-1}2p_{2}^{+}}2\pi\delta(q^{+}-p_{1}^{+}-p_{2}^{+})\frac{1}{D-2}\big|\mathcal{M}\big|^{2}\;.\end{aligned}
\label{eq:crosssectiondef}
\end{equation}
The vectors $\vec{p}_{1}\equiv(p_{1}^{+},\mathbf{p}_{1})$ and $\vec{p}_{2}\equiv(p_{2}^{+},\mathbf{p}_{2})$
describe the $+$ and transverse components of the quark and antiquark,
respectively, and $q^{+}$ is the photon $+$ momentum. The total
$+$ momentum is conserved, as encoded in the delta function. Note
that the factor $1/(D-2)$ accounts for the averaging over the photon
polarization, and that we work for the moment in $D$ dimensions.

In the above formula,  the amplitude $\mathcal{M}$ is defined as:
\begin{equation}
\begin{aligned}_{f}\langle(\mathbf{q})[\vec{p}_{1}]_{s_{1}};(\mathbf{\bar{q}})[\vec{p}_{2}]_{s_{2}}|\hat{F}-1|(\boldsymbol{\gamma})[\vec{q}]_{\lambda}\rangle_{i} & =2\pi\delta(q^{+}-p_{1}^{+}-p_{2}^{+})\mathcal{M}\;.\end{aligned}
\end{equation}
In the l.h.s. the operator $\hat{F}$, which describes the interaction
with the shockwave, acts on the Fock states of the incoming
($i$) photon and outgoing ($f$) quark-antiquark pair. Since these
Fock states are asymptotic, their evolution to and from $x^{+}=0$: the
light-cone time when the scattering takes place, should be calculated
up to a given perturbative order:
\begin{equation}
\begin{aligned}
&_{f}\langle(\mathbf{q})[\vec{p}_{1}]_{s_{1}};(\mathbf{\bar{q}})[\vec{p}_{2}]_{s_{2}}|\hat{F}-1|(\boldsymbol{\gamma})[\vec{q}]_{\lambda}\rangle_{i}\\
&=\langle(\mathbf{q})[\vec{p}_{1}]_{s_{1}};(\mathbf{\bar{q}})[\vec{p}_{2}]_{s_{2}}|\mathcal{U}(+\infty,0)(\hat{F}-1)\mathcal{U}(0,-\infty)|(\boldsymbol{\gamma})[\vec{q}]_{\lambda}\rangle\;,\label{eq:transitionamp}
\end{aligned}
\end{equation}
where $\mathcal{U}$ is the LC-time evolution operator. At lowest non-trivial order, the photon splits into a quark-antiquark
pair before the scattering off the shockwave, while the outgoing quark
pair evolves to asymptotic states without perturbative modifications.
We thus have from the LCPT Feynman rules~\cite{Bjorken:1970ah,Beuf:2016wdz,Altinoluk:2020qet}:
\begin{equation}
\begin{aligned}&\langle(\mathbf{q})[\vec{p}_{1}]_{s_{1}};(\mathbf{\bar{q}})[\vec{p}_{2}]_{s_{2}}|\mathcal{U}(+\infty,0)  =\langle(\mathbf{q})[\vec{p}_{1}]_{s_{1}};(\mathbf{\bar{q}})[\vec{p}_{2}]_{s_{2}}|+\mathcal{O}(g_e,g_s)\;,\\
&\mathcal{U}(0,-\infty)|(\boldsymbol{\gamma})[\vec{q}]_{\lambda}\rangle  =|(\boldsymbol{\gamma})[\vec{q}]_{\lambda}\rangle+\int\mathrm{PS}(\vec{p}_{1}^{\prime},\vec{p}_{2}^{\prime})(2\pi)^{D-1}\delta^{(D-1)}\big(\vec{q}-\vec{p}_{1}^{\prime}-\vec{p}_{2}^{\prime}\big)\\
 &\qquad\qquad\qquad\qquad \times g_{e}e_f\frac{\bar{u}^{s_{1}}(\vec{p}_{1}^{\prime})\cancel{\epsilon}_{\lambda}(\vec{q})u^{s_{2}}(\vec{p}_{2}^{\prime})}{q^{-}-p_{1}^{\prime-}-p_{2}^{\prime-}}|(\mathbf{q})[\vec{p}_{1}^{\prime}];(\mathbf{\bar{q}})[\vec{p}_{2}^{\prime}]\rangle+\mathcal{O}(g_s)\;.
\end{aligned}
\label{eq:timeEVO}
\end{equation}
In the above, we introduced the notation $\mathrm{PS}$ for the measure
of the phase space integrations:
\begin{equation}
\int\mathrm{PS}(\vec{q})=\mu^{4-D}\int\frac{\mathrm{d}^{D-1}\vec{q}\,\theta(q^{+})}{(2\pi)^{D-1}2q^{+}}\;,
\end{equation}
where $\vec{q}\equiv(q^{+},\mathbf{q})$ and where $\theta$ is the
Heaviside step function defined as $\theta(x\geq0)=1$ and $\theta(x<0)=0$.
Combining eqs.   (\ref{eq:transitionamp}) and (\ref{eq:timeEVO}),
 we can reshuffle the terms in the following form:
\begin{equation}
\begin{aligned} &_{f}\langle(\mathbf{q})[\vec{p}_{1}]_{s_{1}};(\mathbf{\bar{q}})[\vec{p}_{2}]_{s_{2}}|\hat{F}-1|(\boldsymbol{\gamma})[\vec{q}]_{\lambda}\rangle_{i}\\
 & =g_{e}e_f\int\mathrm{PS}(\vec{p}_{1}^{\prime},\vec{p}_{2}^{\prime})\times(2\pi)^{D-1}\delta^{(D-1)}\big(\vec{q}-\vec{p}_{1}^{\prime}-\vec{p}_{2}^{\prime}\big)\\
 & \times\frac{\bar{u}^{s_{1}}(\vec{p}_{1}^{\prime})\cancel{\epsilon}_{\lambda}(\vec{q})u^{s_{2}}(\vec{p}_{2}^{\prime})}{q^{-}-p_{1}^{\prime-}-p_{2}^{\prime-}}\times\langle(\mathbf{q})[\vec{p}_{1}];(\mathbf{\bar{q}})[\vec{p}_{2}]|\hat{F}-1|(\mathbf{q})[\vec{p}_{1}^{\prime}];(\mathbf{\bar{q}})[\vec{p}_{2}^{\prime}]\rangle\;.
\end{aligned}
\label{eq:LO}
\end{equation}
The last term in the above expression encodes the scattering of the `bare' $q\bar{q}$
state off the target. In the eikonal approximation, it can be written as:
\begin{equation}
\begin{aligned} & \langle(\mathbf{q})[\vec{p}_{1}];(\mathbf{\bar{q}})[\vec{p}_{2}]|\hat{F}-1|(\mathbf{q})[\vec{p}_{1}^{\prime}];(\mathbf{\bar{q}})[\vec{p}_{2}^{\prime}]\rangle\\
 & =2p_{1}^{+}2\pi\delta(p_{1}^{\prime+}-p_{1}^{+})2p_{2}^{+}2\pi\delta(p_{2}^{\prime+}-p_{2}^{+})\\
 & \times\Big\langle U(\mathbf{p}_{1}^{\prime}-\mathbf{p}_{1})U^{\dagger}(\mathbf{p}_{2}^{\prime}-\mathbf{p}_{2})-(2\pi)^{2(D-2)}\delta^{D-2}(\mathbf{p}_{1}^{\prime}-\mathbf{p}_{1})\delta^{D-2}(\mathbf{p}_{2}^{\prime}-\mathbf{p}_{2})\Big\rangle\;,\\
 & =4p_{1}^{+}p_{2}^{+}2\pi\delta(p_{1}^{\prime+}-p_{1}^{+})2\pi\delta(p_{2}^{\prime+}-p_{2}^{+})\\
 & \times\int_{\mathrm{\mathbf{x}}_{1},\mathbf{x}_{2}}e^{-i\mathbf{x}_{1}\cdot(\mathbf{p}_{1}-\mathbf{p}_{1}^{\prime})}e^{-i\mathbf{x}_{2}\cdot(\mathbf{p}_{2}-\mathbf{p}_{2}^{\prime})}\Big[U_{\mathbf{x}_{1}}U_{\mathbf{x}_{2}}^{\dagger}-1\Big]\;;
\end{aligned}
\label{eq:LOEIK}
\end{equation}
where we introduced the short-hand notation $\int_{\mathbf{x}}=\mu^{D-4}\int\mathrm{d}^{D-2}\mathbf{x}$,
and where
\begin{equation}
U_{\mathbf{x}}=\mathcal{P}\exp\Big(ig_{s}\int\mathrm{d}x^{+}A_{a}^{-}(x^{+},0^{-},\mathbf{x})t^{a}\Big)
\end{equation}
denotes a Wilson line in the fundamental representation going from
$x^{+}=-\infty$ to $x^{+}=+\infty$ with $x^{-}=0$ and transverse
position $\mathbf{x}$. We will use the notation $W_{\mathbf{x}}$
for Wilson lines in the adjoint representation, and suppress the fundamental
color indices. Note that in the eikonal approximation, there is no exchange of spin nor $+$ momentum between the projectile and the target.

Collecting the delta functions in eqs.~\eqref{eq:LO} and~\eqref{eq:LOEIK},
 the phase space integration can be written as follows:
\begin{equation}
\begin{aligned} & \int\mathrm{PS}(\vec{p}_{1}^{\prime},\vec{p}_{2}^{\prime})(2\pi)^{D-1}\delta^{(D-1)}\big(\vec{q}-\vec{p}_{1}^{\prime}-\vec{p}_{2}^{\prime}\big)2\pi\delta(p_{1}^{\prime+}-p_{1}^{+})2\pi\delta(p_{2}^{\prime+}-p_{2}^{+})\\
 & =\frac{2\pi\delta(q^{+}-p_{1}^{+}-p_{2}^{+})}{4p_{1}^{+}p_{2}^{+}}\int_{\mathbf{p}_{1}^{\prime}}\;,
\end{aligned}
\end{equation}
with the convention $\int_{\mathbf{q}}=\mu^{4-D}\int\mathrm{d}^{D-2}\mathbf{q}/(2\pi)^{D-2}$
(note the factor $(2\pi)^{D-2}$ which is not present in the integrations
over transverse coordinate space).

Suppressing the spinor indices, the Dirac structure in eq.~\eqref{eq:LO} can be rewritten in function of good spinors~\cite{Beuf:2016wdz} with the help of
the following intermediary result:
\begin{equation}
\begin{aligned}&\bar{u}(\vec{p}_{1})\cancel{\epsilon}_{\lambda}(\vec{q})v(\vec{p}_{2})\\ & =\bar{u}_{G}(p_{1}^{+})\gamma^{+}\Bigl[\delta^{\lambda\bar{\lambda}}\Bigl(\frac{q^{\bar{\lambda}}}{q^{+}}-\frac{p_{2}^{\bar{\lambda}}}{2p_{2}^{+}}-\frac{p_{1}^{\bar{\lambda}}}{2p_{1}^{+}}\Bigr)-i\sigma^{\lambda\bar{\lambda}}\Bigl(\frac{p_{2}^{\bar{\lambda}}}{2p_{2}^{+}}-\frac{p_{1}^{\bar{\lambda}}}{2p_{1}^{+}}\Bigr)\Bigr]v_{G}(p_{2}^{+})\;,\end{aligned}
\label{eq:spinor2good}
\end{equation}
which holds irregardless of whether we work with quark or antiquark
spinors (since we work with massless quarks).   A very useful feature
of the above formula is that the good spinors only depend on the $+$
component of the momenta,  which means that they are not affected
by the shockwave.   Note as well that,  to arrive at eq.~\eqref{eq:spinor2good},
 we choose the transverse polarization vectors to be $\epsilon_{\lambda}^{i}=\delta^{i\lambda}$.
  We thus get:
\begin{equation}
\begin{aligned} \bar{u}(\vec{p}_{1})\cancel{\epsilon}_{\lambda}(\vec{q})v(\vec{p}_{2})& =-\frac{q^{+}\mathbf{p}_{1}^{\prime\bar{\lambda}}}{2p_{1}^{+}p_{2}^{+}}\bar{u}_{G}^{s_{1}}(p_{1}^{+})\gamma^{+}\bigl[(1-2z)\delta^{\lambda\bar{\lambda}}-i\sigma^{\lambda\bar{\lambda}}\bigr]v_{G}^{s_{2}}(p_{2}^{+})\;,\end{aligned}
\end{equation}
where we defined $z\equiv p_{1}^{+}/q^{+}$ and $\sigma^{ij}\equiv\frac{i}{2}[\gamma^{i},\gamma^{j}]$.
Note that, in our frame, the photon does not have any transverse momentum.

On the other hand, the energy denominator in~\eqref{eq:LO}  gives:
\begin{equation}
\begin{aligned} q^{-}-p_{1}^{\prime-}-p_{2}^{\prime-}=\frac{-q^{+}}{2p_{1}^{+}p_{2}^{+}}\mathbf{p}_{1}^{\prime2}\;.\end{aligned}
\end{equation}
Putting everything together,  we obtain the intermediary result (using
the short-hand $\mathbf{x}_{12}\equiv\mathbf{x}_{1}-\mathbf{x}_{2}$):
\begin{equation}
\begin{aligned}\mathcal{M}_{\mathrm{LO}} & =g_{e}e_f\int_{\mathrm{\mathbf{x}}_{1},\mathbf{x}_{2}}e^{-i\mathbf{p}_{1}\cdot\mathbf{x}_{1}}e^{-i\mathbf{p}_{2}\cdot\mathbf{x}_{2}}\mathrm{Dirac}_{\mathrm{LO}}^{\bar{\lambda}}\int_{\mathbf{p}_{1}^{\prime}}e^{i\mathbf{p}_{1}^{\prime}\cdot\mathbf{x}_{12}}\frac{\mathbf{p}_{1}^{\prime\bar{\lambda}}}{\mathbf{p}_{1}^{\prime2}}\Big[U_{\mathbf{x}_{1}}U_{\mathbf{x}_{2}}^{\dagger}-1\Big]\;,
\end{aligned}
\label{eq:LO2}
\end{equation}
with:
\begin{equation}
\begin{aligned}\mathrm{Dirac}_{\mathrm{LO}}^{\bar{\lambda}} & \equiv\bar{u}_{G}^{s_{1}}(p_{1}^{+})\gamma^{+}\bigl[(1-2z)\delta^{\lambda\bar{\lambda}}-i\sigma^{\lambda\bar{\lambda}}\bigr]v_{G}^{s_{2}}(p_{2}^{+})\;.\end{aligned}
\label{eq:DiracLO}
\end{equation}
Finally,  we can perform the integration over $\mathbf{p}_{1}^{\prime}$,
 which gives:
\begin{equation}
\begin{aligned}\int_{\mathbf{p}_{1}^{\prime}}e^{i\mathbf{p}_{1}^{\prime}\cdot\mathbf{x}_{12}}\frac{\mathbf{p}_{1}^{\prime\bar{\lambda}}}{{\mathbf{p}_{1}^{\prime}}^2} & =-iA^{\bar{\lambda}}(\mathbf{x}_{12})\;,\end{aligned}
\end{equation}
where the Weizs\"acker-Williams field $A^i(\mathbf{x})$ in $D-2$ dimensions is defined by:
\begin{equation}
\begin{aligned}iA^{i}(\mathbf{x})&\equiv\int_{\mathbf{k}}e^{-i\mathbf{k}\cdot\mathbf{x}}\frac{k^{i}}{\mathbf{k}^{2}}  =\frac{-i \mu^{4-D}}{2\pi^{\frac{D}{2}-1}}\frac{x^{i}}{(\mathbf{x}^{2})^{\frac{D}{2}-1}}\Gamma(\frac{D}{2}-1) \overset{D\to4}{=}\frac{-i}{2\pi}\frac{x^{i}}{\mathbf{x}^{2}}\;.
\end{aligned}
\label{eq:WWfield}
\end{equation}
Putting everything together, the LO scattering amplitude finally reads:
\begin{equation}
\begin{aligned}\mathcal{M}_{\mathrm{LO}} & =-ig_{e}e_f\mathrm{Dirac}_{\mathrm{LO}}^{\bar{\lambda}}\int_{\mathrm{\mathbf{x}}_{1},\mathbf{x}_{2}}e^{-i\mathbf{p}_{1}\cdot\mathbf{x}_{1}}e^{-i\mathbf{p}_{2}\cdot\mathbf{x}_{2}}A^{\bar{\lambda}}(\mathbf{x}_{12})\Big[U_{\mathbf{x}_{1}}U_{\mathbf{x}_{2}}^{\dagger}-1\Big]\;.\end{aligned}
\label{eq:LOfinal}
\end{equation}
Hence, the amplitude squared becomes:
\begin{equation}
\begin{aligned}\big|\mathcal{M}_{\mathrm{LO}}\big|^{2} & =4\pi\alpha_{\mathrm{em}}e_f^2\mathrm{Tr}\Big(\mathrm{Dirac}_{\mathrm{LO}}^{\lambda^{\prime}\dagger}\mathrm{Dirac}_{\mathrm{LO}}^{\bar{\lambda}}\Big)\\
 & \times\int_{\mathrm{\mathbf{x}}_{1},\mathbf{x}_{2},\mathrm{\mathbf{x}}_{1^{\prime}},\mathbf{x}_{2^{\prime}}}e^{-i\mathbf{p}_{1}\cdot\mathbf{x}_{11^{\prime}}}e^{-i\mathbf{p}_{2}\cdot\mathbf{x}_{22^{\prime}}}A^{\bar{\lambda}}(\mathbf{x}_{12})A^{\lambda^{\prime}}(\mathbf{x}_{1^{\prime}2^{\prime}})\\
 & \times\Big\langle\mathrm{Tr}\big[U_{\mathbf{x}_{2^{\prime}}}U_{\mathbf{x}_{1^{\prime}}}^{\dagger}-1\big]\big[U_{\mathbf{x}_{1}}U_{\mathbf{x}_{2}}^{\dagger}-1\big]\Big\rangle\;,
\end{aligned}
\label{eq:MLOsquared0}
\end{equation}
where the brackets $\langle\rangle$
denote the average of the semiclassical gluon fields in the target (see also sec.~\ref{sec:kine}).

The Dirac trace is performed as follows:
\begin{equation}
\begin{aligned} & \mathrm{Tr}\Big(\mathrm{Dirac}_{\mathrm{LO}}^{\lambda^{\prime}\dagger}\mathrm{Dirac}_{\mathrm{LO}}^{\bar{\lambda}}\Big) =\mathrm{Tr}\Big(\bar{v}_{G}^{s_{2}}(p_{2}^{+})\gamma^{+}\bigl[(1-2z)\delta^{\lambda\lambda^{\prime}}+i\sigma^{\lambda\lambda^{\prime}}\bigr]u_{G}^{s_{1}}(p_{1}^{+})\\
 &\qquad\qquad\qquad\qquad\qquad\qquad\times\bar{u}_{G}^{s_{1}}(p_{1}^{+})\gamma^{+}\bigl[(1-2z)\delta^{\lambda\bar{\lambda}}-i\sigma^{\lambda\bar{\lambda}}\bigr]v_{G}^{s_{2}}(p_{2}^{+})\Big)\;,\\
 & \qquad\qquad\qquad=4p_{1}^{+}p_{2}^{+}\mathrm{Tr}\Big(\mathcal{P}_{G}\bigl[(1-2z)\delta^{\lambda\lambda^{\prime}}+i\sigma^{\lambda\lambda^{\prime}}\bigr]\mathcal{P}_{G}\bigl[(1-2z)\delta^{\lambda\bar{\lambda}}-i\sigma^{\lambda\bar{\lambda}}\bigr]\Big)\;,
\end{aligned}
\end{equation}
where we used the completeness relations for good spinors~\cite{Beuf:2016wdz}:
\begin{equation}
\sum_{s}u_{G}(p^{+})\bar{u}_{G}(p^{+})\gamma^{+}=2p^{+}\mathcal{P}_{G}\;,
\end{equation}
with the same relation holding for antiquark spinors $v_{G}$,  and
with $\mathcal{P}_{G}=\gamma^{-}\gamma^{+}/2$ the projector on the
good components of the spinor field.

Since $\mathcal{P}_{G}$ commutes with all transverse gamma matrices
and since $\mathcal{P}_{G}\mathcal{P}_{G}=\mathcal{P}_{G}$,  we get:
\begin{equation}
\begin{aligned} & \mathrm{Tr}\Big(\mathrm{Dirac}_{\mathrm{LO}}^{\lambda^{\prime}\dagger}\mathrm{Dirac}_{\mathrm{LO}}^{\bar{\lambda}}\Big)\\
 & =4p_{1}^{+}p_{2}^{+}\Big((1-2z)^{2}\delta^{\lambda^{\prime}\bar{\lambda}}\mathrm{Tr}\big\{\mathcal{P}_{G}\big\}+2i(1-2z)\mathrm{Tr}\big\{\mathcal{P}_{G}\sigma^{\bar{\lambda}\lambda^{\prime}}\big\}+\mathrm{Tr}\big\{\mathcal{P}_{G}\sigma^{\lambda\lambda^{\prime}}\sigma^{\lambda\bar{\lambda}}\big\}\Big)\;.
\end{aligned}
\end{equation}
We finally obtain,  using the identities (\ref{eq:sigmasigma}) and
(\ref{eq:traceP}):
\begin{equation}
\begin{aligned}\mathrm{Tr}\Big(\mathrm{Dirac}_{\mathrm{LO}}^{\lambda^{\prime}\dagger}\mathrm{Dirac}_{\mathrm{LO}}^{\bar{\lambda}}\Big) & =16p_{1}^{+}p_{2}^{+}\delta^{\lambda^{\prime}\bar{\lambda}}\Big(z^{2}+\bar{z}^{2}+\frac{D-4}{2}\Big)\;.\end{aligned}
\label{eq:DiracLOLO}
\end{equation}
Our final result for the LO amplitude squared is then:
\begin{equation}
\begin{aligned}\big|\mathcal{M}_{\mathrm{LO}}\big|^{2} & =64\pi\alpha_{\mathrm{em}}e_f^2N_c p_{1}^{+}p_{2}^{+}\Big(z^{2}+\bar{z}^{2}+\frac{D-4}{2}\Big)\\
 & \times\int_{\mathrm{\mathbf{x}}_{1},\mathbf{x}_{2},\mathrm{\mathbf{x}}_{1^{\prime}},\mathbf{x}_{2^{\prime}}}e^{-i\mathbf{p}_{1}\cdot\mathbf{x}_{11^{\prime}}}e^{-i\mathbf{p}_{2}\cdot\mathbf{x}_{22^{\prime}}}A^{\bar\lambda}(\mathbf{x}_{12})A^{\bar\lambda}(\mathbf{x}_{1^{\prime}2^{\prime}})\\
 & \times\Big\langle Q_{122^{\prime}1^{\prime}}-s_{12}-s_{2^{\prime}1^{\prime}}+1\Big\rangle\;,
\end{aligned}
\label{eq:MLOsquared}
\end{equation}
which leads to the following cross section\footnote{For ease of notation, in this work we suppress the overall summation over light quark flavors in the cross section.} in $D=4$ dimensions:
\begin{equation}
\begin{aligned}\frac{\mathrm{d}\sigma_{\mathrm{LO}}}{\mathrm{d}p_{1}^{+}\mathrm{d}p_{2}^{+}\mathrm{\mathrm{d}}^{2}\mathbf{p}_{1}\mathrm{\mathrm{d}}^{2}\mathbf{p}_{2}} & =\frac{2\alpha_{\mathrm{em}}e^2_f N_{c}}{(2\pi)^{4}}\frac{\delta(1-z-\bar{z})}{(q^{+})^{2}}(z^{2}+\bar{z}^{2})\\
 & \times\int_{\mathrm{\mathbf{x}}_{1},\mathbf{x}_{2},\mathrm{\mathbf{x}}_{1}^{\prime},\mathbf{x}_{2^{\prime}}}e^{-i\mathbf{p}_{1}\cdot\mathbf{x}_{11^{\prime}}}e^{-i\mathbf{p}_{2}\cdot\mathbf{x}_{22^{\prime}}}A^{\bar{\lambda}}(\mathbf{x}_{12})A^{\bar{\lambda}}(\mathbf{x}_{1^{\prime}2^{\prime}})\\
 & \times\Big\langle Q_{122^{\prime}1^{\prime}}-s_{12}-s_{2^{\prime}1^{\prime}}+1\Big\rangle\;.
\end{aligned}
\label{eq:LOXsection}
\end{equation}
In the above two expressions, we introduced the following compact notations for
the dipole and quadrupole:
\begin{equation}
\begin{aligned}
s_{12} & \equiv\frac{1}{N_{c}}\mathrm{Tr}\big( U_{\mathbf{x}_{1}}U_{\mathbf{x}_{2}}^{\dagger}\big)\;,\\
Q_{122^{\prime}1^{\prime}} & \equiv\frac{1}{N_{c}}\mathrm{Tr}\big( U_{\mathbf{x}_{1}}U_{\mathbf{x}_{2}}^{\dagger}U_{\mathbf{x}_{2^{\prime}}}U_{\mathbf{x}_{1^{\prime}}}^{\dagger}\big)\;.
\end{aligned}
\label{eq:dipolequadrupoledef}
\end{equation}

\section{\label{sec:NLO}Next-to-leading order diagrams}
The calculation of the next-to-leading order amplitudes follows largely
the same method as the leading-order case. We will therefore not reproduce
all the intermediate steps but rather give the result for each Feynman
diagram, depicted in figures~\ref{fig:virtual} and~\ref{fig:realdiagrams}.
Of course, all diagrams have a counterpart in which the quark and
antiquark are reversed. We will denote these contributions with an
overline: for example, $\mathrm{\overline{GE}SW}$ is the graph in
which the gluon is radiated by the quark and, after scattering off
the shockwave, absorbed by the antiquark. These \textquoteleft $q\leftrightarrow\bar{q}$
conjugate' amplitudes can be obtained in a straightforward way from
their counterparts applying the following steps:
\begin{enumerate}
\item Interchange \emph{all} the indices $1$ and $2$ except in the Wilson
lines and in the spinors $\bar{u}_{G}(p_{1}^{+})$ and $v_{G}(p_{2}^{+})$.
\item Take the complex conjugate of the part of the Dirac structures sandwiched
between the spinors $\bar{u}_{G}(p_{1}^{+})$ and $v_{G}(p_{2}^{+})$.
\item In LCPT, the vertex for the emission or absorption of a gluon from
the antiquark has an overall minus sign. Add it to the diagrams $\mathrm{\overline{I}SW}$,
$\mathrm{\overline{Q}SW}$ and $\mathrm{\overline{Q}SF}$.
\item Calculate the Wilson line structure separately, there is no simple
rule here.
\end{enumerate}
In tables~\ref{tab:real} and~\ref{tab:virtual}, we list all NLO
contributions to the cross section and their possible contribution
to ultraviolet or infrared divergencies. Their regularization and
either cancellation or renormalization will be the subject of sections
\ref{sec:UV} to~\ref{sec:coll}.

\begin{table}[t]
\begin{centering}
\begin{tabular}{|c|c|c|c|c|}
\hline 
 & \multicolumn{1}{c}{} & \multicolumn{1}{c}{Real NLO} & \multicolumn{1}{c}{} & \tabularnewline
\hline 
\hline 
 & $|\mathcal{M}_{\mathrm{QSF}}|^{2}$ \& $|\mathcal{M}_{\mathrm{\overline{Q}SF}}|^{2}$ & $\mathcal{M}_{\mathrm{\overline{Q}SF}}^{\dagger}\mathcal{M}_{\mathrm{QSF}}$
\& $\mathrm{c.c.}$ & $|\mathcal{M}_{\mathrm{QSW}}|^{2}$ \& $|\mathcal{M}_{\mathrm{\overline{Q}SW}}|^{2}$ & cross\tabularnewline
\hline 
JIMWLK & \cmark & \cmark & \cmark & \cmark\tabularnewline
\hline 
collinear & \cmark &  &  & \tabularnewline
\hline 
soft & ? & \cmark &  & \tabularnewline
\hline 
Sudakov &leading $N_c$ &  &  & \tabularnewline
\hline 
\end{tabular}
\par\end{centering}
\caption{\label{tab:real}Overview of the divergencies and large logarithms
encountered in our calculation, for the different real NLO contributions
to the cross section. In the column \textquoteleft cross', all cross
terms are meant between the gluon emissions before and after the shockwave,
from the quark or from the antiquark, i.e. the interference between
$\mathcal{M}_{\mathrm{QFS}}$, $\mathcal{M}_{\mathrm{\overline{Q}FS}}$,
$\mathcal{M}_{\mathrm{QSW}}$, $\mathcal{M}_{\mathrm{\overline{Q}SW}}$.
Terms involving a real instantaneous gluon emission are strictly finite and do not contribute to the Sudakov logarithms at our accuracy. In this work, we do not attempt to analyze the Sudakov double logarithms beyond the large-$N_c$ limit. Moreover, in our regularization scheme, it is not always possible to unambiguously distinguish soft from rapidity divergences. The question mark indicates this is the case for $|\mathcal{M}_{\mathrm{QSF}}|^{2}$ and $|\mathcal{M}_{\mathrm{\overline{Q}SF}}|^{2}$. The only certainty is that these two diagrams combine with $\mathrm{FSIR}$ (table~\ref{tab:virtual}) into a contribution to the cross section that has rapidity divergences only, see also section~\ref{sec:coll}. }
\end{table}
\begin{table}[t]
\begin{centering}
\begin{tabular}{|c|c|c|c|c|c|c|c|}
\hline 
 & \multicolumn{1}{c}{} & \multicolumn{1}{c}{} & \multicolumn{1}{c}{Virtual NLO} & \multicolumn{1}{c}{} & \multicolumn{1}{c}{} & \multicolumn{1}{c}{} & \tabularnewline
\hline 
\hline 
 & $\mathrm{SESW,sub}$ & $\mathrm{GESW}$ & $\mathrm{GEFS}+\mathrm{IFS}$ & $\mathrm{SESW,UV}$ & $\mathrm{IS}$ & $\mathrm{FSIR}$ & $\mathrm{FSUV}$\tabularnewline
\hline 
ultraviolet &  &  &  & \cmark & \cmark &  & \cmark\tabularnewline
\hline 
JIMWLK & \cmark & \cmark & \cmark & ? & ? & \cmark & ?\tabularnewline
\hline 
collinear &  &  &  &  &  & \cmark & \tabularnewline
\hline 
soft &  &  & \cmark & ? & ? & ? & ?\tabularnewline
\hline 
\end{tabular}
\par\end{centering}
\caption{\label{tab:virtual}Overview of the possible divergencies or large
logarithms encountered in our calculation, for the different virtual
NLO contributions to the cross section. $\mathrm{IS}$ stands for
all the initial-state virtual corrections, which were already obtained
in ref.~\cite{Beuf:2016wdz} and are not recalculated here. $\mathrm{FSIR}$
and $\mathrm{FSUV}$, related to the self-energy corrections in the
final state, are not calculated in sec.~\ref{sec:NLO} neither, but
are introduced in section~\ref{sec:UV}. Just like the real contributions
involving an instantaneously created gluon ($\mathrm{RI}$, $\mathrm{\overline{R}I}$
and interferences), the virtual diagram
$\mathrm{ISW}$ with an instantaneously emitted gluon does not exhibit
any singularity or large logarithm, and hence merely contributes to
the finite part of the cross section. Note that all contributions
in the above table, except for $\mathrm{IS}$, $\mathrm{FSIR}$, and
$\mathrm{FSUV}$, have a $q\leftrightarrow\bar{q}$ counterpart with
the same singular structure. The contributions $\mathrm{SESW,UV}$, $\mathrm{IS}$, and $\mathrm{FSUV}$ exhibit soft- or rapidity singularities, between which we cannot make a distinction, although their combination is finite (see section~\ref{sec:UV}).}
\end{table}

\subsection{\label{subsec:virtual}Virtual corrections}
\begin{figure}[t]
\begin{centering}
\includegraphics[scale=0.5]{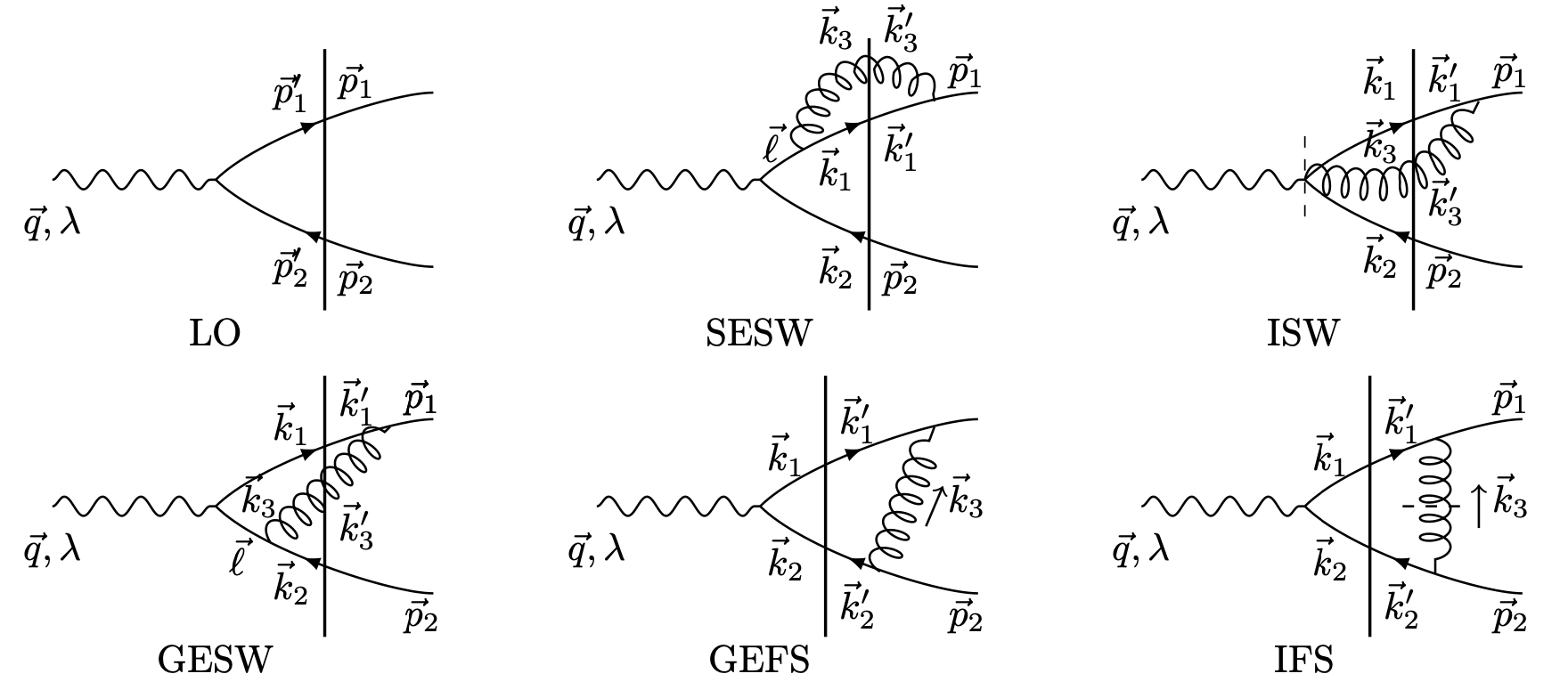}
\par\end{centering}
\caption{\label{fig:virtual}$\mathrm{LO}$: the leading-order Feynman diagram.
The shockwave of semiclassical gluon fields from the hadron target
is depicted by the full vertical line. $\mathrm{SESW}$: self-energy
correction traversing the shockwave. $\mathrm{ISW}$: instantaneous
gluon emission crossing the shockwave. $\mathrm{GESW}$: gluon exchange
crossing the shockwave. $\mathrm{GEFS}$: gluon exchange in the final
state. $\mathrm{IFS}$: instantaneous gluon exchange in the final
state. Virtual corrections before the shockwave are not explicitly
calculated in this work and not shown here, and neither are the self-energy
corrections on the asymptotic final states.}
\end{figure}

\begin{figure}[t]
\begin{centering}
\includegraphics[scale=0.2]{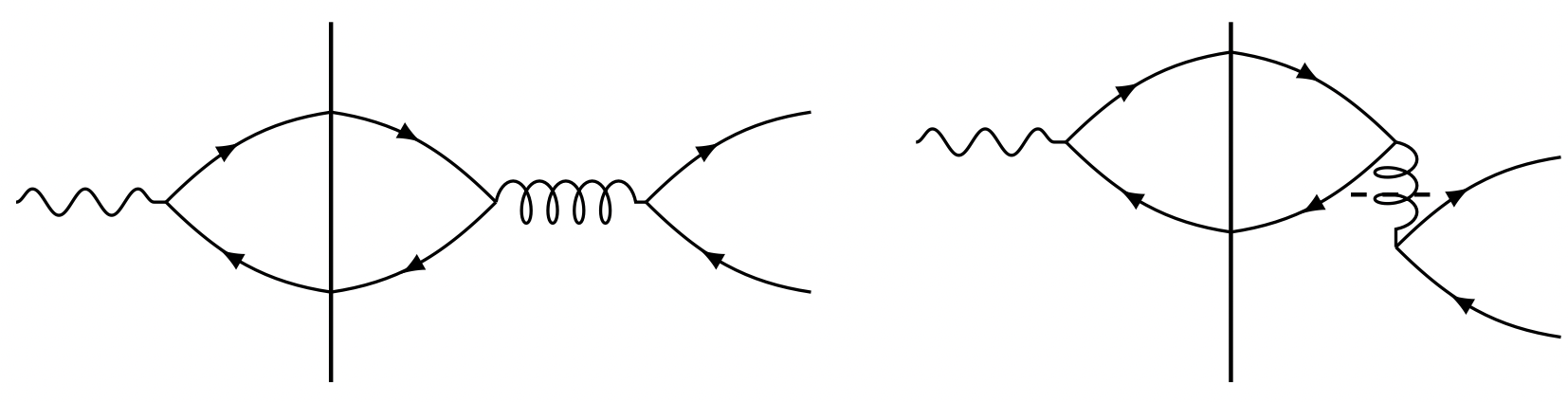}
\par\end{centering}
\caption{\label{fig:s-channel}Two diagrams with a virtual (left) and instantaneous (right) gluon in the $s$-channel after the shockwave. As explained in the main text, these diagrams disappear when considering only the three lightest quarks.}
\end{figure}

\paragraph{ISW: Instantaneous gluon traversing SW}
The amplitude for the instantaneous production of a quark, antiquark,
and gluon from the photon, where the gluon crosses the shockwave and
is absorbed by the outgoing quark (fig.~\ref{fig:virtual}),
is found to be:
\begin{equation}
\begin{aligned}\mathcal{M}_{\mathrm{ISW}} & =i\frac{g_{e}e_fg_{s}^{2}}{2}\int_{0}^{p_{1}^{+}}\frac{\mathrm{d}k_{3}^{+}}{2\pi}\frac{p_{1}^{+}-k_{3}^{+}}{p_{1}^{+}k_{3}^{+}(p_{2}^{+}+k_{3}^{+})}\Big(\frac{k_{3}^{+}}{p_{1}^{+}}\Big)^{D-2}\big(p_{2}^{+}-p_{1}^{+}+k_{3}^{+}\big)\mathrm{Dirac}_{\mathrm{ISW}}^{\eta^{\prime}}(k_{3}^{+})\\
 & \times\int_{\mathbf{x}_{1},\mathbf{x}_{2},\mathbf{x}_{3}}e^{-i\mathbf{p}_{1}\cdot\big(\frac{p_{1}^{+}-k_{3}^{+}}{p_{1}^{+}}\mathbf{x}_{1}+\frac{k_{3}^{+}}{p_{1}^{+}}\mathbf{x}_{3}\big)}e^{-i\mathbf{p}_{2}\cdot\mathbf{x}_{2}}\\
 & \times A^{\eta^{\prime}}(\mathbf{x}_{31})\mathcal{C}\big(\frac{k_{3}^{+}}{p_{1}^{+}}\mathbf{x}_{13}+\mathbf{x}_{21},\frac{k_{3}^{+}}{p_{1}^{+}}\mathbf{x}_{13};\frac{q^{+}(p_{1}^{+}-k_{3}^{+})}{p_{2}^{+}k_{3}^{+}}\big)\\
 & \times\Big[t^{c}U_{\mathbf{x}_{1}}t^{d}U_{\mathbf{x}_{2}}^{\dagger}W_{\mathbf{x}_{3}}^{cd}-C_{F}\Big]\;,
\end{aligned}
\label{eq:instSESWfinal}
\end{equation}
with
\begin{equation}
\begin{aligned}\mathrm{Dirac}_{\mathrm{ISW}}^{\eta^{\prime}}(k_{3}^{+}) & =\bar{u}_{G}^{s_{1}}(p_{1}^{+})\Bigg\{\Bigg[\Big(2\frac{p_{1}^{+}}{k_{3}^{+}}-1\Big)-(D-3)\big(\frac{q^{+}+k_{3}^{+}}{p_{2}^{+}-p_{1}^{+}+k_{3}^{+}}\big)\Bigg]\delta^{\lambda\eta^{\prime}}\\
 & \qquad\qquad+i\sigma^{\lambda\eta^{\prime}}\Bigg[1-\Big(2\frac{p_{1}^{+}}{k_{3}^{+}}-D+3\Big)\big(\frac{q^{+}+k_{3}^{+}}{p_{2}^{+}-p_{1}^{+}+k_{3}^{+}}\big)\Bigg]\Bigg\}\gamma^{+}v_{G}^{s_{2}}(p_{2}^{+})\;,
\end{aligned}
\end{equation}
and where we introduced the generalization of the Coulomb field (see
\cite{Altinoluk:2020qet}) to $D-4$ dimensions,  defined as:
\begin{equation}
\begin{aligned}\mathcal{C}\big(\mathbf{x},\mathbf{y},c\big) & \equiv\int_{\mathbf{k},\mathbf{p}}e^{i\mathbf{p}\cdot\mathbf{x}}e^{i\mathbf{k}\cdot\mathbf{y}}\frac{1}{\mathbf{k}^{2}+c\mathbf{p}^{2}}\\
 & =\mu^{2(4-D)}\frac{\Gamma(D-3)}{4\pi^{D-2}}\frac{c^{\frac{D}{2}-2}}{(\mathbf{x}^{2}+c\mathbf{y}^{2})^{D-3}}\stackrel{D\to4}{=}\frac{1}{(2\pi)^{2}}\frac{1}{\mathbf{x}^{2}+c\mathbf{y}^{2}}\;.
\end{aligned}
\end{equation}

\paragraph{GESW: Gluon exchange traversing the SW}
The amplitude for gluon exchange interacting with the shockwave (fig.~\ref{fig:virtual}) is found to be:
\begin{equation}
\begin{aligned}\mathcal{M}_{\mathrm{GESW}} & =-\frac{ig_{e}e_fg_{s}^{2}}{2}\int_{0}^{p_{1}^{+}}\frac{\mathrm{d}k_{3}^{+}}{2\pi}\frac{k_{3}^{+}}{p_{1}^{+}(p_{2}^{+}+k_{3}^{+})}\mathrm{Dirac}_{\bar{q}\to q}^{\bar{\lambda}\bar{\eta}\eta^{\prime}}(k_{3}^{+})\\
 & \times\int_{\mathbf{x}_{1},\mathbf{x}_{2},\mathbf{x}_{3}}e^{-i\frac{p_{1}^{+}-k_{3}^{+}}{p_{1}^{+}}\mathbf{p}_{1}\cdot\mathbf{x}_{1}}e^{-i\mathbf{p}_{2}\cdot\mathbf{x}_{2}}e^{-i\frac{k_{3}^{+}}{p_{1}^{+}}\mathbf{p}_{1}\cdot\mathbf{x}_{3}}\\
 & \times A^{\eta^{\prime}}(\mathbf{x}_{31})A^{\bar{\eta}}(\mathbf{x}_{32})\mathcal{A}^{\bar{\lambda}}\Big(\frac{p_{2}^{+}\mathbf{x}_{12}+k_{3}^{+}\mathbf{x}_{13}}{p_{2}^{+}+k_{3}^{+}},\frac{k_{3}^{+}}{p_{2}^{+}+k_{3}^{+}}\mathbf{x}_{32};\frac{q^{+}p_{2}^{+}}{k_{3}^{+}(p_{1}^{+}-k_{3}^{+})}\Big)\\
 & \times\Big[t^{c}U_{\mathbf{x}_{1}}t^{d}U_{\mathbf{x}_{2}}^{\dagger}W_{\mathbf{x}_{3}}^{dc}-C_{F}\Big]\;.
\end{aligned}
\label{eq:GESW}
\end{equation}
To arrive at the above expression, we made use of the intermediary
result:
\begin{equation}
\begin{aligned}
\int_{\mathbf{K},\mathbf{P}}\frac{P^{i}}{P^{2}}\frac{K^{j}}{K^{2}+cP^{2}}e^{i\mathbf{K}\cdot\mathbf{r}'}e^{i\mathbf{P}\cdot\mathbf{r}} & =-A^{j}(r^{\prime})\mathcal{A}^{i}(r,r^{\prime},c)
\;,
\end{aligned}
\label{eq:modWWdef-2}
\end{equation}
where $\mathcal{A}$ is the modified Weizs\"acker-Williams field\footnote{Note that we use a slight redefinition of this field with respect
to earlier work~\cite{Altinoluk:2020qet}.} defined as:
\begin{equation}
\begin{aligned}
\mathcal{A}^{i}\big(\mathbf{r},\mathbf{r}^{\prime},\mathcal{\mathcal{C}}\big) & \equiv\frac{-\mu^{4-D}}{64\pi^{D/2}}\frac{r^{i}}{(\mathbf{r}^{2})^{D/2-1}}\Bigg\{32\pi\,\Gamma\big(\frac{D}{2}-1\big)-2^{D}\sqrt{\pi}\mathcal{C}^{\frac{D}{2}-2}\Big(\frac{\mathbf{r}^{\prime2}}{\mathbf{r}^{2}}\Big)^{\frac{D}{2}-2}\Gamma\big(\frac{D}{2}-\frac{3}{2}\big)\\
 &\qquad\qquad\qquad\qquad\quad \times\Big[(\mathbf{r}^{2})^{D-4}\big(\mathcal{C}\mathbf{r}^{\prime2}-\mathbf{r}^{2}\big)\big(\mathcal{C}\mathbf{r}^{\prime2}+\mathbf{r}^{2})^{3-D}\\&\qquad \qquad\qquad\qquad\quad+F_{2,1}\big(\frac{D}{2}-2,D-4,\frac{D}{2}-1,-\mathcal{C}\frac{\mathbf{r}^{\prime2}}{\mathbf{r}^{2}}\big)\Big]\Bigg\}\;.
\end{aligned}
\end{equation}
Luckily, it turns out that we will only need to evaluate $\mathcal{A}$
in finite contributions to the cross section, where we can work in
$D=4$ dimensions for which its expression reduces to:
\begin{equation}
\begin{aligned}
\mathcal{A}^{i}\big(\mathbf{r},\mathbf{r}^{\prime},\mathcal{\mathcal{C}}\big) & =\frac{-1}{2\pi}\frac{r^{i}}{\mathbf{r}^{2}+\mathcal{C}\mathbf{r}^{\prime2}}
\;.
\end{aligned}
\end{equation}
The spinor structure in (\ref{eq:GESW}) has the following form:
\begin{equation}
\begin{aligned}
\mathrm{Dirac}_{\bar{q}\to q}^{\bar{\lambda}\bar{\eta}\eta^{\prime}}(k_{3}^{+}) &= \bar{u}_{G}^{s_{1}}(p_{1}^{+})\Bigg[\Big(2\frac{p_{1}^{+}}{k_{3}^{+}}-1\Big)\delta^{\eta\eta^{\prime}}+i\sigma^{\eta\eta^{\prime}}\Bigg]\Bigg[\Big(1-2\frac{p_{2}^{+}+k_{3}^{+}}{q^{+}}\Big)\delta^{\lambda\bar{\lambda}}+i\sigma^{\lambda\bar{\lambda}}\Bigg]\\
 & \times\Bigg[\Big(1+2\frac{p_{2}^{+}}{k_{3}^{+}}\Big)\delta^{\eta\bar{\eta}}+i\sigma^{\eta\bar{\eta}}\Bigg]\gamma^{+}v_{G}^{s_{2}}(p_{2}^{+})\;,\\
 & =\mathrm{Dirac}_{\bar{q}\to q,(i)}^{\bar{\lambda}\bar{\eta}\eta^{\prime}}+\mathrm{Dirac}_{\bar{q}\to q,(ii)}^{\bar{\lambda}\bar{\eta}\eta^{\prime}}
 \;,
\end{aligned}
\label{eq:Diracqbar2q}
\end{equation}
where we defined:
\begin{equation}
\begin{aligned}
\mathrm{Dirac}_{\bar{q}\to q,(ii)}^{\bar{\lambda}\bar{\eta}\eta^{\prime}} & =4\frac{p_{1}^{+}p_{2}^{+}}{(k_{3}^{+})^{2}}\delta^{\bar{\eta}\eta^{\prime}}\bar{u}_{G}^{s_{1}}(p_{1}^{+})\Bigg[\Big(1-2\frac{p_{2}^{+}+k_{3}^{+}}{q^{+}}\Big)\delta^{\lambda\bar{\lambda}}+i\sigma^{\lambda\bar{\lambda}}\Bigg]\gamma^{+}v_{G}^{s_{2}}(p_{2}^{+})
\;,
\end{aligned}
\label{eq:Diracqbar2q(ii)}
\end{equation}
which is the most singular part of the Dirac structure, scaling like
$1/(k_{3}^{+})^{2}$.

\paragraph{SESW: Self-energy correction traversing the SW}
For the self-energy corrections crossing the shockwave ($\mathrm{SESW}$,
fig.~\ref{fig:virtual}), we obtain:
\begin{equation}
\begin{aligned}
\mathcal{M}_{\mathrm{SESW}} & =i\frac{g_{e}e_fg_{s}^{2}}{2}\int_{0}^{p_{1}^{+}}\frac{\mathrm{d}k_{3}^{+}}{2\pi}\frac{k_{3}^{+}}{(p_{1}^{+})^{2}}\Big[\Big(2\frac{p_{1}^{+}}{k_{3}^{+}}-1\Big)^{2}+(D-3)\Big]\mathrm{Dirac}_{\mathrm{LO}}^{\bar{\lambda}}\\
 & \times\int_{\mathrm{\mathbf{x}}_{1},\mathbf{x}_{2},\mathbf{x}_{3}}e^{-i\mathbf{p}_{1}\cdot\mathbf{x}_{1}}e^{-i\mathbf{p}_{2}\cdot\mathbf{x}_{2}}e^{i\frac{k_{3}^{+}}{p_{1}^{+}}\mathbf{p}_{1}\cdot\mathbf{x}_{13}}A^{\bar{\eta}}(\mathbf{x}_{13})A^{\bar{\eta}}(\mathbf{x}_{13})\\
 & \times\mathcal{A}^{\bar{\lambda}}\Big(\frac{k_{3}^{+}}{p_{1}^{+}}\mathbf{x}_{13}+\mathbf{x}_{21},\frac{k_{3}^{+}}{p_{1}^{+}}\mathbf{x}_{13};\frac{q^{+}(p_{1}^{+}-k_{3}^{+})}{k_{3}^{+}p_{2}^{+}}\Big)\Big[t^{c}U_{\mathbf{x}_{1}}t^{d}U_{\mathbf{x}_{2}}^{\dagger}W_{\mathbf{x}_{3}}^{dc}-C_{F}\Big]\;.
\end{aligned}
\label{eq:SESWunsub}
\end{equation}
The above amplitude contains an ultraviolet divergence,  which
comes from the limit where $\mathbf{x}_{3}\to\mathbf{x}_{1}$. Following
ref.~\cite{Beuf:2016wdz}, we construct a counterterm $\mathcal{M}_{\mathrm{SESW,UV}}$
by taking the $\mathbf{x}_{3}\to\mathbf{x}_{1}$ limit in $\mathcal{M}_{\mathrm{SESW}}$
except in the singular part, and by subtracting an infrared (IR)
divergent contribution $\propto A^{\bar{\eta}}(\mathbf{x}_{13})A^{\bar{\eta}}(\mathbf{x}_{23})$
as follows:
\begin{equation}
\begin{aligned}
\mathcal{M}_{\mathrm{SESW,UV}} & =i\frac{g_{e}e_fg_{s}^{2}}{2}C_{F}\int_{0}^{p_{1}^{+}}\frac{\mathrm{d}k_{3}^{+}}{2\pi}\frac{k_{3}^{+}}{(p_{1}^{+})^{2}}\Big[\Big(2\frac{p_{1}^{+}}{k_{3}^{+}}-1\Big)^{2}+(D-3)\Big]\mathrm{Dirac}_{\mathrm{LO}}^{\bar{\lambda}}\\
 & \times\int_{\mathrm{\mathbf{x}}_{1},\mathbf{x}_{2},\mathbf{x}_{3}}e^{-i\mathbf{p}_{1}\cdot\mathbf{x}_{1}}e^{-i\mathbf{p}_{2}\cdot\mathbf{x}_{2}}A^{\bar{\eta}}(\mathbf{x}_{13})\Big[A^{\bar{\eta}}(\mathbf{x}_{13})-A^{\bar{\eta}}(\mathbf{x}_{23})\Big]A^{\bar{\lambda}}(\mathbf{x}_{21})\\&\times\Big[U_{\mathbf{x}_{1}}U_{\mathbf{x}_{2}}^{\dagger}-1\Big]\;,
\end{aligned}
\label{eq:SESWUV}
\end{equation}
where we used that,  even in $D$ dimensions:
\begin{equation}
\begin{aligned}
\mathcal{A}^{i}\big(\mathbf{r},0,\mathcal{\mathcal{C}}\big) & =A^{i}(\mathbf{r})
\;.
\end{aligned}
\end{equation}
By construction, the counterterm $\mathcal{M}_{\mathrm{SESW,UV}}$
has the same UV pole as $\mathcal{M}_{\mathrm{SESW}}$ while not possessing
any other divergences. To show this is the case,  let
us explicitly evaluate the $\mathbf{x}_{3}$ integration in eq.~\eqref{eq:SESWUV}. We have that:
\begin{equation}
\begin{aligned}
\int_{\mathbf{x}_{3}}A^{i}(\mathbf{x}_{13})A^{i}(\mathbf{x}_{13}) & =\mu^{4-D}\frac{\Gamma(\frac{D}{2}-1)^{2}}{4\pi^{D-2}}\int\frac{\mathrm{d}^{D-2}\mathbf{x}_{3}}{(\mathbf{x}_{13}^{2})^{D-3}}=0
\;,
\end{aligned}
\label{eq:zero}
\end{equation}
since scaleless integrals disappear in dimensional regularization.
This is equivalent to the statement that the above integral contains
two divergences, an UV ($\mathbf{x}_{3}\to\mathbf{x}_{1}$) and an
IR ($\mathbf{x}_{3}\to\infty$) one, which cancel each other.

The second integration in eq.~\eqref{eq:SESWUV} reads:
\begin{equation}
\begin{aligned}
\int_{\mathbf{x}_{3}}A^{i}(\mathbf{x}_{13})A^{i}(\mathbf{x}_{23}) & =\mu^{4-D}\frac{\Gamma\big(\frac{D}{2}-1\big)^{2}}{4\pi^{D-2}}\frac{\pi^{D/2-1}}{\big(\frac{D}{2}-2\big)\Gamma\big(\frac{D}{2}-1\big)}\frac{1}{(\mathbf{x}_{12}^{2})^{\frac{D}{2}-2}}\;,\\
 & =\mu^{4-D}\frac{\Gamma\big(\frac{D}{2}-1\big)}{4\pi^{D/2-1}}\frac{1}{\big(\frac{D}{2}-2\big)}\frac{1}{(\mathbf{x}_{12}^{2})^{\frac{D}{2}-2}}\;,
\end{aligned}
\label{eq:IRcounter}
\end{equation}
which is divergent in the infrared. Combining eqs. (\ref{eq:zero}) and (\ref{eq:IRcounter}),
 the IR pole of the latter cancels the one hidden in the former, and
the overall divergence of $\mathcal{M}_{\mathrm{SESW,UV}}$ can be
interpreted as an UV one.

Since $\mathcal{M}_{\mathrm{SESW,}\mathrm{sub}}\equiv\mathcal{M}_{\mathrm{SESW}}-\mathcal{M}_{\mathrm{SESW,UV}}$
is now free from UV divergences,  we can evaluate it in $D=4$ dimensions:
\begin{equation}
\begin{aligned}
&\mathcal{M}_{\mathrm{SESW,sub}}  =i\frac{g_{e}e_fg_{s}^{2}}{2}\int_{0}^{p_{1}^{+}}\frac{\mathrm{d}k_{3}^{+}}{2\pi}\frac{k_{3}^{+}}{(p_{1}^{+})^{2}}\Big[\Big(2\frac{p_{1}^{+}}{k_{3}^{+}}-1\Big)^{2}+1\Big]\mathrm{Dirac}_{\mathrm{LO}}^{\bar{\lambda}}\\
 & \times\int_{\mathrm{\mathbf{x}}_{1},\mathbf{x}_{2},\mathbf{x}_{3}}e^{-i\mathbf{p}_{1}\cdot\mathbf{x}_{1}}e^{-i\mathbf{p}_{2}\cdot\mathbf{x}_{2}}A^{\bar{\eta}}(\mathbf{x}_{13})\\
 &\times \Bigg\{ e^{i\frac{k_{3}^{+}}{p_{1}^{+}}\mathbf{p}_{1}\cdot\mathbf{x}_{13}}A^{\bar{\eta}}(\mathbf{x}_{13})\mathcal{A}^{\bar{\lambda}}\Big(\frac{k_{3}^{+}}{p_{1}^{+}}\mathbf{x}_{13}+\mathbf{x}_{21},\frac{k_{3}^{+}}{p_{1}^{+}}\mathbf{x}_{13};\frac{q^{+}(p_{1}^{+}-k_{3}^{+})}{k_{3}^{+}p_{2}^{+}}\Big)\\&\quad\times\Big[t^{c}U_{\mathbf{x}_{1}}t^{d}U_{\mathbf{x}_{2}}^{\dagger}W_{\mathbf{x}_{3}}^{dc}-C_{F}\Big] -\Big[A^{\bar{\eta}}(\mathbf{x}_{13})-A^{\bar{\eta}}(\mathbf{x}_{23})\Big]A^{\bar{\lambda}}(\mathbf{x}_{21})C_{F}\Big[U_{\mathbf{x}_{1}}U_{\mathbf{x}_{2}}^{\dagger}-1\Big]\Bigg\}\;.
\end{aligned}
\label{eq:SESWsub}
\end{equation}

\paragraph{GEFS: Gluon exchange in the final state}
We find for the amplitude where the gluon is emitted from the antiquark
and absorbed by the quark in the final state (fig.~\ref{fig:virtual}):
\begin{equation}
\begin{aligned}
\mathcal{M}_{\mathrm{GEFS}} & =-i\frac{g_{e}e_fg_{s}^{2}}{2}\int_{0}^{p_{1}^{+}}\frac{\mathrm{d}k_{3}^{+}}{2\pi}\frac{p_{1}^{+}-k_{3}^{+}}{q^{+}p_{1}^{+}}\mathrm{Dirac}_{\bar{q}\to q}^{\bar{\lambda}\bar{\eta}\eta^{\prime}}(k_{3}^{+})\\
 & \times\int_{\mathrm{\mathbf{x}}_{1},\mathbf{x}_{2}}e^{-i\mathbf{p}_{1}\cdot\big(\frac{p_{1}^{+}-k_{3}^{+}}{p_{1}^{+}}\mathbf{x}_{1}+\frac{k_{3}^{+}}{p_{1}^{+}}\mathbf{x}_{2}\big)}e^{-i\mathbf{p}_{2}\cdot\mathbf{x}_{2}}\\
 & \times A^{\bar{\lambda}}(\mathbf{x}_{12})J^{\eta^{\prime}\bar{\eta}}(k_{3},\mathbf{x}_{12})\times\big[t^{c}U_{\mathbf{x}_{1}}U_{\mathbf{x}_{2}}^{\dagger}t^{c}-C_{F}\big]\;.
\end{aligned}
\label{eq:GEFS}
\end{equation}
with:
\begin{equation}
\begin{aligned}
J^{\eta^{\prime}\bar{\eta}}(k_{3}^{+},\mathbf{x}_{12}) & =\int_{\mathbf{K}}e^{-i\mathbf{K}\cdot\mathbf{x}_{12}}\frac{\mathbf{K}^{\eta^{\prime}}}{\mathbf{K}^{2}-i\epsilon}\frac{\mathbf{K}^{\bar{\eta}}-\frac{k_{3}^{+}q^{+}}{p_{1}^{+}p_{2}^{+}}\mathbf{P}_{\perp}^{\bar{\eta}}}{\big(\mathbf{K}+\frac{p_{1}^{+}-k_{3}^{+}}{p_{1}^{+}}\mathbf{P}_{\perp}\big)^{2}-\frac{p_{2}^{+}+k_{3}^{+}}{p_{2}^{+}}\frac{p_{1}^{+}-k_{3}^{+}}{p_{1}^{+}}\mathbf{P}_{\perp}^{2}-i\epsilon}
\;,
\end{aligned}
\label{eq:J}
\end{equation}
In the above expression, $\mathbf{P}_{\perp}$ is a transverse vector
defined as:
\begin{equation}
\begin{aligned}
\mathbf{P}_{\perp} & \equiv\frac{p_{2}^{+}}{q^{+}}\mathbf{p}_{1}-\frac{p_{1}^{+}}{q^{+}}\mathbf{p}_{2}
\;,
\end{aligned}
\label{eq:P}
\end{equation}
while the loop momentum $\mathbf{K}$ is related to the virtual gluon
transverse momentum through $p_{1}^{+}\mathbf{K}\equiv p_{1}^{+}\mathbf{k}_{3}-k_{3}^{+}\mathbf{p}_{1}$.

\paragraph{IFS: Instantaneous gluon exchange in the final state}
In the case of an instantaneous $q\bar{q}\to q\bar{q}$ final-state
interaction mediated by an instantaneous gluon (fig.~\ref{fig:virtual}) we obtain:
\begin{equation}
\begin{aligned}
\mathcal{M}_{\mathrm{IFS}} & =ig_{e}e_fg_{s}^{2}\int_{0}^{p_{1}^{+}}\frac{\mathrm{d}k_{3}^{+}}{2\pi}\frac{1}{(k_{3}^{+})^{2}}\frac{2(p_{2}^{+}+k_{3}^{+})(p_{1}^{+}-k_{3}^{+})}{q^{+}}\\&\times\bar{u}_{G}^{s_{1}}(p_{1}^{+})\mathbf{\gamma}^{+}\Big[\Big(1-2\frac{p_{2}^{+}+k_{3}^{+}}{q^{+}}\Big)\delta^{\lambda\bar{\lambda}}+i\sigma^{\lambda\bar{\lambda}}\Big]v_{G}^{s_{2}}(p_{2}^{+})\\
 & \times\int_{\mathrm{\mathbf{x}}_{1},\mathbf{x}_{2}}e^{-i\mathbf{p}_{1}\cdot\Big(\frac{p_{1}^{+}-k_{3}^{+}}{p_{1}^{+}}\mathbf{x}_{1}+\frac{k_{3}^{+}}{p_{1}^{+}}\mathbf{x}_{2}\Big)}e^{-i\mathbf{p}_{2}\cdot\mathbf{x}_{2}}A^{\bar{\lambda}}(\mathbf{x}_{12})\\
 & \times\int_{\mathbf{K}}\frac{e^{-i\mathbf{K}\cdot\mathbf{x}_{12}}}{\big(\mathbf{K}+\frac{p_{1}^{+}-k_{3}^{+}}{p_{1}^{+}}\mathbf{P}_{\perp}\big)^{2}-\frac{(p_{2}^{+}+k_{3}^{+})(p_{1}^{+}-k_{3}^{+})}{p_{1}^{+}p_{2}^{+}}\mathbf{P}_{\perp}^{2}-i\epsilon}\times\big[t^{c}U_{\mathbf{x}_{1}}U_{\mathbf{x}_{2}}^{\dagger}t^{c}-C_{F}\big]\;,
\end{aligned}
\label{eq:IFS}
\end{equation}
where the loop momentum is again defined as $p_{1}^{+}\mathbf{K}\equiv p_{1}^{+}\mathbf{k}_{3}-k_{3}^{+}\mathbf{p}_{1}$.

\paragraph{Combining diagrams GEFS and IFS}
The amplitudes (\ref{eq:GEFS}) and (\ref{eq:IFS}) both exhibit an
unphysical $(1/k_{3}^{+})^{2}$ power divergence, which cancels when
summing them. In diagram GEFS, this divergence stems from the Dirac
structure $\mathrm{Dirac}_{\bar{q}\to q,(ii)}^{\bar{\lambda}\bar{\eta}\eta^{\prime}}$.
The subamplitude corresponding to this part of the Dirac structure, which
we denote by $\mathcal{M}_{\mathrm{GEFS},(ii)}$, reads:
\begin{equation}
\begin{aligned}
\mathcal{M}_{\mathrm{GEFS},(ii)} & =-ig_{e}e_fg_{s}^{2}\int_{0}^{p_{1}^{+}}\frac{\mathrm{d}k_{3}^{+}}{2\pi}\frac{2(p_{1}^{+}-k_{3}^{+})}{q^{+}p_{1}^{+}}\frac{p_{1}^{+}p_{2}^{+}}{(k_{3}^{+})^{2}}\\&\times\bar{u}_{G}^{s_{1}}(p_{1}^{+})\Bigg[\Big(1-2\frac{p_{2}^{+}+k_{3}^{+}}{q^{+}}\Big)\delta^{\lambda\bar{\lambda}}+i\sigma^{\lambda\bar{\lambda}}\Bigg]\gamma^{+}v_{G}^{s_{2}}(p_{2}^{+})\\
 & \times\int_{\mathrm{\mathbf{x}}_{1},\mathbf{x}_{2}}e^{-i\mathbf{p}_{1}\cdot\big(\frac{p_{1}^{+}-k_{3}^{+}}{p_{1}^{+}}\mathbf{x}_{1}+\frac{k_{3}^{+}}{p_{1}^{+}}\mathbf{x}_{2}\big)}e^{-i\mathbf{p}_{2}\cdot\mathbf{x}_{2}}A^{\bar{\lambda}}(\mathbf{x}_{12})J^{\eta^{\prime}\eta^{\prime}}(k_{3},\mathbf{x}_{12})\\
 & \times\big[t^{c}U_{\mathbf{x}_{1}}U_{\mathbf{x}_{2}}^{\dagger}t^{c}-C_{F}\big]\;,
\end{aligned}
\end{equation}
and is clearly divergent $\propto1/(k_{3}^{+})^{2}$ in the $k_{3}^{+}\to0$
limit. The situation changes when the above (sub)amplitude is summed
with the $\mathrm{IFS}$ amplitude (\ref{eq:IFS}):
\begin{equation}
\begin{aligned}
&\mathcal{M}_{\mathrm{GEFS},(ii)+\mathrm{IFS}} =i\frac{g_{e}e_fg_{s}^{2}}{\pi}\int_{0}^{p_{1}^{+}}\frac{\mathrm{d}k_{3}^{+}}{k_{3}^{+}}\frac{p_{1}^{+}-k_{3}^{+}}{q^{+}}\\&\times\bar{u}_{G}^{s_{1}}(p_{1}^{+})\mathbf{\gamma}^{+}\Big[\Big(1-2\frac{p_{2}^{+}+k_{3}^{+}}{q^{+}}\Big)\delta^{\lambda\bar{\lambda}}+i\sigma^{\lambda\bar{\lambda}}\Big]v_{G}^{s_{2}}(p_{2}^{+})\\
 & \times\int_{\mathrm{\mathbf{x}}_{1},\mathbf{x}_{2}}e^{-i\mathbf{p}_{1}\cdot\Big(\frac{p_{1}^{+}-k_{3}^{+}}{p_{1}^{+}}\mathbf{x}_{1}+\frac{k_{3}^{+}}{p_{1}^{+}}\mathbf{x}_{2}\Big)}e^{-i\mathbf{p}_{2}\cdot\mathbf{x}_{2}}A^{\bar{\lambda}}(\mathbf{x}_{12})\\
 & \times\int_{\mathbf{K}}\Big(1+\frac{q^{+}}{p_{1}^{+}}\frac{\mathbf{K}\cdot\mathbf{P}_{\perp}}{\mathbf{K}^{2}}\Big)\frac{e^{-i\mathbf{K}\cdot\mathbf{x}_{12}}}{\big(\mathbf{K}+\frac{p_{1}^{+}-k_{3}^{+}}{p_{1}^{+}}\mathbf{P}_{\perp}\big)^{2}-\frac{(p_{2}^{+}+k_{3}^{+})(p_{1}^{+}-k_{3}^{+})}{p_{1}^{+}p_{2}^{+}}\mathbf{P}_{\perp}^{2}-i\epsilon}\\
 & \times\big[t^{c}U_{\mathbf{x}_{1}}U_{\mathbf{x}_{2}}^{\dagger}t^{c}-C_{F}\big]\;,
\end{aligned}
\label{eq:GEFS+IFS}
\end{equation}
and we are left with a logarithmic $k^+\to0$ divergence, which will contribute to JIMWLK.

\paragraph{Final state with gluon in $s$-channel}
Finally, in fig.~\ref{fig:s-channel}, two virtual diagrams are depicted in which, after the scattering off the shockwave, the quark-antiquark pair annihilates and a new pair is created. This process either takes place through an $s$-channel gluon or as an instantaneous $q\bar{q}\to q\bar{q}$ vertex mediated by a fictitious instantaneous gluon. When multiplied with the conjugate leading-order amplitude, these two diagrams contribute to the cross section with a coupling:
\begin{equation}
\begin{aligned}
\big(\sum_f e_f g_e\big)^2 g_s^2\;.
\end{aligned}
\end{equation}
In the above, the summation over quark flavors takes place for the two quark lines separately, since at the level of the cross section the two fermion loops are distinct and only connected by a gluon. This is a unique feature of the two diagrams under consideration; all other contributions to the NLO dijet photoproduction cross section have a coupling $\sum_f e_f^2 g_e^2 g_s^2$. In particular, since in this calculation we only consider the three lightest quarks:
\begin{equation}
\begin{aligned}
\big(\sum_{f=u,d,s} e_f g_e\big)^2 g_s^2=\big(e_u + e_d+ e_s \big)^2 g_e^2g_s^2=\Big(\frac{2}{3}-\frac{1}{3}-\frac{1}{3} \Big)^2 g_e^2g_s^2=0\;,
\end{aligned}
\end{equation}
hence these diagrams can be disregarded.

\subsection{\label{sec:223}Real corrections}
\begin{figure}[t]
\begin{centering}
\includegraphics[clip,scale=0.5]{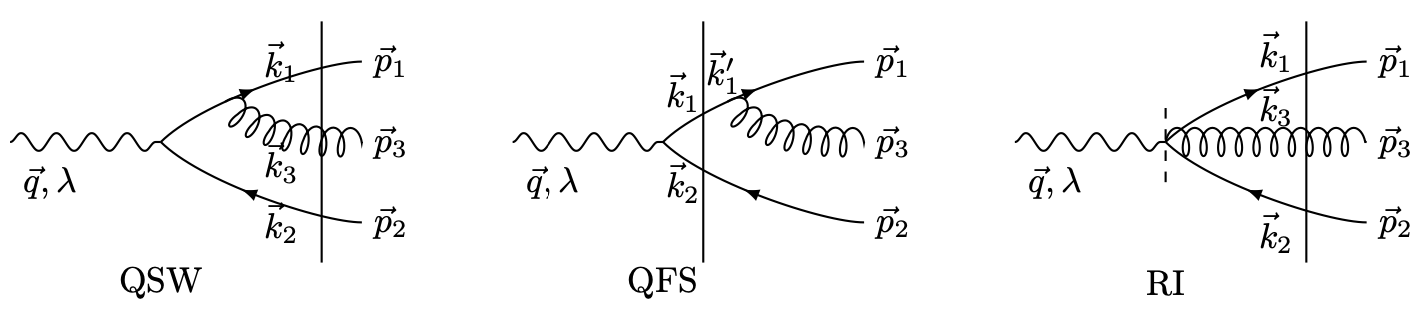}
\par
\end{centering}
\caption{\label{fig:realdiagrams}$\mathrm{QSW}$: real gluon emission from
the quark,  scattering off the shockwave (vertical full line). $\mathrm{QFS}$:
gluon radiated from the quark in the final state. $\mathrm{RI}$:
instantaneously created real gluon scatters off shockwave. }
\end{figure}

\paragraph{QSW: Real gluon scatters off shockwave}
For the diagram where the gluon is emitted from the quark and then
interacts with the shockwave (fig.~\ref{fig:realdiagrams}),
we obtain:
\begin{equation}
\begin{aligned}\mathcal{M}_{\mathrm{QSW}} & =-g_{e}e_fg_{s}\mathrm{Dirac}_{\mathrm{QSW}}^{\bar{\eta}\bar{\lambda}}\frac{p_{3}^{+}}{p_{1}^{+}+p_{3}^{+}}\\
 & \times\int_{\mathrm{\mathbf{x}}_{1},\mathbf{x}_{2},\mathbf{x}_{3},\mathbf{v}}e^{-i\mathbf{p}_{1}\cdot\mathbf{x}_{1}}e^{-i\mathbf{p}_{2}\cdot\mathbf{x}_{2}}e^{-i\mathbf{p}_{3}\cdot\mathbf{x}_{3}}\delta^{(D-2)}\big(\mathbf{v}-\frac{p_{1}^{+}}{p_{1}^{+}+p_{3}^{+}}\mathbf{x}_{1}-\frac{p_{3}^{+}}{p_{1}^{+}+p_{3}^{+}}\mathbf{x}_{3}\big)\\
 & \times A^{\bar{\eta}}(\mathbf{x}_{13})\mathcal{A}^{\bar{\lambda}}\big(\mathbf{v}-\mathbf{x}_{2},\mathbf{x}_{31},\frac{p_{1}^{+}p_{3}^{+}q^{+}}{p_{2}^{+}(p_{1}^{+}+p_{3}^{+})^{2}}\big)\Big[U_{\mathbf{x}_{1}}U_{\mathbf{x}_{3}}^{\dagger}t^{d}U_{\mathbf{x}_{3}}U_{\mathbf{x}_{2}}^{\dagger}-t^{d}\Big]\;,
\end{aligned}
\label{eq:QSWfinal}
\end{equation}
with the Dirac structure:
\begin{equation}
\begin{aligned}\mathrm{Dirac}_{\mathrm{QSW}}^{\bar{\eta}\bar{\lambda}}& =\bar{u}_{G}(p_{1}^{+})\Big[\Big(1+2\frac{p_{1}^{+}}{p_{3}^{+}}\Big)\delta^{\eta\bar{\eta}}-i\sigma^{\eta\bar{\eta}}\Big]\\&\times\Big[\Big(1-2\frac{p_{1}^{+}+p_{3}^{+}}{q^{+}}\Big)\delta^{\lambda\bar{\lambda}}-i\sigma^{\lambda\bar{\lambda}}\Big]\mathbf{\gamma}^{+}v_{G}(p_{2}^{+})
\;.
\end{aligned}
\label{eq:DiracQSW}
\end{equation}

\paragraph{QFS: Gluon emitted from the quark in final state}
The amplitude corresponding to final-state gluon emission from the
quark (fig.~\ref{fig:realdiagrams}) reads:
\begin{equation}
\begin{aligned}
\mathcal{M}_{\mathrm{QFS}} & =ig_{e}e_fg_{s}p_{3}^{+}\mathrm{Dirac}_{\mathrm{QSW}}^{\bar{\eta}\bar{\lambda}}\int_{\mathrm{\mathbf{x}}_{1},\mathbf{x}_{2}}e^{-i(\mathbf{p}_{1}+\mathbf{p}_{3})\cdot\mathbf{x}_{1}}e^{-i\mathbf{p}_{2}\cdot\mathbf{x}_{2}}\\
 & \times A^{\bar{\lambda}}(\mathbf{x}_{12})\frac{(p_{3}^{+}\mathbf{p}_{1}-p_{1}^{+}\mathbf{p}_{3})^{\bar{\eta}}}{\left(p_{3}^{+}\mathbf{p}_{1}-p_{1}^{+}\mathbf{p}_{3}\right)^{2}}\Big[t^{d}U_{\mathbf{x}_{1}}U_{\mathbf{x}_{2}}^{\dagger}-t^{d}\Big]\;,
\end{aligned}
\label{eq:QFS1}
\end{equation}
which can be written as:
\begin{equation}
\begin{aligned}
\mathcal{M}_{\mathrm{QFS}} & =g_{e}e_fg_{s}\frac{p_{3}^{+}}{p_{1}^{+}+p_{3}^{+}}\mathrm{Dirac}_{\mathrm{QSW}}^{\bar{\eta}\bar{\lambda}}\\ & \times\int_{\mathrm{\mathbf{x}}_{1},\mathbf{x}_{2},\mathbf{x}_{3},\mathbf{v}}e^{-i\mathbf{p}_{1}\cdot\mathbf{x}_{1}}e^{-i\mathbf{p}_{2}\cdot\mathbf{x}_{2}}e^{-i\mathbf{p}_{3}\cdot\mathbf{x}_{3}}
\delta^{(D-2)}\big(\mathbf{v}-\frac{p_{1}^{+}}{p_{1}^{+}+p_{3}^{+}}\mathbf{x}_{1}-\frac{p_{3}^{+}}{p_{1}^{+}+p_{3}^{+}}\mathbf{x}_{3}\big)\\&\times A^{\bar{\lambda}}(\mathbf{v}-\mathbf{x}_{2})A^{\bar{\eta}}(\mathbf{x}_{13})\Big[t^{d}U_{\mathbf{v}}U_{\mathbf{x}_{2}}^{\dagger}-t^{d}\Big]\;.
\end{aligned}
\label{eq:QFSfinal}
\end{equation}

\paragraph{RI: Real gluon created instantaneously before the shockwave}
Finally, the real gluon can be radiated instantaneously from the photon
in addition to the quark-antiquark pair (fig.~\ref{fig:realdiagrams}),
yielding the amplitude:
\begin{equation}
\begin{aligned}
\mathcal{M}_{\mathrm{RI}} & =g_{e}e_fg_{s}\frac{p_{1}^{+}p_{3}^{+}(p_{1}^{+}+p_{3}^{+})}{(q^{+})^{3}}\mathrm{Dirac}_{\mathrm{RI}}\\
 & \times\int_{\mathrm{\mathbf{x}}_{1},\mathbf{x}_{2},\mathbf{x}_{3}}e^{-i\mathbf{p}_{1}\cdot\mathbf{x}_{1}}e^{-i\mathbf{p}_{2}\cdot\mathbf{x}_{2}}e^{-i\mathbf{p}_{3}\cdot\mathbf{x}_{3}}\mathcal{C}\big(\frac{p_{1}^{+}}{q^{+}}\mathbf{x}_{21}+\frac{p_{3}^{+}}{q^{+}}\mathbf{x}_{23},\mathbf{x}_{31},\frac{p_{1}^{+}p_{3}^{+}}{q^{+}p_{2}^{+}}\big)\\
 & \times\big[U_{\mathbf{x}_{1}}t^{c}U_{\mathbf{x}_{2}}^{\dagger}W_{\mathbf{x}_{3}}^{cd}-t^{d}\big]\;,
\end{aligned}
\end{equation}
where
\begin{equation}
\begin{aligned}&\mathrm{Dirac}_{\mathrm{RI}}\\& =\bar{u}_{G}(p_{1}^{+})\Bigg[\frac{q^{+}(p_{1}^{+}-p_{2}^{+})}{(p_{1}^{+}+p_{3}^{+})(p_{2}^{+}+p_{3}^{+})}\delta^{\lambda\eta}+\frac{q^{+}(p_{1}^{+}+p_{2}^{+}+2p_{3}^{+})}{(p_{1}^{+}+p_{3}^{+})(p_{2}^{+}+p_{3}^{+})}i\sigma^{\lambda\eta}\Bigg]\gamma^{+}v_{G}(p_{2}^{+})
\;.
\end{aligned}
\end{equation}

\section{\label{sec:UV}UV safety}
The diagrams $\mathrm{SESW}$ and $\mathrm{\overline{SE}SW}$, i.e.
the quark- and antiquark self-energy corrections where the gluon crosses
the shockwave, are UV divergent and need to be regularized by adding
a counterterm, as we demonstrated in eqs.~\eqref{eq:SESWunsub}-\eqref{eq:SESWsub}.
These are the only diagrams which we calculated that exhibit a UV
singularity. There are, however, virtual contributions that we did
not explicitly compute but whose UV poles will cancel with the one
in our counterterm, resulting in a UV-finite total NLO amplitude.
This is what we will now demonstrate.

First, comparing expression~\eqref{eq:SESWUV} for the UV counterterm
with the LO amplitude~\eqref{eq:LOfinal}, we can write:
\begin{equation}
\begin{aligned}
\mathcal{M}_{\mathrm{LO}}+\mathcal{M}_{\mathrm{SESW,UV}} & =-ig_{e}e_f\mathrm{Dirac}_{\mathrm{LO}}^{\bar{\lambda}}\int_{\mathrm{\mathbf{x}}_{1},\mathbf{x}_{2}}e^{-i\mathbf{p}_{1}\cdot\mathbf{x}_{1}}e^{-i\mathbf{p}_{2}\cdot\mathbf{x}_{2}}A^{\bar{\lambda}}(\mathbf{x}_{12})\\
 & \times\Big(1+\frac{\alpha_{s}C_{F}}{2\pi}\mathcal{V}_{\mathrm{SESW,UV}}\Big)\Big[U_{\mathbf{x}_{1}}U_{\mathbf{x}_{2}}^{\dagger}-1\Big]\;,\\
 &=\mathcal{M}_{\mathrm{LO}}\Big(1+\frac{\alpha_{s}C_{F}}{2\pi}\mathcal{V}_{\mathrm{SESW,UV}}\Big)\;,
\end{aligned}
\label{eq:LO+SESWUV}
\end{equation}
where, in the last line, for future convenience we introduced a slight abuse of notation since the factorization of the term $\mathcal{V}_{\mathrm{SESW,UV}}$ is really on the integrand level.
This term reads:
\begin{equation}
\begin{aligned}
\mathcal{V}_{\mathrm{SESW,UV}} & =2\pi\int_{k_{\mathrm{min}}^{+}}^{p_{1}^{+}}\mathrm{d}k_{3}^{+}\frac{k_{3}^{+}}{(p_{1}^{+})^{2}}\Big[\Big(2\frac{p_{1}^{+}}{k_{3}^{+}}-1\Big)^{2}+(D-3)\Big]\\&\times\int_{\mathbf{x}_{3}}A^{\bar{\eta}}(\mathbf{x}_{13})\Big[A^{\bar{\eta}}(\mathbf{x}_{13})-A^{\bar{\eta}}(\mathbf{x}_{23})\Big]\;,\\
 & =-E_{\mathrm{gluon}}(p_{1}^{+},k_{\mathrm{min}}^{+})\frac{\Gamma\big(\frac{D}{2}-1\big)}{2\pi^{D/2-2}}\frac{1}{\big(\frac{D}{2}-2\big)}\frac{\mu^{4-D}}{(\mathbf{x}_{12}^{2})^{\frac{D}{2}-2}}\;.
\end{aligned}
\label{eq:VSESWUV}
\end{equation}
To arrive at the above expression, we made use of eqs.~\eqref{eq:zero}
and~\eqref{eq:IRcounter} and introduced the following phase space
integral over the gluon $+$ momentum:
\begin{equation}
\begin{aligned}
E_{\mathrm{gluon}}(p_{1}^{+},k_{\mathrm{min}}^{+}) & =\int_{k_{\mathrm{min}}^{+}}^{p_{1}^{+}}\mathrm{d}k_{3}^{+}\frac{k_{3}^{+}}{(p_{1}^{+})^{2}}\Big[\Big(2\frac{p_{1}^{+}}{k_{3}^{+}}-1\Big)^{2}+(D-3)\Big]
\;,
\end{aligned}
\label{eq:gluonenergyint}
\end{equation}
where $k_{\mathrm{min}}^{+}$ is a regulator which will be further
specified in subsection (\ref{sec:kine}). Writing $D=4-2\epsilon_{\mathrm{UV}}$,
one can calculate further to obtain:
\begin{equation}
\begin{aligned}
\frac{\Gamma\big(\frac{D}{2}-1\big)}{2\pi^{D/2-2}}\frac{1}{\big(\frac{D}{2}-2\big)}\frac{1}{(\mu^{2}\mathbf{x}_{12}^{2})^{\frac{D}{2}-2}} & =\frac{-1}{2}\Big[\frac{1}{\epsilon_{\mathrm{UV}}}+\gamma_{E}+\mathrm{ln}(\pi\mu^{2}\mathbf{x}_{12}^{2})\Big]+\mathcal{O}(\epsilon_{\mathrm{UV}})
\;,
\end{aligned}
\end{equation}
and:
\begin{equation}
\begin{aligned}
E_{\mathrm{gluon}}(p_{1}^{+},k_{\mathrm{min}}^{+}) & =-4\Big(\ln\frac{k_{\mathrm{min}}^{+}}{p_{1}^{+}}+\frac{3+\epsilon_{\mathrm{UV}}}{4}\Big)
\;,
\end{aligned}
\end{equation}
such that:
\begin{equation}
\begin{aligned}
\mathcal{V}_{\mathrm{SESW,UV}} & =-2\Big[\frac{1}{\epsilon_{\mathrm{UV}}}+\gamma_{E}+\mathrm{ln}(\pi\mu^{2}\mathbf{x}_{12}^{2})\Big]\Big(\ln\frac{k_{\mathrm{min}}^{+}}{p_{1}^{+}}+\frac{3}{4}\Big)-\frac{1}{2}
\;.
\end{aligned}
\end{equation}
Likewise,  the contribution from the antiquark self-energy loop scattering
the shockwave gives:
\begin{equation}
\begin{aligned}
\mathcal{V}_{\mathrm{\mathrm{\overline{SE}SW,UV}}} & =-2\Big[\frac{1}{\epsilon_{\mathrm{UV}}}+\gamma_{E}+\mathrm{ln}(\pi\mu^{2}\mathbf{x}_{12}^{2})\Big]\Big(\ln\frac{k_{\mathrm{min}}^{+}}{p_{2}^{+}}+\frac{3}{4}\Big)-\frac{1}{2}
\;,
\end{aligned}
\label{eq:VSEbarSWUV}
\end{equation}
hence in total:
\begin{equation}
\begin{aligned}
\mathcal{V}_{\mathrm{UV}} & =\mathcal{V}_{\mathrm{\mathrm{SESW,UV}}}+\mathcal{V}_{\mathrm{\mathrm{\overline{SE}SW,UV}}}\\
 & =-2\Big[\frac{1}{\epsilon_{\mathrm{UV}}}+\gamma_{E}+\mathrm{ln}(\pi\mu^{2}\mathbf{x}_{12}^{2})\Big]\Bigg[\frac{3}{2}+\mathrm{ln}\frac{k_{\mathrm{min}}^{+}}{p_{1}^{+}}+\mathrm{ln}\frac{k_{\mathrm{min}}^{+}}{p_{2}^{+}}\Bigg]-1\;.
\end{aligned}
\label{eq:VUV}
\end{equation}

In ref.~\cite{Beuf:2016wdz} (see eq.~144), it was demonstrated
that the loop corrections to the initial state, which we did not explicitly
calculate in this work, can be cast into a similar form factor $\mathcal{V}_{\mathrm{IS}}$
as the one that appears in (\ref{eq:LO+SESWUV}):
\begin{equation}
\begin{aligned}
\mathcal{V}_{\mathrm{IS}} & =\Big[\frac{1}{\epsilon_{\mathrm{UV}}}+\gamma_{E}+\mathrm{ln}(\pi\mu^{2}\mathbf{x}_{12}^{2})\Big]\Big[\frac{3}{2}+\mathrm{ln}\frac{k_{\mathrm{min}}^{+}}{p_{1}^{+}}+\mathrm{ln}\frac{k_{\mathrm{min}}^{+}}{p_{2}^{+}}\Big]+\frac{1}{2}\mathrm{ln}^{2}\frac{p_{1}^{+}}{p_{2}^{+}}-\frac{\pi^{2}}{6}+3
\;.
\end{aligned}
\label{eq:VIS}
\end{equation}
Adding eqs.~\eqref{eq:VUV} and~\eqref{eq:VIS} gives:
\begin{equation}
\begin{aligned}
\mathcal{V}_{\mathrm{UV}}+\mathcal{V}_{\mathrm{IS}}  &=-\Big[\frac{1}{\epsilon_{\mathrm{UV}}}+\gamma_{E}+\mathrm{ln}(\pi\mu^{2}\mathbf{x}_{12}^{2})\Big]\Bigg[\frac{3}{2}+\mathrm{ln}\frac{k_{\mathrm{min}}^{+}}{p_{1}^{+}}+\mathrm{ln}\frac{k_{\mathrm{min}}^{+}}{p_{2}^{+}}\Bigg]\\&+\frac{1}{2}\mathrm{ln}^{2}\frac{p_{1}^{+}}{p_{2}^{+}}-\frac{\pi^{2}}{6}+2\;.
\end{aligned}
\label{eq:VUVVIS}
\end{equation}
Clearly, adding the initial-state loop corrections is not yet sufficient
to cancel the UV poles from the SESW counterterms, and one needs another
UV-divergent contribution. This final ingredient is provided by the
self-energy corrections to the quark and antiquark in the final state.
We did not compute these diagrams explicitly, since they are simply
zero. This is because in dimensional regularization the UV and IR
contributions to a scaleless integral --such as self-energy loops
on an asymptotic massless state-- exactly cancel (cfr. eq.~\eqref{eq:zero}).
This property can, however, be exploited to construct the missing
piece for the UV cancellation in~\eqref{eq:VUVVIS}, writing the
total contribution of the asymptotic (anti)quark leg corrections as
$\mathcal{V}_{\mathrm{FS}}=\mathcal{V}_{\mathrm{FSUV}}+\mathcal{V}_{\mathrm{FSIR}}=0$
with:
\begin{equation}
\begin{aligned}
\mathcal{V}_{\mathrm{FSUV}} & =\Big[\frac{1}{\epsilon_{\mathrm{UV}}}+\gamma_{E}+\mathrm{ln}(\pi\mu^{2}\mathbf{x}_{12}^{2})\Big]\Big[\frac{3}{2}+\mathrm{ln}\frac{k_{\mathrm{min}}^{+}}{p_{1}^{+}}+\mathrm{ln}\frac{k_{\mathrm{min}}^{+}}{p_{2}^{+}}\Big]\;,\\
\mathcal{V}_{\mathrm{FSIR}} & =-\Big[\frac{1}{\epsilon_{\mathrm{IR}}}+\gamma_{E}+\mathrm{ln}(\pi\mu^{2}\mathbf{x}_{12}^{2})\Big]\Big[\frac{3}{2}+\mathrm{ln}\frac{k_{\mathrm{min}}^{+}}{p_{1}^{+}}+\mathrm{ln}\frac{k_{\mathrm{min}}^{+}}{p_{2}^{+}}\Big]\;,
\end{aligned}
\label{eq:VFSIR}
\end{equation}
where we added subscripts to distinguish between positive infinitesimal numbers parameterizing the UV resp. IR pole.

Therefore,  the sum of the leading-order diagram~\eqref{eq:LO}, the initial state
loop corrections~\eqref{eq:VIS}, the UV counterterms from the $\mathrm{SESW}$ and
$\mathrm{\overline{SE}SW}$ diagrams~\eqref{eq:VUV}, and the UV-divergent part of
the final-state corrections~\eqref{eq:VFSIR}, is UV-finite:
\begin{equation}
\begin{aligned}
\mathcal{M}_{\mathrm{LO}+\mathrm{IS}+\mathrm{UV}+\mathrm{FSUV}} & \equiv\mathcal{M}_{\mathrm{LO}}+\mathcal{M}_{\mathrm{SESW,UV}}+\mathcal{M}_{\mathrm{\overline{SE}SW,UV}}+\mathcal{M}_{\mathrm{IS}}+\mathcal{M}_{\mathrm{FSUV}}\\
 & =\mathcal{M}_{\mathrm{LO}} \Big(1+\frac{\alpha_{s}C_{F}}{2\pi}\mathcal{V}_{\mathrm{UV}}+\frac{\alpha_{s}C_{F}}{2\pi}\mathcal{V}_{\mathrm{IS}}+\frac{\alpha_{s}C_{F}}{2\pi}\mathcal{V}_{\mathrm{FSUV}}\Big)\;,\\
 & =\mathcal{M}_{\mathrm{LO}} \Big(1+\frac{\alpha_{s}C_{F}}{2\pi}\big(\frac{1}{2}\mathrm{ln}^{2}\frac{p_{1}^{+}}{p_{2}^{+}}-\frac{\pi^{2}}{6}+2\big)\Big)\;.
\end{aligned}
\label{eq:UVfiniteamplitude}
\end{equation}

After this procedure, what is left from the cancellation of UV divergencies is the finite term above, as well as a new contribution to the virtual
diagrams which stems from the IR-divergent part of the asymptotic leg
corrections~\eqref{eq:VFSIR}:
\begin{equation}
\begin{aligned}
\mathcal{M}_{\mathrm{FSIR}} & =\mathcal{M}_{\mathrm{LO}}\,\frac{\alpha_{s}C_{F}}{2\pi}\mathcal{V}_{\mathrm{FSIR}}
\;,
\end{aligned}
\label{eq:virtualIR}
\end{equation}
which ultimately will cancel with the IR poles from the real
NLO corrections (see section~\ref{sec:coll}).

\section{\label{sec:soft}Soft safety in gluon exchange and interferences}

A second type of infinities that appear in our calculation are soft
divergences, which show up when the gluon momentum $\vec{k}_{3}=(k_{3}^{+},\mathbf{k}_{3})\to0$.
They are distinct from the so-called rapidity divergences associated with $k_{3}^{+}\to0$
but where $\mathbf{k}_{3}$ stays finite. The latter will be regulated with a cutoff $k^+_{\textrm{min}}$
and the remaining logarithms resummed by the JIMWLK equation (see section~\ref{sec:JIMWLK}).

\subsection{Virtual contributions}

When computing the virtual diagrams in section~\ref{subsec:virtual},
we have performed the integration over the transverse loop momentum
$\mathbf{k}_{3}$. Therefore, the information on possible soft singularities
has been lost. To investigate whether a diagram contains a soft divergence,
one needs to rescale the gluon transverse momentum with its $+$ momentum:
$\mathbf{k}_{3}\to\tilde{\mathbf{k}}_{3}=(k_{3}^{+}/p_{1}^{+})\mathbf{k}_{3}$
and then take the $k_{3}^{+}\to0$ limit. The only two virtual diagrams
with a soft singularity are the ones with a gluon exchange in the
final state: $\mathrm{GEFS}$ and $\mathrm{IFS}$ (and their $q\leftrightarrow\bar{q}$
counterparts).

Let us start with the $\mathrm{GEFS}$ diagram and perform a rescaling
$\mathbf{K}=\chi\boldsymbol{\ell}/z$ with $\chi=k_{3}^{+}/p_{2}^{+}$
(writing $\bar{\chi}=1-\chi$) in the transverse integral $J$ (\ref{eq:J}):
\begin{equation}
\begin{aligned}
J^{\eta^{\prime}\bar{\eta}}(k_{3}^{+},\mathbf{x}_{12}) & =\frac{\chi^{2}}{z^{2}}\int_{\mathbf{\boldsymbol{\ell}}}e^{-i\chi\boldsymbol{\ell}\cdot\mathbf{x}_{12}}\frac{(\chi/z)\boldsymbol{\ell}^{\eta^{\prime}}}{(\chi/z)^{2}\boldsymbol{\ell}^{2}}\\&\times\frac{(\chi/z)\boldsymbol{\ell}^{\bar{\eta}}-(\chi/z)\mathbf{P}_{\perp}^{\bar{\eta}}}{\big((\chi/z)\boldsymbol{\ell}+(1-\frac{\bar{z}}{z}\chi)\mathbf{P}_{\perp}\big)^{2}-(1+\chi)(1-\frac{\bar{z}}{z}\chi)\mathbf{P}_{\perp}^{2}-i\epsilon}
\;.
\end{aligned}
\end{equation}
Performing a first-order Taylor expansion in denominator, which tends
to zero when $\chi\to0$, one obtains in the limit:
\begin{equation}
\begin{aligned}
\lim_{k^{+}_3\to0}J^{\eta^{\prime}\bar{\eta}}(k_{3}^{+},\mathbf{x}_{12}) & =\frac{k_{3}^{+}q^{+}}{p_{1}^{+}p_{2}^{+}}\int_{\mathbf{\boldsymbol{\ell}}}\frac{\boldsymbol{\ell}^{\eta^{\prime}}}{\boldsymbol{\ell}^{2}}\frac{\boldsymbol{\ell}^{\bar{\eta}}-\mathbf{P}_{\perp}^{\bar{\eta}}}{2\boldsymbol{\ell}\cdot\mathbf{P}_{\perp}-\mathbf{P}_{\perp}^{2}-i\epsilon}
\;.
\end{aligned}
\label{eq:Jsoft}
\end{equation}
The only other source of powers $1/k_{3}^{+}$ in $\mathcal{M}_{\mathrm{GEFS}}$
(\ref{eq:GEFS}) is the Dirac structure, which needs to scale as $1/(k_{3}^{+})^{2}$
in order to combine with the above integral to give a singularity.
We can therefore pick up only the simple contribution $\mathrm{Dirac}_{\bar{q}\to q,(ii)}^{\bar{\lambda}\bar{\eta}\eta^{\prime}}$
to the full Dirac structure (\ref{eq:Diracqbar2q(ii)}), which gives:
\begin{equation}
\begin{aligned}
\lim_{\mathrm{soft}}\mathcal{M}_{\mathrm{GEFS}} & =-i\frac{g_{e}e_fg_{s}^{2}}{\pi}\int_{0}^{p_{1}^{+}}\frac{\mathrm{d}k_{3}^{+}}{(k_{3}^{+})^{2}}\frac{p_{1}^{+}p_{2}^{+}}{q^{+}}\mathrm{Dirac}_{\mathrm{\overline{LO}}}^{\bar{\lambda}}\\
 & \times\int_{\mathrm{\mathbf{x}}_{1},\mathbf{x}_{2}}e^{-i\mathbf{p}_{1}\cdot\mathbf{x}_{1}}e^{-i\mathbf{p}_{2}\cdot\mathbf{x}_{2}}A^{\bar{\lambda}}(\mathbf{x}_{12})\lim_{k^{+}_3\to0}J^{\eta^{\prime}\eta^{\prime}}(k_{3}^{+},\mathbf{x}_{12})\\
 & \times\big[t^{c}U_{\mathbf{x}_{1}}U_{\mathbf{x}_{2}}^{\dagger}t^{c}-C_{F}\big]\;.
\end{aligned}
\label{eq:GEFSsoft}
\end{equation}
In the above expression, $\mathrm{Dirac}_{\mathrm{\overline{LO}}}^{\bar{\lambda}}$
is the $q\leftrightarrow\bar{q}$ conjugate of the leading-order Dirac
structure eq.~\eqref{eq:DiracLO}), obtained by interchanging $1\leftrightarrow2$
and taking the complex conjugate of the structure between the spinors
(cfr. the algorithm in the beginning of sec.~\ref{sec:NLO}):
\begin{equation}
\begin{aligned}
\mathrm{Dirac}_{\mathrm{\overline{LO}}}^{\bar{\lambda}} & \equiv\bar{u}_{G}^{s_{1}}(p_{1}^{+})\gamma^{+}\bigl[(1-2\bar{z})\delta^{\lambda\bar{\lambda}}+i\sigma^{\lambda\bar{\lambda}}\bigr]v_{G}^{s_{2}}(p_{2}^{+})
\;.
\end{aligned}
\label{eq:DiracLObar}
\end{equation}
Combined with the delta function from the Dirac structure, the soft
limit of the integral $J$ in (\ref{eq:Jsoft}) is thus reduced to
two integrals which both can be analytically solved in dimensional
regularization:
\begin{equation}
\begin{aligned}
I_{1} & =\int_{\boldsymbol{\ell}}\frac{-1}{\Big(-2\boldsymbol{\ell}\cdot\mathbf{P}_{\perp}+\mathbf{P}_{\perp}^{2}\Big)+i\epsilon}\;,\\
I_{2} & =\int_{\boldsymbol{\ell}}\frac{1}{\boldsymbol{\ell}^{2}}\frac{\boldsymbol{\ell}\cdot\mathbf{P}_{\perp}}{\Big(-2\boldsymbol{\ell}\cdot\mathbf{P}_{\perp}+\mathbf{P_{\perp}}^{2}\Big)+i\epsilon}\;.
\end{aligned}
\end{equation}
Let us start with the integral $I_{2}$. Applying the \textquoteleft real'
resp. \textquoteleft complex' Schwinger trick to the first and the
second denominator:
\begin{equation}
\begin{aligned}
\int_{0}^{+\infty}\mathrm{d}\alpha\,e^{-\alpha\Delta} & =\frac{1}{\Delta},\qquad\int_{0}^{+\infty}\mathrm{d}\beta\,e^{i\beta\Delta}=\frac{i}{\Delta+i\epsilon}
\;,
\end{aligned}
\label{eq:Schwingertricks}
\end{equation}
we obtain
\begin{equation}
\begin{aligned}
I_{2} & =i\int_{\boldsymbol{\ell}}\int_{0}^{+\infty}\mathrm{d}\alpha\mathrm{d}\beta\,\boldsymbol{\ell}\cdot\mathbf{P}_{\perp}e^{-\alpha\boldsymbol{\ell}^{2}}e^{i\beta\big(-2\mathbf{P}_{\perp}\cdot\boldsymbol{\ell}+\mathbf{P}_{\perp}^{2}\big)}\\
 & \overset{\boldsymbol{\ell}\to\boldsymbol{\ell}+i\frac{\beta}{\alpha}\mathbf{P}_{\perp}}{=}i\int_{\boldsymbol{\ell}}\int_{0}^{+\infty}\mathrm{d}\alpha\mathrm{d}\beta\,\big(\boldsymbol{\ell}-i\frac{\beta}{\alpha}\mathbf{P}_{\perp}\big)\cdot\mathbf{P}_{\perp}e^{-\alpha\boldsymbol{\ell}^{2}}e^{-\frac{\beta}{\alpha}\big(\beta-i\alpha\big)\mathbf{P}_{\perp}^{2}}\\
 & =\frac{\mu^{4-D}\mathbf{P}_{\perp}^{2}}{(4\pi)^{\frac{D-2}{2}}}\int_{0}^{+\infty}\mathrm{d}\alpha\mathrm{d}\beta\,\frac{\beta}{\alpha^{D/2}}e^{-\frac{\beta}{\alpha}\big(\beta-i\alpha\big)\mathbf{P}_{\perp}^{2}}\;.
\end{aligned}
\end{equation}
Evaluating the integral over $\alpha$, one finds:
\begin{equation}
\begin{aligned}
I_{2} & =\frac{\mu^{4-D}(\mathbf{P}_{\perp}^{2})^{2-\frac{D}{2}}}{(4\pi)^{\frac{D-2}{2}}}\Gamma\big(\frac{D}{2}-1\big)\int_{0}^{+\infty}\mathrm{d}\beta\,\beta^{3-D}e^{i\beta\mathbf{P}_{\perp}^{2}}
\;.
\end{aligned}
\end{equation}
Using the integral representation of the gamma function:
\begin{equation}
\begin{aligned}
\frac{\Gamma(\alpha)}{A^{\alpha}} & =\int_{0}^{\infty}\mathrm{d}t\,t^{\alpha-1}e^{-tA}
\;,
\end{aligned}
\end{equation}
we end up with:
\begin{equation}
\begin{aligned}
I_{2} & =\frac{(-i)^{D-4}\mu^{4-D}(\mathbf{P}_{\perp}^{2})^{\frac{D}{2}-2}}{(4\pi)^{\frac{D-2}{2}}}\Gamma\big(\frac{D}{2}-1\big)\Gamma\big(4-D\big)
\;.
\end{aligned}
\end{equation}
Writing $-i=e^{i(-\frac{\pi}{2}+2n\pi)}$ and $D=4-2\epsilon_{\mathrm{IR}}$:
\begin{equation}
\begin{aligned}
I_{2} & =\frac{1}{8\pi}\Big(\frac{1}{\epsilon_{\mathrm{IR}}}-\gamma_{E}-\ln\Big(\frac{\mathbf{P}_{\perp}^{2}}{4\pi\mu^{2}}\Big)-2i\pi(2n-\frac{1}{2})\Big)+\mathcal{O}(\epsilon_{\mathrm{IR}})
\;.
\end{aligned}
\end{equation}
Likewise,  applying (\ref{eq:Schwingertricks}) to the (finite) integral
$I_{1}$ yields:
\begin{equation}
\begin{aligned}
I_{1} & =-i\int_{\boldsymbol{\ell}}\int_{0}^{+\infty}\mathrm{d}\beta\,e^{i\beta\big(-2\mathbf{P}_{\perp}\cdot\boldsymbol{\ell}+\mathbf{P}_{\perp}^{2}\big)}
\;.
\end{aligned}
\end{equation}
The next step is to rewrite the $D-2$ dimensional $\boldsymbol{\ell}$-integration
as follows:
\begin{equation}
\begin{aligned}
\int_{\boldsymbol{\ell}} & =\frac{\mu^{4-D}}{(2\pi)^{D-2}}\int\mathrm{d}\Omega_{D-3}\int\mathrm{d}\cos\theta\int\mathrm{d}\ell\,\ell^{D-3}\;,\\
 & =\frac{\mu^{4-D}}{(2\pi)^{D-2}}\frac{2\pi^{\frac{D-3}{2}}}{\Gamma\big(\frac{D-3}{2}\big)}\int\mathrm{d}\cos\theta\int\mathrm{d}\ell\,\ell^{D-3}\;.
\end{aligned}
\end{equation}
The $I_{1}$ integral can then be evaluated starting with the integration
over $\cos\theta$, followed by the $\beta$ integral:
\begin{equation}
\begin{aligned}
I_{1} & =-2i\frac{\mu^{4-D}}{(4\pi)^{\frac{D-3}{2}}}\frac{1}{\Gamma\big(\frac{D-3}{2}\big)}\int\mathrm{d}\ell\,\ell^{D-3}\int_{0}^{+\infty}\mathrm{d}\beta J_{0}\big(2\beta\ell|\mathbf{P}_{\perp}|\big)e^{i\beta\mathbf{P}_{\perp}^{2}}\;,\\
 & =\frac{2\mu^{4-D}}{(4\pi)^{\frac{D-3}{2}}}\frac{1}{\Gamma\big(\frac{D-3}{2}\big)}\frac{1}{\mathbf{P}_{\perp}^{2}}\int\mathrm{d}\ell\,\ell^{D-3}\frac{1}{\sqrt{1-4\ell^{2}/\mathbf{P}_{\perp}^{2}}}\;,\\
 & =\frac{2\mu^{4-D}}{(4\pi)^{\frac{D-3}{2}}}\frac{1}{\Gamma\big(\frac{D-3}{2}\big)}\frac{1}{\mathbf{P}_{\perp}^{2}}\frac{i2^{1-D}}{\sqrt{\pi}}\Big(-\frac{1}{\mathbf{P}_{\perp}^{2}}\Big)^{1-D/2}\Gamma\big(\frac{3-D}{2}\big)\Gamma\big(\frac{D-2}{2}\big)\overset{D\to4}{=}\frac{i}{4\pi}\;.
\end{aligned}
\end{equation}
We finally obtain:
\begin{equation}
\begin{aligned}
\lim_{k^{+}_3\to0}J^{\eta^{\prime}\bar{\eta}}(k_{3}^{+},\mathbf{x}_{12}) & =\frac{k_{3}^{+}q^{+}}{p_{1}^{+}p_{2}^{+}}\frac{1}{8\pi}\Big(\frac{1}{\epsilon_{\mathrm{IR}}}-\gamma_{E}-\ln\Big(\frac{\mathbf{P}_{\perp}^{2}}{4\pi\mu^{2}}\Big)-2i\pi(2n-\frac{1}{2})+2i\Big)
\;.
\end{aligned}
\label{eq:Jsoftevaluated}
\end{equation}

The soft limit of diagram $\mathrm{IFS}$ can be extracted in a similar
way.   Rescaling $\mathbf{k}_{3}=\frac{k_{3}^{+}q^{+}}{p_{1}^{+}p_{2}^{+}}\boldsymbol{\ell}-\frac{k_{3}^{+}}{p_{1}^{+}}\mathbf{P}_{\perp}-\frac{k_{3}^{+}}{q^{+}}\mathbf{k}_{\perp}$
in eq.~\eqref{eq:IFS}):
\begin{equation}
\begin{aligned}
& \int_{\mathbf{k}_{3}}\frac{e^{-i\mathbf{k}_{3}\cdot\mathbf{x}_{12}}}{\Big(\mathbf{k}_{3}+\mathbf{P}_{\perp}+\frac{k_{3}^{+}}{q^{+}}\mathbf{k}_{\perp}\Big)^2-\frac{(p_{2}^{+}+k_{3}^{+})(p_{1}^{+}-k_{3}^{+})}{p_{1}^{+}p_{2}^{+}}\mathbf{P}_{\perp}^{2}-i\epsilon}\\
 & =\Big(\frac{k_{3}^{+}q^{+}}{p_{1}^{+}p_{2}^{+}}\Big)^{2}\int_{\boldsymbol{\ell}}\frac{e^{-i\big(\frac{k_{3}^{+}q^{+}}{p_{1}^{+}p_{2}^{+}}\boldsymbol{\ell}-\frac{k_{3}^{+}}{p_{1}^{+}}\mathbf{P}_{\perp}-\frac{k_{3}^{+}}{q^{+}}\mathbf{k}_{\perp}\Big)\cdot\mathbf{x}_{12}}}{\Big(\frac{k_{3}^{+}q^{+}}{p_{1}^{+}p_{2}^{+}}\boldsymbol{\ell}+\bar{\xi}\mathbf{P}_{\perp}\Big)^2-\frac{(p_{2}^{+}+k_{3}^{+})(p_{1}^{+}-k_{3}^{+})}{p_{1}^{+}p_{2}^{+}}\mathbf{P}_{\perp}^{2}-i\epsilon}\;,\\
 & \overset{\lim k_{3}^{+}\to0}{\simeq}\Big(\frac{k_{3}^{+}q^{+}}{p_{1}^{+}p_{2}^{+}}\Big)\int_{\boldsymbol{\ell}}\frac{1}{2\boldsymbol{\ell}\cdot\mathbf{P}_{\perp}-\mathbf{P}_{\perp}^{2}-i\epsilon}=\Big(\frac{k_{3}^{+}q^{+}}{p_{1}^{+}p_{2}^{+}}\Big)I_{1}\;.
\end{aligned}
\end{equation}
Combining this with eqs. (\ref{eq:IFS}), (\ref{eq:GEFSsoft}) and
(\ref{eq:Jsoftevaluated}) yields (writing $\xi=k_{3}^{+}/p_{1}^{+}$):
\begin{equation}
\begin{aligned}
\lim_{\mathrm{soft}}\big(\mathcal{M}_{\mathrm{GEFS}}+\mathcal{M}_{\mathrm{IFS}}\big) & =-i\frac{g_{e}e_fg_{s}^{2}}{\pi}\int\frac{\mathrm{d}\xi}{\xi}\mathrm{Dirac}_{\mathrm{\overline{LO}}}\int_{\mathrm{\mathbf{x}}_{1},\mathbf{x}_{2}}e^{-i\mathbf{p}_{1}\cdot\mathbf{x}_{1}}e^{-i\mathbf{p}_{2}\cdot\mathbf{x}_{2}}A^{\bar{\lambda}}(\mathbf{x}_{12})
 \\&\times\frac{1}{8\pi}\Big(\frac{1}{\epsilon_{\mathrm{IR}}}-\gamma_{E}-\ln\Big(\frac{\mathbf{P}_{\perp}^{2}}{4\pi\mu^{2}}\Big)-2i\pi(2n-\frac{1}{2})\Big)\\
 & \times\big[t^{c}U_{\mathbf{x}_{1}}U_{\mathbf{x}_{2}}^{\dagger}t^{c}-C_{F}\big]\;.
\end{aligned}
\end{equation}
The IR pole will be canceled with certain real NLO corrections, as
will be shown in the next subsection. This cancellation takes place
on the level of the amplitude squared, obtained by multiplying with
$\mathcal{M}_{\mathrm{LO}}^{\dagger}$. Adding as well the $q\leftrightarrow\bar{q}$
diagrams and the complex conjugate (c.c.), we obtain:
\begin{equation}
\begin{aligned}
& \lim_{\mathrm{soft}}\mathcal{M}_{\mathrm{LO}}^{\dagger}\big(\mathcal{M}_{\mathrm{GEFS}}+\mathcal{M}_{\mathrm{IFS}}+\mathcal{M}_{\mathrm{\overline{GE}FS}}+\mathcal{M}_{\mathrm{\overline{I}FS}}\big)+\mathrm{c.c.}\\
 & =64\alpha_{\mathrm{em}} e^2_f \alpha_{s}N_{c}^{2}p_{1}^{+}p_{2}^{+}(z^{2}+\bar{z}^{2})\int\frac{\mathrm{d}\xi}{\xi}\\
 & \times\int_{\mathrm{\mathbf{x}}_{1},\mathbf{x}_{2}}e^{-i\mathbf{p}_{1}\cdot\mathbf{x}_{11^{\prime}}}e^{-i\mathbf{p}_{2}\cdot\mathbf{x}_{22^{\prime}}}A^{\bar{\lambda}}(\mathbf{x}_{12})A^{\bar{\lambda}}(\mathbf{x}_{1^{\prime}2^{\prime}})\times\Big(\frac{1}{\epsilon_{\mathrm{IR}}}-\gamma_{E}-\ln\Big(\frac{\mathbf{P}_{\perp}^{2}}{4\pi\mu^{2}}\Big)\Big)\\
 & \times\Big\langle s_{12}s_{2^{\prime}1^{\prime}}-s_{12}-s_{2^{\prime}1^{\prime}}+1-\frac{1}{N_{c}^{2}}\Big(Q_{122^{\prime}1^{\prime}}-s_{12}-s_{2^{\prime}1^{\prime}}+1\Big)\Big\rangle \;,
\end{aligned}
\label{eq:virtualsoft}
\end{equation}
where we made use of the spinor trace (\ref{eq:DiracLOLO}) with $D=4$,
as well as:
\begin{equation}
\begin{aligned}
\mathrm{Tr}\Big(\mathrm{Dirac}_{\mathrm{LO}}^{\lambda^{\prime}\dagger}\mathrm{Dirac}_{\overline{\mathrm{LO}}}^{\bar{\lambda}}\Big) & =-16p_{1}^{+}p_{2}^{+}\delta^{\bar{\lambda}\lambda^{\prime}}\big(z^{2}+\bar{z}^{2}\big)
\;.
\end{aligned}
\label{eq:DiracLOLObar}
\end{equation}

\subsection{Real contributions}

The IR pole found in the previous subsection stems from a soft virtual
gluon exchange in the final state. It will cancel with real contributions
that have the same topology, i.e. a soft final-state gluon radiated
from the quark in the amplitude and from the antiquark in the complex
conjugate amplitude, or vice versa. The corresponding contribution
to the cross section is:
\begin{equation}
\begin{aligned}
&\int\mathrm{PS}(\vec{p}_3)\mathcal{M}_{\mathrm{\overline{Q}FS}}^{\dagger}\mathcal{M}_{\mathrm{QFS}}  =\int\frac{\mathrm{d}p_{3}^{+}}{4\pi p_{3}^{+}}g_{e}e_f^{2}g_{s}^{2}(p_{3}^{+})^{2}\frac{N_{c}^{2}}{2}\mathrm{Tr}\Big(\mathrm{Dirac}_{\mathrm{\overline{Q}SW}}^{\eta^{\prime}\lambda^{\prime}\dagger}\mathrm{Dirac}_{\mathrm{QSW}}^{\bar{\eta}\bar{\lambda}}\Big)\\
 & \times\int_{\mathrm{\mathbf{x}}_{1}^{\prime},\mathbf{x}_{2^{\prime}},\mathrm{\mathbf{x}}_{1},\mathbf{x}_{2}}e^{-i\mathbf{p}_{1}\cdot\mathbf{x}_{11^{\prime}}}e^{-i\mathbf{p}_{2}\cdot\mathbf{x}_{22^{\prime}}}e^{-i\mathbf{p}_{3}\cdot(\mathbf{x}_{1}-\mathbf{x}_{2^{\prime}})}A^{\lambda^{\prime}}(\mathbf{x}_{1^{\prime}2^{\prime}})A^{\bar{\lambda}}(\mathbf{x}_{12})\\
 & \times\int_{\mathbf{p}_{3}}\frac{(p_{3}^{+}\mathbf{p}_{1}-p_{1}^{+}\mathbf{p}_{3})^{\bar{\eta}}}{\left(p_{3}^{+}\mathbf{p}_{1}-p_{1}^{+}\mathbf{p}_{3}\right)^{2}}\frac{(p_{3}^{+}\mathbf{p}_{2}-p_{2}^{+}\mathbf{p}_{3})^{\eta^{\prime}}}{\left(p_{3}^{+}\mathbf{p}_{2}-p_{2}^{+}\mathbf{p}_{3}\right)^{2}}\\
 & \times\Big\langle s_{12}s_{2^{\prime}1^{\prime}}-s_{12}-s_{2^{\prime}1^{\prime}}+1-\frac{1}{N_{c}^{2}}\Big(Q_{122^{\prime}1^{\prime}}-s_{12}-s_{2^{\prime}1^{\prime}}+1\Big)\Big\rangle\;.
\end{aligned}
\end{equation}
Introducing again $\mathbf{p}_{3}=\frac{p_{3}^{+}}{p_{1}^{+}}\boldsymbol{\ell}$
and taking the $p_{3}^{+}\to0$ limit, the above equation becomes:
\begin{equation}
\begin{aligned}
&\int\mathrm{PS}(\vec{p}_3)\mathcal{M}_{\mathrm{\overline{Q}FS}}^{\dagger}\mathcal{M}_{\mathrm{QFS}}  =64(2\pi)\alpha_{\mathrm{em}} e^2_f \alpha_{s}N_{c}^{2}p_{1}^{+}(p_{2}^{+})^{2}(z^{2}+\bar{z}^{2})\int\frac{\mathrm{d}\xi}{\xi}\\
 & \times\int_{\mathrm{\mathbf{x}}_{1}^{\prime},\mathbf{x}_{2^{\prime}},\mathrm{\mathbf{x}}_{1},\mathbf{x}_{2}}e^{-i\mathbf{p}_{1}\cdot\mathbf{x}_{11^{\prime}}}e^{-i\mathbf{p}_{2}\cdot\mathbf{x}_{22^{\prime}}}A^{\bar{\lambda}}(\mathbf{x}_{1^{\prime}2^{\prime}})A^{\bar{\lambda}}(\mathbf{x}_{12})\int_{\boldsymbol{\ell}}\frac{(\boldsymbol{\ell}-\mathbf{p}_{1})\cdot(p_{1}^{+}\mathbf{p}_{2}-p_{2}^{+}\boldsymbol{\ell})}{(\boldsymbol{\ell}-\mathbf{p}_{1})^{2}\left(p_{1}^{+}\mathbf{p}_{2}-p_{2}^{+}\boldsymbol{\ell}\right)^{2}}\\
 & \times\Big\langle s_{12}s_{2^{\prime}1^{\prime}}-s_{12}-s_{2^{\prime}1^{\prime}}+1-\frac{1}{N_{c}^{2}}\Big(Q_{122^{\prime}1^{\prime}}-s_{12}-s_{2^{\prime}1^{\prime}}+1\Big)\Big\rangle\;,
\end{aligned}
\label{eq:QbarFSQFSalmostsoft}
\end{equation}
where we wrote $\mathrm{d}p_{3}^{+}/p_{3}^{+}\to\mathrm{d}\xi/\xi$
and extracted the leading-power of the Dirac structure:
\begin{equation}
\begin{aligned}
\lim_{p_{3}^{+}\to0}\mathrm{Dirac}_{\mathrm{QSW}}^{\bar{\eta}\bar{\lambda}} & =2\frac{p_{1}^{+}}{p_{3}^{+}}\delta^{\eta\bar{\eta}}\mathrm{Dirac}_{\mathrm{LO}}^{\bar{\lambda}}+\mathcal{O}\big((p_{3}^{+})^{0}\big)
\;,
\end{aligned}
\label{eq:DiracQSWleadingpower}
\end{equation}
which leads to:
\begin{equation}
\begin{aligned}
\lim_{p_{3}^{+}\to0}\mathrm{Dirac}_{\mathrm{QSW}}^{\eta^{\prime}\lambda^{\prime}\dagger}\mathrm{Dirac}_{\mathrm{\overline{Q}SW}}^{\bar{\eta}\bar{\lambda}} & =-64\Big(\frac{p_{1}^{+}p_{2}^{+}}{p_{3}^{+}}\Big)^{2}(z^{2}+\bar{z}^{2})\delta^{\bar{\lambda}\lambda^{\prime}}\delta^{\eta^{\prime}\bar{\eta}}+\mathcal{O}\big((p_{3}^{+})^{0}\big)
\;.
\end{aligned}
\end{equation}
The integral over transverse momentum can be cast into the following
form (where $\mathbf{P}_{\perp}$ is again the momentum vector defined
in (\ref{eq:P})):
\begin{equation}
\begin{aligned}
\int_{\boldsymbol{\ell}}\frac{(\boldsymbol{\ell}-\mathbf{p}_{1})\cdot(p_{1}^{+}\mathbf{p}_{2}-p_{2}^{+}\boldsymbol{\ell})}{(\boldsymbol{\ell}-\mathbf{p}_{1})^{2}\left(p_{1}^{+}\mathbf{p}_{2}-p_{2}^{+}\boldsymbol{\ell}\right)^{2}} & =-\frac{\bar{z}}{q^+}\int_{\boldsymbol{\ell}}\frac{1}{\left(\mathbf{P}_{\perp}+\bar{z}\boldsymbol{\ell}\right)^{2}}-\frac{1}{q^+}\int_{\boldsymbol{\ell}}\frac{\boldsymbol{\ell}\cdot\mathbf{P}_{\perp}}{\boldsymbol{\ell}^{2}\left(\mathbf{P}_{\perp}+\bar{z}\boldsymbol{\ell}\right)^{2}}\;.
\end{aligned}
\end{equation}
The first integral disappears in dimensional regularization because
it is scaleless.   The second one can be computed by applying the
real Schwinger trick (\ref{eq:Schwingertricks}) twice:
\begin{equation}
\begin{aligned}
&-\int_{\boldsymbol{\ell}}\frac{\boldsymbol{\ell}\cdot\mathbf{P}_{\perp}}{q^{+}\boldsymbol{\ell}^{2}\left(\mathbf{P}_{\perp}+\bar{z}\boldsymbol{\ell}\right)^{2}}  =-\frac{1}{p_{2}^{+}}\int_{\boldsymbol{\ell}}\frac{\boldsymbol{\ell}\cdot\mathbf{p}}{\boldsymbol{\ell}^{2}(\boldsymbol{\ell}+\mathbf{p})^{2}}\\
 & =\frac{1}{p_{2}^{+}}\frac{\mathbf{P}_{\perp}^{2}\mu^{4-D}}{(4\pi)^{\frac{D-2}{2}}}\frac{1}{2}(1+e^{i\pi D})(-1)^{-D/2}(\mathbf{P}_{\perp}^{2})^{D/2-3}\Gamma(5-D)\Gamma(\frac{D}{2}-2)\;,\\
 & =-\frac{1}{p_{2}^{+}}\frac{1}{4\pi}\Big(\frac{1}{\epsilon_{\mathrm{IR}}}-\gamma_{E}+2in\pi-\ln\Big(\frac{\mathbf{P}_{\perp}^{2}}{4\pi\mu^{2}}\Big)\Big)\;,
\end{aligned}
\end{equation}
where we wrote $-1=e^{i(\pi+2n\pi)}$. Combining the above result
with eq.~\eqref{eq:Schwingertricks}) and adding the complex conjugate,
we finally obtain:
\begin{equation}
\begin{aligned}
&\int\mathrm{PS}(\vec{p}_3)\mathcal{M}_{\mathrm{\overline{Q}FS}}^{\dagger}\mathcal{M}_{\mathrm{QFS}}+\mathrm{c.c.}  =-64\alpha_{\mathrm{em}} e^2_f \alpha_{s}N_{c}^{2}p_{1}^{+}p_{2}^{+}(z^{2}+\bar{z}^{2})\int\frac{\mathrm{d}\xi}{\xi}\\
 & \times\int_{\mathrm{\mathbf{x}}_{1}^{\prime},\mathbf{x}_{2^{\prime}},\mathrm{\mathbf{x}}_{1},\mathbf{x}_{2}}e^{-i\mathbf{p}_{1}\cdot\mathbf{x}_{11^{\prime}}}e^{-i\mathbf{p}_{2}\cdot\mathbf{x}_{22^{\prime}}}A^{\bar{\lambda}}(\mathbf{x}_{1^{\prime}2^{\prime}})A^{\bar{\lambda}}(\mathbf{x}_{12})\Big(\frac{1}{\epsilon_{\mathrm{IR}}}-\gamma_{E}-\ln\Big(\frac{\mathbf{P}_{\perp}^{2}}{4\pi\mu^{2}}\Big)\Big)\\
 & \times\Big\langle s_{12}s_{2^{\prime}1^{\prime}}-s_{12}-s_{2^{\prime}1^{\prime}}+1-\frac{1}{N_{c}^{2}}\Big(Q_{122^{\prime}1^{\prime}}-s_{12}-s_{2^{\prime}1^{\prime}}+1\Big)\Big\rangle =-\eqref{eq:virtualsoft}.
\end{aligned}
\label{eq:realsoft}
\end{equation}
The above result is exactly the opposite of the soft limit of the
virtual contributions, eq.~\eqref{eq:virtualsoft}. Therefore, the total
cross section is free from soft divergences from contributions with final state gluon exchange topology (including interferences of real final state  emission from the quark or the antiquark). Additional soft divergences, appearing together with collinear divergences, will be discussed in section~\ref{sec:coll}.

\section{\label{sec:JIMWLK}JIMWLK}

\subsection{\label{sec:kine}Kinematics}

So far, among the virtual NLO corrections to the dijet cross section that we have calculated, many have a logarithmically divergent integral over the $+$ momentum $k_3^+$ of the virtual gluon stemming from the $k_3^+\rightarrow 0$ regime. In some cases, like the ${\mathrm{SESW,UV}}$ contribution~\eqref{eq:SESWUV}  or the  $\mathrm{IS}$ contribution~\eqref{eq:VIS}, in which the transverse loop integration can be performed explicitly, one could have used dimensional regularization to deal with these divergences. But in other cases, like the ${\mathrm{GESW}}$ contribution~\eqref{eq:GESW} or the  ${\mathrm{SESW,sub}}$ contribution~\eqref{eq:SESWsub}, the integration over the transverse position of the gluon $\mathbf{x}_{3}$ cannot be performed explicitly due to the presence of a Wilson line at $\mathbf{x}_{3}$. In such cases, dimensional regularization cannot  handle the $k_3^+\rightarrow 0$. For this reason, we introduce the lower cutoff $k_{\mathrm{min}}^{+}$ to regularize all $k_3^+$ loop integrals. Similarly, one encounters divergences in the real NLO corrections at the cross section level from the $p_3^+\rightarrow 0$ regime in the integration over the real gluon $+$ momentum $p_3^+$ (either inside or outside the measured jets, as we will see in sections~\ref{sec:jet} and~\ref{sec:coll}), which we regularize with the same cutoff $k_{\mathrm{min}}^{+}$.

Part of these  $k_3^+\rightarrow 0$ or $p_3^+\rightarrow 0$ divergences are genuine soft divergences which cancel at the dijet cross section level, as we have seen in the previous section~\ref{sec:soft}. However, others are rapidity divergences which survive in the form of single logs of $k_{\mathrm{min}}^{+}$ after regularization. These large logs of $k_{\mathrm{min}}^{+}$ are high-energy leading logs, which can be extracted from the NLO correction to the cross section and resummed into the LO term, via the JIMWLK evolution of the target-averaged color- or Wilson-line operator, as we will now explain.

In the LO cross section~\eqref{eq:LOXsection}, the target-averaged color operator is unevolved and should not yet include high-energy logarithms. We can, therefore, use the notation:
\begin{equation}
\begin{aligned}
 & \Big\langle Q_{122^{\prime}1^{\prime}}-s_{12}-s_{2^{\prime}1^{\prime}}+1\Big\rangle_0\;.
\end{aligned}
\label{eq:LO_op_IC}
\end{equation}
In the simplest scheme for the JIMWLK evolution, which we will use in most of the present study, JIMWLK is viewed as an evolution equation along the $k^+$ axis in logarithmic scale. In this scheme, one defines
\begin{equation}
\begin{aligned}
\Big\langle Q_{122^{\prime}1^{\prime}}-s_{12}-s_{2^{\prime}1^{\prime}}+1\Big\rangle_{Y_f^+}
&\equiv 
 \Big\langle Q_{122^{\prime}1^{\prime}}-s_{12}-s_{2^{\prime}1^{\prime}}+1\Big\rangle_{\ln(k_f^+/k_{\mathrm{min}}^{+})}
\end{aligned}
\label{eq:LO_op_Yfplus}
\end{equation}
as the same  target-averaged operator but now including the resummation of high-energy leading logs associated with gluons with light-cone momentum $k^+$ between the cutoff $k_{\mathrm{min}}^{+}$ and the factorization scale $k_f^+$, in the notations of ref.~\cite{Beuf:2014uia}. The evolution with the factorization scale $k_f^+$, or equivalently with $Y_f^+$, is given by the JIMWLK equation for the LO operator
\begin{equation}
\begin{aligned}
\partial_{Y_f^+}\Big\langle Q_{122^{\prime}1^{\prime}}-s_{12}-s_{2^{\prime}1^{\prime}}+1\Big\rangle_{Y_f^+}
& =
\Big\langle \hat{H}_{\mathrm{JIMWLK}} \big(Q_{122^{\prime}1^{\prime}}-s_{12}-s_{2^{\prime}1^{\prime}}+1\big)\Big\rangle_{Y_f^+}
\;.
\end{aligned}
\label{eq:JIMWLK_eq}
\end{equation}
A more explicit version of this equation, with the action of the JIMWLK Hamiltonian $\hat{H}_{\mathrm{JIMWLK}}$ fully worked out, can be found in ref.~\cite{Dominguez:2011gc}.
Integrating eq.~\eqref{eq:JIMWLK_eq}, one finds
\begin{equation}
\begin{aligned}
\Big\langle Q_{122^{\prime}1^{\prime}}-s_{12}-s_{2^{\prime}1^{\prime}}+1\Big\rangle_{Y_f^+}&=
\Big\langle Q_{122^{\prime}1^{\prime}}-s_{12}-s_{2^{\prime}1^{\prime}}+1\Big\rangle_{0}\\
&+\int_{0}^{Y_f^+} \!\!\!\! \mathrm{d}Y^+\:
\Big\langle \hat{H}_{\mathrm{JIMWLK}} \big(Q_{122^{\prime}1^{\prime}}-s_{12}-s_{2^{\prime}1^{\prime}}+1\big)\Big\rangle_{Y^+}
\;.
\end{aligned}
\label{eq:JIMWLK_eq_int}
\end{equation}
The JIMWLK Hamiltonian is of order $\alpha_s$. Hence, the dependence of a target-averaged operator on $Y^+$ is an effect suppressed by one extra power of $\alpha_s$ in fixed-order perturbation theory.
Therefore, expanding eq.~\eqref{eq:JIMWLK_eq_int} in powers of $\alpha_s$ we can write
\begin{equation}
\begin{aligned}
&\Big\langle Q_{122^{\prime}1^{\prime}}-s_{12}-s_{2^{\prime}1^{\prime}}+1\Big\rangle_{0}
= 
\Big\langle Q_{122^{\prime}1^{\prime}}-s_{12}-s_{2^{\prime}1^{\prime}}+1\Big\rangle_{\ln(k_f^+/k_{\mathrm{min}}^{+})}
\\
&
\qquad\qquad\qquad\qquad\;-
\ln(k_f^+/k_{\mathrm{min}}^{+})
\Big\langle \hat{H}_{\mathrm{JIMWLK}} \big(Q_{122^{\prime}1^{\prime}}-s_{12}-s_{2^{\prime}1^{\prime}}+1\big)\Big\rangle+{\cal O}(\alpha_s^2)
\; ,
\end{aligned}
\label{eq:JIMWLK_eq_int_fixed_order}
\end{equation}
with the scale unspecified for the operator in the second line, since it is not under control at this perturbative order. Hence, inserting eq.~\eqref{eq:JIMWLK_eq_int_fixed_order} into the LO cross section~\eqref{eq:LOXsection}, one substitutes the unevolved target-averaged operator with its evolved (up to the factorization scale $k_f^+$) version, generating an extra NLO term which involves the JIMWLK Hamiltonian.
This new NLO term will subtract the logarithmic dependence on the cutoff   $k_{\mathrm{min}}^{+}$ found in the NLO cross section due to rapidity divergences at $k_3^+\rightarrow 0$ or $p_3^+\rightarrow 0$.

Writing formally the NLO correction to the dijet cross section found from the fixed-order calculation as
\begin{equation}
\begin{aligned}
\mathrm{d}\sigma_{\textrm{NLO}}
& =
\int_{k_{\mathrm{min}}^{+}}^{+\infty}\frac{\mathrm{d}p_{3}^{+}}{p_{3}^{+}}\mathrm{d}\tilde{\sigma}_{\textrm{NLO}} \;,
\end{aligned}
\end{equation}
to separate the gluon $+$ momentum integral from the rest of the
cross section (with other Heaviside or Dirac delta functions constraining $p_{3}^{+}$ included in $\tilde{\sigma}_{\textrm{NLO}}$), one has:
\begin{equation}
\begin{aligned}
\mathrm{d}\sigma_{\textrm{NLO}}
& =
\int_{k_{\mathrm{min}}^{+}}^{k_{f}^{+}}\frac{\mathrm{d}p_{3}^{+}}{p_{3}^{+}}
\hat{H}_{\mathrm{JIMWLK}}\mathrm{d}\sigma_{\mathrm{LO}}\\
&+\int_{k_{\mathrel{\mathrm{min}}}^{+}}^{+\infty}\frac{\mathrm{d}p_{3}^{+}}{p_{3}^{+}}
\Big[\mathrm{d}\tilde{\sigma}_{\textrm{NLO}} -\theta(k_{f}^{+}-p_{3}^{+})
\hat{H}_{\mathrm{JIMWLK}}\mathrm{d}\sigma_{\mathrm{LO}}\Big]
\;.
\end{aligned}
\label{eq:JIMWLKsubtraction}
\end{equation}
By construction, the first term in eq.~\eqref{eq:JIMWLKsubtraction}, extracted from the total NLO correction, identically cancels the second term from  eq.~\eqref{eq:JIMWLK_eq_int_fixed_order} after substituting the left hand side of eq.~\eqref{eq:JIMWLK_eq_int_fixed_order} into the LO cross section~\eqref{eq:LOXsection}. Then, the statement that rapidity divergences are subtracted and resummed thanks to the JIMWLK evolution is equivalent to saying that in the second term of eq.~\eqref{eq:JIMWLKsubtraction}, the cutoff $k_{\mathrm{min}}^{+}$ can be dropped, thanks to cancelations happening at low $p_3^+$ between the terms in the square bracket.
In the rest of this section, we will check this statement, by studying the $p_3^+\rightarrow 0$ (or $k_3^+\rightarrow 0$) limit for each of the NLO contributions to the cross section.

Finally, we should discuss appropriate values for  $k_f^+$ and $k_{\mathrm{min}}^{+}$ (and thus for $Y_f^+$) in order to resum high-energy logarithms via JIMWLK evolution.
The factorization scale $k_f^+$ should be of the order of the $+$ momenta of the measured jets (or at most $q^+$). The cutoff $k_{\mathrm{min}}^{+}$ represents the typical $+$ momentum scale set by the valence (and other large-$x$) partons inside the target, before evolution. Modeling the target before low-$x$ evolution as a collection of partons carrying a fraction of at least $x_0$ of the target momentum $p_{{\scriptscriptstyle A}}^{-}$ and with a  typical transverse mass $Q_0$, one has:
\begin{equation}
\begin{aligned}
k_{\mathrm{min}}^{+}
& =
\frac{Q_0^2}{2x_0\,  p_{{\scriptscriptstyle A}}^{-}} \;.
\end{aligned}
\label{cutoff_kplusmin_scale_choice}
\end{equation}
Moreover, we have chosen a frame in which the photon momentum $q^{\mu}=(q^{+},0,0)$ lies entirely in the light-cone $+$ direction, and the target nucleus $p_{{\scriptscriptstyle A}}^{\mu}=(p_{{\scriptscriptstyle A}}^{+},p_{{\scriptscriptstyle A}}^{-},0)$ mostly in the
light-cone $-$ direction, up to $p_{{\scriptscriptstyle A}}^{+}=M_{{\scriptscriptstyle A}}^2/2p_{{\scriptscriptstyle A}}^{-}$, where $M_{{\scriptscriptstyle A}}$ is the target mass.
Then, the total energy squared of the collision is
\begin{equation}
\begin{aligned}
s
& =
(q+p_{{\scriptscriptstyle A}})^{2}
= 2q^{+}p_{{\scriptscriptstyle A}}^{-}+ M_{{\scriptscriptstyle A}}^2 \simeq 2q^{+}p_{{\scriptscriptstyle A}}^{-}
\end{aligned}
\end{equation}
at high energy. Hence, one can write the cutoff as
\begin{equation}
\begin{aligned}
k_{\mathrm{min}}^{+}
\simeq &
\frac{q^{+} Q_0^2}{x_0\, s}\;,
\end{aligned}
\label{cutoff_kplusmin_scale_choice_2}
\end{equation}
and the range for JIMWLK evolution as
\begin{equation}
\begin{aligned}
Y_f^+
& =
\ln\left(\frac{k_f^+}{k_{\mathrm{min}}^{+} }\right)
=
\ln\left(\frac{k_f^+\, x_0\, s}{q^{+} Q_0^2}\right)\; .
\end{aligned}
\label{Yplusf_scale_choice}
\end{equation}
For this reason, $Y_f^+$ is considered to be a high-energy logarithm. $x_0$ can be taken to be $0.01$, or at most $0.1$. $Q_0$ should be a scale around the transition between perturbative and non-perturbative QCD, or should be related to the initial saturation scale $Q_{s, 0}$ in the case of a large enough nucleus. However, in practice, $Q_0^2/x_0$ can be treated as a parameter in a BK/JIMWLK global fit, together with the shape of the initial condition for the evolution.

The scheme chosen for the JIMWLK resummation, based on an evolution strictly along the $p^+$ axis, is particularly simple to handle. However, this scheme is not unique and, in fact, neither is it optimal as we will discuss in section~\ref{subsec:Sudakov_and_kc_JIMWLK}, in particular for the study of Sudakov logarithms.

In the rest of this section, we start by considering the virtual NLO amplitudes and studying their
$k_{3}^{+}\to0$ limit. After that, we bring them to the level of
the cross section by multiplying them with the complex conjugate of
the LO amplitude $\mathcal{M}_{\mathrm{LO}}^{\dagger}$, constructing
the total virtual contribution to JIMWLK. For the real NLO amplitudes
the procedure is similar, as it turns out to be easiest to take the
$p_{3}^{+}\to0$ limit at the amplitude level. Interestingly, we find that
the thus obtained \textquoteleft virtual' and \textquoteleft real'
contributions to JIMWLK are separately free of subleading-$N_{c}$
terms. In the end, we demonstrate how in the $k_{3}^{+},p_{3}^{+}\to0$
limit, the cross section corresponds to the JIMWLK evolution equations
applied to the Wilson-line structure
\begin{equation}
Q_{122^{\prime}1^{\prime}}-s_{12}-s_{2^{\prime}1^{\prime}}+1
\end{equation}
of the leading-order result, which confirms the resummation of high-energy logs by JIMWLK into the LO term, as presented in this section.


\subsection{Virtual diagrams}

\paragraph{GEFS+IFS}

In the $k_{3}^{+}\to0$ limit, the subamplitude $\mathcal{M}_{\mathrm{GEFS},(ii)+\mathrm{IFS}}$
becomes:
\begin{equation}
\begin{aligned}\lim_{k_{3}^{+}\to0}\mathcal{M}_{\mathrm{GEFS},(ii)+\mathrm{IFS}} & =i\frac{g_{e}e_fg_{s}^{2}}{\pi}\int_{k_{\mathrm{min}}^{+}}^{k_{f}^{+}}\frac{\mathrm{d}k_{3}^{+}}{k_{3}^{+}}\mathrm{Dirac}_{\overline{\mathrm{LO}}}^{\bar{\lambda}}\times\big[t^{c}U_{\mathbf{x}_{1}}U_{\mathbf{x}_{2}}^{\dagger}t^{c}-C_{F}\big]\\
 & \times\int_{\mathrm{\mathbf{x}}_{1},\mathbf{x}_{2}}e^{-i\mathbf{p}_{1}\cdot\mathbf{x}_{1}}e^{-i\mathbf{p}_{2}\cdot\mathbf{x}_{2}}A^{\bar{\lambda}}(\mathbf{x}_{12})\\
 & \times\int_{\mathbf{K}}\Big(\frac{p_{1}^{+}}{q^{+}}+\frac{\mathbf{K}\cdot\mathbf{P}_{\perp}}{\mathbf{K}^{2}}\Big)\frac{e^{-i\mathbf{K}\cdot\mathbf{x}_{12}}}{(\mathbf{K}+\mathbf{P}_{\perp})^{2}-\mathbf{P}_{\perp}^{2}-i\epsilon}\;.
\end{aligned}
\label{eq:GEFS+IFSJIMWLK}
\end{equation}
There is another contribution to JIMWLK due to the $\mathrm{Dirac}_{\bar{q}\to q,(i)}^{\bar{\lambda}\bar{\eta}\eta^{\prime}}$
spinor structure in the amplitude $\mathcal{M}_{\mathrm{GEFS}}$,
which yields:
\begin{equation}
\begin{aligned}\lim_{k_{3}^{+}\to0}\mathcal{M}_{\mathrm{GEFS},(i)} & =-i\frac{g_{e}e_fg_{s}^{2}}{\pi}\int_{k_{\mathrm{min}}^{+}}^{k_{f}^{+}}\frac{\mathrm{d}k_{3}^{+}}{k_{3}^{+}}\frac{p_{1}^{+}-p_{2}^{+}}{2q^{+}}\mathrm{Dirac}_{\overline{\mathrm{LO}}}^{\bar{\lambda}}\times\big[t^{c}U_{\mathbf{x}_{1}}U_{\mathbf{x}_{2}}^{\dagger}t^{c}-C_{F}\big]\\
 & \times\int_{\mathrm{\mathbf{x}}_{1},\mathbf{x}_{2}}e^{-i\mathbf{p}_{1}\cdot\mathbf{x}_{1}}e^{-i\mathbf{p}_{2}\cdot\mathbf{x}_{2}}A^{\bar{\lambda}}(\mathbf{x}_{12})\int_{\mathbf{K}}\frac{e^{-i\mathbf{K}\cdot\mathbf{x}_{12}}}{(\mathbf{K}+\mathbf{P}_{\perp})^{2}-\mathbf{P}_{\perp}^{2}-i\epsilon}\;.
\end{aligned}
\label{eq:GEFS(i)JIMWLK}
\end{equation}
Subamplitudes eq.~\eqref{eq:GEFS+IFSJIMWLK}) and (\ref{eq:GEFS(i)JIMWLK})
nicely combine into:
\begin{equation}
\begin{aligned}\lim_{k_{3}^{+}\to0}\mathcal{M}_{\mathrm{GEFS}} & \equiv\lim_{k_{3}^{+}\to0}\mathcal{M}_{\mathrm{GEFS},(ii)+\mathrm{IFS}}+\mathcal{M}_{\mathrm{GEFS},(i)}\\
 & =i\frac{g_{e}e_fg_{s}^{2}}{2\pi}\int_{k_{\mathrm{min}}^{+}}^{k_{f}^{+}}\frac{\mathrm{d}k_{3}^{+}}{k_{3}^{+}}\mathrm{Dirac}_{\overline{\mathrm{LO}}}^{\bar{\lambda}}\times\big[t^{c}U_{\mathbf{x}_{1}}U_{\mathbf{x}_{2}}^{\dagger}t^{c}-C_{F}\big]\\
 & \times\int_{\mathrm{\mathbf{x}}_{1},\mathbf{x}_{2}}e^{-i\mathbf{p}_{1}\cdot\mathbf{x}_{1}}e^{-i\mathbf{p}_{2}\cdot\mathbf{x}_{2}}A^{\bar{\lambda}}(\mathbf{x}_{12})\int_{\mathbf{K}}\frac{e^{-i\mathbf{K}\cdot\mathbf{x}_{12}}}{\mathbf{K}^{2}}\;.
\end{aligned}
\label{eq:FStotalJIMWLK}
\end{equation}
Finally, with the help of definition (\ref{eq:WWfield}) of the Weizs\"acker-Williams
fields, it is easy to show that:
\begin{equation}
\begin{aligned}\int_{\mathbf{x}_{3}}A^{\eta^{\prime}}(\mathbf{x}_{13})A^{\eta^{\prime}}(\mathbf{x}_{23}) & =-\int_{\mathbf{x}_{3}}\int_{\boldsymbol{\mathbf{\ell}}}\int_{\mathbf{k}}e^{-i\boldsymbol{\ell}\cdot\mathbf{x}_{13}}e^{-i\mathbf{k}\cdot\mathbf{x}_{23}}\frac{\boldsymbol{\ell}\cdot\mathbf{k}}{\boldsymbol{\mathbf{\ell}}^{2}\mathbf{k}^{2}}=\int_{\mathbf{K}}\frac{e^{-i\mathbf{K}\cdot\mathbf{x}_{12}}}{\mathbf{K}^{2}}\;.\end{aligned}
\end{equation}
Multiplying with the complex conjugate of the leading-order amplitude,
we finally obtain:
\begin{equation}
\begin{aligned}&\lim_{k_{3}^{+}\to0}\mathcal{M}_{\mathrm{LO}}^{\dagger}\mathcal{M}_{\mathrm{GEFS}}  =64\pi\alpha_{\mathrm{em}} e^2_f \alpha_{s}N_{c}p_{1}^{+}p_{2}^{+}(z^{2}+\bar{z}^{2})\int_{k_{\mathrm{min}}^{+}}^{k_{f}^{+}}\frac{\mathrm{d}k_{3}^{+}}{k_{3}^{+}}\\
 & \times\int_{\mathrm{\mathbf{x}}_{1}^{\prime},\mathbf{x}_{2^{\prime}},\mathrm{\mathbf{x}}_{1},\mathbf{x}_{2}}e^{-i\mathbf{p}_{1}\cdot\mathbf{x}_{11^{\prime}}}e^{-i\mathbf{p}_{2}\cdot\mathbf{x}_{22^{\prime}}}A^{\bar{\lambda}}(\mathbf{x}_{12})A^{\bar{\lambda}}(\mathbf{x}_{1^{\prime}2^{\prime}})\int_{\mathbf{x}_{3}}A^{\eta^{\prime}}(\mathbf{x}_{13})A^{\eta^{\prime}}(\mathbf{x}_{23})\\
 & \times\Big\langle s_{12}s_{2^{\prime}1^{\prime}}-s_{12}-s_{2^{\prime}1^{\prime}}+1-\frac{1}{N_{c}^{2}}\Big(Q_{122^{\prime}1^{\prime}}-s_{12}-s_{2^{\prime}1^{\prime}}+1\Big)\Big\rangle\\
 & =\lim_{k_{3}^{+}\to0}\mathcal{M}_{\mathrm{LO}}^{\dagger}\mathcal{M}_{\mathrm{\overline{GE}FS}}\;.
\end{aligned}
\label{eq:GEFSJIMWLK}
\end{equation}

\paragraph{GESW}
Since the modified Weizs\"acker-Williams structure is finite in the limit $k_{3}^{+}\to 0$:
\begin{equation}
\begin{aligned}\lim_{k_{3}^{+}\to0}\mathcal{A}^{\bar{\lambda}}\Big(\frac{p_{2}^{+}\mathbf{x}_{12}+k_{3}^{+}\mathbf{x}_{13}}{p_{2}^{+}+k_{3}^{+}},\frac{k_{3}^{+}}{p_{2}^{+}+k_{3}^{+}}\mathbf{x}_{32};\frac{q^{+}p_{2}^{+}}{k_{3}^{+}(p_{1}^{+}-k_{3}^{+})}\Big) & =A^{\bar{\lambda}}(\mathbf{x}_{12})\;,\end{aligned}
\end{equation}
the only contribution to JIMWLK from this diagram comes from the $\mathrm{Dirac}_{q\to\bar{q}(ii)}$ term:
\begin{equation}
\begin{aligned}\lim_{k_{3}^{+}\to0}\mathcal{M}_{\mathrm{GESW}} & =-\frac{ig_{e}e_fg_{s}^{2}}{\pi}\int_{k_{\mathrm{min}}^{+}}^{k_{f}^{+}}\frac{\mathrm{d}k_{3}^{+}}{k_{3}^{+}}\mathrm{Dirac}_{\overline{\mathrm{LO}}}^{\bar{\lambda}}\\
 & \times\int_{\mathbf{x}_{1},\mathbf{x}_{2},\mathbf{x}_{3}}e^{-i\mathbf{p}_{1}\cdot\mathbf{x}_{1}}e^{-i\mathbf{p}_{2}\cdot\mathbf{x}_{2}}A^{\eta^{\prime}}(\mathbf{x}_{31})A^{\eta^{\prime}}(\mathbf{x}_{32})A^{\bar{\lambda}}(\mathbf{x}_{12})\\
 & \times\Big[t^{c}U_{\mathbf{x}_{1}}t^{d}U_{\mathbf{x}_{2}}^{\dagger}W_{\mathbf{x}_{3}}^{dc}-C_{F}\Big]\;.
\end{aligned}
\end{equation}
After multiplying with $\mathcal{M}_{\mathrm{LO}}^{\dagger}$, making
use of eq.~\eqref{eq:DiracLOLObar}), one obtains:
\begin{equation}
\begin{aligned}&\lim_{k_{3}^{+}\to0}\mathcal{M}_{\mathrm{LO}}^{\dagger}\mathcal{M}_{\mathrm{GESW}}  =-128\pi\alpha_{\mathrm{em}} e^2_f \alpha_{s}N_{c}^{2}p_{1}^{+}p_{2}^{+}(z^{2}+\bar{z}^{2})\int_{k_{\mathrm{min}}^{+}}^{k_{f}^{+}}\frac{\mathrm{d}k_{3}^{+}}{k_{3}^{+}}\\
 & \times\int_{\mathrm{\mathbf{x}}_{1}^{\prime},\mathbf{x}_{2^{\prime}},\mathbf{x}_{1},\mathbf{x}_{2},\mathbf{x}_{3}}e^{-i\mathbf{p}_{1}\cdot\mathbf{x}_{11^{\prime}}}e^{-i\mathbf{p}_{2}\cdot\mathbf{x}_{22^{\prime}}}A^{\bar{\lambda}}(\mathbf{x}_{12})A^{\bar{\lambda}}(\mathbf{x}_{1^{\prime}2^{\prime}})A^{\eta^{\prime}}(\mathbf{x}_{31})A^{\eta^{\prime}}(\mathbf{x}_{32})\\
 & \times\Big\langle Q_{322^{\prime}1^{\prime}}s_{13}-s_{13}s_{32}-s_{2^{\prime}1^{\prime}}+1-\frac{1}{N_{c}^{2}}\Big(Q_{122^{\prime}1^{\prime}}-s_{12}-s_{2^{\prime}1^{\prime}}+1\Big)\Big\rangle\;.
\end{aligned}
\label{eq:GESWJIMWLK}
\end{equation}
It is easy to see that the $q\leftrightarrow\bar{q}$ counterpart
of this diagram will give the contribution:
\begin{equation}
\begin{aligned}&\lim_{k_{3}^{+}\to0}\mathcal{M}_{\mathrm{LO}}^{\dagger}\mathcal{M}_{\mathrm{\overline{GE}SW}}  =-128\pi\alpha_{\mathrm{em}} e^2_f \alpha_{s}N_{c}^{2}p_{1}^{+}p_{2}^{+}(z^{2}+\bar{z}^{2})\int_{k_{\mathrm{min}}^{+}}^{k_{f}^{+}}\frac{\mathrm{d}k_{3}^{+}}{k_{3}^{+}}\\
 & \times\int_{\mathrm{\mathbf{x}}_{1}^{\prime},\mathbf{x}_{2^{\prime}},\mathbf{x}_{1},\mathbf{x}_{2},\mathbf{x}_{3}}e^{-i\mathbf{p}_{1}\cdot\mathbf{x}_{11^{\prime}}}e^{-i\mathbf{p}_{2}\cdot\mathbf{x}_{22^{\prime}}}A^{\bar{\lambda}}(\mathbf{x}_{12})A^{\bar{\lambda}}(\mathbf{x}_{1^{\prime}2^{\prime}})A^{\eta^{\prime}}(\mathbf{x}_{31})A^{\eta^{\prime}}(\mathbf{x}_{32})\\
 & \times\Big\langle Q_{2^{\prime}1^{\prime}13}s_{32}-s_{13}s_{32}-s_{2^{\prime}1^{\prime}}+1-\frac{1}{N_{c}^{2}}\Big(Q_{122^{\prime}1^{\prime}}-s_{12}-s_{2^{\prime}1^{\prime}}+1\Big)\Big\rangle\;.
\end{aligned}
\label{eq:GEbarSWJIWMLK}
\end{equation}

\paragraph{SESW }
Taking the $k_{3}^{+}\to0$ limit of (\ref{eq:SESWsub}) is trivial
and yields, after multiplying with $\mathcal{M}_{\mathrm{LO}}^{\dagger}$:
\begin{equation}
\begin{aligned}&\lim_{k_{3}^{+}\to0}\mathcal{M}_{\mathrm{LO}}^{\dagger}\mathcal{M}_{\mathrm{SESW,sub}}  =128\pi\alpha_{\mathrm{em}} e^2_f \alpha_{s}N_{c}^{2}p_{1}^{+}p_{2}^{+}(z^{2}+\bar{z}^{2})\int_{k_{\mathrm{min}}^{+}}^{k_{f}^{+}}\frac{\mathrm{d}k_{3}^{+}}{k_{3}^{+}}\\
 & \times\int_{\mathrm{\mathbf{x}}_{1}^{\prime},\mathbf{x}_{2^{\prime}},\mathrm{\mathbf{x}}_{1},\mathbf{x}_{2},\mathbf{x}_{3}}e^{-i\mathbf{p}_{1}\cdot\mathbf{x}_{11^{\prime}}}e^{-i\mathbf{p}_{2}\cdot\mathbf{x}_{22^{\prime}}}A^{\bar{\lambda}}(\mathbf{x}_{1^{\prime}2^{\prime}})A^{\bar{\lambda}}(\mathbf{x}_{12})A^{\eta^{\prime}}(\mathbf{x}_{31})\\
 & \times\Bigg\{ A^{\eta^{\prime}}(\mathbf{x}_{31})\Big\langle Q_{322^{\prime}1^{\prime}}s_{13}-s_{13}s_{32}-s_{2^{\prime}1^{\prime}}+1-\frac{1}{N_{c}^{2}}\Big(Q_{122^{\prime}1^{\prime}}-s_{12}-s_{2^{\prime}1^{\prime}}+1\Big)\Big\rangle\\
 & \quad-\Big(A^{\eta^{\prime}}(\mathbf{x}_{31})-A^{\eta^{\prime}}(\mathbf{x}_{32})\Big)\Big(1-\frac{1}{N_{c}^{2}}\Big)\Big\langle Q_{122^{\prime}1^{\prime}}-s_{12}-s_{2^{\prime}1^{\prime}}+1\Big\rangle\Bigg\}\;.
\end{aligned}
\label{eq:SESWsubJIMWLK}
\end{equation}
Likewise,  we get for the diagram with a gluon loop on the antiquark:
\begin{equation}
\begin{aligned}&\lim_{k_{3}^{+}\to0}\mathcal{M}_{\mathrm{LO}}^{\dagger}\mathcal{M}_{\mathrm{\overline{SE}SW,sub}}  =128\pi\alpha_{\mathrm{em}} e^2_f \alpha_{s}N_{c}^{2}p_{1}^{+}p_{2}^{+}(z^{2}+\bar{z}^{2})\int_{k_{\mathrm{min}}^{+}}^{k_{f}^{+}}\frac{\mathrm{d}k_{3}^{+}}{k_{3}^{+}}\\
 & \times\int_{\mathbf{x}_{1^{\prime}},\mathbf{x}_{2^{\prime}},\mathbf{x}_{1},\mathbf{x}_{2},\mathbf{x}_{3}}e^{-i\mathbf{p}_{1}\cdot\mathbf{x}_{11^{\prime}}}e^{-i\mathbf{p}_{2}\cdot\mathbf{x}_{22^{\prime}}}A^{\bar{\lambda}}(\mathbf{x}_{1^{\prime}2^{\prime}})A^{\bar{\lambda}}(\mathbf{x}_{12})A^{\eta^{\prime}}(\mathbf{x}_{32})\\
 & \times\Bigg\{ A^{\eta^{\prime}}(\mathbf{x}_{32})\Big\langle Q_{132^{\prime}1^{\prime}}s_{32}-s_{13}s_{32}-s_{2^{\prime}1^{\prime}}+1-\frac{1}{N_{c}^{2}}\Big(Q_{122^{\prime}1^{\prime}}-s_{12}-s_{2^{\prime}1^{\prime}}+1\Big)\Big\rangle\\
 & \quad-\Big(A^{\eta^{\prime}}(\mathbf{x}_{32})-A^{\eta^{\prime}}(\mathbf{x}_{31})\Big)\Big(1-\frac{1}{N_{c}^{2}}\Big)\Big\langle Q_{122^{\prime}1^{\prime}}-s_{12}-s_{2^{\prime}1^{\prime}}+1\Big\rangle\Bigg\}\;.
\end{aligned}
\label{eq:LOSEbarSWsubJIMWLK}
\end{equation}

\paragraph{FSIR}
The last set of virtual diagrams that exhibit a rapidity divergency and hence
contribute to JIMWLK are the IR parts of the self-energy corrections
to the asymptotic (anti)quark, eq.~\eqref{eq:virtualIR}):
\begin{equation}
\begin{aligned}\lim_{k_{3}^{+}\to0}\mathcal{M}_{\mathrm{FSIR}} & =\mathcal{M}_{\mathrm{LO}}\times\frac{\alpha_{s}C_{F}}{2\pi}\lim_{k_{3}^{+}\to0}\mathcal{V}_{\mathrm{FSIR}}\;,\end{aligned}
\end{equation}
where, from eq.~\eqref{eq:VFSIR}):
\begin{equation}
\begin{aligned}\lim_{k_{3}^{+}\to0}\mathcal{V}_{\mathrm{FSIR}} & =2\Big(\frac{1}{\epsilon_{\mathrm{IR}}}+\gamma_{E}+\mathrm{ln}\pi\mathbf{x}_{12}^{2}\mu^{2}\Big)\int_{k_{\mathrm{min}}^{+}}^{k_{f}^{+}}\frac{\mathrm{d}k_{3}^{+}}{k_{3}^{+}}\;,\\
 & =-8\pi\int_{\mathbf{x}_{3}}A^{\eta^{\prime}}(\mathbf{x}_{13})A^{\eta^{\prime}}(\mathbf{x}_{23})\int_{k_{\mathrm{min}}^{+}}^{k_{f}^{+}}\frac{\mathrm{d}k_{3}^{+}}{k_{3}^{+}}\;.
\end{aligned}
\label{eq:FSIRJIMWLK}
\end{equation}
Multiplying with $\mathcal{M}_{\mathrm{LO}}^{\dagger}$:
\begin{equation}
\begin{aligned}&\lim_{k_{3}^{+}\to0}\mathcal{M}_{\mathrm{LO}}^{\dagger}\mathcal{M}_{\mathrm{FSIR}}  =\big|\mathcal{M}_{\mathrm{LO}}\big|^{2}\times\frac{\alpha_{s}C_{F}}{2\pi}\lim_{k_{3}^{+}\to0}\mathcal{V}_{\mathrm{FSIR}}\;,\\
 & =-128\pi\alpha_{\mathrm{em}} e^2_f \alpha_{s}N_{c}^{2}p_{1}^{+}p_{2}^{+}(z^{2}+\bar{z}^{2})\int_{k_{\mathrm{min}}^{+}}^{k_{f}^{+}}\frac{\mathrm{d}k_{3}^{+}}{k_{3}^{+}}\\
 & \times\int_{\mathrm{\mathbf{x}}_{1},\mathbf{x}_{2},\mathrm{\mathbf{x}}_{1}^{\prime},\mathbf{x}_{2^{\prime}}}e^{-i\mathbf{p}_{1}\cdot\mathbf{x}_{11^{\prime}}}e^{-i\mathbf{p}_{2}\cdot\mathbf{x}_{22^{\prime}}}A^{\bar{\lambda}}(\mathbf{x}_{12})A^{\bar{\lambda}}(\mathbf{x}_{1^{\prime}2^{\prime}})\int_{\mathbf{x}_{3}}A^{\eta^{\prime}}(\mathbf{x}_{13})A^{\eta^{\prime}}(\mathbf{x}_{23})\\
 & \times\Big(1-\frac{1}{N_{c}^{2}}\Big)\Big\langle Q_{122^{\prime}1^{\prime}}-s_{12}-s_{2^{\prime}1^{\prime}}+1\Big\rangle\;.
\end{aligned}
\end{equation}

\paragraph{Total virtual contribution to JIMWLK}
Collecting all the above virtual contributions to JIMWLK, we finally
obtain:
\begin{equation}
\begin{aligned}  &\lim_{k_{3}^{+}\to0}\mathcal{M}_{\mathrm{LO}}^{\dagger}\Big(\mathcal{M}_{\mathrm{GEFS}}+\mathcal{M}_{\mathrm{\overline{GE}FS}}+\mathcal{M}_{\mathrm{GESW}}+\mathcal{M}_{\mathrm{\overline{GE}SW}}\\&\qquad\qquad\quad+\mathcal{M}_{\mathrm{SESW,sub}}+\mathcal{M}_{\mathrm{\overline{SE}SW,sub}}+\mathcal{M}_{\mathrm{FSIR}}\Big)\\
 & =128\pi\alpha_{\mathrm{em}} e^2_f \alpha_{s}N_{c}^2 p_{1}^{+}p_{2}^{+}(z^{2}+\bar{z}^{2})\int_{k_{\mathrm{min}}^{+}}^{k_{f}^{+}}\frac{\mathrm{d}k_{3}^{+}}{k_{3}^{+}}\\&\times\int_{\mathrm{\mathbf{x}}_{1}^{\prime},\mathbf{x}_{2^{\prime}},\mathrm{\mathbf{x}}_{1},\mathbf{x}_{2}}e^{-i\mathbf{p}_{1}\cdot\mathbf{x}_{11^{\prime}}}e^{-i\mathbf{p}_{2}\cdot\mathbf{x}_{22^{\prime}}}A^{\bar{\lambda}}(\mathbf{x}_{12})A^{\bar{\lambda}}(\mathbf{x}_{1^{\prime}2^{\prime}})\\
 & \times\int_{\mathbf{x}_{3}}A^{\eta^{\prime}}(\mathbf{x}_{13})A^{\eta^{\prime}}(\mathbf{x}_{23})\Big\langle s_{12}s_{2^{\prime}1^{\prime}}-s_{12}-s_{2^{\prime}1^{\prime}}+1\Big\rangle\\
 & \quad\;\;+A^{\eta^{\prime}}(\mathbf{x}_{13})\Big(A^{\eta^{\prime}}(\mathbf{x}_{13})-A^{\eta^{\prime}}(\mathbf{x}_{23})\Big)\Big\langle Q_{322^{\prime}1^{\prime}}s_{13}-s_{13}s_{32}-s_{2^{\prime}1^{\prime}}+1\Big\rangle\\
 &\quad\;\; +A^{\eta^{\prime}}(\mathbf{x}_{23})\Big(A^{\eta^{\prime}}(\mathbf{x}_{23})-A^{\eta^{\prime}}(\mathbf{x}_{13})\Big)\Big\langle Q_{132^{\prime}1^{\prime}}s_{32}-s_{13}s_{32}-s_{2^{\prime}1^{\prime}}+1\Big\rangle\\
 & \quad\;\;-\Big(A^{\eta^{\prime}}(\mathbf{x}_{13})-A^{\eta^{\prime}}(\mathbf{x}_{23})\Big)\Big(A^{\eta^{\prime}}(\mathbf{x}_{13})-A^{\eta^{\prime}}(\mathbf{x}_{23})\Big)\Big\langle Q_{122^{\prime}1^{\prime}}-s_{12}-s_{2^{\prime}1^{\prime}}+1\Big\rangle\\
 &\quad\;\; -A^{\eta^{\prime}}(\mathbf{x}_{13})A^{\eta^{\prime}}(\mathbf{x}_{23})\Big\langle Q_{122^{\prime}1^{\prime}}-s_{12}-s_{2^{\prime}1^{\prime}}+1\Big\rangle\;.
\end{aligned}
\label{eq:virtualJIMWLK}
\end{equation}
An interesting feature of the above formula is that all subleading-$N_{c}$
contributions have cancelled.

\subsection{Real diagrams}

Taking the $p_{3}^{+}\to0$ limit of the real gluon emission amplitudes
$\mathcal{M}_{\mathrm{QFS}}$ and $\mathcal{M}_{\mathrm{QSW}}$, eq.
(\ref{eq:QFSfinal}) resp. (\ref{eq:QSWfinal}), is very straightforward
due to the simple leading-power behavior (\ref{eq:DiracQSWleadingpower}) of
the Dirac structure:
\begin{equation}
\begin{aligned}\lim_{p_{3}^{+}\to0}\mathcal{M}_{\mathrm{QFS}} & =2g_{e}e_fg_{s}\mathrm{Dirac}_{\mathrm{LO}}^{\bar{\lambda}}\int_{\mathrm{\mathbf{x}}_{1},\mathbf{x}_{2},\mathbf{x}_{3}}e^{-i\mathbf{p}_{1}\cdot\mathbf{x}_{1}}e^{-i\mathbf{p}_{2}\cdot\mathbf{x}_{2}}e^{-i\mathbf{p}_{3}\cdot\mathbf{x}_{3}}\\
 & \times A^{\bar{\lambda}}(\mathbf{x}_{12})A^{\eta}(\mathbf{x}_{13})\Big[t^{d}U_{\mathbf{x}_{1}}U_{\mathbf{x}_{2}}^{\dagger}-t^{d}\Big]\;,
\end{aligned}
\end{equation}
\begin{equation}
\begin{aligned}\lim_{p_{3}^{+}\to0}\mathcal{M}_{\mathrm{QSW}} & =-2g_{e}e_fg_{s}\mathrm{Dirac}_{\mathrm{LO}}^{\bar{\lambda}}\int_{\mathrm{\mathbf{x}}_{1},\mathbf{x}_{2},\mathbf{x}_{3}}e^{-i\mathbf{p}_{1}\cdot\mathbf{x}_{1}}e^{-i\mathbf{p}_{2}\cdot\mathbf{x}_{2}}e^{-i\mathbf{p}_{3}\cdot\mathbf{x}_{3}}\\
 & \times A^{\bar{\lambda}}(\mathbf{x}_{12})A^{\eta}(\mathbf{x}_{13})\Big[U_{\mathbf{x}_{1}}U_{\mathbf{x}_{3}}^{\dagger}t^{d}U_{\mathbf{x}_{3}}U_{\mathbf{x}_{2}}^{\dagger}-t^{d}\Big]\;.
\end{aligned}
\end{equation}
Care should be taken with the amplitudes $\mathcal{M}_{\mathrm{\overline{Q}FS}}$
and $\mathcal{M}_{\mathrm{\overline{Q}SW}}$, which as we remarked
in sec.~\ref{sec:NLO}, receive an additional minus sign from the
LCPT Feynman rules due to the $\bar{q}\to\bar{q}g$ vertex:
\begin{equation}
\begin{aligned}\lim_{p_{3}^{+}\to0}\mathcal{M}_{\mathrm{\overline{Q}FS}} & =2g_{e}e_fg_{s}\mathrm{Dirac}_{\mathrm{\overline{LO}}}^{\bar{\lambda}}\int_{\mathrm{\mathbf{x}}_{1},\mathbf{x}_{2},\mathbf{x}_{3}}e^{-i\mathbf{p}_{1}\cdot\mathbf{x}_{1}}e^{-i\mathbf{p}_{2}\cdot\mathbf{x}_{2}}e^{-i\mathbf{p}_{3}\cdot\mathbf{x}_{3}}\\
 & \times A^{\bar{\lambda}}(\mathbf{x}_{12})A^{\eta}(\mathbf{x}_{23})\Big[U_{\mathbf{x}_{1}}U_{\mathbf{x}_{2}}^{\dagger}t^{d}-t^{d}\Big]\;,
\end{aligned}
\end{equation}
\begin{equation}
\begin{aligned}\lim_{p_{3}^{+}\to0}\mathcal{M}_{\mathrm{\overline{Q}SW}} & =-2g_{e}e_fg_{s}\mathrm{Dirac}_{\mathrm{\overline{LO}}}^{\bar{\lambda}}\int_{\mathrm{\mathbf{x}}_{1},\mathbf{x}_{2},\mathbf{x}_{3}}e^{-i\mathbf{p}_{1}\cdot\mathbf{x}_{1}}e^{-i\mathbf{p}_{2}\cdot\mathbf{x}_{2}}e^{-i\mathbf{p}_{3}\cdot\mathbf{x}_{3}}\\
 & \times A^{\bar{\lambda}}(\mathbf{x}_{12})A^{\eta}(\mathbf{x}_{23})\Big[U_{\mathbf{x}_{1}}U_{\mathbf{x}_{3}}^{\dagger}t^{d}U_{\mathbf{x}_{3}}U_{\mathbf{x}_{2}}^{\dagger}-t^{d}\Big]\;.
\end{aligned}
\end{equation}
With the above expressions at hand, constructing the real part of
the JIMWLK equation is a trivial task, which yields:
\begin{equation}
\begin{aligned} & \lim_{p_{3}^{+}\to0}\int\mathrm{PS}(\vec{p}_3)\big|\mathcal{M}_{\mathrm{QFS}}+\mathcal{M}_{\mathrm{\overline{Q}FS}}+\mathcal{M}_{\mathrm{QSW}}+\mathcal{M}_{\mathrm{\overline{Q}SW}}\big|^{2}\\
 & =128\pi\alpha_{\mathrm{em}} e^2_f \alpha_{s}p_{1}^{+}p_{2}^{+}N_{c}^{2}(z^{2}+\bar{z}^{2})\int_{k_{\mathrm{min}}^{+}}^{k_{f}^{+}}\frac{\mathrm{d}p_{3}^{+}}{p_{3}^{+}}\\
 & \times\int_{\mathrm{\mathbf{x}}_{1}^{\prime},\mathbf{x}_{2^{\prime}},\mathrm{\mathbf{x}}_{1},\mathbf{x}_{2}}e^{-i\mathbf{p}_{1}\cdot\mathbf{x}_{11^{\prime}}}e^{-i\mathbf{p}_{2}\cdot\mathbf{x}_{22^{\prime}}}A^{\lambda}(\mathbf{x}_{12})A^{\lambda}(\mathbf{x}_{1^{\prime}2^{\prime}})\\
 & \times\int_{\mathbf{x}_{3}}\Big(A^{\eta}(\mathbf{x}_{13})A^{\eta}(\mathbf{x}_{1^{\prime}3})+A^{\eta}(\mathbf{x}_{23})A^{\eta}(\mathbf{x}_{2^{\prime}3})\Big)\Big\langle Q_{122^{\prime}1^{\prime}}-s_{12}-s_{2^{\prime}1^{\prime}}+1\Big\rangle\\
 &\quad\; -\Big(A^{\eta}(\mathbf{x}_{13})A^{\eta}(\mathbf{x}_{2^{\prime}3})+A^{\eta}(\mathbf{x}_{23})A^{\eta}(\mathbf{x}_{1^{\prime}3})\Big)\Big\langle s_{12}s_{2^{\prime}1^{\prime}}-s_{12}-s_{2^{\prime}1^{\prime}}+1\Big\rangle\\
 & \quad\;+\Big(A^{\eta}(\mathbf{x}_{13})A^{\eta}(\mathbf{x}_{1^{\prime}3})-A^{\eta}(\mathbf{x}_{23})A^{\eta}(\mathbf{x}_{1^{\prime}3})-A^{\eta}(\mathbf{x}_{13})A^{\eta}(\mathbf{x}_{2^{\prime}3})+A^{\eta}(\mathbf{x}_{23})A^{\eta}(\mathbf{x}_{2^{\prime}3})\Big)\\
 & \qquad\;\times\Big\langle s_{11^{\prime}}s_{2^{\prime}2}-s_{13}s_{32}-s_{2^{\prime}3}s_{31^{\prime}}+1\Big\rangle\\
 & \quad\;-A^{\eta}(\mathbf{x}_{1^{\prime}3})\Big(A^{\eta}(\mathbf{x}_{13})-A^{\eta}(\mathbf{x}_{23})\Big)\Big\langle Q_{322^{\prime}1^{\prime}}s_{13}-s_{13}s_{32}-s_{2^{\prime}1^{\prime}}+1\Big\rangle\\
 & \quad\;-A^{\eta}(\mathbf{x}_{13})\Big(A^{\eta}(\mathbf{x}_{1^{\prime}3})-A^{\eta}(\mathbf{x}_{2^{\prime}3})\Big)\Big\langle Q_{122^{\prime}3}s_{31^{\prime}}-s_{31^{\prime}}s_{2^{\prime}3}-s_{12}+1\Big\rangle\\
 & \quad\;+A^{\eta}(\mathbf{x}_{2^{\prime}3})\Big(A^{\eta}(\mathbf{x}_{13})-A^{\eta}(\mathbf{x}_{23})\Big)\Big\langle Q_{2^{\prime}1^{\prime}13}s_{32}-s_{13}s_{32}-s_{2^{\prime}1^{\prime}}+1\Big\rangle\\
 & \quad\;+A^{\eta}(\mathbf{x}_{23})\Big(A^{\eta}(\mathbf{x}_{1^{\prime}3})-A^{\eta}(\mathbf{x}_{2^{\prime}3})\Big)\Big\langle Q_{31^{\prime}12}s_{2^{\prime}3}-s_{31^{\prime}}s_{2^{\prime}3}-s_{12}+1\Big\rangle\;.
\end{aligned}
\label{eq:JIMWLKreal}
\end{equation}
Just like in the case of the virtual diagrams, all subleading-$N_{c}$
contributions have cancelled.

\subsection{Full JIMWLK limit}

Combining eq.~\eqref{eq:JIMWLKreal}) with eq.~\eqref{eq:virtualJIMWLK})
and its complex conjugate, we obtain:
\begin{equation}
\begin{aligned} & \lim_{p_{3}^{+}\to0}\int\mathrm{PS}(\vec{p}_3)\big|\mathcal{M}_{\mathrm{real}}\big|^{2}+\lim_{k_{3}^{+}\to0}\big|\mathcal{M}_{\mathrm{virtual}}\big|^{2}\\
 & =64\pi\alpha_{\mathrm{em}} e_f^2 p_{1}^{+}p_{2}^{+}N_{c}(z^{2}+\bar{z}^{2})\int_{\mathrm{\mathbf{x}}_{1}^{\prime},\mathbf{x}_{2^{\prime}},\mathrm{\mathbf{x}}_{1},\mathbf{x}_{2}}e^{-i\mathbf{p}_{1}\cdot\mathbf{x}_{11^{\prime}}}e^{-i\mathbf{p}_{2}\cdot\mathbf{x}_{22^{\prime}}}A^{\lambda}(\mathbf{x}_{12})A^{\lambda}(\mathbf{x}_{1^{\prime}2^{\prime}})\\
 & \times\Bigg(\frac{\alpha_{s}N_{c}}{(2\pi)^{2}}\int_{k_{\mathrm{min}}^{+}}^{k_{f}^{+}}\frac{\mathrm{d}k_{3}^{+}}{k_{3}^{+}}\Bigg)\int_{\mathbf{x}_{3}}\Bigg\{-\mathcal{K}_{1}(\mathbf{x}_{1},\mathbf{x}_{2},\mathbf{x}_{2^{\prime}},\mathbf{x}_{1^{\prime}};\mathbf{x}_{3})\times \langle Q_{122^{\prime}1^{\prime}}\rangle\\
 &\qquad +\mathcal{A}(\mathbf{x}_{1},\mathbf{x}_{2},\mathbf{x}_{2^{\prime}},\mathbf{x}_{1^{\prime}};\mathbf{x}_{3})\times \langle s_{11^{\prime}}s_{2^{\prime}2}\rangle +\mathcal{B}(\mathbf{x}_{1},\mathbf{x}_{2},\mathbf{x}_{2^{\prime}},\mathbf{x}_{1^{\prime}};\mathbf{x}_{3})\times \langle s_{12}s_{2^{\prime}1^{\prime}}\rangle \\
 &\qquad +\mathcal{K}_{2}(\mathbf{x}_{1};\mathbf{x}_{2},\mathbf{x}_{1^{\prime}};\mathbf{x}_{3})\times \langle Q_{322^{\prime}1^{\prime}}s_{13}\rangle +\mathrm{c.c.}\\
 &\qquad +\mathcal{K}_{2}(\mathbf{x}_{2};\mathbf{x}_{1},\mathbf{x}_{2^{\prime}};\mathbf{x}_{3})\times \langle Q_{2^{\prime}1^{\prime}13}s_{32}\rangle +\mathrm{c.c.}+2\frac{\mathbf{x}_{12}^{2}}{\mathbf{x}_{13}^{2}\mathbf{x}_{23}^{2}}\big\langle s_{12}-s_{13}s_{32}\big\rangle+\mathrm{c.c.}\Bigg\}\;.
\end{aligned}
\label{eq:fullJIMWLK}
\end{equation}
The above expression is the JIMWLK equation for $\big|\mathcal{M}_{\mathrm{LO}}\big|^{2}$,
eq.~\eqref{eq:MLOsquared}), consisting in the evolution of the quadrupole
$Q_{122^{\prime}1^{\prime}}$ (eq. 4 in Ref.~\cite{Dominguez:2011gc})
and, in the last line, of the dipole $s_{12}$ and its complex conjugate.
We have therefore proven what we asserted in eq.~\eqref{eq:JIMWLKsubtraction}),
namely that the part of the cross section $\mathrm{d}\sigma$ enhanced
by large logarithms $\ln k_{f}^{+}/k_{\mathrm{min}}^{+}$ takes the
form $\hat{H}_{\mathrm{JIMWLK}}\mathrm{d}\sigma_{\mathrm{LO}}$.

The structures $\mathcal{A}\;,\mathcal{B}\;,\mathcal{K}_{1}$ and $\mathcal{K}_{2}$
in eq.~\eqref{eq:fullJIMWLK}) are defined according to the notation
in~\cite{Dominguez:2011gc}, and read:
\begin{equation}
\begin{aligned}\mathcal{A}(\mathbf{x}_{1},\mathbf{x}_{2},\mathbf{x}_{2^{\prime}},\mathbf{x}_{1^{\prime}};\mathbf{x}_{3}) & =2(2\pi)^{2}\Big(A^{\eta}(\mathbf{x}_{13})A^{\eta}(\mathbf{x}_{1^{\prime}3})-A^{\eta}(\mathbf{x}_{23})A^{\eta}(\mathbf{x}_{1^{\prime}3})\\
 & \qquad\qquad-A^{\eta}(\mathbf{x}_{13})A^{\eta}(\mathbf{x}_{2^{\prime}3})+A^{\eta}(\mathbf{x}_{23})A^{\eta}(\mathbf{x}_{2^{\prime}3})\Big)\;,\\
 & =\frac{\mathbf{x}_{1^{\prime}2}^{2}}{\mathbf{x}_{1^{\prime}3}^{2}\mathbf{x}_{23}^{2}}+\frac{\mathbf{x}_{2^{\prime}1}^{2}}{\mathbf{x}_{2^{\prime}3}^{2}\mathbf{x}_{13}^{2}}-\frac{\mathbf{x}_{1^{\prime}1}^{2}}{\mathbf{x}_{1^{\prime}3}^{2}\mathbf{x}_{13}^{2}}-\frac{\mathbf{x}_{2^{\prime}2}^{2}}{\mathbf{x}_{2^{\prime}3}^{2}\mathbf{x}_{23}^{2}}\;,
\end{aligned}
\end{equation}
\begin{equation}
\begin{aligned}\mathcal{\mathcal{B}}(\mathbf{x}_{1},\mathbf{x}_{2},\mathbf{x}_{2^{\prime}},\mathbf{x}_{1^{\prime}};\mathbf{x}_{3}) & =2(2\pi)^{2}\Big(A^{\eta}(\mathbf{x}_{13})A^{\eta}(\mathbf{x}_{23})+A^{\eta}(\mathbf{x}_{1^{\prime}3})A^{\eta}(\mathbf{x}_{2^{\prime}3})\\
 &\qquad\qquad -A^{\eta}(\mathbf{x}_{13})A^{\eta}(\mathbf{x}_{2^{\prime}3})-A^{\eta}(\mathbf{x}_{23})A^{\eta}(\mathbf{x}_{1^{\prime}3})\Big)\;,\\
 & =\frac{\mathbf{x}_{12^{\prime}}^{2}}{\mathbf{x}_{13}^{2}\mathbf{x}_{2^{\prime}3}^{2}}+\frac{\mathbf{x}_{1^{\prime}2}^{2}}{\mathbf{x}_{1^{\prime}3}^{2}\mathbf{x}_{23}^{2}}-\frac{\mathbf{x}_{12}^{2}}{\mathbf{x}_{13}^{2}\mathbf{x}_{23}^{2}}-\frac{\mathbf{x}_{1^{\prime}2^{\prime}}^{2}}{\mathbf{x}_{1^{\prime}3}^{2}\mathbf{x}_{2^{\prime}3}^{2}}\;,
\end{aligned}
\end{equation}
\begin{equation}
\begin{aligned}&\mathcal{K}_{1}(\mathbf{x}_{1},\mathbf{x}_{2},\mathbf{x}_{2^{\prime}},\mathbf{x}_{1^{\prime}};\mathbf{x}_{3})  =-2(2\pi)^{2}\Big(A^{\eta}(\mathbf{x}_{13})A^{\eta}(\mathbf{x}_{1^{\prime}3})+A^{\eta}(\mathbf{x}_{23})A^{\eta}(\mathbf{x}_{2^{\prime}3})\\
 &\qquad\qquad\qquad -A^{\eta}(\mathbf{x}_{13})A^{\eta}(\mathbf{x}_{13})+A^{\eta}(\mathbf{x}_{13})A^{\eta}(\mathbf{x}_{23})-A^{\eta}(\mathbf{x}_{23})A^{\eta}(\mathbf{x}_{23})\\
 &\qquad\qquad\qquad -A^{\eta}(\mathbf{x}_{1^{\prime}3})A^{\eta}(\mathbf{x}_{1^{\prime}3})+A^{\eta}(\mathbf{x}_{2^{\prime}3})A^{\eta}(\mathbf{x}_{1^{\prime}3})-A^{\eta}(\mathbf{x}_{2^{\prime}3})A^{\eta}(\mathbf{x}_{2^{\prime}3})\Big)\;,\\
 & \qquad\qquad\qquad \qquad\qquad\;=\frac{\mathbf{x}_{11^{\prime}}^{2}}{\mathbf{x}_{13}^{2}\mathbf{x}_{1^{\prime}3}^{2}}+\frac{\mathbf{x}_{12}^{2}}{\mathbf{x}_{13}^{2}\mathbf{x}_{23}^{2}}+\frac{\mathbf{x}_{22^{\prime}}^{2}}{\mathbf{x}_{23}^{2}\mathbf{x}_{2^{\prime}3}^{2}}+\frac{\mathbf{x}_{1^{\prime}2^{\prime}}^{2}}{\mathbf{x}_{1^{\prime}3}^{2}\mathbf{x}_{2^{\prime}3}^{2}}\;,
\end{aligned}
\end{equation}
\begin{equation}
\begin{aligned}\mathcal{K}_{2}(\mathbf{x}_{1};\mathbf{x}_{2},\mathbf{x}_{1^{\prime}};\mathbf{x}_{3};\mathbf{x}_{3}) & =2(2\pi)^{2}\Big(A^{\eta}(\mathbf{x}_{13})-A^{\eta}(\mathbf{x}_{1^{\prime}3})\Big)\Big(A^{\eta}(\mathbf{x}_{13})-A^{\eta}(\mathbf{x}_{23})\Big)\;,\\
 & =\frac{\mathbf{x}_{12}^{2}}{\mathbf{x}_{23}^{2}\mathbf{x}_{13}^{2}}+\frac{\mathbf{x}_{11^{\prime}}^{2}}{\mathbf{x}_{13}^{2}\mathbf{x}_{1^{\prime}3}^{2}}-\frac{\mathbf{x}_{21^{\prime}}^{2}}{\mathbf{x}_{23}^{2}\mathbf{x}_{1^{\prime}3}^{2}}\;.
\end{aligned}
\end{equation}

\section{\label{sec:jet}Jet definition}
In the previous sections, we were always concerned with partonic scattering
amplitudes, hiding any hadronic physics inside the nonperturbative
CGC averages over the Wilson lines. The aim of this work, however,
is to compute the NLO \emph{dijet} photoproduction cross section,
not the diquark one, which implies that at a certain point we need
to quantify what we mean by a jet.

At leading order, transforming the $\gamma A\to q\bar{q}X$ cross
section to the $\gamma A\to\mathrm{dijet}X$ one is trivial and
done by simply identifying the quark and antiquark partonic momenta
$\vec{p}_{1}$ and $\vec{p}_{2}$ with the jet momenta $\vec{p}_{j1}$ and $\vec{p}_{j2}$.
This same trivial identification can be done for the virtual NLO contributions
to the cross section. It is implicitly assumed that the outgoing momenta
are sufficiently separated to prevent the quark and antiquark being
grouped within the same jet.

For the real NLO corrections, the situation is different. In this
case, two steps are needed to go from the tree-level $\gamma A\to q\bar{q}gX$ partonic cross section to the real NLO correction to the dijet cross section. First, a jet algorithm needs to be applied in order to determine in what part of the phase space the three partons are considered to form three separate jets, and in what part of the phase space two of the partons are clustered into a single jet. Second, in the case of the three jet configuration, in order to obtain the corresponding contribution to the inclusive dijet cross section at NLO, two jets should be identified with the measured jets with momenta $\vec{p}_{j1}$ and $\vec{p}_{j2}$ while the third jet should be integrated over.

The jet definition/algorithm that we use in this NLO calculation is as follows: if two partons $i$ and $j$ in the final state are such that
\begin{equation}
\begin{aligned}
\frac{(p_{i}^{+}+p_{j}^{+})}{\left|\mathbf{p}_{i}+\mathbf{p}_{j}\right|}
\left|\frac{\mathbf{p}_{i}}{p_{i}^{+}}-\frac{\mathbf{p}_{j}}{p_{j}^{+}}\right| & < R
 \;,
\end{aligned}
\label{eq:jetdef_1}
\end{equation}
then they are considered to be part of the same jet with momentum $(p_{i}^{+}+p_{j}^{+},  \mathbf{p}_{i}+\mathbf{p}_{j}) $. Otherwise, the partons $i$ and $j$ are considered to form separate jets. This definition depends on the jet radius parameter $R$, satisfying $0<R<1$.

In the present study, for simplicity and in order to be able to perform analytic calculations as far as possible, we are using our jet definition in the narrow-jet limit, meaning that formally $R\rightarrow 0$. In practice, this means that terms in $\log(R)$ or terms independent of $R$ are kept in the cross section, whereas terms suppressed as positive powers of $R$ are neglected. In diagrams without a collinear divergence, one can identify each of the three partons with a separate jet, because the contribution from the merging of two partons into one jet takes place only in a parametrically small part of the phase space, suppressed by powers of $R$.
Among the real NLO corrections, only $|\mathcal{M}_{\mathrm{QFS}}|^{2}$ has a quark-gluon collinear divergence  and only $|\mathcal{M}_{\mathrm{\overline{Q}FS}}|^{2}$ has an antiquark-gluon collinear divergence. In the narrow-jet limit, it is then sufficient to distinguish when the quark and the gluon (resp. the antiquark and gluon) are clustered into a single jet or not by the jet algorithm in the $|\mathcal{M}_{\mathrm{QFS}}|^{2}$ contribution (resp. in the $|\mathcal{M}_{\mathrm{\overline{Q}FS}}|^{2}$ contribution).

As a remark, when the left hand side of eq.~\eqref{eq:jetdef_1} is much smaller than $1$, one has the equivalence
\begin{equation}
\begin{aligned}
\frac{(p_{i}^{+}+p_{j}^{+})}{\left|\mathbf{p}_{i}+\mathbf{p}_{j}\right|}
\left|\frac{\mathbf{p}_{i}}{p_{i}^{+}}-\frac{\mathbf{p}_{j}}{p_{j}^{+}}\right| & \sim \sqrt{{\Delta y_{ij}}^2 + {\Delta \phi_{ij}}^2}
 \;,
\end{aligned}
\label{eq:jetdef_small_R_1}
\end{equation}
where $\Delta y_{ij}$ and $\Delta \phi_{ij}$ are the difference in rapidity and in azimuthal angle between the two partons. For this reason, our jet algorithm and the Cambridge/Aachen algorithm (see refs.~\cite{Dokshitzer:1997in,Wobisch:1998wt}) become equivalent in the narrow-jet limit $R\rightarrow 0$.

Moreover, in the case of three-jet configurations, we always consider the quark jet to be the first measured jet, the antiquark jet to be the second measured jet, and the gluon jet to be the jet which is unmeasured and integrated over.
Hence, the results that we present correspond specifically to the flavor-tagged inclusive quark-antiquark dijet cross section at NLO. Other contributions to the full inclusive dijet cross section at NLO, for example with a gluon jet and a quark jet,  do not involve any conceptual difficulty (in particular, they do not contain any divergence whatsoever), and could be obtained in a straightforward way from our intermediate results on the quark-antiquark-gluon partonic cross section.

From the $|\mathcal{M}_{\mathrm{QFS}}|^{2}$ contribution to the $q\bar{q}g$ partonic cross section
\begin{equation}
\begin{aligned}
\frac{\mathrm{d}\sigma_{q\bar{q}g}^{|{\mathrm{QFS}}|^2}}{
\mathrm{d}p_{1}^{+}\mathrm{\mathrm{d}}^{D-2}\mathbf{p}_{1}
\mathrm{d}p_{2}^{+}\mathrm{\mathrm{d}}^{D-2}\mathbf{p}_{2}
\mathrm{d}p_{3}^{+}\mathrm{\mathrm{d}}^{D-2}\mathbf{p}_{3}
}
 & =
\frac{\theta(p_{1}^{+})}{(2\pi)^{D-1}2p_{1}^{+}}
\frac{\theta(p_{2}^{+})}{(2\pi)^{D-1}2p_{2}^{+}}
\frac{\theta(p_{3}^{+})}{(2\pi)^{D-1}2p_{3}^{+}}
\\
&
\times
\frac{2\pi\delta(q^{+}-p_{1}^{+}-p_{2}^{+}-p_{3}^{+})}{2q^{+}}
\frac{1}{(D-2)}|\mathcal{M}_{\mathrm{QFS}}|^{2}\;,
\end{aligned}
\label{eq:qqbarg_QFSsquare}
\end{equation}
applying our jet definition in the narrow-jet limit, we thus obtain two contributions to the dijet cross section. The first one, in which each of the three partons is forming a jet and in which the gluon jet is integrated over, reads
\begin{equation}
\begin{aligned}
&\frac{\mathrm{d}\sigma_{\textrm{dijet}}^{|{\mathrm{QFS}}|^2;\textrm{ out}}}{
\mathrm{d}p_{j1}^{+}\mathrm{\mathrm{d}}^{D-2}\mathbf{p}_{j1}
\mathrm{d}p_{j2}^{+}\mathrm{\mathrm{d}}^{D-2}\mathbf{p}_{j2}
}\\
& = \,
 \int\mathrm{PS}(\vec{p}_3)\;
\Big[1-
 \theta_{\mathrm{in}}(\vec{p}_{j1},\vec{p}_{3})
 \Big]
\frac{\mathrm{d}\sigma_{q\bar{q}g}^{|{\mathrm{QFS}}|^2}}{
\mathrm{d}p_{j1}^{+}\mathrm{\mathrm{d}}^{D-2}\mathbf{p}_{j1}
\mathrm{d}p_{j2}^{+}\mathrm{\mathrm{d}}^{D-2}\mathbf{p}_{j2}
\mathrm{d}p_{3}^{+}\mathrm{\mathrm{d}}^{D-2}\mathbf{p}_{3}
}
 \\
 &= 
\frac{\theta(p_{j1}^{+})}{(2\pi)^{D-1}2p_{j1}^{+}}
\frac{\theta(p_{j2}^{+})}{(2\pi)^{D-1}2p_{j2}^{+}}
 \int\mathrm{PS}(\vec{p}_3)\;
\frac{2\pi\delta(q^{+}-p_{j1}^{+}-p_{j2}^{+}-p_{3}^{+})}{2q^{+}}
\\
 & \times
\Big[1-
 \theta_{\mathrm{in}}(\vec{p}_{j1},\vec{p}_{3})
 \Big]
 \frac{1}{(D-2)}|\mathcal{M}_{\mathrm{QFS}}|^{2}
 \bigg|_{\vec{p}_{1} = \vec{p}_{j1} ;\,  \vec{p}_{2} = \vec{p}_{j2}}  \;.
\end{aligned}
\label{eq:QFSsquare_out}
\end{equation}
We have excluded the region in which the quark and the gluon are merged into one jet thanks to $\theta_{\mathrm{in}}(\vec{p}_{i},\vec{p}_{j})$, which is defined as
\begin{equation}
\begin{aligned}\theta_{\mathrm{in}}(\vec{p}_{i},\vec{p}_{j})
& =
\theta\Bigg((\mathbf{p}_{i}+\mathbf{p}_{j})^{2}R^{2}
-(p_{i}^{+}+p_{j}^{+})^{2}
\Big(\frac{\mathbf{p}_{i}}{p_{i}^{+}}
-\frac{\mathbf{p}_{j}}{p_{j}^{+}}\Big)^{2}
\Bigg)
\end{aligned}
\label{eq:jetdef_2}
\end{equation}
and enforces the condition \eqref{eq:jetdef_1}. The second contribution,  in which the quark and the gluon are combined into one jet of momentum $\vec{p}_{j1} = \vec{p}_{1}+\vec{p}_{3}$ whereas the antiquark forms a jet of momentum $\vec{p}_{j2} = \vec{p}_{2}$, reads
\begin{equation}
\begin{aligned}
&\frac{\mathrm{d}\sigma_{\textrm{dijet}}^{|{\mathrm{QFS}}|^2;\textrm{ in}}}{
\mathrm{d}p_{j1}^{+}\mathrm{\mathrm{d}}^{D-2}\mathbf{p}_{j1}
\mathrm{d}p_{j2}^{+}\mathrm{\mathrm{d}}^{D-2}\mathbf{p}_{j2}
}=
  \frac{\theta(p_{j1}^{+})}{(2\pi)^{D-1}2p_{j1}^{+}}
\frac{\theta(p_{j2}^{+})}{(2\pi)^{D-1}2p_{j2}^{+}}
 \frac{2\pi\delta(q^{+}-p_{j1}^{+}-p_{j2}^{+})}{2q^{+}}
\\
&\times\int\mathrm{PS}(\vec{p}_3)
   \theta_{\mathrm{in}}(\vec{p}_{j1}\!-\!\vec{p}_{3},\vec{p}_{3})
   \theta(p_{j1}^{+}\!-\!p_{3}^{+})
\frac{p_{j1}^{+}}{(p_{j1}^{+}\!-\!p_{3}^{+})}
\frac{1}{(D\!-\!2)}|\mathcal{M}_{\mathrm{QFS}}|^{2}
 \bigg|_{\vec{p}_{1} = \vec{p}_{j1}\!-\!\vec{p}_{3};\vec{p}_{2} = \vec{p}_{j2}} .
\end{aligned}
\label{eq:QFSsquare_in}
\end{equation}

For the  $|\mathcal{M}_{\mathrm{\overline{Q}FS}}|^{2}$ contribution, which is the square of the gluon emission by the antiquark in the final state, one obtains two contributions to the NLO dijet cross section in a similar way. Their expressions are the same as eqs.~\eqref{eq:QFSsquare_out} and \eqref{eq:QFSsquare_in}, up to the exchange of the role of the quark and the antiquark, meaning in particular $\vec{p}_{1}\leftrightarrow \vec{p}_{2}$  and  $\vec{p}_{j1}\leftrightarrow \vec{p}_{j2}$.

Finally, for all the contributions to the $q\bar{q}g$ partonic cross section other than
$|\mathcal{M}_{\mathrm{QFS}}|^{2}$ and $|\mathcal{M}_{\mathrm{\overline{Q}FS}}|^{2}$,
which are collinear safe, we go from the partonic to the dijet cross section by integrating over the gluon momentum and identifying the produced quark and antiquark with the measured jets, as
\begin{equation}
\begin{aligned}
\frac{\mathrm{d}\sigma_{\textrm{dijet}}^{\textrm{NLO real; coll. safe diags.}}}{
\mathrm{d}p_{j1}^{+}\mathrm{\mathrm{d}}^{D-2}\mathbf{p}_{j1}
\mathrm{d}p_{j2}^{+}\mathrm{\mathrm{d}}^{D-2}\mathbf{p}_{j2}
}
 & =
 \int\mathrm{PS}(\vec{p}_3)\;
\frac{\mathrm{d}\sigma_{q\bar{q}g}^{\textrm{coll. safe  diags.}}}{
\mathrm{d}p_{j1}^{+}\mathrm{\mathrm{d}}^{D-2}\mathbf{p}_{j1}
\mathrm{d}p_{j2}^{+}\mathrm{\mathrm{d}}^{D-2}\mathbf{p}_{j2}
\mathrm{d}p_{3}^{+}\mathrm{\mathrm{d}}^{D-2}\mathbf{p}_{3}
}
 \;.
\end{aligned}
\label{eq:coll_safe_NLO_real_dijet}
\end{equation}

\section{\label{sec:coll}Collinear and soft safety in final state fragmentation}


As already mentioned in the previous section, the real contributions $|\mathrm{QFS}|^{2}$ and $|\mathrm{\overline{Q}FS}|^{2}$ to the dijet cross section have collinear divergences. Such divergences come from the transverse integration, and we are handling them with dimensional regularization. In addition, these diagrams also lead to soft divergences at the dijet cross section level. In our regularization scheme, using dimensional regularization for transverse integrals and the cutoff $k^+_{\textrm{min}}$ for the $p_3^+$ integral, it is difficult to distinguish unambiguously between genuine soft divergences and rapidity divergences, since both arise from the $p_3^+\rightarrow 0$ regime. In this section, calculating the real contributions from $|\mathrm{QFS}|^{2}$ to the dijet cross section, we will show that most divergences cancel in the sum of the real corrections $|\mathrm{QFS}|^{2}$ and $|\mathrm{\overline{Q}FS}|^{2}$ on the one hand, and of the virtual correction $\mathrm{FSIR}$ both in the amplitude and the complex conjugate amplitude on the other hand. The only leftover divergence in the resulting sum is the rapidity divergence associated with the JIMWLK evolution, already discussed in
section~\ref{sec:JIMWLK}. This is an evidence for the cancelation of both collinear and soft divergences between these contributions. In order to calculate the real correction $|\mathrm{QFS}|^{2}$ to the dijet cross section, following the discussion in section~\ref{sec:jet}, we first split it into the \emph{in} contribution~\eqref{eq:QFSsquare_in}, in which the quark and the gluon belong to the same jet, and the~\emph{out}  contribution~\eqref{eq:QFSsquare_out}, in which they form separate jets.

The amplitude for the diagram $\mathrm{QFS}$ is given in eq.~\eqref{eq:QFS1}. Squaring it and summing over the colors, helicities and the photon polarization, one finds
\begin{equation}
\begin{aligned}
\big|\mathcal{M}_{\mathrm{QFS}}\big|^{2}
&= 
 (4\pi)^{2}\alpha_{\mathrm{em}} e^2_f \alpha_{s}C_{F}N_{c} 8p_{1}^{+}p_{2}^{+}(p_{3}^{+})^{2}\\
&\times \bigg[2\Big(\frac{p_{1}^{+}+p_{3}^{+}}{q^{+}}\Big)^{2}
 +2\Big(\frac{p_{2}^{+}}{q^{+}}\Big)^{2}+(D-4)\bigg]
 \bigg[\Big(1+2\frac{p_{1}^{+}}{p_{3}^{+}}\Big)^{2}+(D-3)\bigg]
 \\
 &  \times
  \frac{1}{(p_{3}^{+}\mathbf{p}_{1}-p_{1}^{+}\mathbf{p}_{3})^{2}}
 \int_{\mathrm{\mathbf{x}}_{1},\mathbf{x}_{2},\mathrm{\mathbf{x}}_{1}^{\prime},\mathbf{x}_{2^{\prime}}}e^{-i(\mathbf{p}_{1}+\mathbf{p}_{3})\cdot\mathbf{x}_{11^{\prime}}}e^{-i\mathbf{p}_{2}\cdot\mathbf{x}_{22^{\prime}}}A^{\bar{\lambda}}(\mathbf{x}_{12})A^{\bar{\lambda}}(\mathbf{x}_{1^{\prime}2^{\prime}})
 \\
 &  \times
  \Big\langle Q_{122^{\prime}1^{\prime}}-s_{12}-s_{2^{\prime}1^{\prime}}+1\Big\rangle\;.
\end{aligned}
\label{eq:QFSsquared}
\end{equation}


\subsection{Contribution $|\mathrm{QFS}|^{2}$; in}
When the gluon and the quark are merged into a single jet by the jet algorithm, we have by definition $\vec{p}_{1}+\vec{p}_{3}=\vec{p}_{j1}$. Introducing the transverse momentum of the gluon relative to the one of jet $j1$
\begin{equation}
\begin{aligned}
\mathbf{P}_{3} &\equiv  \mathbf{p}_{3} - \frac{p_{3}^{+}}{p_{j1}^{+}}  \mathbf{p}_{j1}
\;,
\end{aligned}
\label{eq:def_P3_shifted_mom}
\end{equation}
one can rewrite the denominator of eq.~\eqref{eq:QFSsquared} as
\begin{equation}
\begin{aligned}
(p_{3}^{+}\mathbf{p}_{1}-p_{1}^{+}\mathbf{p}_{3})^{2} = & (p_{j1}^{+})^{2} {\mathbf{P}_{3}^{2}}
\;,
\end{aligned}
\label{eq:denom_QFS2_in}
\end{equation}
Then, comparing with the LO squared amplitude \eqref{eq:MLOsquared}, eq.~\eqref{eq:QFSsquared} becomes:
\begin{equation}
\begin{aligned}
\big|\mathcal{M}_{\mathrm{QFS}}\big|^{2} \bigg|_{\textrm{in}}
& =
\big|\mathcal{M}_{\mathrm{LO}}\big|^{2}\times4\pi\alpha_{s}C_{F}
\frac{(p_{3}^{+})^{2}(p_{j1}^{+}\!-\!p_{3}^{+})}{ (p_{j1}^{+})^{3} {\mathbf{P}_{3}^{2}}}
\Big[\Big(1\!+\!2\frac{(p_{j1}^{+}\!-\!p_{3}^{+})}{p_{3}^{+}}\Big)^{2}\!+\!(D\!-\!3)\Big]\;.
\end{aligned}
\label{eq:QFS2_ampl_square_factorization_in}
\end{equation}
Moreover, with the notation \eqref{eq:def_P3_shifted_mom},
\begin{equation}
\begin{aligned}
\theta_{\mathrm{in}}(\vec{p}_{1},\vec{p}_{3})
 & =\,
 \theta_{\mathrm{in}}(\vec{p}_{j1}-\vec{p}_{3},\vec{p}_{3})
 =
 \theta\Bigg( \mathbf{p}_{j1}^{2}R^{2}-\frac{(p_{j1}^{+})^4}{(p_{j1}^{+}-p_{3}^{+})^2\, (p_{3}^{+})^2} {\mathbf{P}_{3}^{2}}\Bigg)\;.
\end{aligned}
\label{eq:theta_in_in}
\end{equation}
Inserting eqs.~\eqref{eq:QFS2_ampl_square_factorization_in} and \eqref{eq:theta_in_in} into eq.~\eqref{eq:QFSsquare_in}, one finds the inside jet radiation contribution from $|\mathrm{QFS}|^{2}$ to factorize as
\begin{equation}
\begin{aligned}
&
\frac{\mathrm{d}\sigma_{\textrm{dijet}}^{|{\mathrm{QFS}}|^2;\textrm{ in}}}{
\mathrm{d}p_{j1}^{+}\mathrm{\mathrm{d}}^{D-2}\mathbf{p}_{j1}
\mathrm{d}p_{j2}^{+}\mathrm{\mathrm{d}}^{D-2}\mathbf{p}_{j2}
}
  =
 \frac{\mathrm{d}\sigma^{\textrm{dijet}}_{\mathrm{LO}}}{
\mathrm{d}p_{j1}^{+}\mathrm{\mathrm{d}}^{D-2}\mathbf{p}_{j1}
\mathrm{d}p_{j2}^{+}\mathrm{\mathrm{d}}^{D-2}\mathbf{p}_{j2}
}
\times
 \frac{\alpha_{s}C_{F}}{2\pi}
 \mathcal{V}_{|{\mathrm{QFS}}|^2}^{\textrm{in}}
 \; ,
\end{aligned}
\label{eq:QFSsquare_in_2}
\end{equation}
 where
\begin{equation}
\begin{aligned}
  \mathcal{V}_{|{\mathrm{QFS}}|^2}^{\textrm{in}}
  & =   
 (2\pi)(4\pi)  \int\mathrm{PS}(\vec{p}_3)\;
   \theta(p_{j1}^{+}-p_{3}^{+})\;
\frac{1}{ (p_{j1}^{+})^{2} {\mathbf{P}_{3}^{2}}}
\\
 &\times
  \theta\Bigg(\frac{(p_{j1}^{+}-p_{3}^{+})^2 (p_{3}^{+})^2}{(p_{j1}^{+})^4} \mathbf{p}_{j1}^{2}R^{2}- {\mathbf{P}_{3}^{2}}\Bigg)\;
\Big[(2p_{j1}^{+}-p_{3}^{+})^{2}+(D-3) (p_{3}^{+})^2\Big]
 \; .
\end{aligned}
\label{eq:QFSsquare_in_3}
\end{equation}
In eq.~\eqref{eq:QFSsquare_in_3}, the transverse integral is straightforward to calculate in dimensional regularization, and yields
 \begin{equation}
\begin{aligned}
 &
 4\pi \mu^{4-D}\int\frac{\mathrm{d}^{D-2}\mathbf{P}_{3}}{(2\pi)^{D-2}}
 \frac{1}{{\mathbf{P}_{3}^{2}}}
 \theta\Bigg(\frac{(p_{j1}^{+}-p_{3}^{+})^2\, (p_{3}^{+})^2}{(p_{j1}^{+})^4} \mathbf{p}_{j1}^{2}R^{2}- {\mathbf{P}_{3}^{2}}\Bigg)
 \\
&=   - \frac{1}{\epsilon_{\textrm{coll}}} \frac{1}{\Gamma(1-\epsilon)}
 \left[\frac{(p_{j1}^{+}-p_{3}^{+})^2 (p_{3}^{+})^2}{(p_{j1}^{+})^4} \frac{\mathbf{p}_{j1}^{2}R^{2}}{4\pi \mu^2 }
 \right]^{-\epsilon}
 \\
&= 
-\frac{1}{\epsilon_{\mathrm{coll}}}+\gamma_{E}+\ln\frac{\mathbf{p}_{j1}^{2}}{4\pi\mu^{2}}+2\ln R
+2\ln\frac{p_{3}^{+}}{p_{j1}^{+}}+2\ln\left(\frac{p_{j1}^{+}-p_{3}^{+}}{p_{j1}^{+}}\right)
+\mathcal{O}(\epsilon)\;.
\end{aligned}
\label{eq:QFS2_in_perp_integ}
\end{equation}
Note that we perform the $D\rightarrow 4$ expansion before taking the integration over $p_3^+$, since in our scheme, only the transverse integral (and thus the UV and collinear divergences) is regulated with dimensional regularization, whereas the $p_3^+$-integral is regulated with the lower cutoff $k_{\textrm{min}}^+$.

Thus,  eq.~\eqref{eq:QFSsquare_in_3} becomes:
\begin{equation}
\begin{aligned}
  \mathcal{V}_{|{\mathrm{QFS}}|^2}^{\textrm{in}}
   & =
  \int_{k_{\textrm{min}}^+}^{p_{j1}^{+}}
  \frac{\mathrm{d}p_{3}^{+}}{2p_{3}^{+}} \; \frac{1}{ (p_{j1}^{+})^{2}}
   \Big[4 (p_{j1}^{+})^{2} - 4 p_{j1}^{+}p_{3}^{+} +2(1-\epsilon) (p_{3}^{+})^2
\Big]
\;
\\
 & \times
 \bigg[-\frac{1}{\epsilon_{\mathrm{coll}}}+\gamma_{E}+\ln\frac{\mathbf{p}_{j1}^{2}}{4\pi\mu^{2}}
+2\ln R+2\ln\frac{p_{3}^{+}}{p_{j1}^{+}}+2\ln\left(\frac{p_{j1}^{+}-p_{3}^{+}}{p_{j1}^{+}}\right)
+\mathcal{O}(\epsilon)
 \bigg]
\\
   & =
 \left(-\frac{3}{2}+2\ln\frac{p_{j1}^{+}}{k_{\mathrm{min}}^{+}}\right)
 \left(-\frac{1}{\epsilon_{\mathrm{coll}}}+\gamma_{E}+\ln\frac{\mathbf{p}_{j1}^{2}}{4\pi\mu^{2}}+2\ln R\right)\\
 &-2\left(\ln\frac{p_{j1}^{+}}{k_{\mathrm{min}}^{+}}\right)^{2}
 -\frac{2\pi^{2}}{3}+\frac{13}{2}+\mathcal{O}(\epsilon)
 \; .
\end{aligned}
\label{eq:QFSsquare_in_4}
\end{equation}
As explained in section~\ref{sec:kine}, $k_{\mathrm{min}}^{+}$ plays a double role in our calculation. On the one hand it is used as a regulator for the integrals in $p^+_3$ or $k^+_3$. On the other hand, it is used to specify the physical scale set by the target at which the low-$x$ evolution is starting, or equivalently to encode the dependence on the total energy $\sqrt{s}$ of the collision. The JIMWLK evolution resums single high-energy logarithms, written as logarithms of $k_{\mathrm{min}}^+$,
and unlike the jet radius parameter $R$ does not depend on details of the process. Hence, in the result \eqref{eq:QFSsquare_in_4}, the term in $\ln^2k_{\mathrm{min}}^+$ and the term in
$\ln R\,  \ln k_{\mathrm{min}}^+$ can definitely not be subtracted and resummed by JIMWLK. Instead, these two terms in eq.~\eqref{eq:QFSsquare_in_4} should be understood as manifestations of the standard soft-collinear double logarithmic divergence for final state gluon radiation in our hybrid regularization scheme, having nothing to do with rapidity divergences and low-$x$ evolution. In the remainder of this section, we will see how these $\ln^2k_{\mathrm{min}}^+$ and
$\ln R\,  \ln k_{\mathrm{min}}^+$ terms as well as the collinear $1/\epsilon_{\mathrm{coll}}$ pole are canceled by other contributions, leaving only the expected single high-energy logarithm to be subtracted and resummed by JIMWLK, following section~\ref{sec:JIMWLK}.

 For the diagram $|{\mathrm{\overline{Q}FS}}|^2$, the contribution from the regime in which the antiquark and the gluon are merged into the same jet can be calculated in the same way, leading to
\begin{equation}
\begin{aligned}
  \mathcal{V}_{|{\mathrm{\overline{Q}FS}}|^2}^{\textrm{in}}
   & =
 \left(-\frac{3}{2}+2\ln\frac{p_{j2}^{+}}{k_{\mathrm{min}}^{+}}\right)
 \left(-\frac{1}{\epsilon_{\mathrm{coll}}}+\gamma_{E}+\ln\frac{\mathbf{p}_{j2}^{2}}{4\pi\mu^{2}}+2\ln R\right)
\\& -2\left(\ln\frac{p_{j2}^{+}}{k_{\mathrm{min}}^{+}}\right)^{2}
 -\frac{2\pi^{2}}{3}+\frac{13}{2}+\mathcal{O}(\epsilon)
 \; .
\end{aligned}
\label{eq:QbarFSsquare_in_4}
\end{equation}


\subsection{Contribution $|\mathrm{QFS}|^{2}$; out}
Let us now consider the other contribution to the dijet cross section from the $|\mathrm{QFS}|^{2}$ diagram, in which the quark and the gluon are forming separate jets according to our jet definition.
In that case, the quark and antiquark momenta are identified with the momenta of the measured jets, as
$\vec{p}_{1}=\vec{p}_{j1}$ and $\vec{p}_{2}=\vec{p}_{j2}$. Using again the notation \eqref{eq:def_P3_shifted_mom}, the expression \eqref{eq:denom_QFS2_in} for the denominator in eq.~\eqref{eq:QFSsquared} stays valid. By contrast, the $\mathbf{x}_{11^{\prime}}$ dependent phase factor in eq.~\eqref{eq:QFSsquared} now becomes:
\begin{equation}
\begin{aligned}
e^{-i(\mathbf{p}_{1}+\mathbf{p}_{3})\cdot\mathbf{x}_{11^{\prime}}}
& =
e^{-i(\mathbf{p}_{j1}+\mathbf{p}_{3})\cdot\mathbf{x}_{11^{\prime}}}
=
e^{-i\mathbf{p}_{j1}\cdot\mathbf{x}_{11^{\prime}}} e^{-i\mathbf{P}_{3}\cdot\mathbf{x}_{11^{\prime}}}
e^{-i\frac{p_{3}^{+}}{p_{j1}^{+}}  \mathbf{p}_{j1}\cdot\mathbf{x}_{11^{\prime}}}
\;.
\end{aligned}
\label{eq:phases_QFS2_out}
\end{equation}
With all this, the squared amplitude $|\mathcal{M}_{\mathrm{QFS}}\big|^{2}$  from eq.~\eqref{eq:QFSsquared} can be written in the case of three separate jets as
\begin{equation}
\begin{aligned}
\big|\mathcal{M}_{\mathrm{QFS}}\big|^{2}\bigg|_{\textrm{out}}
&=
\big|\mathcal{M}_{\mathrm{LO}}\big|^{2}\times 4\pi\alpha_{s}C_{F}
\frac{(p_{3}^{+})^{2}}{ (p_{j1}^{+})^{2}}
\Big[\Big(1+2\frac{p_{j1}^{+}}{p_{3}^{+}}\Big)^{2}+(D-3)\Big]
\frac{1}{ {\mathbf{P}_{3}^{2}}} e^{-i\mathbf{P}_{3}\cdot\mathbf{x}_{11^{\prime}}}
\\
& \times
\frac{\bigg[2\Big(\frac{p_{j1}^{+}+p_{3}^{+}}{q^{+}}\Big)^{2}
 +2\Big(\frac{p_{j2}^{+}}{q^{+}}\Big)^{2}+(D-4)\bigg]}{\bigg[2\Big(\frac{p_{j1}^{+}}{q^{+}}\Big)^{2}
 +2\Big(\frac{p_{j2}^{+}}{q^{+}}\Big)^{2}+(D-4)\bigg]}
e^{-i\frac{p_{3}^{+}}{p_{j1}^{+}}  \mathbf{p}_{j1}\cdot\mathbf{x}_{11^{\prime}}}
\;,
\end{aligned}
\label{eq:QFS2_ampl_square_factorization_out}
\end{equation}
with $\big|\mathcal{M}_{\mathrm{LO}}\big|^{2}$ written as in eq.~\eqref{eq:MLOsquared}.
Note that eq.~\eqref{eq:QFS2_ampl_square_factorization_out} is, once again, a slight abuse of notation, since the factorization actually happens happens at the integrand level and the integrations over $\mathbf{x}_{1}$ and $\mathbf{x}_{1^{\prime}}$ are hidden inside $|\mathcal{M}_{\mathrm{LO}}|^{2}$.

Moreover, from the definition \eqref{eq:jetdef_2}, one has now
\begin{equation}
\begin{aligned}
1-\theta_{\mathrm{in}}(\vec{p}_{j1},\vec{p}_{3})
& =
\theta\Bigg(
(p_{j1}^{+}+p_{3}^{+})^{2}
\Big(\frac{\mathbf{p}_{3}}{p_{3}^{+}}
-\frac{\mathbf{p}_{j1}}{p_{j1}^{+}}\Big)^{2}
-
(\mathbf{p}_{j1}+\mathbf{p}_{3})^{2}R^{2}
\Bigg)
\\
& =
\theta\Bigg(
{\mathbf{P}_{3}^{2}}
-
\frac{(p_{3}^{+})^2}{(p_{j1}^{+}+p_{3}^{+})^{2}}
\Big(\frac{(p_{j1}^{+}+p_{3}^{+})}{p_{j1}^{+}}  \mathbf{p}_{j1}  +\mathbf{P}_{3}\Big)^{2}R^{2}
\Bigg)
\end{aligned}
\label{eq:jetdef_out}
\end{equation}
In the narrow jet $R\rightarrow 0$ limit, this theta function changes values from $0$ to $1$ at a parametrically small value of $|\mathbf{P}_{3}|$, which scales as $R$. Hence, in that limit, $\mathbf{P}_{3}$ is negligible compared to $\mathbf{p}_{j1}$, so that
\begin{equation}
\begin{aligned}
1-\theta_{\mathrm{in}}(\vec{p}_{j1},\vec{p}_{3})
& \rightarrow
\theta\Bigg(
{\mathbf{P}_{3}^{2}}
-
\frac{(p_{3}^{+})^2}{(p_{j1}^{+})^2}
 \mathbf{p}_{j1}^{2} R^{2}
\Bigg)
\;.
\end{aligned}
\label{eq:jetdef_out_narrow}
\end{equation}

Inserting the expressions \eqref{eq:QFS2_ampl_square_factorization_out} and \eqref{eq:jetdef_out_narrow} into eq.~\eqref{eq:QFSsquare_out}, we can write the contribution to the dijet cross section from $|\mathrm{QFS}|^{2}$ with three separate jets as
\begin{equation}
\begin{aligned}
&\frac{\mathrm{d}\sigma_{\textrm{dijet}}^{|{\mathrm{QFS}}|^2;\textrm{ out}}}{
\mathrm{d}p_{j1}^{+}\mathrm{\mathrm{d}}^{D-2}\mathbf{p}_{j1}
\mathrm{d}p_{j2}^{+}\mathrm{\mathrm{d}}^{D-2}\mathbf{p}_{j2}
}\\
 & =
\frac{\theta(p_{j1}^{+})}{(2\pi)^{D-1}2p_{j1}^{+}}
\frac{\theta(p_{j2}^{+})}{(2\pi)^{D-1}2p_{j2}^{+}}
\frac{1}{2q^{+}}
 \frac{1}{(D-2)}|\mathcal{M}_{\mathrm{LO}}|^{2}
4\pi\alpha_{s}C_{F}
 \int\mathrm{PS}(\vec{p}_3)
\\
 & \times
 2\pi\delta(q^{+}-p_{j1}^{+}-p_{j2}^{+}-p_{3}^{+})
\theta\Bigg({\mathbf{P}_{3}^{2}}-\frac{(p_{3}^{+})^2}{(p_{j1}^{+})^2}\mathbf{p}_{j1}^{2} R^{2}
\Bigg)
\frac{1}{ {\mathbf{P}_{3}^{2}}} e^{-i\mathbf{P}_{3}\cdot\mathbf{x}_{11^{\prime}}}
e^{-i\frac{p_{3}^{+}}{p_{j1}^{+}}  \mathbf{p}_{j1}\cdot\mathbf{x}_{11^{\prime}}}
\\
& \times
\frac{\bigg[2\Big(\frac{p_{j1}^{+}+p_{3}^{+}}{q^{+}}\Big)^{2}
 +2\Big(\frac{p_{j2}^{+}}{q^{+}}\Big)^{2}+(D-4)\bigg]}{\bigg[2\Big(\frac{p_{j1}^{+}}{q^{+}}\Big)^{2}
 +2\Big(\frac{p_{j2}^{+}}{q^{+}}\Big)^{2}+(D-4)\bigg]}
 \frac{(p_{3}^{+})^{2}}{ (p_{j1}^{+})^{2}}
\Big[\Big(1+2\frac{p_{j1}^{+}}{p_{3}^{+}}\Big)^{2}+(D-3)\Big]
  \;.
\end{aligned}
\label{eq:QFSsquare_out_2}
\end{equation}
In eq.~\eqref{eq:QFSsquare_out_2}, the integral over $\mathbf{P}_{3}$ is actually finite, since the phase prevents UV divergences to occur, and the theta function is cutting off the collinear regime, preventing the gluon to belong to the quark jet. Hence, dimensional regularization is actually not necessary here and one can take $D= 4-2\epsilon \rightarrow 4$. The transverse integral in eq.~\eqref{eq:QFSsquare_out_2} can be calculated as
 \begin{equation}
\begin{aligned}
 &
 4\pi \mu^{4-D}\int\frac{\mathrm{d}^{D-2}\mathbf{P}_{3}}{(2\pi)^{D-2}}
 \frac{1}{{\mathbf{P}_{3}^{2}}} e^{-i\mathbf{P}_{3}\cdot\mathbf{x}_{11^{\prime}}}
 \theta\Bigg({\mathbf{P}_{3}^{2}}-\frac{(p_{3}^{+})^2}{(p_{j1}^{+})^2}\mathbf{p}_{j1}^{2} R^{2}
\Bigg)
 \\
& =
2\int_{0}^{+\infty} \frac{\mathrm{d}|\mathbf{P}_{3}|}{|\mathbf{P}_{3}|} \textrm{J}_0\big(|\mathbf{P}_{3}| |\mathbf{x}_{11^{\prime}}|\big) \theta\bigg(|\mathbf{P}_{3}|-\frac{p_{3}^{+}}{p_{j1}^{+}}|\mathbf{p}_{j1}| R
\bigg)\;
 \\
& =
-2\ln R -\ln\left(\frac{\mathbf{p}_{j1}^{2}\mathbf{x}_{11^{\prime}}^{2}}{{c_{0}^{2}}}\right)
-2\ln\frac{p_{3}^{+}}{p_{j1}^{+}}
+\mathcal{O}(R^2)
\;,
\end{aligned}
\label{eq:QFS2_out_perp_integ}
\end{equation}
in the narrow-jet limit $R\rightarrow 0$. In eq.~\eqref{eq:QFS2_out_perp_integ}, $\textrm{J}_0$ is the Bessel function of the first kind, and $c_0\equiv 2 e^{-\gamma_E}$.
At this stage, eq.~\eqref{eq:QFSsquare_out_2} becomes
\begin{equation}
\begin{aligned}
&\frac{\mathrm{d}\sigma_{\textrm{dijet}}^{|{\mathrm{QFS}}|^2;\textrm{ out}}}{
\mathrm{d}p_{j1}^{+}\mathrm{\mathrm{d}}^{2}\mathbf{p}_{j1}
\mathrm{d}p_{j2}^{+}\mathrm{\mathrm{d}}^{2}\mathbf{p}_{j2}
}\\
 & =
\frac{\theta(p_{j1}^{+})}{(2\pi)^{3}2p_{j1}^{+}}
\frac{\theta(p_{j2}^{+})}{(2\pi)^{3}2p_{j2}^{+}}
\frac{1}{2q^{+}}
 \frac{1}{2}|\mathcal{M}_{\mathrm{LO}}|^{2}
\frac{\alpha_{s}C_{F}}{2\pi}
 \int_{k^+_{\textrm{min}}}^{+\infty}\frac{\mathrm{d}p_{3}^{+}}{p_{3}^{+}}
 e^{-i\frac{p_{3}^{+}}{p_{j1}^{+}}  \mathbf{p}_{j1}\cdot\mathbf{x}_{11^{\prime}}}
\\
 & \times
 2\pi\delta(q^{+}-p_{j1}^{+}-p_{j2}^{+}-p_{3}^{+})
 \left[-2\ln R -\ln\left(\frac{\mathbf{p}_{j1}^{2}\mathbf{x}_{11^{\prime}}^{2}}{{c_{0}^{2}}}\right)
-2\ln\frac{p_{3}^{+}}{p_{j1}^{+}}\right]
\\
& \times
\frac{\big[(p_{j1}^{+}+p_{3}^{+})^{2}
 +(p_{j2}^{+})^{2}\big]}{\big[(p_{j1}^{+})^{2}
 +(p_{j2}^{+})^{2}\big]}
\bigg[2+2\frac{p_{3}^{+}}{p_{j1}^{+}}+\frac{(p_{3}^{+})^{2}}{ (p_{j1}^{+})^{2}}
\bigg]
  \;.
\end{aligned}
\label{eq:QFSsquare_out_3}
\end{equation}
In this expression, we could simply use the delta function in order to perform the $p_3^+$ integration. However, the result would be proportional to $1/(q^{+}-p_{j1}^{+}-p_{j2}^{+})$, which could diverge depending on the kinematics of the jets, unless the cutoff $k^+_{\textrm{min}}$ is taken into account. In order to obtain a well behaved result and to extract the sensitivity of the expression \eqref{eq:QFSsquare_out_3} on the cutoff $k^+_{\textrm{min}}$, let us split it into several pieces.
First, let us take the integrand at $p_3^+\rightarrow 0$, including in the delta function. The corresponding piece of eq.~\eqref{eq:QFSsquare_out_3} can be written as
\begin{equation}
\begin{aligned}
\frac{\mathrm{d}\sigma_{\textrm{dijet}}^{|{\mathrm{QFS}}|^2;\textrm{ out; soft}}}{
\mathrm{d}p_{j1}^{+}\mathrm{\mathrm{d}}^{2}\mathbf{p}_{j1}
\mathrm{d}p_{j2}^{+}\mathrm{\mathrm{d}}^{2}\mathbf{p}_{j2}
}
 & =
\frac{\mathrm{d}\sigma_{\textrm{dijet}}^{{\mathrm{LO}}}}{
\mathrm{d}p_{j1}^{+}\mathrm{\mathrm{d}}^{2}\mathbf{p}_{j1}
\mathrm{d}p_{j2}^{+}\mathrm{\mathrm{d}}^{2}\mathbf{p}_{j2}
} \times
\frac{\alpha_{s}C_{F}}{2\pi}
\mathcal{V}_{|{\mathrm{QFS}}|^2}^{\textrm{out; soft}}
\end{aligned}
\label{eq:QFSsquare_out_soft_1}
\end{equation}
where
\begin{equation}
\begin{aligned}
 \mathcal{V}_{|{\mathrm{QFS}}|^2}^{\textrm{out; soft}}
 & =
2 \int_{k^+_{\textrm{min}}}^{p_{j1}^{+}}\frac{\mathrm{d}p_{3}^{+}}{p_{3}^{+}}
  \left[-2\ln R -\ln\left(\frac{\mathbf{p}_{j1}^{2}\mathbf{x}_{11^{\prime}}^{2}}{{c_{0}^{2}}}\right)
-2\ln\frac{p_{3}^{+}}{p_{j1}^{+}}\right]
\\
 & =
 2\left(\ln\frac{p_{j1}^{+}}{k_{\mathrm{min}}^{+}}\right)^{2}
  -4 \ln\left(\frac{p_{j1}^{+}}{k_{\mathrm{min}}^{+}}\right) \ln R
-2   \ln\left(\frac{p_{j1}^{+}}{k_{\mathrm{min}}^{+}}\right)
\ln\left(\frac{\mathbf{p}_{j1}^{2}\mathbf{x}_{11^{\prime}}^{2}}{{c_{0}^{2}}}\right)
  \; ,
\end{aligned}
\label{eq:QFSsquare_out_soft_2}
\end{equation}
introducing $p_{j1}^{+}$ as an upper bound in this soft contribution. The first two terms in eq.~\eqref{eq:QFSsquare_out_soft_2}: in $\ln^2k_{\mathrm{min}}^+$ and
$\ln R\,  \ln k_{\mathrm{min}}^+$, precisely cancel the ones found in from the in-jet radiation contribution \eqref{eq:QFSsquare_in_4}. As discussed earlier, these terms cannot be associated with high-energy evolution but are, instead, manifestations of soft-collinear divergences in our regularization scheme. Their cancellation is a crucial requirement for the consistency of our hybrid regularization scheme in which we combine transverse dimensional regularization with a lower cutoff for $p_{3}^{+}$.

Subtracting the term \eqref{eq:QFSsquare_out_soft_1} from  \eqref{eq:QFSsquare_out_3} is sufficient to remove any divergence from the $p_{3}^{+}$ integral, or more precisely any logarithmic sensitivity on $k_{\mathrm{min}}^{+}$, so that one can remove the cutoff $k_{\mathrm{min}}^{+}$ in the leftover piece.
Nevertheless, the finite $p_{3}^{+}$ integral can still produce a potentially large logarithm. The reason for this is that the $p_{3}^{+}\rightarrow 0$ approximation of the integrand of eq.~\eqref{eq:QFSsquare_out_3} in order to obtain \eqref{eq:QFSsquare_out_soft_1} is not always a good approximation for small but finite  $p_{3}^{+}\ll p_{j1}^{+}\;, p_{j2}^{+}$. Indeed, in this regime, the smallness of  $p_{3}^{+}$ compared to  $p_{j1}^{+}$ can be compensated for very large values of $|\mathbf{p}_{j1}\cdot\mathbf{x}_{11^{\prime}}|$, making the phase factor in eq.~\eqref{eq:QFSsquare_out_3} non trivial even at small $p_{3}^{+}$. In order to have control on this issue, let us extract as well from eq.~\eqref{eq:QFSsquare_out_3} the contribution
\begin{equation}
\begin{aligned}
\frac{\mathrm{d}\sigma_{\textrm{dijet}}^{|{\mathrm{QFS}}|^2;\textrm{ out; phase}}}{
\mathrm{d}p_{j1}^{+}\mathrm{\mathrm{d}}^{2}\mathbf{p}_{j1}
\mathrm{d}p_{j2}^{+}\mathrm{\mathrm{d}}^{2}\mathbf{p}_{j2}
}
 & =
\frac{\mathrm{d}\sigma_{\textrm{dijet}}^{{\mathrm{LO}}}}{
\mathrm{d}p_{j1}^{+}\mathrm{\mathrm{d}}^{2}\mathbf{p}_{j1}
\mathrm{d}p_{j2}^{+}\mathrm{\mathrm{d}}^{2}\mathbf{p}_{j2}
} \times
\frac{\alpha_{s}C_{F}}{2\pi}
 \mathcal{V}_{|{\mathrm{QFS}}|^2}^{\textrm{out; phase}}
\end{aligned}
\label{eq:QFSsquare_out_phase_1}
\end{equation}
where
\begin{equation}
\begin{aligned}
 \mathcal{V}_{|{\mathrm{QFS}}|^2}^{\textrm{out; phase}}
 & =
 2\int_{0}^{p_{j1}^{+}}\frac{\mathrm{d}p_{3}^{+}}{p_{3}^{+}}
 \left[e^{-i\frac{p_{3}^{+}}{p_{j1}^{+}}  \mathbf{p}_{j1}\cdot\mathbf{x}_{11^{\prime}}}-1\right]
  \left[-2\ln R -\ln\left(\frac{\mathbf{p}_{j1}^{2}\mathbf{x}_{11^{\prime}}^{2}}{{c_{0}^{2}}}\right)
-2\ln\frac{p_{3}^{+}}{p_{j1}^{+}}\right]
\\
& =
 2\int_{0}^{1}\frac{\mathrm{d}\xi}{\xi}
 \Big[e^{-i\xi  \mathbf{p}_{j1}\cdot\mathbf{x}_{11^{\prime}}}-1\Big]
  \left[-2\ln R -\ln\left(\frac{\mathbf{p}_{j1}^{2}\mathbf{x}_{11^{\prime}}^{2}}{{c_{0}^{2}}}\right)
-2\ln\xi\right]
  \; .
\end{aligned}
\label{eq:QFSsquare_out_phase_2}
\end{equation}
The above contribution will be further studied in section~\ref{sec:corrlimit}, in the back-to-back jets limit.

Finally, we call \emph{regular} the leftover part of eq.~\eqref{eq:QFSsquare_out_3} obtained after subtracting
the terms \eqref{eq:QFSsquare_out_soft_1} and \eqref{eq:QFSsquare_out_phase_1}, as
\begin{equation}
\begin{aligned}
\mathrm{d}\sigma_{\textrm{dijet}}^{|{\mathrm{QFS}}|^2;\textrm{ out; reg}}&=\mathrm{d}\sigma_{\textrm{dijet}}^{|{\mathrm{QFS}}|^2;\textrm{ out}}-\mathrm{d}\sigma_{\textrm{dijet}}^{|{\mathrm{QFS}}|^2;\textrm{ out; soft}}-\mathrm{d}\sigma_{\textrm{dijet}}^{|{\mathrm{QFS}}|^2;\textrm{ out; phase}}\;,
\end{aligned}
\label{eq:QFSsquare_out_reg_1}
\end{equation}
since in this contribution, the integration over $p_{3}^{+}$ cannot produce a divergent result even without cutoff and cannot produce further large logarithms. It can be written explicitly as
\begin{equation}
\begin{aligned}
&\frac{\mathrm{d}\sigma_{\textrm{dijet}}^{|{\mathrm{QFS}}|^2;\textrm{ out; reg}}}{
\mathrm{d}p_{j1}^{+}\mathrm{\mathrm{d}}^{2}\mathbf{p}_{j1}
\mathrm{d}p_{j2}^{+}\mathrm{\mathrm{d}}^{2}\mathbf{p}_{j2}
}
=
\frac{\theta(p_{j1}^{+})}{(2\pi)^{3}2p_{j1}^{+}}
\frac{\theta(p_{j2}^{+})}{(2\pi)^{3}2p_{j2}^{+}}
\frac{1}{2q^{+}}
 \frac{1}{2}|\mathcal{M}_{\mathrm{LO}}|^{2}\\
  &\qquad\qquad\times\frac{\alpha_{s}C_{F}}{2\pi}
 \int_{0}^{+\infty}
   \frac{\mathrm{d}\xi}{\xi}
 e^{-i\xi  \mathbf{p}_{j1}\cdot\mathbf{x}_{11^{\prime}}}
 \left[-2\ln R -\ln\left(\frac{\mathbf{p}_{j1}^{2}\mathbf{x}_{11^{\prime}}^{2}}{{c_{0}^{2}}}\right)
-2\ln\xi\right]
\\
 &\qquad\qquad\times
\bigg\{
2\pi\delta(q^{+}-(1+\xi)p_{j1}^{+}-p_{j2}^{+})
\bigg[1+
\frac{(2\xi+\xi^2)(p_{j1}^{+})^{2}}{(p_{j1}^{+})^{2}
 +(p_{j2}^{+})^{2}}
 \bigg]
\big(1+(1+\xi)^2\big)
\\
&\qquad\qquad\qquad-4\pi \delta(q^{+}-p_{j1}^{+}-p_{j2}^{+}) \theta(1 -\xi)
\bigg\}
\;.
\end{aligned}
\label{eq:QFSsquare_out_reg_2}
\end{equation}

For the diagram $|{\mathrm{\overline{Q}FS}}|^2$, the contribution from the regime in which the antiquark and the gluon form separate jet can be split in the same way into three contributions, and one obtains
\begin{equation}
\begin{aligned}
 \mathcal{V}_{|{\mathrm{\overline{Q}FS}}|^2}^{\textrm{out; soft}}
 = &\mathcal{V}_{|{\mathrm{QFS}}|^2}^{\textrm{out; soft}} (1\leftrightarrow2)\; ,
\end{aligned}
\label{eq:QbarFSsquare_out_soft_2}
\end{equation}
\begin{equation}
\begin{aligned}
 \mathcal{V}_{|{\mathrm{\overline{Q}FS}}|^2}^{\textrm{out; phase}}
 = \mathcal{V}_{|{\mathrm{QFS}}|^2}^{\textrm{out; phase}}(1\leftrightarrow2) \;,
\end{aligned}
\label{eq:QbarFSsquare_out_phase_2}
\end{equation}
and
\begin{equation}
\begin{aligned}
\frac{\mathrm{d}\sigma_{\textrm{dijet}}^{|{\mathrm{\overline{Q}FS}}|^2;\textrm{ out; reg}}}{
\mathrm{d}p_{j1}^{+}\mathrm{\mathrm{d}}^{2}\mathbf{p}_{j1}
\mathrm{d}p_{j2}^{+}\mathrm{\mathrm{d}}^{2}\mathbf{p}_{j2}}&=\frac{\mathrm{d}\sigma_{\textrm{dijet}}^{|{\mathrm{QFS}}|^2;\textrm{ out; reg}}}{
\mathrm{d}p_{j1}^{+}\mathrm{\mathrm{d}}^{2}\mathbf{p}_{j1}
\mathrm{d}p_{j2}^{+}\mathrm{\mathrm{d}}^{2}\mathbf{p}_{j2}}(1\leftrightarrow2)\;.
\end{aligned}
\label{eq:QbarFSsquare_out_reg_2}
\end{equation}


\subsection{Cancellation of collinear and soft divergences}

As already discussed, soft and collinear divergences manifest themselves in various ways when $|{\mathrm{QFS}}|^2$ is calculated using transverse dimensional regularization and a cutoff $k_{\mathrm{min}}^{+}$.
In addition to the $1/\epsilon$ collinear pole in the inside jet radiation contribution \eqref{eq:QFSsquare_in_4}, both eqs.~\eqref{eq:QFSsquare_in_4} and \eqref{eq:QFSsquare_out_soft_2} contain terms of soft-collinear origin, namely in $\ln^2k_{\mathrm{min}}^+$ and in $\ln R\,  \ln k_{\mathrm{min}}^+$. It is then difficult to disentangle single logarithms $\ln k_{\mathrm{min}}^+$ associated with either soft or rapidity divergences.

Adding eqs.~\eqref{eq:QFSsquare_in_4} and \eqref{eq:QFSsquare_out_soft_2} together, the terms in $\ln^2k_{\mathrm{min}}^+$ and in $\ln R\,  \ln k_{\mathrm{min}}^+$ cancel, and one obtains
\begin{equation}
\begin{aligned}
  \mathcal{V}_{|{\mathrm{QFS}}|^2}^{\textrm{in}} +  \mathcal{V}_{|{\mathrm{QFS}}|^2}^{\textrm{out; soft}}
   & =
 \left[\frac{3}{2}- 2\ln\left(\frac{p_{j1}^{+}}{k_{\mathrm{min}}^{+}}\right)\right]
 \left[\frac{1}{\epsilon_{\mathrm{coll}}}+\gamma_{E}
+\ln\left(\pi\mu^{2}\mathbf{x}_{11^{\prime}}^{2}\right)\right]
   \\
   &
-\frac{3}{2}
\ln\left(\frac{\mathbf{p}_{j1}^{2}\mathbf{x}_{11^{\prime}}^{2}}{{c_{0}^{2}}}\right)
-3\ln R-\frac{2\pi^{2}}{3}+\frac{13}{2}
 +\mathcal{O}(\epsilon)
 \; .
\end{aligned}
\label{eq:QFSsquare_in_plus_soft_out}
\end{equation}
The other two contributions \eqref{eq:QFSsquare_out_phase_2} and \eqref{eq:QFSsquare_out_reg_2} from $|{\mathrm{QFS}}|^2$ do not contain any divergence.
Similarly, for the diagram $|{\mathrm{\overline{Q}FS}}|^2$, one finds
\begin{equation}
\begin{aligned}
  \mathcal{V}_{|{\mathrm{\overline{Q}FS}}|^2}^{\textrm{in}} +  \mathcal{V}_{|{\mathrm{\overline{Q}FS}}|^2}^{\textrm{out; soft}}
   & =
 \left[\frac{3}{2}- 2\ln\left(\frac{p_{j2}^{+}}{k_{\mathrm{min}}^{+}}\right)\right]
 \left[\frac{1}{\epsilon_{\mathrm{coll}}}+\gamma_{E}
+\ln\left(\pi\mu^{2}\mathbf{x}_{22^{\prime}}^{2}\right)\right]
   \\
   &
-\frac{3}{2}
\ln\left(\frac{\mathbf{p}_{j2}^{2}\mathbf{x}_{22^{\prime}}^{2}}{{c_{0}^{2}}}\right)
-3\ln R-\frac{2\pi^{2}}{3}+\frac{13}{2}
 +\mathcal{O}(\epsilon)
 \; .
\end{aligned}
\label{eq:QbarFSsquare_in_plus_soft_out}
\end{equation}

In the NLO virtual corrections to DIS dijets, the only contribution containing a collinear divergence is $\mathrm{FSIR}$, see eq.~\eqref{eq:VFSIR}, which factorizes as well from the LO cross section at integrand level. That contribution correspond to the IR part of the self energy diagrams for the quark and for the antiquark in the  final state. With the identification of the momenta of partons and jets, it can be written as
\begin{equation}
\begin{aligned}
\mathcal{V}_{\mathrm{FSIR}}
& =
\left[\frac{1}{\epsilon_{\mathrm{coll}}}+\gamma_{E}+\mathrm{ln}(\pi\mu^{2}\mathbf{x}_{12}^{2})\right]
\left[-\frac{3}{2}+\mathrm{ln}\left(\frac{p_{j1}^{+}}{k_{\mathrm{min}}^{+}}\right)
+\mathrm{ln}\left(\frac{p_{j2}^{+}}{k_{\mathrm{min}}^{+}}\right)  \right]
\;.
\end{aligned}
\label{eq:VFSIR_1}
\end{equation}
Eq.~\eqref{eq:VFSIR_1} is corresponds to a virtual correction in the amplitude. We should also consider that virtual correction in the complex conjugate amplitude, which is
\begin{equation}
\begin{aligned}
\mathcal{V}_{\mathrm{FSIR}^{\dagger}}
& =
\left[\frac{1}{\epsilon_{\mathrm{coll}}}+\gamma_{E}+\mathrm{ln}(\pi\mu^{2}\mathbf{x}_{1'2'}^{2})\right]
\left[-\frac{3}{2}+\mathrm{ln}\left(\frac{p_{j1}^{+}}{k_{\mathrm{min}}^{+}}\right)
+\mathrm{ln}\left(\frac{p_{j2}^{+}}{k_{\mathrm{min}}^{+}}\right)  \right]
\;.
\end{aligned}
\label{eq:VFSIR_dag_1}
\end{equation}
The only difference between the two is the dependence on the transverse coordinates $\mathbf{x}_{1}$ and $\mathbf{x}_{2}$ of the Wilson lines in the amplitude, or on the transverse coordinates $\mathbf{x}_{1'}$ and $\mathbf{x}_{2'}$ of the Wilson lines in the amplitude.

In the sum of the contributions \eqref{eq:QFSsquare_in_plus_soft_out}, \eqref{eq:QbarFSsquare_in_plus_soft_out}, \eqref{eq:VFSIR_1} and \eqref{eq:VFSIR_dag_1}, the collinear poles cancel, and one finds
\begin{equation}
\begin{aligned}
  \mathcal{V}_{{\mathrm{jet}}}
  \equiv &
  \mathcal{V}_{|{\mathrm{QFS}}|^2}^{\textrm{in}}
  +  \mathcal{V}_{|{\mathrm{QFS}}|^2}^{\textrm{out; soft}}
  + \mathcal{V}_{|{\mathrm{\overline{Q}FS}}|^2}^{\textrm{in}}
  +  \mathcal{V}_{|{\mathrm{\overline{Q}FS}}|^2}^{\textrm{out; soft}}
  +\mathcal{V}_{\mathrm{FSIR}}  + \mathcal{V}_{\mathrm{FSIR}^{\dagger}}
 \\
& =
 -2  \ln\left(\frac{p_{j1}^{+}}{k_{\mathrm{min}}^{+}}\right)
 \ln\left(\frac{\mathbf{x}_{11^{\prime}}^{2}}{|\mathbf{x}_{12}|\, |\mathbf{x}_{1'2'}|}\right)
  -2  \ln\left(\frac{p_{j2}^{+}}{k_{\mathrm{min}}^{+}}\right)
 \ln\left(\frac{\mathbf{x}_{22^{\prime}}^{2}}{|\mathbf{x}_{12}|\, |\mathbf{x}_{1'2'}|}\right)
\\
 &
 -3 \ln\left(\frac{|\mathbf{p}_{j1}|\, |\mathbf{p}_{j2}|\, |\mathbf{x}_{12}|\, |\mathbf{x}_{1'2'}|}{{c_{0}^{2}}}\right)
 -6  \ln R
-\frac{4\pi^{2}}{3}+13
\; .
\end{aligned}
\label{eq:total_VFS}
\end{equation}
In this expression, all soft and collinear divergences have canceled~\cite{FaridJyvaskyla}. Only single logarithmic dependence on $k_{\mathrm{min}}^{+}$ remains, which now correspond to high-energy evolution. This can be shown as follows. Introducing the factorization scale $k_f^+$,
one can isolate the dependence on the cutoff $k_{\mathrm{min}}^{+}$ as
\begin{equation}
\begin{aligned}
  \mathcal{V}_{{\mathrm{jet}}}
& =
 2  \ln\left(\frac{k_f^{+}}{k_{\mathrm{min}}^{+}}\right)
 \ln\left(\frac{\mathbf{x}_{12}^{2} \mathbf{x}_{1'2'}^{2}}{\mathbf{x}_{11^{\prime}}^{2} \mathbf{x}_{22^{\prime}}}\right)
 -2  \ln\left(\frac{p_{j1}^{+}}{k_f^{+}}\right)
 \ln\left(\frac{\mathbf{x}_{11^{\prime}}^{2}}{|\mathbf{x}_{12}|\, |\mathbf{x}_{1'2'}|}\right)\\
  &-2  \ln\left(\frac{p_{j2}^{+}}{k_f^{+}}\right)
 \ln\left(\frac{\mathbf{x}_{22^{\prime}}^{2}}{|\mathbf{x}_{12}|\, |\mathbf{x}_{1'2'}|}\right)
 -3 \ln\left(\frac{|\mathbf{p}_{j1}|\, |\mathbf{p}_{j2}|\, |\mathbf{x}_{12}|\, |\mathbf{x}_{1'2'}|}{{c_{0}^{2}}}\right)\\
 &-6  \ln R
-\frac{4\pi^{2}}{3}+13
\; .
\end{aligned}
\label{eq:total_VFS_2}
\end{equation}
The first term in eq.~\eqref{eq:total_VFS_2} amounts to a multiplication of the $\textrm{LO}$ cross section
by
\begin{equation}
\begin{aligned}
&\frac{\alpha_s C_F}{2\pi} 2  \ln\left(\frac{k_f^{+}}{k_{\mathrm{min}}^{+}}\right)
 \ln\left(\frac{\mathbf{x}_{12}^{2} \mathbf{x}_{1'2'}^{2}}{\mathbf{x}_{11^{\prime}}^{2} \mathbf{x}_{22^{\prime}}}\right)\\
  &=
 2\alpha_s N_c \left(1-\frac{1}{N_c^2}\right) \ln\left(\frac{k_f^{+}}{k_{\mathrm{min}}^{+}}\right)
 \int_{\mathbf{x}_{3}}
 \Big(
 A^{\eta}(\mathbf{x}_{13})A^{\eta}(\mathbf{x}_{1^{\prime}3})
 \\
 &+A^{\eta}(\mathbf{x}_{23})A^{\eta}(\mathbf{x}_{2^{\prime}3})
 -A^{\eta}(\mathbf{x}_{13})A^{\eta}(\mathbf{x}_{23})
 -A^{\eta}(\mathbf{x}_{1^{\prime}3})A^{\eta}(\mathbf{x}_{2^{\prime}3})
 \Big)
\; .
\end{aligned}
\label{eq:JIMWLK_term_in_VFS}
\end{equation}
This corresponds indeed to the part of the JIMWLK evolution proportional to $\alpha_s C_F$ times the $\textrm{LO}$ cross section. Hence, in our resummation scheme for high-energy leading logs, the first term in eq.~\eqref{eq:total_VFS_2} is the part which is subtracted and resummed by the  JIMWLK evolution.


\section{\label{sec:crosssection}Inclusive dijet cross section}


In this section, we present the full differential NLO cross section for $\gamma A\to\mathrm{dijet}X$
in the CGC. It is given by the sum:
\begin{equation}
\begin{aligned}\mathrm{d}\sigma_{\mathrm{NLO}}^{\gamma A\to\mathrm{dijet}X} & =\mathrm{d}\sigma_{\mathrm{LO}}+\mathrm{d}\sigma_{\mathrm{jet}}+\mathrm{d}\sigma_{\mathrm{softGE}}+\mathrm{d}\sigma^{\mathrm{Sudakov}}_{\mathrm{real}}+\mathrm{d}\sigma_{\mathrm{virtual}}^{\mathrm{finite}}+\mathrm{d}\sigma_{\mathrm{real}}^{\mathrm{finite}}\;,\end{aligned}
\label{eq:totalcrosssection}
\end{equation}
where it is understood that the rapidity divergences are absorbed into JIMWLK according to eq. \eqref{eq:JIMWLKsubtraction}. The cross section is written in terms of different contributions which are each separately soft- and collinear safe. We will shortly discuss each of the terms, and provide their explicit expressions in the next subsections.

The first part of the cross section is the leading-order one, given
in eq.~\eqref{eq:LOXsection}), where the outgoing partons are identified
with jets: $\vec{p}_{1,2}\to\vec{p}_{j1,j2}$.

Second, following the discussion in sections~\ref{sec:jet} and~\ref{sec:coll}, the real NLO corrections due to final-state gluon emission contain soft-collinear singularities. These singularities are cured when applying the jet algorithm and summing the contributions due to gluon emission inside the jet, soft gluon emission outside the jet, and the IR part of the virtual self-energy corrections to the final state:
\begin{equation}
\begin{aligned}
&\frac{\mathrm{d}\sigma_{\mathrm{jet}}}{
\mathrm{d}p_{j1}^{+}\mathrm{\mathrm{d}}^{2}\mathbf{p}_{j1}
\mathrm{d}p_{j2}^{+}\mathrm{\mathrm{d}}^{2}\mathbf{p}_{j2}
}
  =
 \frac{\mathrm{d}\sigma^{\textrm{dijet}}_{\mathrm{LO}}}{
\mathrm{d}p_{j1}^{+}\mathrm{\mathrm{d}}^{D-2}\mathbf{p}_{j1}
\mathrm{d}p_{j2}^{+}\mathrm{\mathrm{d}}^{D-2}\mathbf{p}_{j2}
}
\times
 \frac{\alpha_{s}C_{F}}{2\pi}
 \mathcal{V}_{\mathrm{jet}} \; .
\end{aligned}
\end{equation}
The explicit expression for $\mathcal{V}_{\mathrm{jet}}$ is given in eq~\eqref{eq:total_VFS_2}.

Third, as explained in section~\ref{sec:soft}, the contribution to the cross section due to diagram $\mathrm{GEFS},(ii)+\mathrm{IFS}$ and its $q\leftrightarrow\bar{q}$ counterpart contains a soft divergence. This singularity cancels with the one found in the interference between real final-state emission from the quark in the amplitude, and from the antiquark in the complex conjugate amplitude or vice versa. Summing both virtual and real contributions, one ends up with the following finite contribution to the cross section which we might call $\mathrm{softGE}$ for soft final-state gluon exchange:
\begin{equation}
\begin{aligned}
&\frac{\mathrm{d}\sigma^{}_{\mathrm{softGE}}}{\mathrm{d}p_{j1}^{+}\mathrm{\mathrm{d}}^{D-2}\mathbf{p}_{j1}\mathrm{d}p_{j2}^{+}\mathrm{\mathrm{d}}^{D-2}\mathbf{p}_{j2}}  =\frac{1}{2q^{+}}\frac{\theta(p_{j1}^{+})}{(2\pi)^{D-1}2p_{j1}^{+}}\frac{\theta(p_{j2}^{+})}{(2\pi)^{D-1}2p_{j2}^{+}}\\
 & \times\frac{1}{D-2}\Bigg[2\pi\delta(q^{+}-p_{j1}^{+}-p_{j2}^{+})\Big( \mathcal{M}_{\mathrm{LO}}^{\dagger}\mathcal{M}_{\mathrm{GEFS},(ii)+\mathrm{IFS}}+q\leftrightarrow\bar{q}+\mathrm{c.c.}\Big)\\
 &\qquad+\int\mathrm{PS}(\vec{p}_{3})2\pi\delta(q^{+}-p_{j1}^{+}-p_{j2}^{+}-p_{3}^+)\Big(\mathcal{M}_{\mathrm{\overline{Q}FS}}^{\dagger}\mathcal{M}_{\mathrm{QFS}}+\mathrm{c.c.}\Big) \Bigg]_{\vec{p}_{1,2}\to\vec{p}_{j1,j2}}\;.
 \end{aligned}
 \label{eq:virtualsoft}
\end{equation}

Fourth, there are terms due to the final-state gluon radiation outside the jet, that in certain kinematics will become enhanced by a large Sudakov double logarithm (see sections~\ref{sec:coll} and~\ref{sec:corrlimit}):
\begin{equation}
\begin{aligned}
\frac{\mathrm{d}\sigma^{\mathrm{Sudakov}}_{\mathrm{real}}}{
\mathrm{d}p_{j1}^{+}\mathrm{\mathrm{d}}^{2}\mathbf{p}_{j1}
\mathrm{d}p_{j2}^{+}\mathrm{\mathrm{d}}^{2}\mathbf{p}_{j2}
}
 & =
\frac{\mathrm{d}\sigma_{\textrm{dijet}}^{{\mathrm{LO}}}}{
\mathrm{d}p_{j1}^{+}\mathrm{\mathrm{d}}^{2}\mathbf{p}_{j1}
\mathrm{d}p_{j2}^{+}\mathrm{\mathrm{d}}^{2}\mathbf{p}_{j2}
} \times
\frac{\alpha_{s}C_{F}}{2\pi}
\Big(\mathcal{V}_{|{\mathrm{QFS}}|^2}^{\textrm{out; phase}}+\mathcal{V}_{|{\mathrm{\overline{Q}FS}}|^2}^{\textrm{out; phase}}\Big)\;,
\end{aligned}
\end{equation}
The explicit formulae for $\mathcal{V}_{|{\mathrm{QFS}}|^2}^{\textrm{out; phase}}$ and $\mathcal{V}_{|{\mathrm{\overline{Q}FS}}|^2}^{\textrm{out; phase}}$ are given in eqs.~\eqref{eq:QFSsquare_out_phase_2} and \eqref{eq:QbarFSsquare_out_phase_2}.

The final contributions to eq.~\eqref{eq:totalcrosssection} are due to purely finite virtual and real NLO corrections:
\begin{equation}
\begin{aligned}&\frac{\mathrm{d}\sigma^{\mathrm{finite}}_{\mathrm{virtual}}}{\mathrm{d}p_{j1}^{+}\mathrm{\mathrm{d}}^{2}\mathbf{p}_{j1}\mathrm{d}p_{j2}^{+}\mathrm{\mathrm{d}}^{2}\mathbf{p}_{j2}}  =\frac{1}{2q^{+}}\frac{\theta(p_{j1}^{+})}{(2\pi)^{3}2p_{j1}^{+}}\frac{\theta(p_{j2}^{+})}{(2\pi)^{3}2p_{j2}^{+}}2\pi\delta(q^{+}-p_{j1}^{+}-p_{j2}^{+})\\
 & \times\frac{1}{2}\Bigg[\Big(\mathcal{M}_{\mathrm{LO}}^{\dagger}\mathcal{M}_{\mathrm{SESW,sub}}+\mathcal{M}_{\mathrm{LO}}^{\dagger}\mathcal{M}_{\mathrm{GESW}}+\mathcal{M}_{\mathrm{LO}}^{\dagger}\mathcal{M}_{\mathrm{ISW}}\\
 &+\mathcal{M}_{\mathrm{LO}}^{\dagger}\mathcal{M}_{\mathrm{GEFS},(i)}+q\leftrightarrow\bar{q}\Big) +\mathcal{M}_{\mathrm{LO}}^{\dagger}\mathcal{M}_{\mathrm{IS}+\mathrm{UV}+\mathrm{FSUV}}+\mathrm{c.c.}\Bigg]_{\vec{p}_{1,2}\to\vec{p}_{j1,j2}}\;,
\end{aligned}
\label{eq:virtualfinite}
\end{equation}
and
\begin{equation}
\begin{aligned}&\frac{\mathrm{d}\sigma^{\mathrm{finite}}_{\mathrm{real}}}{\mathrm{d}p_{j1}^{+}\mathrm{\mathrm{d}}^{2}\mathbf{p}_{j1}\mathrm{d}p_{j2}^{+}\mathrm{\mathrm{d}}^{2}\mathbf{p}_{j2}}\\  &=\frac{\mathrm{d}\sigma_{\textrm{dijet}}^{|{\mathrm{QFS}}|^2;\textrm{out;reg}}}{
\mathrm{d}p_{j1}^{+}\mathrm{\mathrm{d}}^{2}\mathbf{p}_{j1}
\mathrm{d}p_{j2}^{+}\mathrm{\mathrm{d}}^{2}\mathbf{p}_{j2}
}+\frac{\mathrm{d}\sigma_{\textrm{dijet}}^{|{\mathrm{\overline{Q}FS}}|^2;\textrm{out;reg}}}{
\mathrm{d}p_{j1}^{+}\mathrm{\mathrm{d}}^{2}\mathbf{p}_{j1}
\mathrm{d}p_{j2}^{+}\mathrm{\mathrm{d}}^{2}\mathbf{p}_{j2}
} 
+\frac{1}{2q^{+}}\frac{\theta(p_{j1}^{+})}{(2\pi)^{3}2p_{j1}^{+}}\frac{\theta(p_{j2}^{+})}{(2\pi)^{3}2p_{j2}^{+}}\frac{1}{4p_{3}^{+}}\\
&\times\int_{\mathbf{p}_{3}}\bigg\{|\mathcal{M}_{\mathrm{QSW}}|^{2}+\mathcal{M}_{\mathrm{QSW}}^{\dagger}\mathcal{M}_{\mathrm{QFS}}+\mathcal{M}_{\mathrm{QFS}}^{\dagger}\mathcal{M}_{\mathrm{QSW}}\\
 &\qquad\qquad +|\mathcal{M}_{\mathrm{\overline{Q}SW}}|^{2}+\mathcal{M}_{\mathrm{\overline{Q}SW}}^{\dagger}\mathcal{M}_{\mathrm{\overline{Q}FS}}+\mathcal{M}_{\mathrm{\overline{Q}FS}}^{\dagger}\mathcal{M}_{\mathrm{\overline{Q}SW}}\\
 &\qquad\qquad +\mathcal{M}_{\mathrm{\overline{Q}SW}}^{\dagger}\mathcal{M}_{\mathrm{QSW}}+\mathcal{M}_{\mathrm{\overline{Q}SW}}^{\dagger}\mathcal{M}_{\mathrm{QFS}}+\mathcal{M}_{\mathrm{\overline{Q}FS}}^{\dagger}\mathcal{M}_{\mathrm{QSW}}+\mathrm{c.c.}\\
 &\qquad\qquad +\mathcal{M}_{\mathrm{RI}}^{\dagger}\Big(\mathcal{M}_{\mathrm{QSW}}+\mathcal{M}_{\mathrm{QFS}}\Big)+\mathrm{c.c.}\\
 &\qquad\qquad +\mathcal{M}_{\mathrm{RI}}^{\dagger}\Big(\mathcal{M}_{\mathrm{\overline{Q}FS}}+\mathcal{M}_{\mathrm{\overline{Q}SW}}\Big)+\mathrm{c.c.}+\big|\mathcal{M}_{\mathrm{RI}}\big|^{2}\bigg\}\bigg|_{\vec{p}_{1,2}\to\vec{p}_{j1,j2},\,p_{3}^{+}=q^{+}-p_{j1}^{+}-p_{j2}^{+}>0}\;.
\end{aligned}
\label{eq:realfinite}
\end{equation}

\subsection{\label{subsec:softGE}SoftGE}
As discussed in section~\ref{sec:NLO}, the amplitudes due to gluon
exchange in the final state exhibit an unphysical power-divergence
in $k_{3}^{+}$. This is cured by summing the problematic part of
the amplitude, $\mathcal{M}_{\mathrm{GEFS},(ii)}$, with the amplitude
$\mathcal{M}_{\mathrm{IFS}}$ corresponding to the final-state exchange
of an instantaneous gluon:
\begin{equation}
\begin{aligned}&\mathcal{M}_{\mathrm{LO}}^{\dagger}\mathcal{M}_{\mathrm{GEFS},(ii)+\mathrm{IFS}}\\ & =64(2\pi)\alpha_{\mathrm{em}} e^2_f \alpha_{s}N_{c}^{2}p_{1}^{+}p_{2}^{+}\int_{k_{f}^{+}}^{p_{1}^{+}}\frac{\mathrm{d}k_{3}^{+}}{k_{3}^{+}}\frac{p_{1}^{+}-k_{3}^{+}}{q^{+}}\Bigg[(z^{2}+\bar{z}^{2})+(1-2z)\frac{k_{3}^{+}}{q^{+}}\Bigg]\\
 & \times\int_{\mathbf{x}_{1^{\prime}},\mathbf{x}_{2^{\prime}},\mathbf{x}_{1},\mathbf{x}_{2}}e^{-i\mathbf{p}_{1}\cdot\mathrm{\mathbf{x}}_{11^{\prime}}}e^{-i\mathbf{p}_{2}\cdot\mathbf{x}_{22^{\prime}}}e^{i\frac{k_{3}^{+}}{p_{1}^{+}}\mathbf{p}_{1}\cdot\mathbf{x}_{12}}A^{\lambda}(\mathbf{x}_{1^{\prime}2^{\prime}})A^{\lambda}(\mathbf{x}_{12})\\
 & \times\int_{\mathbf{K}}\Big(1+\frac{\mathbf{K}\cdot\mathbf{P}_{\perp}}{z\mathbf{K}^{2}}\Big)\frac{e^{-i\mathbf{K}\cdot\mathbf{x}_{12}}}{(\mathbf{K}+\frac{p_{1}^{+}-k_{3}^{+}}{p_{1}^{+}}\mathbf{P}_{\perp})^{2}-\frac{(p_{2}^{+}+k_{3}^{+})(p_{1}^{+}-k_{3}^{+})}{p_{1}^{+}p_{2}^{+}}\mathbf{P}_{\perp}^{2}-i\epsilon}\\
 & \times\Big\langle s_{2^{\prime}1^{\prime}}s_{12}-s_{12}-s_{2^{\prime}1^{\prime}}+1-\frac{1}{N_{c}^{2}}\Big(Q_{2^{\prime}1^{\prime}12}-s_{12}-s_{2^{\prime}1^{\prime}}-1\Big)\Big\rangle\;,
\end{aligned}
\label{eq:LOGEFSIFS}
\end{equation}
where we used that:
\begin{equation}
\begin{aligned}\mathrm{Tr}\Big(\mathrm{Dirac}_{\mathrm{LO}}^{\lambda^{\prime}\dagger}\mathrm{Dirac}_{\bar{q}\to q,(ii)}^{\bar{\lambda}\bar{\eta}\eta^{\prime}}\Big) & =-64\frac{(p_{1}^{+}p_{2}^{+})^{2}}{(k_{3}^{+})^{2}}\delta^{\bar{\eta}\eta^{\prime}}\delta^{\bar{\lambda}\lambda^{\prime}}\Bigg[(z^{2}+\bar{z}^{2})+(1-2z)\frac{k_{3}^{+}}{q^{+}}\Bigg]\;.\end{aligned}
\label{eq:DiracLOqbarqii}
\end{equation}

Due to its simple Wilson-line structure, the $q\leftrightarrow\bar{q}$
conjugate contributions to the cross section can be simply obtained
from the above result by exchanging the quark and antiquark indices:
\begin{equation}
\begin{aligned}\mathcal{M}_{\mathrm{LO}}^{\dagger}\mathcal{M}_{\mathrm{\overline{GE}FS},(ii)+\mathrm{\overline{I}FS}} & =\mathcal{M}_{\mathrm{LO}}^{\dagger}\mathcal{M}_{\mathrm{GEFS},(ii)+\mathrm{IFS}}(1\leftrightarrow2)\;,\\\mathrm{Tr}\Big(\mathrm{Dirac}_{\mathrm{LO}}^{\lambda^{\prime}\dagger}\mathrm{Dirac}_{q\to\bar{q}}^{\bar{\lambda}\bar{\eta}\eta^{\prime}}\Big) & =\mathrm{Tr}\Big(\mathrm{Dirac}_{\mathrm{LO}}^{\lambda^{\prime}\dagger}\mathrm{Dirac}_{\bar{q}\to q}^{\bar{\lambda}\bar{\eta}\eta^{\prime}}\Big)(1\leftrightarrow2)\;.
\end{aligned}
\end{equation}

The terms due to interference between final-state emission from the
quark and antiquark read:
\begin{equation}
\begin{aligned} & \int_{\mathbf{p}_{3}}\Big[\mathcal{M}_{\mathrm{\overline{Q}FS}}^{\dagger}\mathcal{M}_{\mathrm{QFS}}+\mathcal{M}_{\mathrm{QFS}}^{\dagger}\mathcal{M}_{\mathrm{\overline{Q}FS}}\Big]\\
 & =-4(2\pi)^{2}\alpha_{\mathrm{em}} e^2_f \alpha_{s}\mathrm{Tr}\Big(\mathrm{Dirac}_{\mathrm{\overline{Q}SW}}^{\eta^{\prime}\lambda^{\prime}\dagger}\mathrm{Dirac}_{\mathrm{QSW}}^{\bar{\eta}\bar{\lambda}}\Big)\frac{p_{3}^{+}}{p_{j1}^{+}+p_{3}^{+}}\frac{p_{3}^{+}}{p_{j2}^{+}+p_{3}^{+}}\\
 & \times\Bigg\{\int_{\mathbf{v},\mathbf{v}^{\prime}}\int_{\mathbf{x}_{1},\mathbf{x}_{1^{\prime}},\mathbf{x}_{2},\mathbf{x}_{2^{\prime}},\mathbf{x}_{3}}e^{-i\mathbf{p}_{j1}\cdot\mathbf{x}_{11^{\prime}}}e^{-i\mathbf{p}_{j2}\cdot\mathbf{x}_{22^{\prime}}}\\
 & \times\delta^{(2)}\big(\mathbf{v}-\frac{p_{j1}^{+}}{p_{j1}^{+}+p_{3}^{+}}\mathbf{x}_{1}-\frac{p_{3}^{+}}{p_{j1}^{+}+p_{3}^{+}}\mathbf{x}_{3}\big)\delta^{(2)}\big(\mathbf{v}^{\prime}-\frac{p_{j2}^{+}}{p_{j2}^{+}+p_{3}^{+}}\mathbf{x}_{2^{\prime}}-\frac{p_{3}^{+}}{p_{j2}^{+}+p_{3}^{+}}\mathbf{x}_{3}\big)\\
 & \times A^{\bar{\eta}}(\mathbf{x}_{13})A^{\eta^{\prime}}(\mathbf{x}_{2^{\prime}3})A^{\bar{\lambda}}(\mathbf{v}-\mathbf{x}_{2})A^{\lambda^{\prime}}(\mathbf{v}^{\prime}-\mathbf{x}_{1^{\prime}})\\
 & \times\Big\langle s_{v^{\prime}1^{\prime}}s_{v2}-s_{v2}-s_{v^{\prime}1^{\prime}}+1-\frac{1}{N_{c}^{2}}\Big(Q_{v2v^{\prime}1^{\prime}}-s_{v2}-s_{v^{\prime}1^{\prime}}+1\Big)\Big\rangle+\mathrm{c}.\mathrm{c}.\Bigg\}\;,
\end{aligned}
\label{eq:QbarFSQFS}
\end{equation}
with the Dirac trace, calculated with the help of identity (\ref{eq:DiracTraceBoss}):
\begin{equation}
\begin{aligned} & \mathrm{Tr}\Big(\mathrm{Dirac}_{\mathrm{\overline{Q}SW}}^{\eta^{\prime}\lambda^{\prime}\dagger}\mathrm{Dirac}_{\mathrm{QSW}}^{\bar{\eta}\bar{\lambda}}\Big)\\
 & =-32\frac{p_{1}^{+}p_{2}^{+}}{q^{+}p_{3}^{+}}\bigg[\frac{\big(2p_{1}^{+}p_{2}^{+}+(p_{1}^{+}+p_{2}^{+})p_{3}^{+}\big)\big((p_{1}^{+})^{2}+(p_{2}^{+})^{2}+(p_{1}^{+}+p_{2}^{+})p_{3}^{+}\big)}{q^{+}p_{3}^{+}}\delta^{\bar{\lambda}\lambda^{\prime}}\delta^{\bar{\eta}\eta^{\prime}}\\
 & \qquad-(p_{1}^{+}-p_{2}^{+})^{2}\epsilon^{\bar{\lambda}\lambda^{\prime}}\epsilon^{\bar{\eta}\eta^{\prime}}\bigg]\;.
\end{aligned}
\label{eq:DiracQbarSWQSW}
\end{equation}

\subsection{\label{subsec:finitevirtualcross}Finite virtual}

\paragraph{SESW}
The virtual contributions to the cross section due to the gluon self-energy
correction scattering off the shockwave, and subtracting the UV divergence,
read:
\begin{equation}
\begin{aligned}&\mathcal{M}_{\mathrm{LO}}^{\dagger}\mathcal{M}_{\mathrm{SESW,sub}}  =-32(2\pi)\alpha_{\mathrm{em}} e^2_f \alpha_{s}N_{c}^{2}p_{1}^{+}p_{2}^{+}(z^{2}+\bar{z}^{2})\\
 & \times\int_{k_{f}^{+}}^{p_{1}^{+}}\frac{\mathrm{d}k_{3}^{+}}{k_{3}^{+}}\Big[1+\Big(1-\frac{k_{3}^{+}}{p_{1}^{+}}\Big)^{2}\Big]\\;int_{\mathbf{x}_{1^{\prime}},\mathbf{x}_{2^{\prime}},\mathrm{\mathbf{x}}_{1},\mathbf{x}_{2},\mathbf{x}_{3}}e^{-i\mathbf{p}_{1}\cdot\mathrm{\mathbf{x}}_{11^{\prime}}}e^{-i\mathbf{p}_{2}\cdot\mathbf{x}_{22^{\prime}}}A^{\bar{\lambda}}(\mathbf{x}_{1^{\prime}2^{\prime}})A^{\bar{\eta}}(\mathbf{x}_{13})\\
 & \times\Bigg\{ e^{i\frac{k_{3}^{+}}{p_{1}^{+}}\mathbf{p}_{1}\cdot\mathbf{x}_{13}}A^{\bar{\eta}}(\mathbf{x}_{13})\mathcal{A}^{\bar{\lambda}}\Big(\frac{k_{3}^{+}}{p_{1}^{+}}\mathbf{x}_{13}+\mathbf{x}_{21},\frac{k_{3}^{+}}{p_{1}^{+}}\mathbf{x}_{13};\frac{q^{+}(p_{1}^{+}-k_{3}^{+})}{k_{3}^{+}p_{2}^{+}}\Big)\\
 & \qquad\times\Big\langle Q_{322^{\prime}1^{\prime}}s_{13}-s_{13}s_{32}-s_{2^{\prime}1^{\prime}}+1-\frac{1}{N_{c}^{2}}\Big(Q_{122^{\prime}1^{\prime}}-s_{12}-s_{2^{\prime}1^{\prime}}+1\Big)\Big\rangle\\
 & -\Big(A^{\bar{\eta}}(\mathbf{x}_{13})-A^{\bar{\eta}}(\mathbf{x}_{23})\Big)A^{\bar{\lambda}}(\mathbf{x}_{21})\Big(1-\frac{1}{N_{c}^{2}}\Big)\Big\langle Q_{122^{\prime}1^{\prime}}-s_{12}-s_{2^{\prime}1^{\prime}}+1\Big\rangle\Bigg\}\;,
\end{aligned}
\label{eq:LOSESWsub}
\end{equation}
and:
\begin{equation}
\begin{aligned}&\mathcal{M}_{\mathrm{LO}}^{\dagger}\mathcal{M}_{\mathrm{\overline{SE}SW,sub}}  =+32(2\pi)\alpha_{\mathrm{em}} e^2_f \alpha_{s}N_{c}^{2}p_{1}^{+}p_{2}^{+}(z^{2}+\bar{z}^{2})\\
 & \times\int_{k_{f}^{+}}^{p_{2}^{+}}\frac{\mathrm{d}k_{3}^{+}}{k_{3}^{+}}\Big[1+\Big(1-\frac{k_{3}^{+}}{p_{2}^{+}}\Big)^{2}\Big]\\;int_{\mathbf{x}_{1^{\prime}},\mathbf{x}_{2^{\prime}},\mathrm{\mathbf{x}}_{1},\mathbf{x}_{2},\mathbf{x}_{3}}e^{-i\mathbf{p}_{1}\cdot\mathrm{\mathbf{x}}_{11^{\prime}}}e^{-i\mathbf{p}_{2}\cdot\mathbf{x}_{22^{\prime}}}A^{\bar{\lambda}}(\mathbf{x}_{1^{\prime}2^{\prime}})A^{\bar{\eta}}(\mathbf{x}_{23})\\
 & \times\Bigg\{ e^{i\frac{k_{3}^{+}}{p_{2}^{+}}\mathbf{p}_{2}\cdot\mathbf{x}_{23}}A^{\bar{\eta}}(\mathbf{x}_{23})\mathcal{A}^{\bar{\lambda}}\Big(\frac{k_{3}^{+}}{p_{2}^{+}}\mathbf{x}_{23}+\mathbf{x}_{12},\frac{k_{3}^{+}}{p_{2}^{+}}\mathbf{x}_{23};\frac{q^{+}(p_{2}^{+}-k_{3}^{+})}{k_{3}^{+}p_{1}^{+}}\Big)\\
 & \qquad\times\Big\langle Q_{132^{\prime}1^{\prime}}s_{32}-s_{13}s_{32}-s_{2^{\prime}1^{\prime}}+1-\frac{1}{N_{c}^{2}}\Big(Q_{122^{\prime}1^{\prime}}-s_{12}-s_{2^{\prime}1^{\prime}}+1\Big)\Big\rangle\\
 & -\Big(A^{\bar{\eta}}(\mathbf{x}_{23})-A^{\bar{\eta}}(\mathbf{x}_{13})\Big)A^{\bar{\lambda}}(\mathbf{x}_{12})\Big(1-\frac{1}{N_{c}^{2}}\Big)\Big\langle Q_{122^{\prime}1^{\prime}}-s_{12}-s_{2^{\prime}1^{\prime}}+1\Big\rangle\Bigg\}\;.
\end{aligned}
\label{eq:LOSEbarSWsub}
\end{equation}

\paragraph{GESW}
We find for the contributions due to a gluon being exchanged between
the antiquark and quark, traveling through the shockwave:
\begin{equation}
\begin{aligned}&\mathcal{M}_{\mathrm{LO}}^{\dagger}\mathcal{M}_{\mathrm{GESW}}  =2\pi\alpha_{\mathrm{em}} e^2_f \alpha_{s}N_{c}^{2}\int_{k_{f}^{+}}^{p_{1}^{+}}\frac{\mathrm{d}k_{3}^{+}}{k_{3}^{+}}\frac{(k_{3}^{+})^{2}}{p_{1}^{+}(p_{2}^{+}+k_{3}^{+})}\mathrm{Tr}\Big(\mathrm{Dirac}_{\mathrm{LO}}^{\lambda^{\prime}\dagger}\mathrm{Dirac}_{\bar{q}\to q}^{\bar{\lambda}\bar{\eta}\eta^{\prime}}\Big)\\
 &\qquad \times\int_{\mathbf{x}_{1^{\prime}},\mathbf{x}_{2^{\prime}},\mathbf{x}_{1},\mathbf{x}_{2},\mathbf{x}_{3}}e^{-i\mathbf{p}_{1}\cdot\mathrm{\mathbf{x}}_{11^{\prime}}}e^{-i\mathbf{p}_{2}\cdot\mathbf{x}_{22^{\prime}}}e^{i\frac{k_{3}^{+}}{p_{1}^{+}}\mathbf{p}_{1}\cdot\mathbf{x}_{13}}A^{\lambda^{\prime}}(\mathbf{x}_{1^{\prime}2^{\prime}})\\
 &\qquad \times A^{\eta^{\prime}}(\mathbf{x}_{13})A^{\bar{\eta}}(\mathbf{x}_{23})\mathcal{A}^{\bar{\lambda}}\Big(\frac{p_{2}^{+}\mathbf{x}_{12}+k_{3}^{+}\mathbf{x}_{13}}{p_{2}^{+}+k_{3}^{+}},\frac{k_{3}^{+}}{p_{2}^{+}+k_{3}^{+}}\mathbf{x}_{32};\frac{q^{+}p_{2}^{+}}{k_{3}^{+}(p_{1}^{+}-k_{3}^{+})}\Big)\\
 & \qquad\times\Big\langle Q_{322^{\prime}1^{\prime}}s_{13}-s_{13}s_{32}-s_{2^{\prime}1^{\prime}}+1-\frac{1}{N_{c}^{2}}\Big(Q_{122^{\prime}1^{\prime}}-s_{12}-s_{2^{\prime}1^{\prime}}+1\Big)\Big\rangle\;,
\end{aligned}
\label{eq:LOGESW}
\end{equation}
where the Dirac trace is easily calculated applying identity (\ref{eq:DiracTraceBoss}),
with the following result:
\begin{equation}
\begin{aligned}\mathrm{Tr}\Big(\mathrm{Dirac}_{\mathrm{LO}}^{\lambda^{\prime}\dagger}\mathrm{Dirac}_{\bar{q}\to q}^{\bar{\lambda}\bar{\eta}\eta^{\prime}}\Big) & =32\frac{p_{1}^{+}p_{2}^{+}}{(q^{+}k_{3}^{+})^{2}}\Bigg[\Big((k_{3}^{+})^{2}+k_{3}^{+}(p_{2}^{+}-p_{1}^{+})-2p_{1}^{+}p_{2}^{+}\Big)\\
 & \times\Big((p_{1}^{+})^{2}+(p_{2}^{+})^{2}+k_{3}^{+}(p_{2}^{+}-p_{1}^{+})\Big)\delta^{\bar{\lambda}\lambda^{\prime}}\delta^{\bar{\eta}\eta^{\prime}}\\
 & +q^{+}k_{3}^{+}(k_{3}^{+}+p_{2}^{+}-p_{1}^{+})^{2}\epsilon^{\bar{\lambda}\lambda^{\prime}}\epsilon^{\bar{\eta}\eta^{\prime}}\Bigg]\;.
\end{aligned}
\label{eq:DiracLOqbarq}
\end{equation}

The $q\leftrightarrow\bar{q}$ conjugate, where the gluon is emitted
from the quark and, after interaction with the shockwave, absorbed
by the antiquark, is given by:
\begin{equation}
\begin{aligned}&\mathcal{M}_{\mathrm{LO}}^{\dagger}\mathcal{M}_{\mathrm{\overline{GE}SW}}  =2\pi\alpha_{\mathrm{em}} e^2_f \alpha_{s}N_{c}^{2}\int_{k_{f}^{+}}^{p_{2}^{+}}\frac{\mathrm{d}k_{3}^{+}}{k_{3}^{+}}\frac{(k_{3}^{+})^{2}}{p_{2}^{+}(p_{1}^{+}+k_{3}^{+})}\mathrm{Tr}\Big(\mathrm{Dirac}_{\mathrm{LO}}^{\lambda^{\prime}\dagger}\mathrm{Dirac}_{q\to\bar{q}}^{\bar{\lambda}\bar{\eta}\eta^{\prime}}\Big)\\
 &\qquad \times\int_{\mathbf{x}_{1^{\prime}},\mathbf{x}_{2^{\prime}},\mathbf{x}_{1},\mathbf{x}_{2},\mathbf{x}_{3}}e^{-i\mathbf{p}_{1}\cdot\mathrm{\mathbf{x}}_{11^{\prime}}}e^{-i\mathbf{p}_{2}\cdot\mathbf{x}_{22^{\prime}}}e^{i\frac{k_{3}^{+}}{p_{2}^{+}}\mathbf{p}_{2}\cdot\mathbf{x}_{23}}A^{\lambda^{\prime}}(\mathbf{x}_{1^{\prime}2^{\prime}})\\
 &\qquad \times A^{\eta^{\prime}}(\mathbf{x}_{32})A^{\bar{\eta}}(\mathbf{x}_{31})\mathcal{A}^{\bar{\lambda}}\Big(\frac{p_{1}^{+}\mathbf{x}_{21}+k_{3}^{+}\mathbf{x}_{23}}{p_{1}^{+}+k_{3}^{+}},\frac{k_{3}^{+}}{p_{1}^{+}+k_{3}^{+}}\mathbf{x}_{31};\frac{q^{+}p_{1}^{+}}{k_{3}^{+}(p_{2}^{+}-k_{3}^{+})}\Big)\\
 & \qquad\times\Big\langle Q_{132^{\prime}1^{\prime}}s_{32}-s_{13}s_{32}-s_{2^{\prime}1^{\prime}}+1-\frac{1}{N_{c}^{2}}\Big(Q_{122^{\prime}1^{\prime}}-s_{12}-s_{2^{\prime}1^{\prime}}+1\Big)\Big\rangle\;,
\end{aligned}
\label{eq:LOGEbarSW}
\end{equation}
and the Dirac trace yields:
\begin{equation}
\begin{aligned}&\mathrm{Tr}\Big(\mathrm{Dirac}_{\mathrm{LO}}^{\lambda^{\prime}\dagger}\mathrm{Dirac}_{q\to\bar{q}}^{\bar{\lambda}\bar{\eta}\eta^{\prime}}\Big)\\  &=-32\frac{p_{1}^{+}p_{2}^{+}}{(q^{+}k_{3}^{+})^{2}}\Bigg[\Big(k_{3}^{+}(p_{1}^{+}-p_{2}^{+})-2p_{1}^{+}p_{2}^{+}\Big)
  \Big((k_{3}^{+})^{2}+k_{3}^{+}(p_{1}^{+}-p_{2}^{+})-2p_{1}^{+}p_{2}^{+}\Big)\delta^{\bar{\lambda}\lambda^{\prime}}\delta^{\bar{\eta}\eta^{\prime}}\\
 & \qquad\qquad\qquad\quad-q^{+}(k_{3}^{+})^{2}(k_{3}^{+}+p_{1}^{+}-p_{2}^{+})\epsilon^{\bar{\lambda}\lambda^{\prime}}\epsilon^{\bar{\eta}\eta^{\prime}}\Bigg]\;..
\end{aligned}
\label{eq:DiracLOqqbar}
\end{equation}

\paragraph{ISW}
The following contributions are due to the instantaneous splitting
of the photon into a gluon, quark and antiquark. After traveling
through the shockwave, the gluon is absorbed by the quark:
\begin{equation}
\begin{aligned}&\mathcal{M}_{\mathrm{LO}}^{\dagger}\mathcal{M}_{\mathrm{ISW}} \\& =-32(2\pi)\alpha_{\mathrm{em}} e^2_f \alpha_{s}N_{c}^{2}p_{1}^{+}p_{2}^{+}\int_{k_{f}^{+}}^{p_{1}^{+}}\mathrm{d}k_{3}^{+}\frac{p_{1}^{+}-k_{3}^{+}}{q^{+}p_{1}^{+}}\Big(\frac{p_{1}^{+}}{p_{2}^{+}+k_{3}^{+}}+\frac{p_{2}^{+}(p_{1}^{+}-k_{3}^{+})}{(p_{1}^{+})^{2}}\Big)\\
 & \times\int_{\mathbf{x}_{1^{\prime}},\mathbf{x}_{2^{\prime}},\mathbf{x}_{1},\mathbf{x}_{2},\mathbf{x}_{3}}e^{-i\mathbf{p}_{1}\cdot\mathrm{\mathbf{x}}_{11^{\prime}}}e^{-i\mathbf{p}_{2}\cdot\mathbf{x}_{22^{\prime}}}e^{i\mathbf{p}_{1}\cdot\frac{k_{3}^{+}}{p_{1}^{+}}\mathbf{x}_{13}}\\
 & \times A^{\lambda^{\prime}}(\mathbf{x}_{1^{\prime}2^{\prime}})A^{\lambda^{\prime}}(\mathbf{x}_{31})\mathcal{C}\big(\frac{k_{3}^{+}}{p_{1}^{+}}\mathbf{x}_{13}+\mathbf{x}_{21},\frac{k_{3}^{+}}{p_{1}^{+}}\mathbf{x}_{13};\frac{q^{+}(p_{1}^{+}-k_{3}^{+})}{p_{2}^{+}k_{3}^{+}}\big)\\
 & \times\Big\langle Q_{322^{\prime}1^{\prime}}s_{13}-s_{13}s_{32}-s_{2^{\prime}1^{\prime}}+1-\frac{1}{N_{c}^{2}}\Big(Q_{122^{\prime}1^{\prime}}-s_{12}-s_{2^{\prime}1^{\prime}}+1\Big)\Big\rangle\;,
\end{aligned}
\label{eq:LOISW}
\end{equation}
where we used the following result for the Dirac trace:
\begin{equation}
\begin{aligned}\mathrm{Tr}\Big(\mathrm{Dirac}_{\mathrm{LO}}^{\lambda^{\prime}\dagger}\mathrm{Dirac}_{\mathrm{ISW}}^{\eta^{\prime}}\Big) & =32p_{1}^{+}p_{2}^{+}\delta^{\lambda^{\prime}\eta^{\prime}}\frac{(p_{1}^{+})^{3}+p_{2}^{+}(p_{1}^{+}-k_{3}^{+})(p_{2}^{+}+k_{3}^{+})}{k_{3}^{+}q^{+}(k_{3}^{+}+p_{2}^{+}-p_{1}^{+})}\;.\end{aligned}
\label{eq:DiracLOILSW}
\end{equation}
Likewise, when the gluon is absorbed by the antiquark, one obtains:
\begin{equation}
\begin{aligned}&\mathcal{M}_{\mathrm{LO}}^{\dagger}\mathcal{M}_{\mathrm{\overline{I}SW}} \\& =32(2\pi)\alpha_{\mathrm{em}} e^2_f \alpha_{s}N_{c}^{2}p_{1}^{+}p_{2}^{+}\int_{k_{f}^{+}}^{p_{2}^{+}}\mathrm{d}k_{3}^{+}\frac{p_{2}^{+}-k_{3}^{+}}{q^{+}p_{2}^{+}}\Big(\frac{p_{2}^{+}}{p_{1}^{+}+k_{3}^{+}}+\frac{p_{1}^{+}(p_{2}^{+}-k_{3}^{+})}{(p_{2}^{+})^{2}}\Big)\\
 & \times\int_{\mathbf{x}_{1^{\prime}},\mathbf{x}_{2^{\prime}},\mathbf{x}_{1},\mathbf{x}_{2},\mathbf{x}_{3}}e^{-i\mathbf{p}_{1}\cdot\mathrm{\mathbf{x}}_{11^{\prime}}}e^{-i\mathbf{p}_{2}\cdot\mathbf{x}_{22^{\prime}}}e^{i\mathbf{p}_{2}\cdot\frac{k_{3}^{+}}{p_{2}^{+}}\mathbf{x}_{23}}\\
 & \times A^{\lambda^{\prime}}(\mathbf{x}_{1^{\prime}2^{\prime}})A^{\lambda^{\prime}}(\mathbf{x}_{32})\mathcal{C}\big(\frac{k_{3}^{+}}{p_{2}^{+}}\mathbf{x}_{23}+\mathbf{x}_{12},\frac{k_{3}^{+}}{p_{2}^{+}}\mathbf{x}_{23};\frac{q^{+}(p_{2}^{+}-k_{3}^{+})}{p_{1}^{+}k_{3}^{+}}\big)\\
 & \times\Big\langle Q_{132^{\prime}1^{\prime}}s_{32}-s_{13}s_{32}-s_{2^{\prime}1^{\prime}}+1-\frac{1}{N_{c}^{2}}\Big(Q_{122^{\prime}1^{\prime}}-s_{12}-s_{2^{\prime}1^{\prime}}+1\Big)\Big\rangle\;,
\end{aligned}
\label{eq:LOIbarSW}
\end{equation}
where we used that:
\begin{equation}
\begin{aligned}\mathrm{Tr}\Big(\mathrm{Dirac}_{\mathrm{LO}}^{\lambda^{\prime}\dagger}\mathrm{Dirac}_{\mathrm{\overline{I}SW}}^{\eta^{\prime}}\Big) & =-32p_{1}^{+}p_{2}^{+}\delta^{\lambda^{\prime}\eta^{\prime}}\frac{(p_{2}^{+})^{3}+p_{1}^{+}(p_{2}^{+}-k_{3}^{+})(p_{1}^{+}+k_{3}^{+})}{k_{3}^{+}q^{+}(k_{3}^{+}+p_{1}^{+}-p_{2}^{+})}\;.\end{aligned}
\label{eq:DiracLOILbarSW}
\end{equation}

\paragraph{GEFS,(i)}
The following contribution is what is left after subtracting $\mathcal{M}_{\mathrm{GEFS},(ii)}$ from the amplitude due to final-state gluon exchange:
\begin{equation}
\begin{aligned}\mathcal{M}_{\mathrm{LO}}^{\dagger}\mathcal{M}_{\mathrm{GEFS},(i)} & =(2\pi)\alpha_{\mathrm{em}} e^2_f \alpha_{s}N_{c}^{2}\int_{k_{f}^{+}}^{p_{1}^{+}}\mathrm{d}k_{3}^{+}\frac{p_{1}^{+}-k_{3}^{+}}{q^{+}p_{1}^{+}}\mathrm{Tr}\Big(\mathrm{Dirac}_{\mathrm{LO}}^{\lambda^{\prime}\dagger}\mathrm{Dirac}_{\bar{q}\to q,(i)}^{\bar{\lambda}\bar{\eta}\eta^{\prime}}\Big)\\
 & \times\int_{\mathbf{x}_{1^{\prime}},\mathbf{x}_{2^{\prime}},\mathbf{x}_{1},\mathbf{x}_{2}}e^{-i\mathbf{p}_{1}\cdot\mathrm{\mathbf{x}}_{11^{\prime}}}e^{-i\mathbf{p}_{2}\cdot\mathbf{x}_{22^{\prime}}}e^{i\frac{k_{3}^{+}}{p_{1}^{+}}\mathbf{p}_{1}\cdot\mathbf{x}_{12}}\\
 & \times A^{\lambda^{\prime}}(\mathbf{x}_{1^{\prime}2^{\prime}})A^{\bar{\lambda}}(\mathbf{x}_{12})J^{\eta^{\prime}\bar{\eta}}(k_{3},\mathbf{x}_{12})\\
 & \times\Big\langle s_{2^{\prime}1^{\prime}}s_{12}-s_{12}-s_{2^{\prime}1^{\prime}}+1-\frac{1}{N_{c}^{2}}\Big(Q_{2^{\prime}1^{\prime}12}-s_{12}-s_{2^{\prime}1^{\prime}}-1\Big)\Big\rangle\;,
\end{aligned}
\label{eq:LOGEFS}
\end{equation}
with the Dirac trace:
\begin{equation}
\begin{aligned}\mathrm{Tr}\Big(\mathrm{Dirac}_{\mathrm{LO}}^{\lambda^{\prime}\dagger}\mathrm{Dirac}_{\bar{q}\to q,(i)}^{\bar{\lambda}\bar{\eta}\eta^{\prime}}\Big) & =32\frac{p_{1}^{+}p_{2}^{+}}{(q^{+}k_{3}^{+})^{2}}\Bigg[\Big((k_{3}^{+})^{2}+k_{3}^{+}(p_{2}^{+}-p_{1}^{+})\Big)\\
 & \times\Big((p_{1}^{+})^{2}+(p_{2}^{+})^{2}+k_{3}^{+}(p_{2}^{+}-p_{1}^{+})\Big)\delta^{\bar{\lambda}\lambda^{\prime}}\delta^{\bar{\eta}\eta^{\prime}}\\
 & +q^{+}k_{3}^{+}(k_{3}^{+}+p_{2}^{+}-p_{1}^{+})^{2}\epsilon^{\bar{\lambda}\lambda^{\prime}}\epsilon^{\bar{\eta}\eta^{\prime}}\Bigg]\;.
\end{aligned}
\label{eq:DiracLOqbarqi}
\end{equation}

The $q\leftrightarrow\bar{q}$ conjugate contributions to the cross section can be simply obtained
from the above result by exchanging the quark and antiquark indices:
\begin{equation}
\begin{aligned}\mathcal{M}_{\mathrm{LO}}^{\dagger}\mathcal{M}_{\mathrm{\overline{GE}FS},(i)} & =\mathcal{M}_{\mathrm{LO}}^{\dagger}\mathcal{M}_{\mathrm{GEFS},(i)}(1\leftrightarrow2)\;,\\
\mathrm{Tr}\Big(\mathrm{Dirac}_{\mathrm{LO}}^{\lambda^{\prime}\dagger}\mathrm{Dirac}_{q\to\bar{q}}^{\bar{\lambda}\bar{\eta}\eta^{\prime}}\Big) & =\mathrm{Tr}\Big(\mathrm{Dirac}_{\mathrm{LO}}^{\lambda^{\prime}\dagger}\mathrm{Dirac}_{\bar{q}\to q}^{\bar{\lambda}\bar{\eta}\eta^{\prime}}\Big)(1\leftrightarrow2)\;.
\end{aligned}
\end{equation}

\paragraph{IS+UV+FSUV}

Finally, the contributions due to the initial-state corrections, ultraviolet
counterterms, and ultraviolet part of the self-energy corrections
to the asymptotic final states read:
\begin{equation}
\begin{aligned}\mathcal{M}_{\mathrm{LO}}^{\dagger}\mathcal{M}_{\mathrm{IS}+\mathrm{UV}+\mathrm{FSUV}} & =\big|\mathcal{M}_{\mathrm{LO}}\big|^{2}\frac{\alpha_{s}C_{F}}{2\pi}\big(\frac{1}{2}\mathrm{ln}^{2}\frac{z}{\bar{z}}-\frac{\pi^{2}}{6}+2\big)\;.\end{aligned}
\end{equation}

\subsection{\label{subsec:realcross}Real terms}
\paragraph{Final-state gluon radiation outside jet}
The two terms in the first line of eq.~\eqref{eq:realfinite} are due to what we call the `regular' (see section~\ref{sec:coll}) parts of $|{\mathrm{QFS}}|^2$ and $|{\mathrm{\overline{Q}FS}}|^2$, where the gluon is emitted outside the jet:
\begin{equation}
\begin{aligned}
&\frac{\mathrm{d}\sigma_{\textrm{dijet}}^{|{\mathrm{QFS}}|^2;\textrm{ out; reg}}}{
\mathrm{d}p_{j1}^{+}\mathrm{\mathrm{d}}^{2}\mathbf{p}_{j1}
\mathrm{d}p_{j2}^{+}\mathrm{\mathrm{d}}^{2}\mathbf{p}_{j2}
}
=
\frac{\theta(p_{j1}^{+})}{(2\pi)^{3}2p_{j1}^{+}}
\frac{\theta(p_{j2}^{+})}{(2\pi)^{3}2p_{j2}^{+}}
\frac{1}{2q^{+}}
 \frac{1}{2}|\mathcal{M}_{\mathrm{LO}}|^{2}\\
 & \times\frac{\alpha_{s}C_{F}}{2\pi}
 \int_{0}^{+\infty}
   \frac{\mathrm{d}\xi}{\xi}
 e^{-i\xi  \mathbf{p}_{j1}\cdot\mathbf{x}_{11^{\prime}}}
 \left[-2\ln R -\ln\left(\frac{\mathbf{p}_{j1}^{2}\mathbf{x}_{11^{\prime}}^{2}}{{c_{0}^{2}}}\right)
-2\ln\xi\right] \\
& \times
\bigg\{2\pi\delta(q^{+}-(1+\xi)p_{j1}^{+}-p_{j2}^{+})
\bigg[1+
\frac{(2\xi+\xi^2)(p_{j1}^{+})^{2}}{(p_{j1}^{+})^{2}
 +(p_{j2}^{+})^{2}}
 \bigg]
\big(1+(1+\xi)^2\big)\;
\\
&\quad
-4\pi \delta(q^{+}-p_{j1}^{+}-p_{j2}^{+}) \theta(1 -\xi)
\bigg\}
\;.
\end{aligned}
\label{eq:QFSsquare_out_reg_2}
\end{equation}
and
\begin{equation}
\begin{aligned}
&\frac{\mathrm{d}\sigma_{\textrm{dijet}}^{|{\mathrm{\overline{Q}FS}}|^2;\textrm{ out; reg}}}{
\mathrm{d}p_{j1}^{+}\mathrm{\mathrm{d}}^{2}\mathbf{p}_{j1}
\mathrm{d}p_{j2}^{+}\mathrm{\mathrm{d}}^{2}\mathbf{p}_{j2}
}
  =
\frac{\theta(p_{j1}^{+})}{(2\pi)^{3}2p_{j1}^{+}}
\frac{\theta(p_{j2}^{+})}{(2\pi)^{3}2p_{j2}^{+}}
\frac{1}{2q^{+}}
 \frac{1}{2}|\mathcal{M}_{\mathrm{LO}}|^{2}
\\
 &\times\frac{\alpha_{s}C_{F}}{2\pi}
 \int_{0}^{+\infty}
   \frac{\mathrm{d}\xi}{\xi}
 e^{-i\xi  \mathbf{p}_{j2}\cdot\mathbf{x}_{22^{\prime}}}
 \left[-2\ln R -\ln\left(\frac{\mathbf{p}_{j2}^{2}\mathbf{x}_{22^{\prime}}^{2}}{{c_{0}^{2}}}\right)
-2\ln\xi\right]
\\
& \times
\bigg\{
2\pi\delta(q^{+}-p_{j1}^{+}-(1+\xi)p_{j2}^{+})
\bigg[1+
\frac{(2\xi+\xi^2)(p_{j2}^{+})^{2}}{(p_{j1}^{+})^{2}
 +(p_{j2}^{+})^{2}}
 \bigg]
\big(1+(1+\xi)^2\big)\;
\\
&\quad
-4\pi \delta(q^{+}-p_{j1}^{+}-p_{j2}^{+}) \theta(1 -\xi)
\bigg\}
\;.
\end{aligned}
\label{eq:QbarFSsquare_out_reg_2}
\end{equation}
\paragraph{QFS and QSW}
The next terms are due to gluon emission before the shockwave and
the interference with the final-state emission:
\begin{equation}
\begin{aligned} & \int_{\mathbf{p}_{3}}\Big[|\mathcal{M}_{\mathrm{QSW}}|^{2}+\mathcal{M}_{\mathrm{QSW}}^{\dagger}\mathcal{M}_{\mathrm{QFS}}+\mathcal{M}_{\mathrm{QFS}}^{\dagger}\mathcal{M}_{\mathrm{QSW}}\Big]\\
 & =2(2\pi)^{2}\alpha_{\mathrm{em}} e^2_f \alpha_{s}N_{c}^{2}\mathrm{Tr}\Big(\mathrm{Dirac}_{\mathrm{QSW}}^{\bar{\eta}\bar{\lambda}}\mathrm{Dirac}_{\mathrm{QSW}}^{\eta^{\prime}\lambda^{\prime}\dagger}\Big)\Big(\frac{p_{3}^{+}}{p_{j1}^{+}+p_{3}^{+}}\Big)^{2}\\
 & \times\int_{\mathbf{v},\mathbf{v}^{\prime}}\int_{\mathbf{x}_{1},\mathbf{x}_{1^{\prime}},\mathbf{x}_{2},\mathbf{x}_{2^{\prime}},\mathbf{x}_3}e^{-i\mathbf{p}_{j1}\cdot\mathbf{x}_{11^{\prime}}}e^{-i\mathbf{p}_{j2}\cdot\mathbf{x}_{22^{\prime}}}\delta^{(2)}\big(\mathbf{v}-\frac{p_{j1}^{+}}{p_{j1}^{+}+p_{3}^{+}}\mathbf{x}_{1}-\frac{p_{3}^{+}}{p_{j1}^{+}+p_{3}^{+}}\mathbf{x}_{3}\big)\\
 & \times\delta^{(2)}\big(\mathbf{v}^{\prime}-\frac{p_{j1}^{+}}{p_{j1}^{+}+p_{3}^{+}}\mathbf{x}_{1^{\prime}}-\frac{p_{3}^{+}}{p_{j1}^{+}+p_{3}^{+}}\mathbf{x}_{3}\big)A^{\bar{\eta}}\big(\mathbf{x}_{13}\big)A^{\eta^{\prime}}\big(\mathbf{x}_{1^{\prime}3} \big)\\
 & \times\Bigg\{\mathcal{A}^{\lambda^{\prime}}\big(\mathbf{v}^{\prime}-\mathbf{x}_{2^{\prime}},\mathbf{x}_{31^{\prime}},\frac{p_{j1}^{+}p_{3}^{+}q^{+}}{p_{j2}^{+}(p_{j1}^{+}+p_{3}^{+})^{2}}\big)\mathcal{A}^{\bar{\lambda}}\big(\mathbf{v}-\mathbf{x}_{2},\mathbf{x}_{31},\frac{p_{j1}^{+}p_{3}^{+}q^{+}}{p_{j2}^{+}(p_{j1}^{+}+p_{3}^{+})^{2}}\big)\\
 & \qquad\times\Big\langle s_{11^{\prime}}s_{2^{\prime}2}-s_{32}s_{13}-s_{31^{\prime}}s_{2^{\prime}3}+1-\frac{1}{N_{c}^{2}}\Big(Q_{122^{\prime}1^{\prime}}-s_{12}-s_{2^{\prime}1^{\prime}}+1\Big)\Big\rangle\\
 &\quad -\mathcal{A}^{\lambda^{\prime}}\big(\mathbf{v}^{\prime}-\mathbf{x}_{2^{\prime}},\mathbf{x}_{31^{\prime}},\frac{p_{j1}^{+}p_{3}^{+}q^{+}}{p_{j2}^{+}(p_{j1}^{+}+p_{3}^{+})^{2}}\big)A^{\bar{\lambda}}(\mathbf{v}-\mathbf{x}_{2})\\
 & \qquad\times\Big\langle Q_{v22^{\prime}3}s_{31^{\prime}}-s_{31^{\prime}}s_{2^{\prime}3}-s_{v2}+1-\frac{1}{N_{c}^{2}}\Big(Q_{v22^{\prime}1^{\prime}}-s_{v2}-s_{2^{\prime}1^{\prime}}+1\Big)\Big\rangle+\mathrm{c.c}\;.\Bigg\}\;.
\end{aligned}
\label{eq:QSWQFS}
\end{equation}
Likewise, when the gluon is radiated from the antiquark:
\begin{equation}
\begin{aligned} & \int_{\mathbf{p}_{3}}\Big[|\mathcal{M}_{\mathrm{\overline{Q}SW}}|^{2}+\mathcal{M}_{\mathrm{\overline{Q}SW}}^{\dagger}\mathcal{M}_{\mathrm{\overline{Q}FS}}+\mathcal{M}_{\mathrm{\overline{Q}FS}}^{\dagger}\mathcal{M}_{\mathrm{\overline{Q}SW}}\Big]\\
 & =2(2\pi)^{2}\alpha_{\mathrm{em}} e^2_f \alpha_{s}N_{c}^{2}\mathrm{Tr}\Big(\mathrm{Dirac}_{\mathrm{\overline{Q}SW}}^{\eta^{\prime}\lambda^{\prime}\dagger}\mathrm{Dirac}_{\mathrm{\overline{Q}SW}}^{\bar{\eta}\bar{\lambda}}\Big)\Big(\frac{p_{3}^{+}}{p_{2}^{+}+p_{3}^{+}}\Big)^{2}\\
 & \times\int_{\mathbf{v},\mathbf{v}^{\prime}}\int_{\mathbf{x}_{1},\mathbf{x}_{1^{\prime}},\mathbf{x}_{2},\mathbf{x}_{2^{\prime}},\mathbf{x}_3}e^{-i\mathbf{p}_{j1}\cdot\mathbf{x}_{11^{\prime}}}e^{-i\mathbf{p}_{j2}\cdot\mathbf{x}_{22^{\prime}}}\delta^{(2)}\big(\mathbf{v}-\frac{p_{j2}^{+}}{p_{j2}^{+}+p_{3}^{+}}\mathbf{x}_{2}-\frac{p_{3}^{+}}{p_{j2}^{+}+p_{3}^{+}}\mathbf{x}_{3}\big)\\
 & \times\delta^{(2)}\big(\mathbf{v}^{\prime}-\frac{p_{j2}^{+}}{p_{j2}^{+}+p_{3}^{+}}\mathbf{x}_{2^{\prime}}-\frac{p_{3}^{+}}{p_{j2}^{+}+p_{3}^{+}}\mathbf{x}_{3}\big)A^{\bar{\eta}}(\mathbf{x}_{23})A^{\eta^{\prime}}(\mathbf{x}_{2^{\prime}3})\\
 & \times\Bigg\{\mathcal{A}^{\lambda^{\prime}}\big(\mathbf{v}^{\prime}-\mathbf{x}_{1^{\prime}},\mathbf{x}_{32^{\prime}},\frac{p_{j2}^{+}p_{3}^{+}q^{+}}{p_{j1}^{+}(p_{j2}^{+}+p_{3}^{+})^{2}}\big)\mathcal{A}^{\bar{\lambda}}\big(\mathbf{v}-\mathbf{x}_{1},\mathbf{x}_{32},\frac{p_{j2}^{+}p_{3}^{+}q^{+}}{p_{j1}^{+}(p_{j2}^{+}+p_{3}^{+})^{2}}\big)\\
 & \qquad\times\Big\langle s_{11^{\prime}}s_{2^{\prime}2}-s_{2^{\prime}3}s_{31^{\prime}}-s_{13}s_{32}+1-\frac{1}{N_{c}^{2}}\big(Q_{122^{\prime}1^{\prime}}-s_{12}-s_{2^{\prime}1^{\prime}}+1\big)\Big\rangle\\
 & \quad-\mathcal{A}^{\lambda^{\prime}}\big(\mathbf{v}^{\prime}-\mathbf{x}_{1^{\prime}},\mathbf{x}_{32^{\prime}},\frac{p_{j2}^{+}p_{3}^{+}q^{+}}{p_{j1}^{+}(p_{j2}^{+}+p_{3}^{+})^{2}}\big)A^{\bar{\lambda}}(\mathbf{v}-\mathbf{x}_{1})\\
 & \qquad\times\Big\langle Q_{31^{\prime}1v}s_{2^{\prime}3}-s_{2^{\prime}3}s_{31^{\prime}}-s_{1v}+1-\frac{1}{N_{c}^{2}}\big(Q_{2^{1v\prime}1^{\prime}}-s_{1v}-s_{2^{\prime}1^{\prime}}+1\big)\Big\rangle+\mathrm{c.c.}\Bigg\}\;.
\end{aligned}
\label{eq:QbarSWQbarFS}
\end{equation}
The contributions to the cross section due to the interference between
gluon radiation from the quark and from the antiquark read:
\begin{equation}
\begin{aligned} & \int_{\mathbf{p}_{3}}\Big[\mathcal{M}_{\mathrm{\overline{Q}SW}}^{\dagger}\mathcal{M}_{\mathrm{QSW}}+\mathcal{M}_{\mathrm{\overline{Q}SW}}^{\dagger}\mathcal{M}_{\mathrm{QFS}}+\mathcal{M}_{\mathrm{\overline{Q}FS}}^{\dagger}\mathcal{M}_{\mathrm{QSW}}+\mathrm{c.c.}\Big]\\
 & =-2(2\pi)^{2}\alpha_{\mathrm{em}} e^2_f \alpha_{s}N_{c}^{2}\mathrm{Tr}\Big(\mathrm{Dirac}_{\mathrm{\overline{Q}SW}}^{\eta^{\prime}\lambda^{\prime}\dagger}\mathrm{Dirac}_{\mathrm{QSW}}^{\bar{\eta}\bar{\lambda}}\Big)\frac{p_{3}^{+}}{p_{j1}^{+}+p_{3}^{+}}\frac{p_{3}^{+}}{p_{j2}^{+}+p_{3}^{+}}\\
 & \times\int_{\mathbf{v},\mathbf{v}^{\prime}}\int_{\mathbf{x}_{1},\mathbf{x}_{1^{\prime}},\mathbf{x}_{2},\mathbf{x}_{2^{\prime}},\mathbf{x}_3}e^{-i\mathbf{p}_{j1}\cdot\mathbf{x}_{11^{\prime}}}e^{-i\mathbf{p}_{j2}\cdot\mathbf{x}_{22^{\prime}}}\delta^{(2)}\big(\mathbf{v}^{\prime}-\frac{p_{j2}^{+}}{p_{j2}^{+}+p_{3}^{+}}\mathbf{x}_{2^{\prime}}-\frac{p_{3}^{+}}{p_{j2}^{+}+p_{3}^{+}}\mathbf{x}_{3}\big)\\
 & \times\delta^{(2)}\big(\mathbf{v}-\frac{p_{j1}^{+}}{p_{j1}^{+}+p_{3}^{+}}\mathbf{x}_{1}-\frac{p_{3}^{+}}{p_{j1}^{+}+p_{3}^{+}}\mathbf{x}_{3}\big)A^{\eta^{\prime}}(\mathbf{x}_{2^{\prime}3})A^{\bar{\eta}}(\mathbf{x}_{13})\\
 & \times\Bigg\{\mathcal{A}^{\lambda^{\prime}}\big(\mathbf{v}^{\prime}-\mathbf{x}_{1^{\prime}},\mathbf{x}_{32^{\prime}},\frac{p_{j2}^{+}p_{3}^{+}q^{+}}{p_{j1}^{+}(p_{j2}^{+}+p_{3}^{+})^{2}}\big)\mathcal{A}^{\bar{\lambda}}\big(\mathbf{v}-\mathbf{x}_{2},\mathbf{x}_{31},\frac{p_{j1}^{+}p_{3}^{+}q^{+}}{p_{j2}^{+}(p_{j1}^{+}+p_{3}^{+})^{2}}\big)\\
 & \qquad\times\Big\langle s_{11^{\prime}}s_{2^{\prime}2}-s_{2^{\prime}3}s_{31^{\prime}}-s_{13}s_{32}+1-\frac{1}{N_{c}^{2}}\big(Q_{122^{\prime}1^{\prime}}-s_{12}-s_{2^{\prime}1^{\prime}}+1\big)\Big\rangle\\
 & \quad-\mathcal{A}^{\lambda^{\prime}}\big(\mathbf{v}^{\prime}-\mathbf{x}_{1^{\prime}},\mathbf{x}_{32^{\prime}},\frac{p_{j2}^{+}p_{3}^{+}q^{+}}{p_{j1}^{+}(p_{j2}^{+}+p_{3}^{+})^{2}}\big)A^{\bar{\lambda}}(\mathbf{v}-\mathbf{x}_{2})\\
 & \qquad\times\Big\langle Q_{v22^{\prime}3}s_{31^{\prime}}-s_{v2}-s_{2^{\prime}3}s_{31^{\prime}}+1-\frac{1}{N_{c}^{2}}\Big(Q_{v22^{\prime}1^{\prime}}-s_{v2}-s_{2^{\prime}1^{\prime}}+1\Big)\Big\rangle\\
 & \quad-A^{\lambda^{\prime}}(\mathbf{v}^{\prime}-\mathbf{x}_{1^{\prime}})\mathcal{A}^{\bar{\lambda}}\big(\mathbf{v}-\mathbf{x}_{2},\mathbf{x}_{31},\frac{p_{j1}^{+}p_{3}^{+}q^{+}}{p_{j2}^{+}(p_{j1}^{+}+p_{3}^{+})^{2}}\big)\\
 & \qquad\times\Big\langle Q_{v^{\prime}1^{\prime}13}s_{32}-s_{13}s_{32}-s_{v^{\prime}1^{\prime}}+1-\frac{1}{N_{c}^{2}}\Big(Q_{v^{\prime}1^{\prime}12}-s_{12}-s_{v^{\prime}1^{\prime}}+1\Big)\Big\rangle\\
 &\quad +\mathrm{c.c.}\Bigg\}\;.
\end{aligned}
\label{eq:QbarSWQSW}
\end{equation}
The Dirac traces appearing in the above two expressions are easily
calculated using the identities (\ref{eq:sigmasigma}) and (\ref{eq:traceP}),
and read:
\begin{equation}
\begin{aligned}&\mathrm{Tr}\Big(\mathrm{Dirac}_{\mathrm{QSW}}^{\eta^{\prime}\lambda^{\prime}\dagger}\mathrm{Dirac}_{\mathrm{QSW}}^{\bar{\eta}\bar{\lambda}}\Big) \\ &=32p_{1}^{+}p_{2}^{+}\Bigg[\frac{(p_{1}^{+})^{2}+(p_{1}^{+}+p_{3}^{+})^{2}}{(p_{3}^{+})^{2}}\frac{(p_{2}^{+})^{2}+(p_{1}^{+}+p_{3}^{+})^{2}}{(q^{+})^{2}}\delta^{\bar{\lambda}\lambda^{\prime}}\delta^{\bar{\eta}\eta^{\prime}}\\
 & \qquad\qquad\quad+\frac{(2p_{1}^{+}+p_{3}^{+})(p_{1}^{+}-p_{2}^{+}+p_{3}^{+})}{p_{3}^{+}q^{+}}\epsilon^{\bar{\lambda}\lambda^{\prime}}\epsilon^{\bar{\eta}\eta^{\prime}}\Bigg]\;,
\end{aligned}
\label{eq:DiracQSWQSW}
\end{equation}
\begin{equation}
\begin{aligned}\mathrm{Tr}\Big(\mathrm{Dirac}_{\mathrm{\overline{Q}SW}}^{\eta^{\prime}\lambda^{\prime}\dagger}\mathrm{Dirac}_{\mathrm{\overline{Q}SW}}^{\bar{\eta}\bar{\lambda}}\Big) & =\mathrm{Tr}\Big(\mathrm{Dirac}_{\mathrm{QSW}}^{\eta^{\prime}\lambda^{\prime}\dagger}\mathrm{Dirac}_{\mathrm{QSW}}^{\bar{\eta}\bar{\lambda}}\Big)(1\leftrightarrow2)\;.\end{aligned}
\label{eq:DiracQbarSWQbarSW}
\end{equation}

\paragraph{RI: instantaneous gluon emission}
We find for the contribution due to the instantaneous emission of
a gluon:
\begin{equation}
\begin{aligned}&\int_{\mathbf{p}_{3}}\big|\mathcal{M}_{\mathrm{RI}}\big|^{2}  =4(2\pi)^{2}\alpha_{\mathrm{em}} e^2_f \alpha_{s}\frac{p_{1}^{+2}p_{3}^{+2}(p_{1}^{+}+p_{3}^{+})^{2}}{(q^{+})^{6}}\\&\mathrm{Tr}\big|\mathrm{Dirac}_{\mathrm{RI}}\big|^{2}\\;int_{\mathbf{x}_{1^{\prime}},\mathbf{x}_{2^{\prime}},\mathrm{\mathbf{x}}_{1},\mathbf{x}_{2},\mathbf{x}_{3}}e^{-i\mathbf{p}_{1}\cdot\mathbf{x}_{11^{\prime}}}e^{-i\mathbf{p}_{2}\cdot\mathbf{x}_{22^{\prime}}}\\
 & \times\mathcal{C}\big(\frac{p_{1}^{+}}{q^{+}}\mathbf{x}_{21}+\frac{p_{3}^{+}}{q^{+}}\mathbf{x}_{23},\mathbf{x}_{31},\frac{p_{1}^{+}p_{3}^{+}}{q^{+}p_{2}^{+}}\big)\mathcal{C}\big(\frac{p_{1}^{+}}{q^{+}}\mathbf{x}_{2^{\prime}1^{\prime}}+\frac{p_{3}^{+}}{q^{+}}\mathbf{x}_{2^{\prime}3},\mathbf{x}_{31^{\prime}},\frac{p_{1}^{+}p_{3}^{+}}{q^{+}p_{2}^{+}}\big)\\
 & \times\Big\langle s_{11^{\prime}}s_{2^{\prime}2}-s_{2^{\prime}3}s_{31^{\prime}}-s_{13}s_{32}+1-\frac{1}{N_{c}^{2}}\big(Q_{122^{\prime}1^{\prime}}-s_{12}-s_{2^{\prime}1^{\prime}}+1\big)\Big\rangle\;,
\end{aligned}
\label{eq:RIRI}
\end{equation}
with the Dirac trace:
\begin{equation}
\begin{aligned}\mathrm{Tr}\big|\mathrm{Dirac}_{\mathrm{RI}}\big|^{2} & =\frac{16(q^{+})^{2}p_{1}^{+}p_{2}^{+}}{(p_{1}^{+}+p_{3}^{+})^{2}(p_{2}^{+}+p_{3}^{+})^{2}}\Big((p_{1}^{+}-p_{2}^{+})^{2}+(p_{1}^{+}+p_{2}^{+}+2p_{3}^{+})^{2}\Big)\;.\end{aligned}
\label{eq:DiracRIRI}
\end{equation}

\paragraph{Interference terms}
Finally, the interference terms due to instantaneous gluon emission
in the complex conjugate amplitude and radiation from a quark in the
amplitude read:
\begin{equation}
\begin{aligned} & \int_{\mathbf{p}_{3}}\mathcal{M}_{\mathrm{RI}}^{\dagger}\Big(\mathcal{M}_{\mathrm{QSW}}+\mathcal{M}_{\mathrm{QFS}}\Big)=2(2\pi)^{2}\alpha_{\mathrm{em}} e^2_f \alpha_{s}N_{c}^{2}\frac{p_{1}^{+}(p_{3}^{+})^{2}}{(q^{+})^{3}}\mathrm{Tr}\Big(\mathrm{Dirac}_{\mathrm{RI}}^{\dagger}\mathrm{Dirac}_{\mathrm{QSW}}^{\bar{\eta}\bar{\lambda}}\Big)\\
 & \times\int_{\mathbf{x}_{1^{\prime}},\mathbf{x}_{2^{\prime}},\mathbf{x}_{1},\mathbf{x}_{2},\mathbf{x}_{3},\mathbf{v}}e^{-i\mathbf{p}_{1}\cdot\mathbf{x}_{11^{\prime}}}e^{-i\mathbf{p}_{2}\cdot\mathbf{x}_{22^{\prime}}}\delta^{(2)}\big(\mathbf{v}-\frac{p_{1}^{+}}{p_{1}^{+}+p_{3}^{+}}\mathbf{x}_{1}-\frac{p_{3}^{+}}{p_{1}^{+}+p_{3}^{+}}\mathbf{x}_{3}\big)\\
 & \times\mathcal{C}\big(\frac{p_{1}^{+}}{q^{+}}\mathbf{x}_{2^{\prime}1^{\prime}}+\frac{p_{3}^{+}}{q^{+}}\mathbf{x}_{2^{\prime}3},\mathbf{x}_{31^{\prime}},\frac{p_{1}^{+}p_{3}^{+}}{q^{+}p_{2}^{+}}\big)A^{\bar{\eta}}(\mathbf{x}_{13})\\
 & \times\Bigg\{-\mathcal{A}^{\bar{\lambda}}\big(\mathbf{v}-\mathbf{x}_{2},\mathbf{x}_{31},\frac{p_{1}^{+}p_{3}^{+}q^{+}}{p_{2}^{+}(p_{1}^{+}+p_{3}^{+})^{2}}\big)\\
 & \times\Big\langle s_{11^{\prime}}s_{2^{\prime}2}-s_{32}s_{13}-s_{31^{\prime}}s_{2^{\prime}3}+1-\frac{1}{N_{c}^{2}}\Big(Q_{122^{\prime}1^{\prime}}-s_{12}-s_{2^{\prime}1^{\prime}}+1\Big)\Big\rangle\\
 & +A^{\bar{\lambda}}(\mathbf{v}-\mathbf{x}_{2})\Big\langle Q_{v22^{\prime}3}s_{31^{\prime}}-s_{v2}-s_{31^{\prime}}s_{2^{\prime}3}+1-\frac{1}{N_{c}^{2}}\Big(Q_{v22^{\prime}1^{\prime}}-s_{v2}-s_{2^{\prime}1^{\prime}}+1\Big)\Big\rangle\Bigg\}\;,
\end{aligned}
\label{eq:RIQSWQFS}
\end{equation}
with the Dirac trace:
\begin{equation}
\begin{aligned}\mathrm{Tr}\Big(\mathrm{Dirac}_{\mathrm{RI}}^{\dagger}\mathrm{Dirac}_{\mathrm{QSW}}^{\bar{\eta}\bar{\lambda}}\Big) & =-32p_{1}^{+}p_{2}^{+}\delta^{\bar{\lambda}\bar{\eta}}\frac{1}{p_{3}^{+}}\Big(\frac{(p_{1}^{+}+p_{3}^{+})^{2}}{p_{2}^{+}+p_{3}^{+}}+\frac{p_{1}^{+}p_{2}^{+}}{p_{1}^{+}+p_{3}^{+}}\Big)\;.
\end{aligned}
\label{eq:DiracRIQSW}
\end{equation}
Likewise, for the interference between instantaneous emission with
gluon radiation from the antiquark, we find:
\begin{equation}
\begin{aligned} & \int_{\mathbf{p}_{3}}\mathcal{M}_{\mathrm{RI}}^{\dagger}\Big(\mathcal{M}_{\mathrm{\overline{Q}SW}}+\mathcal{M}_{\mathrm{\overline{Q}FS}}\Big)\\&=-2(2\pi)\alpha_{\mathrm{em}} e^2_f \alpha_{s}N_{c}^{2}\frac{p_{1}^{+}(p_{3}^{+})^{2}(p_{1}^{+}+p_{3}^{+})}{(q^{+})^{3}(p_{2}^{+}+p_{3}^{+})}\mathrm{Dirac}_{\mathrm{RI}}^{\dagger}\mathrm{Dirac}_{\mathrm{\bar{Q}SW}}^{\bar{\eta}\bar{\lambda}}\\
 & \times\int_{\mathbf{x}_{1^{\prime}},\mathbf{x}_{2^{\prime}},\mathbf{x}_{1},\mathbf{x}_{2},\mathbf{x}_{3},\mathbf{v}}e^{-i\mathbf{p}_{1}\cdot\mathbf{x}_{11^{\prime}}}e^{-i\mathbf{p}_{2}\cdot\mathbf{x}_{22^{\prime}}}\delta^{(2)}\big(\mathbf{v}-\frac{p_{2}^{+}}{p_{2}^{+}+p_{3}^{+}}\mathbf{x}_{2}-\frac{p_{3}^{+}}{p_{2}^{+}+p_{3}^{+}}\mathbf{x}_{3}\big)\\
 & \times\mathcal{C}\big(\frac{p_{1}^{+}}{q^{+}}\mathbf{x}_{2^{\prime}1^{\prime}}+\frac{p_{3}^{+}}{q^{+}}\mathbf{x}_{2^{\prime}3^{\prime}},\mathbf{x}_{3^{\prime}1^{\prime}},\frac{p_{1}^{+}p_{3}^{+}}{q^{+}p_{2}^{+}}\big)A^{\bar{\eta}}(\mathbf{x}_{23})\\
 & \times\Bigg\{-\mathcal{A}^{\bar{\lambda}}\big(\mathbf{v}-\mathbf{x}_{1},\mathbf{x}_{32},\frac{p_{2}^{+}p_{3}^{+}q^{+}}{p_{1}^{+}(p_{2}^{+}+p_{3}^{+})^{2}}\big)\\
 & \times\Big\langle s_{2^{\prime}2}s_{11^{\prime}}-s_{13}s_{32}-s_{2^{\prime}3}s_{31^{\prime}}+1-\frac{1}{N_{c}^{2}}\Big(Q_{122^{\prime}1^{\prime}}-s_{12}-s_{2^{\prime}1^{\prime}}+1\Big)\Big\rangle\\
 & +A^{\bar{\lambda}}(\mathbf{v}-\mathbf{x}_{1})\Big\langle Q_{31^{\prime}1v}s_{2^{\prime}3}-s_{1v}-s_{2^{\prime}3}s_{31^{\prime}}+1-\frac{1}{N_{c}^{2}}\Big(Q_{2^{1v\prime}1^{\prime}}-s_{1v}-s_{2^{\prime}1^{\prime}}+1\Big)\Big\rangle\Bigg\}\;,
\end{aligned}
\label{eq:RIQbarSWQbarFS}
\end{equation}
with the Dirac trace:
\begin{equation}
\begin{aligned}\mathrm{Tr}\Big(\mathrm{Dirac}_{\mathrm{RI}}^{\dagger}\mathrm{Dirac}_{\mathrm{\overline{Q}SW}}^{\bar{\eta}\bar{\lambda}}\Big) & =32p_{1}^{+}p_{2}^{+}\delta^{\bar{\lambda}\bar{\eta}}\frac{1}{p_{3}^{+}}\Big(\frac{(p_{2}^{+}+p_{3}^{+})^{2}}{p_{1}^{+}+p_{3}^{+}}+\frac{p_{1}^{+}p_{2}^{+}}{p_{2}^{+}+p_{3}^{+}}\Big)\;.
\end{aligned}
\label{eq:DiracRIQbarSW}
\end{equation}

\section{\label{sec:corrlimit}Correlation limit}
In the so-called `correlation limit' (ref.~\cite{Dominguez:2010xd,Dominguez:2011wm}),
the two outgoing jets are back-to-back in the transverse plane. In
this kinematic configuration, we recover a scale $\mathbf{k}_{\perp}=\mathbf{p}_{j1}+\mathbf{p}_{j2}$
set by the vector sum of the transverse jet momenta, which is small
with respect to the center-of-mass energy $\sqrt{s}$ and the large transverse momentum
of the jets $\mathbf{P}_{\perp}^{2}\sim\mathbf{p}_{j1}^{2}\sim\mathbf{p}_{j2}^{2}$. $\mathbf{k}_\perp^2$ is typically of the order of the saturation scale $Q_{s}^2$ but
can in principle even become nonperturbative:
\begin{equation}
s\gg\mathbf{P}_{\perp}^{2}\gg\mathbf{k}_{\perp}^{2}\sim Q_{s}^{2}\;.
\end{equation}

The emergence of the large ratio $\mathbf{P}_{\perp}^{2}/\mathbf{k}_{\perp}^{2}\gg1$
implies the appearance of large `Sudakov' logarithms,
which need to be resummed on top of the rapidity logarithms $Y_f^+\propto\ln s$. Sudakov resummation is governed by the Collins-Soper-Sterman (CSS~\cite{Collins:1984kg,Collins:1988ig})
evolution equations, which are embedded in the framework of transverse momentum dependent (TMD) factorization (ref.~\cite{Collins:2011zzd,Angeles-Martinez:2015sea}). In TMD factorization, the cross section can be written as the convolution
of a hard part with TMD parton distribution functions and fragmentation functions (TMD PDFs and TMD FFs), which evolve according to CSS.

It has been known since a long time~\cite{Dominguez:2011wm} that, in the correlation limit, the leading-order CGC
cross section~\eqref{eq:LOXsection} can be written as the product of a hard part with a gluon TMD PDF, consistent with what one would find in TMD factorization at tree level. One part of the proof that this correspondence holds at higher orders, is to demonstrate that the Sudakov logarithms in our NLO cross section have the right form and can be absorbed into CSS. It turns out that this first step is already highly non-trivial, and we will limit ourselves in this work to the Sudakov double logarithms at large $N_c$.

To double leading logarithmic (DLL) accuracy, the TMD factorization formula for our process $\gamma A\to \mathrm{dijet}+X$ is:
\begin{equation}
\begin{aligned}\frac{\mathrm{d}\sigma_{\mathrm{DLL}}^{\mathrm{TMD}}}{\mathrm{d}z\mathrm{d}\bar{z}\mathrm{\mathrm{d}}^{2}\mathbf{P}_{\perp}\mathrm{\mathrm{d}}^{2}\mathbf{k}_{\perp}}= & \frac{\mathrm{d}\sigma_{\mathrm{LO}}^{\mathrm{TMD}}}{\mathrm{d}z\mathrm{d}\bar{z}\mathrm{\mathrm{d}}^{2}\mathbf{P}_{\perp}\mathrm{\mathrm{d}}^{2}\mathbf{k}_{\perp}}\times e^{-\frac{1}{2}S_{A}(\mathbf{b}-\mathbf{b}^\prime,\,\mathbf{P}_{\perp})}\;,\end{aligned}
\end{equation}
with some slight abuse of notation, since the product is really at the integrand level, and where $S_A$ is the perturbative Sudakov factor which reads at LO and in the DLL approximation:
\begin{equation}
S_{A}(\mathbf{b}-\mathbf{b}^\prime,\,\mathbf{P}_{\perp})=\frac{\alpha_{s}N_{c}}{2\pi}\ln^{2}\frac{\mathbf{P}_{\perp}^{2}(\mathbf{b}-\mathbf{b}^\prime)^{2}}{c_0^2}\;,\label{eq:doubleSudakovlog}
\end{equation}
In the above formulas, $\mathbf{b}-\mathbf{b}^\prime$ is the transverse coordinate conjugate to $\mathbf{k}_\perp$.

We will show in this section that our NLO cross section, in the correlation limit, takes the form
\begin{equation}
\begin{aligned}\mathrm{d}\sigma^{\mathrm{NLO}}_{\gamma A\to \mathrm{dijet}+X}&\overset{\mathrm{corr.\;lim.}}{=} \mathrm{d}\sigma_{\mathrm{LO}}^{\mathrm{TMD}}\Big(1-\frac{\alpha_{s}N_{c}}{4\pi}\ln^{2}\frac{\mathbf{P}_{\perp}^{2}(\mathbf{b}-\mathbf{b}^\prime)^{2}}{c_0^2}+\mathcal{O}(\mathrm{single\;logs})\Big)\\&\quad\;+\mathcal{O}(\mathrm{finite\;terms})\;,\end{aligned}
\label{eq:DLL}
\end{equation}
hence proving agreement between the CGC and the TMD calculations, at least to DLL accuracy. The Sudakov double logarithm comes, at least at leading $N_c$, from soft-collinear gluon radiation just outside the jet, i.e. from the contributions $\mathcal{V}_{|{\mathrm{QFS}}|^2}^{\textrm{out; phase}}$ (eqs.~\eqref{eq:QFSsquare_out_phase_1} and \eqref{eq:QFSsquare_out_phase_2} and its $q\leftrightarrow\bar{q}$ counterpart. However, as was remarked very recently in~\cite{FaridJyvaskyla}, taken at face value the double logarithm in this contribution comes with the wrong sign (see sec.~\ref{sec:wrongsign}). As we will demonstrate in subsection~\ref{subsec:Sudakov_and_kc_JIMWLK}, this wrong sign is compensated for by the mismatch between naive and kinematically-improved low-$x$ resummation.

Finally, we should remark that in the seminal paper ref.~\cite{Mueller:2013wwa}, eq.~\ref{eq:DLL} was already inferred from an analysis of Higgs hadroproduction combined with kinematical arguments. While physically insightful, the approach to the dijet case in~\cite{Mueller:2013wwa} has some limitations. The calculation presented in this section relies on the complete NLO dijet calculation instead, which requires a systematic treatment of the low-$x$ resummation. As such, our treatment constitutes an important step towards reaching full NLO accuracy in the back-to-back limit.

\subsection{Leading order}

Introducing the vector sum of the transverse momenta of the outgoing
(anti)quark:
\begin{equation}
\begin{aligned}\mathbf{k}_{\perp} & =\mathbf{p}_{1}+\mathbf{p}_{2}\;,\end{aligned}
\end{equation}
one can rewrite the transverse coordinates in function of $\mathbf{b}$
and $\mathbf{r}$ which are conjugate to $\mathbf{k}_{\perp}$ and
to $\mathbf{P}_{\perp}$, respectively (remember that $z=p_{1}^{+}/q^{+}$
and $\bar{z}=p_{2}^{+}/q^{+}$):
\begin{equation}
\begin{aligned}\mathbf{x}_{1} & =\mathbf{b}+\bar{z}\mathbf{r},\quad\mathbf{x}_{2}=\mathbf{b}-z\mathbf{r}\;.\end{aligned}
\label{eq:corrlimit}
\end{equation}
After this coordinate transform, the squared of the LO amplitude (\ref{eq:MLOsquared})
becomes:
\begin{equation}
\begin{aligned}\big|\mathcal{M}_{\mathrm{LO}}\big|^{2} & =16(4\pi)\alpha_{\mathrm{em}}e_f^2 p_{1}^{+}p_{2}^{+}(z^{2}+\bar{z}^{2})N_{c}\\
 & \times\int_{\mathrm{\mathbf{r}},\mathrm{\mathbf{r}}^{\prime},\mathrm{\mathbf{b}},\mathbf{b}^{\prime}}e^{-i\mathbf{P}_{\perp}\cdot(\mathbf{r}-\mathbf{r}^{\prime})}e^{-i\mathbf{k}_{\perp}\cdot(\mathbf{b}-\mathbf{b}^{\prime})}A^{\lambda^{\prime}}(\mathbf{r})A^{\lambda^{\prime}}(\mathbf{r}^{\prime})\\
 & \times\mathrm{Tr}\Big\langle Q_{122^{\prime}1^{\prime}}-s_{12}-s_{2^{\prime}1^{\prime}}+1\Big\rangle\;.
\end{aligned}
\label{eq:LOsquaredcorr0}
\end{equation}
Taking the correlation limit $\mathbf{P}_{\perp}\gg\mathbf{k}_{\perp}$
then implies $\mathbf{b},\mathbf{b}^{\prime}\gg\mathbf{r},\mathbf{r}^{\prime}$,
which we will use to perform a Taylor expansion of the Wilson lines.
At the lowest non-trivial order, the only nonzero contribution will
come from the quadrupole since all $\mathcal{O}(\mathbf{r})$ corrections
on dipoles evaluated in either two unprimed or two primed coordinates
disappear:
\begin{equation}
\begin{aligned}\mathrm{Tr}\Big[\partial_{\mathbf{r}}^{i}(U_{\mathbf{x}_{i}})U_{\mathbf{b}}^{\dagger}\Big]\Big|_{\mathbf{\mathbf{r}}=0} & =\mathrm{Tr}\Big[\partial_{\mathbf{r}}^{i}\partial_{\mathbf{r}}^{j}(U_{\mathbf{x}_{i}})U_{\mathbf{b}}^{\dagger}\Big]\Big|_{\mathbf{\mathbf{r}}=0}=0\;.\end{aligned}
\end{equation}
It is then easy to see that:
\begin{equation}
\mathrm{Tr}\Big\langle Q_{122^{\prime}1^{\prime}}-s_{12}-s_{2^{\prime}1^{\prime}}+1\Big\rangle\simeq\mathbf{r}^{i}\mathbf{r}^{\prime j}\frac{\mathrm{Tr}}{N_{c}}\Big\langle U_{\mathbf{b}}\big(\partial^{i}U_{\mathbf{b}}^{\dagger}\big)\big(\partial^{j}U_{\mathbf{b}^{\prime}}\big)U_{\mathbf{b}^{\prime}}^{\dagger}\Big\rangle
\end{equation}
up to higher orders in the Taylor expansion. eq.~\eqref{eq:LOsquaredcorr0})
then simplifies to:
\begin{equation}
\begin{aligned}\big|\mathcal{M}_{\mathrm{LO}}\big|^{2} & \overset{\mathrm{TMD}}{=}16\frac{\alpha_{\mathrm{em}}e_f^2}{\pi}p_{1}^{+}p_{2}^{+}(z^{2}+\bar{z}^{2})\int_{\mathrm{\mathbf{r}},\mathrm{\mathbf{r}}^{\prime}}e^{-i\mathbf{P}_{\perp}\cdot(\mathbf{r}-\mathbf{r}^{\prime})}\frac{\mathbf{r}\cdot\mathbf{r}^{\prime}}{\mathbf{r}^{2}\mathbf{r}^{\prime2}}\mathbf{r}^{i}\mathbf{r}^{\prime j}\\
 & \times\int_{\mathbf{b},\mathbf{b}^{\prime}}e^{-i\mathbf{k}_{\perp}\cdot(\mathbf{b}-\mathbf{b}^{\prime})}\mathrm{Tr}\Big\langle U_{\mathbf{b}}\big(\partial^{i}U_{\mathbf{b}}^{\dagger}\big)\big(\partial^{j}U_{\mathbf{b}^{\prime}}\big)U_{\mathbf{b}^{\prime}}^{\dagger}\Big\rangle\;.
\end{aligned}
\label{eq:LOcorrelation}
\end{equation}
In the thus obtained Wilson-line structure, one can recognize the
so-called \textquoteleft hadron correlator' which is parameterized
by the unpolarized $\mathcal{F}_{\mathrm{WW}}$ and linearly-polarized
$\mathcal{H}_{\mathrm{WW}}$ Weizs\"acker-Williams gluon TMD (\cite{Mulders:2000sh}):
\begin{equation}
\begin{aligned} & \int_{\mathbf{b},\mathbf{b}^{\prime}}e^{-i\mathbf{k}_{\perp}\cdot(\mathbf{b}-\mathbf{b}^{\prime})}\mathrm{Tr}\Big\langle U_{\mathbf{b}}\big(\partial^{i}U_{\mathbf{b}}^{\dagger}\big)\big(\partial^{j}U_{\mathbf{b}^{\prime}}\big)U_{\mathbf{b}^{\prime}}^{\dagger}\Big\rangle\\
 & =g_{s}^{2}(2\pi)^{3}\frac{1}{4}\Bigg[\frac{\delta^{ij}}{2}\mathcal{F}_{\mathrm{WW}}(x_{{\scriptscriptstyle A}},\mathbf{k}_{\perp})+\Big(\frac{\mathbf{k}_{\perp}^{i}\mathbf{k}_{\perp}^{j}}{\mathbf{k}_{\perp}^{2}}-\frac{\delta^{ij}}{2}\Big)\mathcal{H}_{\mathrm{WW}}(x_{{\scriptscriptstyle A}},\mathbf{k}_{\perp})\Bigg]\;.
\end{aligned}
\label{eq:hadroncorrelator}
\end{equation}
Finally, using eq.~\eqref{eq:crosssectiondef}), we end up with (note
that at LO and at the cross section level, the parton momenta can
be identified with jet momenta):
\begin{equation}
\begin{aligned}\frac{\mathrm{d}\sigma_{\mathrm{LO}}^{\mathrm{TMD}}}{\mathrm{d}z\mathrm{d}\bar{z}\mathrm{\mathrm{d}}^{2}\mathbf{P}_{\perp}\mathrm{\mathrm{d}}^{2}\mathbf{k}_{\perp}} & =\frac{2\alpha_{\mathrm{em}}e_f^2}{(2\pi)^{7}}(z^{2}+\bar{z}^{2})2\pi\delta(z+\bar{z}-1)\int_{\mathrm{\mathbf{r}},\mathrm{\mathbf{r}}^{\prime}}e^{-i\mathbf{P}_{\perp}\cdot(\mathbf{r}-\mathbf{r}^{\prime})}\frac{\mathbf{r}\cdot\mathbf{r}^{\prime}}{\mathbf{r}^{2}\mathbf{r}^{\prime2}}\mathbf{r}^{i}\mathbf{r}^{\prime j}\\
 & \times\int_{\mathbf{b},\mathbf{b}^{\prime}}e^{-i\mathbf{k}_{\perp}\cdot(\mathbf{b}-\mathbf{b}^{\prime})}\mathrm{Tr}\Big\langle U_{\mathbf{b}}\big(\partial^{i}U_{\mathbf{b}}^{\dagger}\big)\big(\partial^{j}U_{\mathbf{b}^{\prime}}\big)U_{\mathbf{b}^{\prime}}^{\dagger}\Big\rangle\;,\\
 & =\frac{\alpha_{\mathrm{em}} e^2_f \alpha_{s}}{\mathbf{P}_{\perp}^{4}}(z^{2}+\bar{z}^{2})\delta(1-z-\bar{z})\mathcal{F}_{\mathrm{WW}}(x_{{\scriptscriptstyle A}},\mathbf{k}_{\perp})\;,
\end{aligned}
\end{equation}
which is the same expression as one would find in a leading-order
TMD factorization approach~\cite{Dominguez:2011wm}. Note that at
the present lowest perturbative order, one needs an extra scale such
as a nonzero photon virtuality $Q^{2}$ or heavy-quark mass to be
sensitive to the linearly-polarized gluon TMD $\mathcal{H}_{\mathrm{WW}}$
\cite{Dominguez:2011br,Akcakaya:2012si,Marquet:2017xwy}.


\subsection{\label{sec:wrongsign}Sudakov double logs in the NLO cross section}

The contributions $ \mathcal{V}_{|{\mathrm{QFS}}|^2}^{\textrm{out; phase}}$ and  $\mathcal{V}_{|{\mathrm{\overline{Q}FS}}|^2}^{\textrm{out; phase}} $ (see eqs.~\eqref{eq:QFSsquare_out_phase_1}, \eqref{eq:QFSsquare_out_phase_2} and \eqref{eq:QbarFSsquare_out_phase_2}) factorize from the LO cross section (at integrand level), and can produce double large logarithms. In the correlation limit, since $\mathbf{p}_{j1}\rightarrow \mathbf{P}_{\perp}$ and $\mathbf{x}_{11^{\prime}}\rightarrow \mathbf{b}-\mathbf{b}^{\prime}$, $ \mathcal{V}_{|{\mathrm{QFS}}|^2}^{\textrm{out; phase}}$ becomes
\begin{equation}
\begin{aligned}
 \mathcal{V}_{|{\mathrm{QFS}}|^2}^{\textrm{out; phase}}
&\underset{\mathrm{corr.\;lim.}}{=} 
 2\int_{0}^{1}\frac{\mathrm{d}\xi}{\xi}
 \Big[e^{-i\xi  \mathbf{P}_{\perp}\cdot(\mathbf{b}-\mathbf{b}^{\prime})}-1\Big]\\
 &\quad\;\times \left[-2\ln R -\ln\left(\frac{\mathbf{P}_{\perp}^{2}(\mathbf{b}-\mathbf{b}^{\prime})^{2}}{{c_{0}^{2}}}\right)
-2\ln\xi\right]
  \; ,
\end{aligned}
\label{eq:QFSsquare_out_phase_corr_lim_1}
\end{equation}
and $\mathcal{V}_{|{\mathrm{\overline{Q}FS}}|^2}^{\textrm{out; phase}} $ has the same expression in the correlation limit.
For simplicity let us restrict ourselves to the case in which the azimuthal angle of $\mathbf{P}_{\perp}$ is integrated over. Then, we have
\begin{equation}
\begin{aligned}
& \mathcal{V}_{|{\mathrm{QFS}}|^2}^{\textrm{out; phase}}\\
&\underset{\mathrm{corr.\;lim.}}{=} 
 2\int_{0}^{1}\frac{\mathrm{d}\xi}{\xi}
 \Big[\textrm{J}_0\big(\xi  |\mathbf{P}_{\perp}|\, |\mathbf{b}-\mathbf{b}^{\prime}|\big)-1\Big]
  \left[-2\ln R -\ln\left(\frac{\xi^2\, \mathbf{P}_{\perp}^{2}(\mathbf{b}-\mathbf{b}^{\prime})^{2}}{{c_{0}^{2}}}\right)
\right]
\\
& \quad\;=
\frac{1}{2} \left(\ln\left(\frac{\mathbf{P}_{\perp}^{2}(\mathbf{b}-\mathbf{b}^{\prime})^{2}}{{c_{0}^{2}}}\right)\right)^2
+2 \ln R\, \ln\left(\frac{\mathbf{P}_{\perp}^{2}(\mathbf{b}-\mathbf{b}^{\prime})^{2}}{{c_{0}^{2}}}\right)
\\&\quad\;+{\cal O}\left(\frac{1}{\sqrt{|\mathbf{P}_{\perp}|\, |\mathbf{b}-\mathbf{b}^{\prime}|}}\right)
  \; ,
\end{aligned}
\label{eq:QFSsquare_out_phase_corr_lim_2}
\end{equation}
Hence, in the correlation limit, the NLO corrections from eqs.~\eqref{eq:QFSsquare_out_phase_1}, \eqref{eq:QFSsquare_out_phase_2} and \eqref{eq:QbarFSsquare_out_phase_2}) reduce to the $\textrm{LO}$ cross section with the extra factor
\begin{equation}
\begin{aligned}
&\frac{\alpha_s\, C_F}{2\pi} \bigg[\mathcal{V}_{|{\mathrm{QFS}}|^2}^{\textrm{out; phase}}
+\mathcal{V}_{|{\mathrm{\overline{Q}FS}}|^2}^{\textrm{out; phase}}
\bigg]
\overset{\mathrm{corr.\;lim.}}{=} 
\frac{\alpha_s\, C_F}{2\pi}\, \left(\ln\left(\frac{\mathbf{P}_{\perp}^{2}(\mathbf{b}-\mathbf{b}^{\prime})^{2}}{{c_{0}^{2}}}\right)\right)^2
\\&-\frac{2\alpha_s\, C_F}{\pi} \ln \left(\frac{1}{R}\right) \ln\left(\frac{\mathbf{P}_{\perp}^{2}(\mathbf{b}-\mathbf{b}^{\prime})^{2}}{{c_{0}^{2}}}\right)
+{\cal O}\left(\frac{\alpha_s\, C_F}{\sqrt{|\mathbf{P}_{\perp}|\, |\mathbf{b}-\mathbf{b}^{\prime}|}}\right)
\end{aligned}
\label{eq:QFSsquare_plus_QbarFSsquare_out_phase_corr_lim_1}
\end{equation}
in its integrand. These logarithms are indeed of the Sudakov type. However, the coefficient of the double log is positive, whereas the total Sudakov double log term should be negative.

The only other class of diagrams in our calculation which can give a Sudakov double logarithm are the diagrams involving soft divergences and which give a contribution with the same Wilson-line operator as at LO.
In the final-state exchange diagrams discussed in section~\ref{sec:soft}, such as GEFS or the real interference between $\mathrm{QFS}$ and $\mathrm{\overline{Q}FS}$,
 the leading-$N_c$ operators do not include quadrupoles (and drop in the correlation limit), and only subleading $N_c$ terms are proportional to the color operator appearing at LO.

In the case of high-energy logs  in section~\ref{sec:JIMWLK}, there is a cancelation of the subleading-$N_c$ terms between the $C_F$ terms from $|\mathrm{QFS}|^2$ and $|\mathrm{\overline{Q}FS}|^2$, the final state diagrams discussed in section~\ref{sec:soft}, and the real initial state emission diagrams $|\mathrm{QSW}+\mathrm{\overline{Q}SW}|^2$.
However, since our projectile, a photon, does not carry color charge, the initial state emission diagrams $|\mathrm{QSW}+\mathrm{\overline{Q}SW}|^2$ cannot have soft divergences. Hence, the cancelation of subleading $N_c$ terms occurring for high-energy logarithms cannot fully extend to the case of Sudakov double logarithms.

Due to the complexity of the diagrams from section~\ref{sec:soft}, we are not attempting in this study to calculate their possible subleading-$N_c$ contribution to Sudakov logarithms, and we stay instead at large-$N_c$. Hence, we can actually  replace $C_F$ by $N_c/2$ in eq.~\eqref{eq:QFSsquare_plus_QbarFSsquare_out_phase_corr_lim_1} at this accuracy.

In our calculation of the NLO dijet cross section in the correlation limit, we thus obtain Sudakov double logarithms with the expected coefficient at large $N_c$ but the wrong sign. We will now discuss the effect on this result from the kinematical improvement of the high-energy leading log resummation.

\subsection{\label{subsec:Sudakov_and_kc_JIMWLK}Sudakov double logs from the mismatch of naive and kinematically consistent low-$x$ resummation}


In order to perform the subtraction and resummation of low-$x$ (or high-energy) logs for the NLO dijet cross section, we have used the simplest scheme for the JIMWLK resummation. As explained in section~\ref{sec:kine}, in this scheme the evolution takes place along the $p^+$ axis, from a cutoff $k^+_{\textrm{min}}$ to a factorization scale $k^+_f$. The JIMWLK evolution then resums multiple gluon emissions within this interval, which are strongly ordered in $p^+$ only.

However, such a simple scheme for the low-$x$ resummation is known to fail beyond low-$x$ leading logarithmic accuracy. Indeed, such a scheme amounts to an approximation of infinite collision energy $\sqrt{s}$ in order to simplify the kinematics, which then leads to serious issues like NLO cross sections which can go negative and unstable low-$x$ evolution equations at next-to-leading log accuracy. The main ingredient for a resolution of these issues is an improvement of the kinematics in the high-energy evolution equation~\cite{Ciafaloni:1987ur,Andersson:1995ju,Kwiecinski:1996td,Salam:1998tj,Motyka:2009gi,Beuf:2014uia,Iancu:2015vea,Ducloue:2019ezk,Hatta:2016ujq}, also known as consistency constraint or kinematical constraint, which can be also understood as an all order resummation of leading collinear logarithms within the high-energy evolution equation, either BFKL~\cite{Ciafaloni:1987ur,Andersson:1995ju,Kwiecinski:1996td,Salam:1998tj,Motyka:2009gi} or BK~\cite{Motyka:2009gi,Beuf:2014uia,Iancu:2015vea,Ducloue:2019ezk}. The case of the JIMWLK equation has been much less studied so far, due to its complexity, although the same issue is present and can be addressed following a similar strategy~\cite{Hatta:2016ujq}.

The main reason for this failure is that the strong $p^+$ ordering of successive emissions is necessary but not sufficient for large high-energy leading logs to arise in higher-order calculations for a large but finite collision energy $\sqrt{s}$. Instead, the necessary and sufficient condition is that successive emissions should be strongly ordered both in $p^+$ and in $p^-$, in opposite directions. The JIMWLK equation in the scheme that we have used thus leads to an oversubtraction of high-energy logarithms, in the domain satisfying the $p^+$ ordering but violating the $p^-$ ordering. Note that the situation would have been similar if we had chosen a scheme for JIMWLK based on $p^-$ ordering only, except that the oversubtraction would have happened in a different kinematic domain.

The contribution to be subtracted from the NLO correction and resummed into the LL evolution of the LO term should thus be defined with two conditions, which require two factorization scales. Schematically, in addition to the condition $p_3^+<k_f^+$ for the gluon of momentum $p_3$ to participate to the high-energy evolution, one should also include a condition of the type
$p_3^->k_f^-$. Note that both factorization scales $k_f^+$ and $k_f^-$ are used to specify the endpoint of the high-energy evolution of the target. They should thus be chosen in relation to the kinematics of the observable (the dijet in our case), and be independent of the kinematics of the target and of the collision energy $\sqrt{s}$. First, $k_f^+$ should be smaller or equal to the $+$ momenta of both jets. A natural choice for $k_f^+$ is then:
\begin{equation}
\begin{aligned}
k_f^+ & =  \frac{p_{j1}^{+} p_{j2}^{+}}{q^+}
\;.
\end{aligned}
\label{kfplus_choice}
\end{equation}
The other factorization scale, $k_f^-$, should be a typical $-$ momentum scale set by the dijet final state. Using the dijet mass squared
\begin{equation}
\begin{aligned}
2 p_{j1}^{\mu}p_{j2\, \mu}
& =
\frac{\big(p_{j2}^{+} \mathbf{p}_{j1}   -p_{j1}^{+} \, \mathbf{p}_{j2}    \big)^2}{p_{j1}^{+} p_{j2}^{+}}
= \frac{(q^+)^2\, \mathbf{P}_{\perp}^2}{p_{j1}^{+} p_{j2}^{+}}
\label{dijet_mass}
\end{aligned}
\end{equation}
and $q^+$, a natural choice for $k_f^-$  can be written as:
\begin{equation}\begin{aligned}
k_f^- & =  \frac{2 p_{j1}^{\mu}p_{j2\, \mu} }{2q^+} = \frac{q^+\, \mathbf{P}_{\perp}^2}{2p_{j1}^{+} p_{j2}^{+}}= \frac{\mathbf{P}_{\perp}^2}{2k_f^{+}}
\;.
\end{aligned}
\label{kfminus_choice}
\end{equation}

The BK and JIMWLK evolution equation are usually written in mixed space, with $+$ momenta and transverse positions. Hence, one has no direct access to the $p^-$ of the gluon, which complicates the imposition of a condition of the type $p_3^->k_f^-$. In practice, one is led to build a proxy for the $p^-$ of the gluon out of the available mixed-space variables, so that  $k_f^-$ becomes a lower bound on a combination of $+$ momenta and transverse positions involved in the gluon emission~\cite{Motyka:2009gi,Beuf:2014uia,Iancu:2015vea,Ducloue:2019ezk,Hatta:2016ujq}.
In the present section, we do not need to enter into such technicalities. We can stay at a quite schematic level in order to understand the interplay between Sudakov double logs and the kinematical improvement of the JIMWLK equation.

Sudakov logarithms are known to exponentiate, and thus to appear at higher orders in terms containing the same Wilson-line operator (or the same parton distribution) as the LO term. Hence, we will now focus on the part in the JIMWLK evolution that involves this same LO operator. From the results of section~\ref{sec:JIMWLK}, in our naive scheme for JIMWLK, at the cross section level this contribution amounts to multiplying the LO cross section with the factor
\begin{equation}
\begin{aligned}
&2\alpha_s N_c\, \ln\left(\frac{k^+_f}{k^+_{\textrm{min}}}\right)
\int_{\mathbf{x}_3}
\Big[A^{\eta}(\mathbf{x}_{1^{\prime}3})A^{\eta}(\mathbf{x}_{13})
+A^{\eta}(\mathbf{x}_{2^{\prime}3})A^{\eta}(\mathbf{x}_{23})
\\&\qquad\qquad\qquad\quad\qquad\qquad-A^{\eta}(\mathbf{x}_{13})A^{\eta}(\mathbf{x}_{23})
-A^{\eta}(\mathbf{x}_{1^{\prime}3})A^{\eta}(\mathbf{x}_{2^{\prime}3})
\Big]
\\
 &=
2\alpha_s N_c\, \int_{k^+_{\textrm{min}}}^{k^+_f} \frac{\mathrm{d} p_3^+}{p_3^+}
\int_{\mathbf{P}_3} \frac{1}{\mathbf{P}_3^2}
\Big[e^{i\mathbf{P}_3\cdot \mathbf{x}_{1^{\prime}1}}+e^{i\mathbf{P}_3\cdot \mathbf{x}_{2^{\prime}2}}
-e^{i\mathbf{P}_3\cdot \mathbf{x}_{12}}-e^{i\mathbf{P}_3\cdot \mathbf{x}_{1^{\prime}2^{\prime}}}\Big]
\; ,
\end{aligned}
\label{LOlike_naive_JIMWLK_term}
\end{equation}
using the momentum representation \eqref{eq:WWfield} of the Weizs\"acker-Williams fields. The expression \eqref{LOlike_naive_JIMWLK_term} also corresponds to the leading-$N_c$ part of eq.~\eqref{eq:JIMWLK_term_in_VFS}.

For the contribution \eqref{LOlike_naive_JIMWLK_term}, we can implement the extra condition $p_3^->k_f^-$ corresponding to kinematical improvement directly in momentum space as
\begin{equation}\begin{aligned}
\frac{\mathbf{P}_3^2}{2p_3^+}
>&
k_f^- = \frac{\mathbf{P}_{\perp}^2}{2k_f^{+}}
\;.
\end{aligned}
\label{kc_cond}
\end{equation}

In other sections, we have obtained the NLO dijet cross section using a naive scheme for high-energy log subtraction (and resummation). Using instead a kinematically improved scheme for this subtraction, we would obtain the same result for the NLO cross section, plus an extra term: the difference between the naive and the kinematically improved integrated JIMWLK evolutions of the LO cross section. The contribution to this difference which is proportional to the LO operator would then be
\begin{equation}\begin{aligned}
 &
2\alpha_s N_c\, \int_{k^+_{\textrm{min}}}^{k^+_f} \frac{\mathrm{d} p_3^+}{p_3^+}
\int_{\mathbf{P}_3} \frac{1}{\mathbf{P}_3^2}
\Big[e^{i\mathbf{P}_3\cdot \mathbf{x}_{1^{\prime}1}}+e^{i\mathbf{P}_3\cdot \mathbf{x}_{2^{\prime}2}}
-e^{i\mathbf{P}_3\cdot \mathbf{x}_{12}}-e^{i\mathbf{P}_3\cdot \mathbf{x}_{1^{\prime}2^{\prime}}}\Big]
 \\
 &\hspace{2cm}\times
\left[1 - \theta\left(\frac{\mathbf{P}_3^2}{2p_3^+} \! -\! \frac{\mathbf{P}_{\perp}^2}{2k^+_f} \right)
\right]
\;,
\end{aligned}
\label{naive_kc_evol_diff_1}
\end{equation}
with the term $1$ in the bracket corresponding to the naive evolution with $p^+$ ordering only, whereas the theta function implements the condition \eqref{kc_cond}, meaning imposing as well the $p^-$ ordering of the gluons participating to the evolution with respect to the $k_f^-$ scale set by the dijet kinematics.
Making all the bounds explicit as theta functions, the expression \eqref{naive_kc_evol_diff_1} becomes:
\begin{equation}\begin{aligned}
 &
2\alpha_s N_c\, \int_{0}^{+\infty} \frac{\mathrm{d} p_3^+}{p_3^+}
\int_{\mathbf{P}_3} \frac{1}{\mathbf{P}_3^2}
\Big[e^{i\mathbf{P}_3\cdot \mathbf{x}_{1^{\prime}1}}+e^{i\mathbf{P}_3\cdot \mathbf{x}_{2^{\prime}2}}
-e^{i\mathbf{P}_3\cdot \mathbf{x}_{12}}-e^{i\mathbf{P}_3\cdot \mathbf{x}_{1^{\prime}2^{\prime}}}\Big]
 \\
 &\hspace{2cm}\times\,
 \theta\left(k^+_f \! -\! p_3^+\right)
\theta\left(p_3^+ \! -\! k^+_{\textrm{min}}\right)
\theta\left(\frac{p_3^+}{k^+_f} \mathbf{P}_{\perp}^2 \! -\!\mathbf{P}_3^2 \right)
\label{naive_kc_evol_diff_2}
\;.
\end{aligned}\end{equation}
In terms of $p_3^+$, there is one upper but two lower bounds. $k^+_{\textrm{min}}$ was first introduced as a cutoff to regulate the $p_3^+$-integral. In the high-energy resummation, $k^+_{\textrm{min}}$ also plays the role of the starting point of the evolution of the target, and allows us to implement the dependence of the cross section on the energy of the collision as $k^+_{\textrm{min}} \propto 1/s$. According to both of these interpretations for $k^+_{\textrm{min}}$, the lower bound $p_3^+> k^+_{\textrm{min}}$ is less restrictive than the other one in the expression~\eqref{naive_kc_evol_diff_2}, and can be dropped. Then, eq.~\eqref{naive_kc_evol_diff_2} becomes
\begin{equation}\begin{aligned}
 \eqref{naive_kc_evol_diff_2}
& =
2\alpha_s N_c\, \int_{\mathbf{P}_3} \frac{1}{\mathbf{P}_3^2}
 \theta\left(\mathbf{P}_{\perp}^2 \! -\!\mathbf{P}_3^2 \right)
\ln\left(\frac{\mathbf{P}_{\perp}^2}{\mathbf{P}_3^2}   \right)
\\&\times\Big[e^{i\mathbf{P}_3\cdot \mathbf{x}_{1^{\prime}1}}+e^{i\mathbf{P}_3\cdot \mathbf{x}_{2^{\prime}2}}
-e^{i\mathbf{P}_3\cdot \mathbf{x}_{12}}-e^{i\mathbf{P}_3\cdot \mathbf{x}_{1^{\prime}2^{\prime}}}\Big]
 \\
& =
\frac{2\alpha_s N_c}{\pi} \int_{0}^{|\mathbf{P}_{\perp}|} \frac{\mathrm{d} |\mathbf{P}_3|}{|\mathbf{P}_3|}
\ln\left(\frac{|\mathbf{P}_{\perp}|}{ |\mathbf{P}_3|}   \right)\\
&\times\Big[ \textrm{J}_0\left(|\mathbf{P}_3|\, |\mathbf{x}_{1^{\prime}1}|\right)
+  \textrm{J}_0\left(|\mathbf{P}_3|\, |\mathbf{x}_{2^{\prime}2}|\right)
- \textrm{J}_0\left(|\mathbf{P}_3|\, |\mathbf{x}_{12}|\right)
- \textrm{J}_0\left(|\mathbf{P}_3|\, |\mathbf{x}_{1^{\prime}2^{\prime}}|\right)
\Big]
\label{naive_kc_evol_diff_3}
\;.
\end{aligned}\end{equation}

In the correlation limit, one has $\mathbf{x}_{12}\rightarrow 0$, $\mathbf{x}_{1^{\prime}2^{\prime}}\rightarrow 0$,  $\mathbf{x}_{1^{\prime}1}\rightarrow \mathbf{b}^{\prime}-\mathbf{b}$ and
$\mathbf{x}_{2^{\prime}2}\rightarrow \mathbf{b}^{\prime}-\mathbf{b}$, so that the expression~\eqref{naive_kc_evol_diff_3} becomes
\begin{equation}\begin{aligned}
 &
\frac{4\alpha_s N_c}{\pi} \int_{0}^{|\mathbf{P}_{\perp}|\, |\mathbf{b}^{\prime}\! -\!\mathbf{b}|} \frac{\mathrm{d} \tau}{\tau}
\Big[\ln\left({|\mathbf{P}_{\perp}|\, |\mathbf{b}^{\prime}-\mathbf{b}|}\right)
-\ln\left({ \tau} \right)
\Big]
\Big[ \textrm{J}_0\left(\tau\right)
- 1
\Big]
 \\
& =
-\frac{\alpha_s N_c}{2\pi} \left\{
\ln^2 \left(\frac{\mathbf{P}_{\perp}^2\, (\mathbf{b}^{\prime}-\mathbf{b})^2}{{c_0}^2}\right)
+ O\left(\frac{1}{\sqrt{|\mathbf{P}_{\perp}|\, |\mathbf{b}^{\prime}-\mathbf{b}|}}\right)
\right\}
\end{aligned}\end{equation}
for $|\mathbf{P}_{\perp}||\mathbf{b}^{\prime}-\mathbf{b}|\gg 1 $.
This is a Sudakov double log type term, this time with the expected negative sign. Combining this contribution together with the leading $N_c$ term from eq.~\eqref{eq:QFSsquare_plus_QbarFSsquare_out_phase_corr_lim_1}, one finally obtains the following total double logarithmic contribution
\begin{equation}\begin{aligned}
 &
-\frac{\alpha_s N_c}{4\pi}
\ln^2 \left(\frac{\mathbf{P}_{\perp}^2\, (\mathbf{b}^{\prime}-\mathbf{b})^2}{{c_0}^2}\right)
\end{aligned}\end{equation}
at leading-$N_c$, which is indeed the expected Sudakov double log term from eq.~\ref{eq:DLL}.

Several remarks are in order:
\begin{itemize}
  \item The calculation outlined in this subsection provides an extra motivation for the kinematical improvement of high-energy evolution equations like JIMWLK: it allows one to obtain the correct Sudakov double logarithms as a leftover in the NLO cross section after high-energy resummation. The situation can be summarized as follows. In the naive scheme for JIMWLK resummation based on $p^+$ ordering only, any gluon radiation with smaller $p^+$ than the dijet scale $k_f^+$ is treated as part of the evolution of the target. By contrast, including kinematical improvement allows one to split such small-$p^+$ gluon radiation into two parts: the true contribution to the evolution of the target with smaller $p^+$ but larger $p^-$ than the dijet, and the soft radiation with both $p^+$ and $p^-$ smaller than the dijet. Sudakov logarithms originate from the soft regime only, which is thus distinct from the true regime contributing to the evolution of the target. But the naive scheme in $p^+$ without kinematical improvement for the target evolution misses this fact, and leads to oversubtracting high-energy logs out of the soft regime. This motivates future studies in order to understand the practical implementation of the proposal of ref.~\cite{Hatta:2016ujq} for the kinematical improvement of JIMWLK, or to construct other prescriptions.

  \item We have focused on the leading-$N_c$ contribution only. Subleading-$N_c$ terms cancel in the naive version of the BK and JIMWLK equation, but this cancelation can be broken when kinematical improvement is included, as can be seen from ref.~\cite{Beuf:2014uia} in the BK case. Hence, at this stage, we have no control on a possible subleading-$N_c$ correction to the coefficient of the Sudakov double log.

  \item If we had used a scheme for JIMWLK resummation based on $p^-$ ordering only, the situation would have been more favorable before kinematical improvement. In that case, only gluon radiation with larger $p^-$ than the dijet (and any $p^+$) would have been treated as part of the evolution of the target. This would not overlap with the soft regime characterized by both smaller $p^+$ and $p^-$ than the dijet. Hence, Sudakov logs should be obtained correctly in that case even without kinematical improvement of JIMWLK, since the oversubtraction of high-energy logs happens in the regime of larger $p^+$ and larger $p^-$ than the dijet, which is well separated from the soft regime. A formulation of the BK equation as evolution along $p^-$ was proposed in ref.~\cite{Ducloue:2019ezk}. However, it is crucially based on specific properties of BK, and for the moment such a scheme does not exist for JIMWLK. 

\end{itemize}


\subsection{Beyond the double leading logarithmic approximation}
In principle, having obtained the full NLO cross section in general
kinematics, one should be able to extend the notion of the correlation
limit and write down a TMD-factorized NLO cross section for back-to-back
dijets. All virtual diagrams would contribute since they preserve
the leading-order kinematics. The most important real NLO corrections
stem from gluon emissions inside or close to the jets, yielding Sudakov double-
and single logarithms as well as finite terms. 

A first step is to extend the calculation above to the Sudakov single logarithms, which is left for future work.
In the second step, where all the virtual diagrams need to be analyzed to obtain the finite contributions to the NLO cross section, we immediately encounter some difficulties. Indeed, TMD factorization
is obvious for all diagrams with initial- or final-state loop corrections
(i.e. $\mathrm{IS+UV+FSUV}$, $\mathrm{GEFS}$, $\mathrm{IFS}$),
at least in the sense that up to power corrections one can write the
Wilson lines as the WW gluon TMDs (\ref{eq:hadroncorrelator}) which
decouple from the hard part. Note that the rationale for this power
expansion in our approach comes from the phases $e^{i\mathbf{P}_{\perp}\cdot\mathbf{r}}e^{i\mathbf{k}_{\perp}\cdot\mathbf{b}}$
which imply that $\mathbf{r}\ll\mathbf{b}$ when $\mathbf{P}_{\perp}\gg\mathbf{k}_{\perp}$.
However, in all virtual graphs where the gluon scatters off the shockwave
($\mathrm{SESW}$, $\mathrm{GESW}$, $\mathrm{ISW}$) this procedure
is compromised due to the phase $e^{i\mathbf{P}_{\perp}\cdot(\mathbf{r}+\xi\mathbf{x}_{13})}e^{i\mathbf{k}_{\perp}\cdot\mathbf{b}}$,
which now enforces $|\mathbf{r}+\xi\mathbf{x}_{13}|\ll\mathbf{b}$.
With this condition, the Wilson-line structure
\begin{equation}
Q_{322^{\prime}1^{\prime}}s_{13}-s_{13}s_{32}-s_{2^{\prime}1^{\prime}}+1-\frac{1}{N_{c}^{2}}\Big(Q_{122^{\prime}1^{\prime}}-s_{12}-s_{2^{\prime}1^{\prime}}+1\Big)
\end{equation}
cannot be cast in any form resembling a TMD. Such a form can only
be established when one requires that $\mathbf{r}\ll\mathbf{b}$ and
$\mathbf{x}_{13}\ll\mathbf{b}$ separately, which yields:
\begin{equation}
\begin{aligned} & \Big\langle Q_{322^{\prime}1^{\prime}}s_{13}-s_{13}s_{32}-s_{2^{\prime}1^{\prime}}+1-\frac{1}{N_{c}^{2}}\Big(Q_{122^{\prime}1^{\prime}}-s_{12}-s_{2^{\prime}1^{\prime}}+1\Big)\Big\rangle\\
 & \simeq\Big(\frac{2C_{F}}{N_{c}}\mathbf{r}^{i}\mathbf{r}^{\prime j}-\mathbf{x}_{13}^{i}\mathbf{r}^{\prime j}\Big)\frac{\mathrm{Tr}}{N_{c}}\Big\langle U_{\mathbf{b}}\big(\partial^{i}U_{\mathbf{b}}^{\dagger}\big)\big(\partial^{j}U_{\mathbf{b}^{\prime}}\big)U_{\mathbf{b}^{\prime}}^{\dagger}\Big\rangle\;,
\end{aligned}
\end{equation}
and would bring the virtual NLO contributions in a TMD-factorized
form. Unfortunately, it is not clear to us whether such an expansion
can be justified.

\section{Conclusions}
Making use of the dipole picture of the CGC effective theory and of LCPT, we have calculated the cross section for the inclusive production of two jets in the scattering of a real photon with a target proton or nucleus at low~$x$. The computation was performed at next-to-leading order in $\alpha_s$, while resumming the multiple rescatterings of the partons off the semiclassical gluon fields in the target to all orders in the eikonal approximation. Using dimensional regularization, we explicitly showed the cancellation of ultraviolet singularities between different virtual NLO corrections on the amplitude level. Likewise, we demonstrated how both soft and collinear divergences, which appear in final-state radiation, cancel through an intricate interplay between the jet definition and certain virtual diagrams. Moreover, we regularized rapidity divergences with the standard cutoff method and absorbed the resulting large logarithms into the JIMWLK equations. This hybrid scheme, i.e. the dimensional regularization of UV and soft-collinear divergences combined with a cutoff for rapidity divergences, is commonly used in higher-order CGC calculations. However, it has the drawback that it obfuscates the distinction between `genuine' soft divergences on the one hand, and rapidity divergences on the other. The former are typically associated with initial- and final state radiation, while the latter are the hallmark of low-$x$ physics and related to the highly boosted target. The fact that, as we show in section~\ref{sec:coll}, not only all soft and collinear singularities cancel in the final-state, but the only large logarithms left are those removed by JIMWLK, is a powerful confirmation of the consistency of this scheme.

After having obtained the NLO dijet cross section, we have explored the back-to-back limit to investigate whether our result could be cast in a form consistent with TMD factorization. At leading order, the overlap of the CGC and the TMD frameworks for processes such as this has been demonstrated already some time ago~\cite{Dominguez:2010xd,Dominguez:2011wm}. However, a full next-to-leading order matching is much more involved, since it constitutes an analysis of all Sudakov double and single logarithms as well as finite NLO contributions. A first demonstration that Sudakov double logarithms arise in the hadroproduction of a Higgs boson at low-$x$ was performed in~\cite{Mueller:2013wwa}, where also the precise form of these double logs in dijet production was inferred based on kinematical arguments. In this work, we revisited the analysis of the Sudakov double logarithms in the dijet case, this time based on the full NLO calculation. We argue that a kinematical improvement of the JIMWLK resummation is crucial to obtain the correct result for the Sudakov double logarithms. We have also identified a class of virtual contributions that, at least at first sight, break TMD factorization on the finite or potentially single-logarithmic level. Before drawing definite conclusions a more thorough study of the correlation limit of our process is needed, which is left for future work. The inclusive dijet electroproduction process has been studied within TMD factorization in refs.~\cite{delCastillo:2020omr, delCastillo:2021znl}.

While this work was in progress, the NLO calculation of the inclusive dijet \emph{electro}production cross section appeared in ref.~\cite{Caucal:2021ent}. This calculation was performed in a covariant formulation of the CGC rather than in LCPT, and using a different UV subtraction scheme. In the photoproduction limit $Q^2\to 0$, our cross section and the $\gamma_{T}^{*}A\to\mathrm{dijet}X$ cross section in~\cite{Caucal:2021ent} should coincide. The largest difference between both results stems from the treatment of the jet. Indeed, the final result eq.~7.16 of~\cite{Caucal:2021ent} for the dijet cross section is still sensitive to both single- and double logarithms in the rapidity renormalization scale $z_f=k^+_f/q^+$ (or in the rapidity cutoff $k^+_{\mathrm{min}}$ if the JIMWLK subtraction is performed after the application of the jet algorithm.) Since these logarithms cannot be absorbed into JIMWLK and have a soft origin, not a rapidity one, they are unphysical and still need to cancel with soft gluon emission outside the jet, as demonstrated in section~\ref{sec:coll}. Other differences between our results are relatively minor and mainly related to the precise $+$ momentum due to the Dirac traces. In appendix~\ref{sec:BNL}, we cast our partonic cross section in the same notations and conventions as~\cite{Caucal:2021ent} to facilitate a detailed comparison.

\section*{Acknowledgements}
The work of PT is supported by a postdoctoral fellowship fundamental
research of the Research Foundation - Flanders (FWO) no. 1233422N.
PT thanks Dani\"el Boer, Francesco G. Celiberto, Miguel Echevarr\'ia, and Cristian Pisano for
numerous stimulating discussions. TA is supported in part by the National Science Centre (Poland) under the research grant no. 2018/31/D/ST2/00666 (SONATA 14).
GB is supported in part by the National Science Centre (Poland) under the research grant no. 2020/38/E/ST2/00122 (SONATA BIS 10).
The work of TA and GB has been performed in the framework of MSCA RISE 823947 ``Heavy ion collisions: collectivity and precision in saturation physics''  (HIEIC). This work has received funding from the European Un\-ion's Horizon 2020 research and innovation programme under grant agreement No. 824093.

\appendix

\section{Appendices}
\subsection{Gamma matrices in dimensional regularization}
In this appendix,  we establish the gamma matrix identities in $D-2$
transverse Euclidean dimensions.
	
From the definition $\{\gamma^{\mu},\gamma^{\nu}\}=2g^{\mu\nu}$,
 we immediately obtain:
\begin{equation}
\{\gamma^{i},\gamma^{j}\}=-2\delta^{ij}\;,\label{eq:gammacommutatorD-2}
\end{equation}
from which the following identity trivially follows:
\begin{equation}
\begin{aligned}\gamma^{i}\gamma^{i} & =-(D-2)\;.\end{aligned}
\end{equation}
Repeatedly applying identity (\ref{eq:gammacommutatorD-2}) allows
us to write:
\begin{equation}
\begin{aligned}\gamma^{i}\gamma^{j}\gamma^{k}\gamma^{l} & =\gamma^{k}\gamma^{l}\gamma^{i}\gamma^{j}+2\delta^{il}\gamma^{k}\gamma^{j}-2\delta^{ik}\gamma^{l}\gamma^{j}+2\delta^{jl}\gamma^{i}\gamma^{k}-2\delta^{jk}\gamma^{i}\gamma^{l}\;.\end{aligned}
\end{equation}
With the help of the above relation, it is straightforward to work
out the commutation relation for the Dirac sigma, defined as $\sigma^{ij}=(i/2)[\gamma^{i},\gamma^{j}]$,
in $D-2$ dimensions:
\begin{equation}
\begin{aligned}\big[\sigma^{ij},\sigma^{kl}\big] & =2i\delta^{il}\sigma^{kj}-2i\delta^{ik}\sigma^{lj}+2i\delta^{jl}\sigma^{ik}-2i\delta^{jk}\sigma^{il}\;,\end{aligned}
\label{eq:sigmacommutator}
\end{equation}
as well as the contraction:
\begin{equation}
\begin{aligned}\sigma^{ij}\sigma^{il} & =(D-3)\delta^{jl}+i(D-4)\sigma^{jl}\;.\end{aligned}
\label{eq:sigmasigma}
\end{equation}
When evaluating Dirac traces, one can use the fact that $\gamma^{+}$
(and $\gamma^{-}$) commute with the transverse gamma matrices, and
hence also with $\sigma^{ij}$, and then apply the completeness relation
\begin{equation}
\begin{aligned}u_{G}^{s}(q^{+})\bar{u}_{G}^{s}(q^{+})\gamma^{+} & =2q^{+}\mathcal{P}_{G}\;,\end{aligned}
\end{equation}
where $\mathcal{P}_{G}=\gamma^{-}\gamma^{+}/2$ is the projector on
good spinor states.

The above definitions allow one to easily demonstrate the following
identities:
\begin{equation}
\begin{aligned}\mathrm{Tr}\big(\mathcal{P}_{G}\big) & =2\;,\\
\mathrm{Tr}\big(\mathcal{P}_{G}\sigma^{ij}\big) & =0\;,\\
\mathrm{Tr}\big(\mathcal{P}_{G}\sigma^{ij}\sigma^{kl}\big) & =2(g^{ik}g^{jl}-g^{il}g^{jk})\overset{D\to4}{=}2\epsilon^{ij}\epsilon^{kl}\;,\\
\mathrm{Tr}\big(\mathcal{P}_{G}\sigma^{ij}\sigma^{kl}\sigma^{im}\sigma^{kn}\big) & \overset{D\to4}{=}\mathrm{Tr}\big(\mathcal{P}_{G}\sigma^{kl}\sigma^{ij}\sigma^{im}\sigma^{kn}\big)\overset{D\to4}{=}2\delta^{jm}\delta^{ln}\;.
\end{aligned}
\label{eq:traceP}
\end{equation}
Moreover, using the above relations as well as the commutation relation
(\ref{eq:sigmacommutator}), it is straightforward to prove the following
relation:
\begin{equation}
\begin{aligned} & \mathrm{Tr}\Big\{\mathcal{P}_{G}\big(A\delta^{\lambda\lambda^{\prime}}+i\sigma^{\lambda\lambda^{\prime}}\big)\big(B\delta^{\eta\eta^{\prime}}+i\sigma^{\eta\eta^{\prime}}\big)\big(C\delta^{\lambda\bar{\lambda}}+i\sigma^{\lambda\bar{\lambda}}\big)\big(D\delta^{\eta\bar{\eta}}+i\sigma^{\eta\bar{\eta}}\big)\Big\}\\
 & \overset{D\to4}{=}2\Big[(AC-1)(BD-1)\delta^{\bar{\eta}\eta^{\prime}}\delta^{\bar{\lambda}\lambda^{\prime}}+(A-C)(D-B)\sigma^{\bar{\eta}\eta^{\prime}}\sigma^{\bar{\lambda}\lambda^{\prime}}\Big]\;.
\end{aligned}
\label{eq:DiracTraceBoss}
\end{equation}
\subsection{\label{sec:BNL}Cross section in the notation of Caucal-Salazar-Venugopalan}

In this appendix, we provide our cross section (section~\ref{sec:crosssection})
cast in the notations and conventions of ref.~\cite{Caucal:2021ent},
to facilitate comparison with their result for the $\gamma_{T}^{*}A\to q\bar{q}X$
and $\gamma_{T}^{*}A\to q\bar{q}gX$ NLO impact factors in the limit
of vanishing photon virtuality $Q^{2}\to0$.

In~\cite{Caucal:2021ent}, the indices $x$, $y$, and $z$ are used
instead of $1$, $2$, and $3$ for the coordinates of the quark,
antiquark, and gluon, respectively, such that e.g. $\mathbf{r}_{xy}=\mathbf{x}_{12}$.
Plus momenta are always written as fractions with the photon $+$
momentum $q^{+}$, i.e. $z_{q}\equiv p_{1}^{+}/q^{+}$, $z_{\bar{q}}\equiv p_{2}^{+}/q^{+}$,
and $z_{g}\equiv k_{3}^{+}/q^{+}$ (or $z_{g}\equiv p_{3}^{+}/q^{+}$
in real diagrams). The coordinate vectors
\begin{equation}
\begin{aligned}\mathbf{R}_{\mathrm{SE}} & \equiv-\frac{k_{3}^{+}}{p_{1}^{+}}\mathbf{x}_{13}+\mathbf{x}_{12}\;,\\
\mathbf{R}_{\mathrm{V}} & \equiv\frac{k_{3}^{+}}{p_{2}^{+}+k_{3}^{+}}\mathbf{x}_{23}+\mathbf{x}_{12}=\frac{p_{2}^{+}\mathbf{x}_{12}+k_{3}^{+}\mathbf{x}_{13}}{p_{2}^{+}+k_{3}^{+}}\;,
\end{aligned}
\end{equation}
and
\begin{equation}
\begin{aligned}X_{V}^{2} & \equiv\frac{p_{2}^{+}}{q^{+}}\frac{p_{1}^{+}-k_{3}^{+}}{q^{+}}\mathbf{x}_{12}^{2}+\frac{k_{3}^{+}}{q^{+}}\frac{p_{1}^{+}-k_{3}^{+}}{q^{+}}\mathbf{x}_{31}^{2}+\frac{k_{3}^{+}p_{2}^{+}}{(q^{+})^{2}}\mathbf{x}_{32}^{2}\;,\end{aligned}
\end{equation}
appear in the virtual diagrams $\mathrm{SESW}$, $\mathrm{GESW}$
and $\mathrm{ISW}$. The following short-hand notations are used for
the Wilson-line structures:
\begin{equation}
\begin{aligned}
\Xi_{\mathrm{LO}} & \equiv\Big\langle Q_{2^{\prime}1^{\prime}12}-s_{12}-s_{2^{\prime}1^{\prime}}+1\Big\rangle\;,\\
\frac{2}{N_{c}}\Xi_{\mathrm{NLO,1}} & \equiv\Big\langle  Q_{322^{\prime}1^{\prime}}s_{13}-s_{13}s_{32}-s_{2^{\prime}1^{\prime}}+1-\frac{1}{N_{c}^{2}}\Big(Q_{122^{\prime}1^{\prime}}-s_{12}-s_{2^{\prime}1^{\prime}}+1\Big)\Big\rangle\;,\\
\frac{2}{N_{c}}\Xi_{\mathrm{NLO,3}} & \equiv\Big\langle  s_{2^{\prime}1^{\prime}}s_{12}-s_{12}-s_{2^{\prime}1^{\prime}}+1-\frac{1}{N_{c}^{2}}\Big(Q_{122^{\prime}1^{\prime}}-s_{12}-s_{2^{\prime}1^{\prime}}+1\Big)\Big\rangle\;,
\end{aligned}
\end{equation}
while the compact notation

\begin{equation}
\begin{alignedat}{1}\int\mathrm{d}\Pi_{\mathrm{LO}}\equiv & \int_{\mathrm{\mathbf{x}}_{1}^{\prime},\mathbf{x}_{2}^{\prime},\mathrm{\mathbf{x}}_{1},\mathbf{x}_{2}}e^{-i\mathbf{p}_{1}\cdot\mathrm{\mathbf{x}}_{11^{\prime}}}e^{-i\mathbf{p}_{2}\cdot\mathbf{x}_{22^{\prime}}}\end{alignedat}
\end{equation}
is used for the transverse integral.

With the above notations, it is easy to show that the modified Weizs\"acker-Williams-
and Coulomb fields that appear in $\mathrm{SESW}$, $\mathrm{GESW}$,
and $\mathrm{ISW}$ can be cast in the following form:
\begin{equation}
\begin{aligned}\mathcal{A}^{\bar{\lambda}}\Big(\frac{k_{3}^{+}}{p_{1}^{+}}\mathbf{x}_{13}+\mathbf{x}_{21},\frac{k_{3}^{+}}{p_{1}^{+}}\mathbf{x}_{13};\frac{q^{+}(p_{1}^{+}-k_{3}^{+})}{k_{3}^{+}p_{2}^{+}}\Big) & =\frac{1}{2\pi}z_{q}z_{\bar{q}}\frac{\mathbf{R}_{\mathrm{SE}}^{\bar{\lambda}}}{X_{V}^{2}}\\
\mathcal{A}^{\bar{\lambda}}\Big(\frac{p_{2}^{+}\mathbf{x}_{12}+k_{3}^{+}\mathbf{x}_{13}}{p_{2}^{+}+k_{3}^{+}},\frac{k_{3}^{+}}{p_{2}^{+}+k_{3}^{+}}\mathbf{x}_{32};\frac{q^{+}p_{2}^{+}}{k_{3}^{+}(p_{1}^{+}-k_{3}^{+})}\Big) & =-\frac{1}{2\pi}(z_{q}-z_{g})(z_{\bar{q}}+z_{g})^{2}\frac{\mathbf{R}_{\mathrm{V}}^{\bar{\lambda}}}{X_{V}^{2}}\;,\\
\mathcal{C}\big(\frac{k_{3}^{+}}{p_{1}^{+}}\mathbf{x}_{13}+\mathbf{x}_{21},\frac{k_{3}^{+}}{p_{1}^{+}}\mathbf{x}_{13};\frac{q^{+}(p_{1}^{+}-k_{3}^{+})}{p_{2}^{+}k_{3}^{+}}\big) & =\frac{1}{(2\pi)^{2}}z_{q}z_{\bar{q}}\frac{1}{X_{V}^{2}}\;.
\end{aligned}
\end{equation}
Finally, with the indices $J\equiv j1$ and $K\equiv j2$ for the
jet initiated by the quark resp. antiquark, we can the contribution
to the cross section due to $\mathrm{SESW,sub}$ as follows:

\begin{equation}
\begin{aligned}\frac{\mathrm{d}\sigma_{\mathrm{SESW,sub}}}{\mathrm{d}\eta_{J}\mathrm{d}\eta_{K}\mathrm{\mathrm{d}}^{2}\mathbf{p}_{J}\mathrm{\mathrm{d}}^{2}\mathbf{p}_{K}} & =\frac{\alpha_{\mathrm{em}}e_{f}^{2}N_{c}}{(2\pi)^{6}}\delta(1-z_{J}-z_{K})\int\mathrm{d}\Pi_{\mathrm{LO}}2z_{J}^{2}z_{K}^{2}\frac{1}{\mathbf{r}_{x^{\prime}y^{\prime}}^{2}}\\
 & \times\frac{\alpha_{s}}{\pi}\int_{z_{\mathrm{min}}}^{z_{J}}\frac{\mathrm{d}z_{g}}{z_{g}}\int\frac{\mathrm{d}^{2}\mathbf{z}_{\perp}}{\pi}(z_{J}^{2}+z_{K}^{2})\Big(1-\frac{z_{g}}{z_{J}}+\frac{z_{g}^{2}}{2z_{J}^{2}}\Big)\\
 & \times\Bigg\{ e^{-i\frac{z_{g}}{z_{J}}\mathbf{p}_{J}\cdot\mathbf{r}_{zx}}\frac{1}{\mathbf{r}_{xz}^{2}}\frac{\mathbf{R}_{\mathrm{SE}}\cdot\mathbf{r}_{x^{\prime}y^{\prime}}}{X_{V}^{2}}\Xi_{\mathrm{NLO,1}}\\
 & +\frac{1}{z_{J}z_{K}}\Big(\frac{1}{\mathbf{r}_{xz}^{2}}-\frac{\mathbf{r}_{xz}\cdot\mathbf{r}_{yz}}{\mathbf{r}_{xz}^{2}\mathbf{r}_{yz}^{2}}\Big)\frac{\mathbf{r}_{xy}\cdot\mathbf{r}_{x^{\prime}y^{\prime}}}{\mathbf{r}_{xy}^{2}}C_{F}\Xi_{\mathrm{LO}}\Bigg\}\;,
\end{aligned}
\end{equation}
in complete agreement with eq. B.1 in~\cite{Caucal:2021ent} up to
the replacements $\frac{z_{g}}{2z_{J}^{2}}\to\frac{z_{g}^{2}}{2z_{J}^{2}}$
and $\mathbf{r}_{xy}\to\mathbf{r}_{x^{\prime}y^{\prime}}$ in the
third line, which are likely typos.

Likewise, for the contribution due to $\mathrm{GESW}$:
\begin{equation}
\begin{aligned}&\frac{\mathrm{d}\sigma_{\mathrm{GESW}}}{\mathrm{d}\eta_{J}\mathrm{d}\eta_{K}\mathrm{\mathrm{d}}^{2}\mathbf{p}_{J}\mathrm{\mathrm{d}}^{2}\mathbf{p}_{K}}  =\frac{\alpha_{\mathrm{em}}e_{f}^{2}N_{c}}{(2\pi)^{6}}\delta(1-z_{J}-z_{K})\int\mathrm{d}\Pi_{\mathrm{LO}}2z_{J}^{2}z_{K}^{2}\frac{1}{\mathbf{r}_{x^{\prime}y^{\prime}}^{2}}\\
 & \times\frac{\alpha_{s}}{\pi}\int_{z_{\mathrm{min}}}^{z_{J}}\frac{\mathrm{d}z_{g}}{z_{g}}\int\frac{\mathrm{d}^{2}\mathbf{z}_{\perp}}{\pi}e^{-i\frac{z_{g}}{z_{J}}\mathbf{p}_{J}\cdot\mathbf{r}_{zx}}\frac{(z_{J}-z_{g})(z_{K}+z_{g})}{2z_{J}^{2}z_{K}}\\
 & \times\Bigg[\Big(z_{g}^{2}+z_{g}(z_{K}-z_{J})-2z_{J}z_{K}\Big)\Big(z_{J}^{2}+z_{K}^{2}+z_{g}(z_{K}-z_{J})\Big)\frac{\mathbf{r}_{zx}\cdot\mathbf{r}_{zy}}{\mathbf{r}_{zx}^{2}\mathbf{r}_{zy}^{2}}\frac{\mathbf{R}_{\mathrm{V}}\cdot\mathbf{r}_{x^{\prime}y^{\prime}}}{X_{V}^{2}}\\
 & \quad-z_{g}(z_{g}+z_{K}-z_{J})^{2}\frac{\mathbf{r}_{zx}\times\mathbf{r}_{zy}}{\mathbf{r}_{zx}^{2}\mathbf{r}_{zy}^{2}}\frac{\mathbf{R}_{\mathrm{V}}\times\mathbf{r}_{x^{\prime}y^{\prime}}}{X_{V}^{2}}\Bigg]\Xi_{\mathrm{NLO,1}}\;.
\end{aligned}
\end{equation}
The term $\mathbf{R}_{\mathrm{V}}\cdot\mathbf{r}_{x^{\prime}y^{\prime}}$
is equal to a factor $z_{K}+z_{g}$ times the one in eq. B.1, assuming
the Bessel function in the fourth line should read $K_{1}(QX_{\mathrm{V}})$.
With the same assumption, for the term $\mathbf{R}_{\mathrm{V}}\times\mathbf{r}_{x^{\prime}y^{\prime}}$,
the discrepancy is bigger with a prefactor:
\begin{equation}
\begin{aligned}\mathrm{term}_{\mathbf{R}_{\mathrm{V}}\times\mathbf{r}_{x^{\prime}y^{\prime}}}^{\mathrm{TABM}} & =\frac{(z_{K}+z_{g})(z_{g}+z_{K}-z_{J})}{1+z_{g}-2z_{J}(z_{K}+z_{g})}\mathrm{term}_{\mathbf{R}_{\mathrm{V}}\times\mathbf{r}_{x^{\prime}y^{\prime}}}^{\mathrm{CSV}}\;.\end{aligned}
\end{equation}
Finally, for $\mathrm{ISW}$:
\begin{equation}
\begin{aligned}\frac{\mathrm{d}\sigma_{\mathrm{ISW}}}{\mathrm{d}\eta_{J}\mathrm{d}\eta_{K}\mathrm{\mathrm{d}}^{2}\mathbf{p}_{J}\mathrm{\mathrm{d}}^{2}\mathbf{p}_{K}} & =\frac{\alpha_{\mathrm{em}}e_{f}^{2}N_{c}}{(2\pi)^{6}}\delta(1-z_{J}-z_{K})\int\mathrm{d}\Pi_{\mathrm{LO}}2z_{J}^{2}z_{K}^{2}\frac{1}{\mathbf{r}_{x^{\prime}y^{\prime}}^{2}}\\
 & \times\frac{\alpha_{s}}{\pi}\int_{z_{\mathrm{min}}}^{z_{J}}\frac{\mathrm{d}z_{g}}{z_{g}}\int\frac{\mathrm{d}^{2}\mathbf{z}_{\perp}}{\pi}e^{-i\frac{z_{g}}{z_{J}}\mathbf{p}_{J}\cdot\mathbf{r}_{zx}}\Xi_{\mathrm{NLO,1}}\\
 & \times-\frac{z_{g}(z_{J}-z_{g})}{2z_{J}}\Big(\frac{z_{J}}{z_{K}+z_{g}}+\frac{z_{K}(z_{J}-z_{g})}{z_{J}^{2}}\Big)\frac{\mathbf{r}_{zx}\cdot\mathbf{r}_{x^{\prime}y^{\prime}}}{\mathbf{r}_{zx}^{2}\mathbf{r}_{x^{\prime}y^{\prime}}^{2}}\frac{1}{X_{V}^{2}}\;,
\end{aligned}
\end{equation}
which is a factor $1/2$ times the result in eq. B.1 if we assume
that a factor $-1$ was forgotten in front of the term $z_{g}(z_{g}-z_{J})^{2}z_{K}/z_{J}^{3}$
in the second line.

Likewise, the notations $\mathbf{k}_{\perp}\equiv\mathbf{p}_{1}$,
$\mathbf{p}_{\perp}\equiv\mathbf{p}_{2}$, and $\mathbf{k}_{g\perp}\equiv\mathbf{k}_{3}\mathrm{or}\mathbf{p}_{3}$ are employed for the
transverse momentum of the quark, antiquark, and virtual resp. real gluon. The vector sum of the quark- and gluon momenta is written as $\boldsymbol{\Delta}_{\perp}\equiv\mathbf{p}_{1}+\mathbf{p}_{2}$, while
$\mathbf{P}_{\perp}$ denotes the same momentum combination as in our work.
With the notation:
\begin{equation}
\begin{aligned}\boldsymbol{\Delta}_{\mathrm{V3}}^{2} & \equiv-\frac{p_{2}^{+}+k_{3}^{+}}{p_{2}^{+}}\frac{p_{1}^{+}-k_{3}^{+}}{p_{1}^{+}}\mathbf{P}_{\perp}^{2}\;,\end{aligned}
\end{equation}
and making use of the identity:
\begin{equation}
\begin{aligned}\int_{\mathbf{K}}\frac{e^{i\mathbf{K}\cdot\mathbf{x}}}{\mathbf{K}^{2}+\mathbf{Q}^{2}} & =\frac{1}{2\pi}K_{0}\big(\big|\mathbf{x}\big|\big|\mathbf{Q}\big|\big)\;,\end{aligned}
\label{eq:BesselK0}
\end{equation}
we can rewrite the integral $J$ (\ref{eq:J}) as follows:
\begin{equation}
\begin{aligned}\delta^{\eta^{\prime}\bar{\eta}}J^{\eta^{\prime}\bar{\eta}}(k_{3}^{+},\mathbf{x}_{12}) & =e^{i(1-\frac{z_{g}}{z_{q}})\mathbf{P}_{\perp}\cdot\mathbf{r}_{xy}}\frac{1}{2\pi}K_{0}\big(r_{xy}\mathbf{\Delta_{\mathrm{V3}}}\big)\\&+\frac{z_{g}}{2z_{K}(z_{J}-z_{g})}\frac{1}{2\pi}\mathcal{J}_{\odot}\big(\mathbf{r}_{xy},\big(1-\frac{z_{g}}{z_{q}}\big)\mathbf{P}_{\perp},\mathbf{\Delta_{\mathrm{V3}}}\big)\;,\\
\epsilon^{\bar{\eta}\eta^{\prime}}J^{\eta^{\prime}\bar{\eta}}(k_{3}^{+},\mathbf{x}_{12}) & =-i\frac{z_{g}}{(z_{J}-z_{g})z_{K}}\frac{1}{2\pi}\mathcal{J}_{\otimes}\big(\mathbf{r}_{xy},\big(1-\frac{z_{g}}{z_{q}}\big)\mathbf{P}_{\perp},\mathbf{\Delta_{\mathrm{V3}}}\big)\;.
\end{aligned}
\end{equation}
The contributions to the cross section due to the virtual diagrams
$\mathrm{GEFS},(i)$ and $\mathrm{GEFS},(ii)+\mathrm{IFS}$ can then
be written as follows:
\begin{equation}
\begin{aligned}&\frac{\mathrm{d}\sigma_{\mathrm{\mathrm{GEFS},}(i)}}{\mathrm{d}\eta_{J}\mathrm{d}\eta_{K}\mathrm{\mathrm{d}}^{2}\mathbf{p}_{J}\mathrm{\mathrm{d}}^{2}\mathbf{p}_{K}}  =\frac{\alpha_{\mathrm{em}}e_{f}^{2}N_{c}}{(2\pi)^{6}}\delta(1-z_{J}-z_{K})\int\mathrm{d}\Pi_{\mathrm{LO}}2z_{J}^{2}z_{K}^{2}\frac{1}{\mathbf{r}_{x^{\prime}y^{\prime}}^{2}}\\
 & \times\frac{\alpha_{s}}{\pi}\int_{z_{\mathrm{min}}}^{z_{J}}\frac{\mathrm{d}z_{g}}{z_{g}}\frac{1}{\mathbf{r}_{xy}^{2}}\Xi_{\mathrm{NLO,3}}\frac{1}{z_{J}z_{K}}\frac{z_{J}-z_{g}}{z_{J}z_{g}}\\
 & \times\Bigg\{\Big(z_{g}^{2}+z_{g}(z_{K}-z_{J})\Big)\Big(z_{J}^{2}+z_{K}^{2}+z_{g}(z_{K}-z_{J})\Big)\\
 & \quad\times\Big(e^{i(\mathbf{P}_{\perp}+z_{g}\boldsymbol{\Delta}_{\perp})\cdot\mathbf{r}_{xy}}(\mathbf{r}_{xy}\cdot\mathbf{r}_{x^{\prime}y^{\prime}})K_{0}\big(r_{xy}\mathbf{\Delta_{\mathrm{V3}}}\big)\\
 & \qquad+\frac{z_{g}}{2z_{K}(z_{J}-z_{g})}e^{i\frac{z_{g}}{z_{J}}\mathbf{p}_{J}\cdot\mathbf{r}_{xy}}(\mathbf{r}_{xy}\cdot\mathbf{r}_{x^{\prime}y^{\prime}})\mathcal{J}_{\odot}\big(\mathbf{r}_{xy},\big(1-\frac{z_{g}}{z_{q}}\big)\mathbf{P}_{\perp},\mathbf{\Delta_{\mathrm{V3}}}\big)\Big)\\
 & \quad-i\frac{z_{g}^{2}(z_{g}+z_{K}-z_{J})^{2}}{(z_{J}-z_{g})z_{K}}e^{i\frac{z_{g}}{z_{J}}\mathbf{p}_{J}\cdot\mathbf{r}_{xy}}(\mathbf{r}_{xy}\times\mathbf{r}_{x^{\prime}y^{\prime}})\mathcal{J}_{\otimes}\big(\mathbf{r}_{xy},\big(1-\frac{z_{g}}{z_{q}}\big)\mathbf{P}_{\perp},\mathbf{\Delta_{\mathrm{V3}}}\big)\Bigg\}\;,
\end{aligned}
\end{equation}
and:
\begin{equation}
\begin{aligned}&\frac{\mathrm{d}\sigma_{\mathrm{\mathrm{GEFS},}(ii)+\mathrm{IFS}}}{\mathrm{d}\eta_{J}\mathrm{d}\eta_{K}\mathrm{\mathrm{d}}^{2}\mathbf{p}_{J}\mathrm{\mathrm{d}}^{2}\mathbf{p}_{K}}  =\frac{\alpha_{\mathrm{em}}e_{f}^{2}N_{c}}{(2\pi)^{6}}\delta(1-z_{J}-z_{K})\int\mathrm{d}\Pi_{\mathrm{LO}}2z_{J}^{2}z_{K}^{2}\frac{1}{\mathbf{r}_{x^{\prime}y^{\prime}}^{2}}\\
 & \times\frac{\alpha_{s}}{\pi}\int_{z_{\mathrm{min}}}^{z_{J}}\frac{\mathrm{d}z_{g}}{z_{g}}\frac{1}{\mathbf{r}_{xy}^{2}}\Xi_{\mathrm{NLO,3}}2\frac{z_{J}-z_{g}}{z_{J}z_{K}}\Big(z_{J}^{2}+z_{K}^{2}+z_{g}(z_{K}-z_{J})\Big)\\
 & \times\bigg\{ e^{i(\mathbf{P}_{\perp}+z_{g}\boldsymbol{\Delta}_{\perp})\cdot\mathbf{r}_{xy}}(\mathbf{r}_{xy}\cdot\mathbf{r}_{x^{\prime}y^{\prime}})K_{0}\big(r_{xy}\mathbf{\Delta_{\mathrm{V3}}}\big)\\
 & \quad-\frac{1}{2(z_{J}-z_{g})}e^{i\frac{z_{g}}{z_{J}}\mathbf{p}_{J}\cdot\mathbf{r}_{xy}}(\mathbf{r}_{xy}\cdot\mathbf{r}_{x^{\prime}y^{\prime}})\mathcal{J}_{\odot}\big(\mathbf{r}_{xy},\big(1-\frac{z_{g}}{z_{q}}\big)\mathbf{P}_{\perp},\mathbf{\Delta_{\mathrm{V3}}}\big)\bigg\}\;.
\end{aligned}
\end{equation}
We completely agree with eq. B.4 in~\cite{Caucal:2021ent} for the
terms involving $K_{0}$ and $\mathcal{J}_{\odot}$ (up to $-\frac{z_{g}}{2z_{J}z_{K}}\to-\frac{z_{g}^{2}}{2z_{J}z_{K}}$
which is likely a typo). However, we do not agree for the prefactor
of $\mathcal{J}_{\otimes}$.

For the real partonic contribution due to $|\mathcal{M_\mathrm{QFS}}|^2$,
we have:
\begin{equation}
\begin{aligned}\frac{\mathrm{d}\sigma_{\mathrm{QFS}^{2}}}{\mathrm{d}\eta_{q}\mathrm{d}\eta_{\bar{q}}\mathrm{d}\eta_{g}\mathrm{\mathrm{d}}^{2}\mathbf{k}_{\perp}\mathrm{\mathrm{d}}^{2}\mathbf{p}_{\perp}} & =\frac{\alpha_{\mathrm{em}}e_{f}^{2}N_{c}}{(2\pi)^{6}}\delta(1-z_{q}-z_{\bar{q}}-z_{g})\alpha_{s}\int\mathrm{d}\Pi_{\mathrm{LO}}\\
 & \times C_{F}\Xi_{\mathrm{LO}}8z_{q}z_{\bar{q}}\Big(z_{\bar{q}}^{2}-(1-z_{\bar{q}})^{2}\Big)\Big(1+\frac{z_{g}}{z_{q}}+\frac{z_{g}^{2}}{2z_{q}^{2}}\Big)\\
 & \times\frac{\mathbf{r}_{xy}\cdot\mathbf{r}_{x^{\prime}y^{\prime}}}{\mathbf{r}_{xy}^{2}\mathbf{r}_{x^{\prime}y^{\prime}}^{2}}\frac{e^{-i\mathbf{k}_{g\perp}\cdot\mathbf{r}_{xx^{\prime}}}}{(\mathbf{k}_{g\perp}-z_{g}/z_{q}\mathbf{k}_{\perp})^{2}}\;,
\end{aligned}
\end{equation}
which exactly corresponds to B.5, and for the interference term  $\mathcal{M^\dagger_\mathrm{\overline{Q}FS}}\mathcal{M_\mathrm{QFS}}$:
\begin{equation}
\begin{aligned}&\frac{\mathrm{d}\sigma_{\mathrm{\overline{Q}FS}^\dagger\mathrm{QFS}}}{\mathrm{d}\eta_{q}\mathrm{d}\eta_{\bar{q}}\mathrm{d}\eta_{g}\mathrm{\mathrm{d}}^{2}\mathbf{k}_{\perp}\mathrm{\mathrm{d}}^{2}\mathbf{p}_{\perp}\mathrm{\mathrm{d}}^{2}\mathbf{k}_{g\perp}}  =\frac{\alpha_{\mathrm{em}}e_{f}^{2}N_{c}}{(2\pi)^{8}}\delta(1-z_{q}-z_{\bar{q}}-z_{g})\alpha_{s}\\
&\times\int\mathrm{d}\Pi_{\mathrm{LO}}\Xi_{\mathrm{NLO,3}}8z_{q}z_{\bar{q}}e^{-i\mathbf{k}_{g\perp}\cdot\mathbf{r}_{xx^{\prime}}}
  \Bigg\{(z_{q}+z_{\bar{q}}-2z_{q}z_{\bar{q}})\Big(1+\frac{z_{g}}{2z_{q}}+\frac{z_{g}}{2z_{\bar{q}}}\Big)\\
 &\qquad\qquad \qquad\times\frac{-\mathbf{r}_{xy}\cdot\mathbf{r}_{x^{\prime}y^{\prime}}}{\mathbf{r}_{xy}^{2}\mathbf{r}_{x^{\prime}y^{\prime}}^{2}}\frac{\big(\mathbf{k}_{g\perp}-\frac{z_{g}}{z_{q}}\mathbf{k}_{\perp}\big)\cdot\big(\mathbf{k}_{g\perp}-\frac{z_{g}}{z_{\bar{q}}}\mathbf{p}_{\perp}\big)}{\big(\mathbf{k}_{g\perp}-\frac{z_{g}}{z_{q}}\mathbf{k}_{\perp}\big)^{2}\big(\mathbf{k}_{g\perp}-\frac{z_{g}}{z_{\bar{q}}}\mathbf{p}_{\perp}\big)^{2}}\\
 & \qquad\qquad\qquad-\frac{z_{g}(z_{q}-z_{\bar{q}})^{2}}{2z_{q}z_{\bar{q}}}\frac{-\mathbf{r}_{xy}\times\mathbf{r}_{x^{\prime}y^{\prime}}}{\mathbf{r}_{xy}^{2}\mathbf{r}_{x^{\prime}y^{\prime}}^{2}}\frac{\big(\mathbf{k}_{g\perp}-\frac{z_{g}}{z_{q}}\mathbf{k}_{\perp}\big)\times\big(\mathbf{k}_{g\perp}-\frac{z_{g}}{z_{\bar{q}}}\mathbf{p}_{\perp}\big)}{\big(\mathbf{k}_{g\perp}-\frac{z_{g}}{z_{q}}\mathbf{k}_{\perp}\big)^{2}\big(\mathbf{k}_{g\perp}-\frac{z_{g}}{z_{\bar{q}}}\mathbf{p}_{\perp}\big)^{2}}\Bigg\}\;,
\end{aligned}
\end{equation}
which agrees with B.7 up to the difference $1+\frac{z_{g}}{z_{q}}+\frac{z_{g}}{z_{\bar{q}}}\to1+\frac{z_{g}}{2z_{q}}+\frac{z_{g}}{2z_{\bar{q}}}$
in the third line and obvious typos in the transverse momentum structures.

For the remaining real contributions, the authors
of~\cite{Caucal:2021ent} do not provide an explicit evaluation of
the Dirac traces hence we cannot compare further.

\end{document}